\documentclass{aa} 
\usepackage{graphicx}
\usepackage{txfonts}
\usepackage[colorlinks=true,allcolors=blue]{hyperref}
\usepackage{schemata}
\usepackage{tikz}
\usepackage{aalongtable}
\usepackage{lscape}
\usepackage{mathtools}
\usepackage{amsmath}
\usepackage{xcolor}
\usepackage{tabularx}
\usepackage{xtab,booktabs}
\usepackage{subcaption}
\usepackage{multirow}
\usepackage{array}

\begin{document} 

\title{CARMENES input catalogue of M dwarfs}
\subtitle{IX. Multiplicity from close spectroscopic binaries to ultra-wide systems}
\titlerunning{M dwarf multiplicity from close spectroscopic binaries to ultra-wide systems}

\author{
    C.\,Cifuentes\inst{\ref{cab-vil}}
    \and J.\,A.\,Caballero\inst{\ref{cab-vil}}
    \and J.\,Gonz\'alez-Payo\inst{\ref{ucm},\ref{unie}}
    \and P.\,J.\,Amado\inst{\ref{iaa}}
    \and V.\,J.\,S.\,B\'ejar\inst{\ref{iac},\ref{ull}}
    \and A.\,J.\,Burgasser\inst{\ref{ucsd}}
    \and M.\,Cort\'es-Contreras\inst{\ref{ucm}}
    \and N.\,Lodieu\inst{\ref{iac},\ref{ull}}    
    \and D.\,Montes\inst{\ref{ucm}}
    \and A.\,Quirrenbach\inst{\ref{lsw}}
    \and A.\,Reiners\inst{\ref{iag}}
    \and I.\,Ribas\inst{\ref{ice},\ref{ieec}} 
    \and J.\,Sanz-Forcada\inst{\ref{cab-vil}}
    \and W.\,Seifert\inst{\ref{lsw}} 
    \and M.\,R.\,Zapatero~Osorio\inst{\ref{cab-vil}}
}
    
\institute{ 
        Centro de Astrobiolog\'ia, CSIC-INTA, Camino Bajo del Castillo s/n, Campus European Space Astronomy Centre, 28692 Villanueva de la Ca\~nada, Madrid, Spain\label{cab-vil}, \email{ccifuentes@cab.inta-csic.es} 
    \and
        Departamento de F\'isica de la Tierra y Astrof\'isica \& IPARCOS-UCM (Instituto de F\'isica de Part\'iculas y del Cosmos de la UCM), Facultad de Ciencias F\'isicas, Universidad Complutense de Madrid, 28040 Madrid, Spain\label{ucm}
    \and
        UNIE Universidad, Departamento de Ciencia y Tecnolog\'ia, Arapiles 14, 28015 Madrid\label{unie}
    \and
        Instituto de Astrof\'isica de Andaluc\'ia (CSIC), Glorieta de la Astronom\'ia s/n, 18008 Granada, Spain\label{iaa}
    \and 
        Instituto de Astrof\'isica de Canarias, V\'ia L\'actea s/n, 38205 San Crist\'obal de La Laguna, Tenerife, Spain\label{iac}
    \and  
        Departamento de Astrof\'isica, Universidad de La Laguna, Astrof\'isico Francisco S\'anchez s/n, 38206 La Laguna, Tenerife, Spain\label{ull}
    \and
        Department of Astronomy and Astrophysics, University of California, San Diego, 9500 Gilman Drive, La Jolla, CA 92093, USA\label{ucsd}
    \and
        Landessternwarte, Zentrum f\"ur Astronomie der Universit\"at Heidelberg, K\"onigstuhl 12, 69117 Heidelberg, Germany \label{lsw}    
    \and
        Institut f\"ur Astrophysik und Geophysik, Georg-August-Universit\"at-G\"ottingen, Friedrich-Hund-Platz 1, 37077 G\"ottingen, Germany\label{iag}
    \and      
        Institut de Ci\`encies de l'Espai (CSIC-IEEC), Can Magrans s/n, Campus UAB, 08193 Bellaterra, Barcelona, Spain\label{ice}
        \and
        Institut d'Estudis Espacials de Catalunya (IEEC), 08034 Barcelona, Spain\label{ieec}
}

\date{Received 8 October 2024 / Accepted 28 {November} 2024}

 
\abstract
{
Multiplicity studies greatly benefit from focusing on M dwarfs because they are often paired in a variety of configurations with both stellar and substellar objects, including exoplanets.
}
{
We aim to address the observed multiplicity of M dwarfs by conducting a systematic analysis using the latest available astrophotometric data.
}
{
For every star in a sample of 2214 M dwarfs from the CARMENES catalogue, we investigated the existence of resolved and unresolved physical companions in the literature and in all-sky surveys, especially in {\em Gaia} DR3 data products.
We covered a very wide range of separations, from known spectroscopic binaries in tight arrangements ($\sim$0.01\,au) to remarkably separated ultra-wide pairs ($\sim$10$^5$\,au).
}
{
We identified 835 M dwarfs in 720 multiple systems, predominantly binaries. Thus, we propose 327 new binary candidates based on {\em Gaia} data.
If these candidates are finally confirmed, we expect the multiplicity fraction of M dwarfs to be {40.3$^{+2.1}_{-2.0}$\,\%}.
When only considering the systems already identified, the multiplicity fraction is reduced to 27.8$^{+1.9}_{-1.8}$\,\%. This result is in line with most of the values published in the literature.
We also identified M-dwarf multiple systems with FGK, white dwarf, ultra-cool dwarf, and exoplanet companions, as well as those in young stellar kinematic groups. We studied their physical separations, orbital periods, binding energies, and mass ratios.
}
{
We argue that based on reliable astrometric data and spectroscopic investigations from the literature (even when considering detection biases), the multiplicity fraction of M dwarfs could still be significantly underestimated. This calls for further high-resolution follow-up studies to validate these findings.
}

\maketitle


\section{Introduction}
\label{section:introduction}

Stellar multiplicity is a natural consequence of the stellar formation process \citep[][also see \citealt{Duc13} and \citealt{Off23} for reviews]{Cha03b,Goo07,Bate12,Tok18}.
The frequency of multiple systems is known to increase with the primary stellar mass \citep[][]{Lad06,Par14,Off23}.
The observational evidence shows that multiplicity is 
greater that 80\,\% for OBA-type stars \citep{Kou07,Mas09,Chi12}, 
around 50\,\% for solar-type stars \citep{Abt76,Duq91,Rag10}, 
and 10--30\,\% for {very} low-mass stars and brown dwarfs \citep{Bur03,Bur07,Bou03,Joe08,Fon18}. 
In the case of M dwarfs, several studies in the last three decades have pointed to a multiplicity of 20--30\,\% \citep[e.g.][]{Jan12,War15,Cor17a,Win19a,Cla24}.

If  all stars are indeed born together with their siblings in groups, it is natural to question how single stars came to be.
Many of them may not have remained together as they evolved, while a significant fraction could also remain undetected.
The dynamical interplay between the components turns into a competition for attaining stable orbits \citep[][but see \citealt{Kin12b}]{Ell14,Sad17}.
It comes as no surprise that young stars are usually found to be part of multiple systems \citep[e.g.][]{Lei93,Rei93,Kou07,Sha17}.
The lifetimes of young systems are too short for them to have settled down into stable configurations \citep[we refer e.g. to the {simulations} carried out by][]{All11}.
For them, it is often unclear whether a given group of stars can be treated as a young trapezia-like architecture, a small stellar kinematic group, or a mature mini-cluster \cite[][]{Mam10,Tok22,GP23}. 

\begin{table*}[]
\caption[]{Multiplicity fraction for M dwarfs calculated in this work and published in the literature.}
\label{tab:literature}
\centering
\begin{tabular}{lccccccc}
\hline\hline 
\noalign{\smallskip}
Reference       &       Spectral        &       Sample  &       $d_{\rm lim}$$^b$       &       $s^c$         &       MF & MF+        &       Methodology$^d$ \\
& range$^a$ & size & [pc] & [au] & [\%] & [\%] & \\ 
   \noalign{\smallskip}\hline \noalign{\smallskip}
        &       M0--M9  &       {2214}  &       $\sim$10--33    &       $\lesssim$ 10$^5$  &       27.8$^{+1.9}_{-1.8}$ & 40.3$^{+2.1}_{-2.0}$                     &               \\          \noalign{\smallskip}
This work       &       M0.0--M4.5      &       {2038} &        $\sim$24--33    &       $\lesssim$ 10$^5$ &        {28.2}$^{+2.0}_{-1.9}$ & {40.5} $^{+ 2.1}_{- 2.1 }$             &       Meta (Sect.~\ref{section:analysis})  \\       \noalign{\smallskip}
 &      M5.0--M9.5      &       {176} &  $\sim$10--24   &       $\lesssim$ 10$^5$  &       22.7$^{+6.7}_{-5.6}$ & 38.1$^{+7.4}_{-6.8}$     &               \\
\noalign{\smallskip} \hline \noalign{\smallskip}
\cite{Cla24}    &       M0--M9  &       1125    &       15      &       \ldots  &       \multicolumn{2}{c}{23.5 $\pm$ 2.0}      &       SI, Meta        \\          \noalign{\smallskip}
\cite{Sus22}    &       M       &       1550    &       15      &       $\lesssim$ 10$^4$  &       \multicolumn{2}{c}{22.9 $\pm$ 2.8}      &       Meta    \\          \noalign{\smallskip}
\cite{Rey21}    &       M       &       249             &       10      & \ldots  & \multicolumn{2}{c}{22.9$^e$}  &       Meta    \\          \noalign{\smallskip}
\cite{Win19a}   &       M       &       1120    &       25      &       $\lesssim$ 10$^4$  &       \multicolumn{2}{c}{26.8 $\pm$ 1.4}      &       WI      \\          \noalign{\smallskip}
\cite{Cor17a}   &       M0--M5  &       425     &       14 (86\,\%)     &       $\sim$1.4--65.6 &       \multicolumn{2}{c}{19.5 $\pm$ 2.3$^f$}  &       LI      \\          \noalign{\smallskip}
\cite{War15}    &       K7--M6  &       245     &       15      &       $\sim$3--10\,000        &       \multicolumn{2}{c}{23.5 $\pm$ 3.2}      &       AO, WI  \\          \noalign{\smallskip}
\cite{Jod13}    &       K5--M4  &       451     &       25      &       $\lesssim$ 80      &       \multicolumn{2}{c}{20.3$^{+6.9}_{-5.2}$}        &       LI      \\          \noalign{\smallskip}
\cite{Jan12}    &       M0--M5  &       761     &       52      &       $\sim$3--227    &       \multicolumn{2}{c}{27 $\pm$ 3}        &       LI      \\          \noalign{\smallskip}
\cite{Ber10}    &       M0--M6  &       108     &       52      &       $\sim$3--180    &       \multicolumn{2}{c}{32 $\pm$ 6}        &       LI      \\          \noalign{\smallskip}
\cite{Law08}    &       M4.5--M6.0      &       108     &       $\lesssim$ 20      &       $\lesssim$ 80   &       \multicolumn{2}{c}{13.6$^{+6.5}_{-4.0}$}        &       LI      \\          \noalign{\smallskip}
\cite{Rei97a}   &       K2--M6  &       106     &       8       &       $\lesssim$ 1800    &       \multicolumn{2}{c}{32}  &       SI, WI, RV      \\          \noalign{\smallskip}
\cite{Lei97}    &       M0--M6  &       34      &       5       &       $\sim$1--100    &       \multicolumn{2}{c}{26 $\pm$ 9}        &       SI      \\          \noalign{\smallskip}
\cite{Sim96}    &       M       &       66      &       8       &       $\sim$100--1400 &       \multicolumn{2}{c}{40}  &       SI, WI, RV  \\          \noalign{\smallskip}
\cite{Fis92}    &       M       &       62      &       20      &       $\lesssim$ 10$^4$  &       \multicolumn{2}{c}{42 $\pm$ 9}  &       SI, WI  \\          \noalign{\smallskip}
\cite{Hen91}    &       M       &       74      &       8       &       \ldots  &       \multicolumn{2}{c}{31.3\,\%}    &       SI      \\          
\noalign{\smallskip}
\cite{Hen90}    &       M       &       27      &       5.2     &       $\lesssim$ 1000            &       \multicolumn{2}{c}{38 $\pm$ 9}  &       SI      \\          \noalign{\smallskip}
\noalign{\smallskip}
\hline
\end{tabular}
\tablefoot{
\tablefoottext{a}{`M' should be read as `all the M dwarfs within the volume limited by $d_{\rm lim}$', when no specific limitation on the spectral classification of the sample is given.}
\tablefoottext{b}{The volume-complete samples limited by these distances are motivated in Sect.~\ref{section:sample}.}
\tablefoottext{c}{Not every star in the sample has been studied in the literature with the same level of detail. In Sect.~\ref{ssection:fraction} we distinguish between three possible categories depending on the resolution, from those only known (at most) {\em Gaia} to those studied in a higher detail.}
\tablefoottext{d}{
Methods referred to data acquisition.
    SI: Speckle interferometry; 
        LI: Lucky imaging; 
        WI: Wide-field imaging; 
        AO: Adaptive optics; 
        RV: Radial-velocity;
    Meta: Literature meta-analysis.}
\tablefoottext{e}{\cite{Rey21} identified at least 94 multiple systems with M dwarfs in their 10-parsec sample, which covered a wide parameter space in mass ratios, magnitude differences, angular separations, inclinations, and orientations. 
They calculate a MF for all the range of masses of 27.4 $\pm$ 2.3\,\%.
While the authors did not provide a specific MF for M-dwarfs for direct comparison, we derived from their results (52 multiple M dwarfs out of 249 in total) a specific value, given the meticulous completeness of their study (see especially their {Fig}.~2).
For this we employed the data of their first update \citep{Rey22}.
\tablefoottext{f}{The authors indicated that the percentage may increase to at least 36\,\% by including the pairs at $\rho < $ 0.2\,arcsec and $\rho > $ 5\,arcsec.}
}}
\end{table*}

Systems that contain more than two components are, in principle, unstable \citep{Har72,Goo05}.
However, the dynamical evolution is able to produce a hierarchical arrangement of binaries within binaries or nested orbits, which leads to stability \citep{Eva68,Bon03,Tok14,Pow23}.
In their seminal work, \cite{Pov67} numerically simulated the formation of runaway stars in few-body clusters, where the high kinetic energy of the runaways is balanced by the binding energy of the binaries formed during close interactions.
The loss of angular momentum during the shrinkage of the closest pair is transferred to the third component, which can result in an ejection from the original compact arrangement, a process that unfolds during the first tens to hundreds of millions of years\footnote{{Throughout the manuscript we use the symbol `a' for annus (Julian year) instead of `yr'. We refer to \url{http://exoterrae.eu/annus.html} for further details.}}
\citep[][and especially \citealt{Reip12}]{Del03,Tok06,Moe10,Kou10}. 
A consequence of the momentum transfer is that a large fraction of close binaries are part of hierarchical triple systems \citep[e.g.][]{Cza22}, a fact that different investigations noted some time ago \citep[e.g.][and see some examples by \citealt{Cif21}]{Maz90,Bat03,Tok06,Pri06,Bas06,Cab07,Kou10,Rap13}.
This means that in many instances of wide binaries, one of the components is (or will be) further resolved as a very compact binary itself. 
The Proxima--$\alpha$~Cen~AB triple system constitutes the nearest example of this kind.
Luckily, from a mathematical perspective, these configurations can be treated dynamically as two-body problems \citep[e.g.][]{Eva68}. 
However, the formation of close binaries in triple systems has also been linked to the Kozai-Lidov mechanism \citep{Koz62,Lid62}, but this has been challenged by the discovery of many wide tertiaries with isotropic orientations and low eccentricities that are inconsistent with the mechanism's predictions \citep{Hwa23}.

While an upper limit for binary separation is not formally established, wide binary systems, particularly those exceeding 0.1--0.2\,pc, are extremely fragile and can be easily disrupted by the Galactic field \citep{Wei87,Cab09,Tok17,GP23}.
Increasingly refined data from surveys such as Hipparcos \citep{Per97}, Tycho-2 \citep{Hog00}, and its successor {\em Gaia} \citep{Gaia16pru} have shown that stellar systems with very large orbital separations (of up to 1\,pc or greater) do exist \citep[][]{Cab10a,Sha11,Oh17,And17}.
The conclusion of these analyses is that there is no strict cut-off in the semimajor axis ($a$) of wide binary systems, as previously theoretically predicted by \citet{Was87}; rather, a {cut-off in} binding energy is more likely (min~$|U_g^*| \approx 10^{33}$\,J; \citealt{Cab10a}). 
Proxima Centauri serves as an example of a star that remains bound to its system despite having a value of $a$ roughly similar to the Hill radius of $\alpha$~Centauri AB \citep{Mat93,Wer06,Reip12,Ker17}.

Momentum transfer is not the only mechanism to account for the large separations of the widest binary systems.
Considering two modes of binary breakup, namely, core splitting and stellar ejection, \cite{Sad17} predicted that the majority of wide binaries break apart, but with some systems becoming tighter within several million astronomical units {(au)}.
\cite{Goo05} noted that these ejections in hierarchical systems would also produce a significant population of close binaries that are typically only detectable using advanced techniques.
Some binaries are so closely packed that they are disguised as single objects by direct imaging.
While very close binaries are undetectable by direct imaging, they can be recognised in the Doppler shift of the spectral lines (spectroscopic binaries), in the periodic eclipsing of their light (eclipsing binaries), or in the measurable change in their motion (astrometric binaries).
There are also compact unresolved triples with hierarchical orbits \citep[][even existing within a space of only a few au; see  \citealt{Moh24}]{Maz01,Bar21}.
The lack of detailed observations regarding many individual stars carries an important observational bias, leaving some very close pairs unrecognised as such.
\cite{Kro91} found that previous studies had underestimated the number of low-mass stars and proposed that assuming independent component masses for binary systems reconciles discrepancies between different luminosity function samples. 
\cite{Pis91} noted that the impact of photometrically unresolved binaries on the luminosity function depends on the mass ratio, $q = \mathcal{M}_2 / \mathcal{M}_1$, and is strongest when both components have similar masses.

Stars in multiple systems offer the precious opportunity to directly measure fundamental parameters, such as masses, radii, or both \citep[e.g.][]{Pop80,Mat00,Zap04,Tor10,Schw19}.
However, close companions are capable of influencing every stage of the stellar evolution.
Stars that have a common origin must have the same age and chemical composition.
Therefore, wide binaries are assumed to be coeval \citep{Har94,Whi01,Jor05,Sta06,Mak08,Kra09}, as well as co-chemical \citep{Giz97b,Gra01,Des04,Des06,Kra09,Haw20}.
Sufficiently resolved pairs can be useful to prove this assumption and serve as pieces in the puzzle of the Galactic formation.
Fitting these pieces together and reassembling the original configuration is the goal of Galactic archaeology studies, which benefits from wide pair systems \citep{And19,Haw20}.
Important applications of wide binaries are the calibration of metallicities of M dwarfs \citep[e.g.][]{Bon05,Bea06,Lep07b,Roj10,Mon18,Marf21},  age-metallicity relation \citep[e.g.][]{Reb16,Zha24}, age-magnetic activity relation \citep[e.g.][]{Gar11,Cha12,Kim21}, and even investigations into the dark matter in the Milky Way \citep[e.g.][]{Yoo04,Cha04}.
The statistics regarding the frequency of multiple systems, their primary-companion mass ratios and their physical separations can also set meaningful constraints for models of stellar formation and evolution.
For instance, they can help determine whether a primordial population composed exclusively of multiples is as likely as predicted \citep{Har94,Whi01,Par09,Reg11,Cla12,Lei13,Rei14,Par14}.

Studies of stellar multiplicity take on particular significance for M dwarfs because they potentially rely on a vast sample of study. 
The nearest stars represent a valuable sample for multiplicity studies because they allow accurate photometric and astrometric measurements. 
Their multiplicity characteristics carry the imprints of the formation and evolution of our Galaxy.
Although intrinsically small and faint \citep[$\mathcal{M} \lesssim 0.62\,{\rm M}_\odot$, $\mathcal{L} \lesssim 0.076\,{\rm L}_\odot$,][and references therein]{Cif20}, M dwarfs make up the majority of the stars in the Galaxy \citep{Hen94,Hen06,Rei04,Boc10,Win15,Rey21,Gol23,Kir24}. 
M dwarfs have also gained importance in the last two decades because of the search for Earth-like planets, especially in their habitable zone \citep{Sca07,Kop13}, either with space missions for transit surveys (CoRoT,  \citealt{Auv09}, \textit{Kepler,}  \citealt{Bor10}, TESS,  \citealt{Ric14}) or with the radial-velocity (RV) method \citep{Bon13a,Fou18,Rei18a,Rib23}.
Among the notable high-resolution ground-based spectrographs that undertake RV searches is the Calar Alto high-Resolution search for M dwarfs with Exoearths with Near-infrared and optical \'Echelle Spectrographs\footnote{\url{https://carmenes.caha.es}} \citep[CARMENES,][]{Qui14}.

Table~\ref{tab:literature} displays the multiplicity fraction (MF) of M dwarfs, namely, the proportion of these stars that are the most massive components of a multiple system (Sect.~\ref{ssection:fraction}), along with the values reported in the literature by different authors during the past three decades, including the present work.
We indicate (when possible) the relevant constrains of the studies: spectral range, sample size, completeness volume, search separations, and methodology.
Not all the portions of the spectral range of M dwarfs have been studied equally well regarding their multiplicity.
Also, the detection limits (again Sect.~\ref{ssection:fraction}) are not consistent between studies, leading to different proportions of undetected binaries in very compact arrangements.
While it has not been included in the list due to greater difficulty in comparison, we mention other systematic efforts in multiplicity investigations, such as the pioneering infrared imaging of 55 low-mass binaries by \cite{Skr89} or the \textit{Hubble} Space Telescope snapshot high-resolution images of 225 stars by \cite{Die12}.
Many of the publications in Table~\ref{tab:literature} predate {\em Gaia}, but some of them made predictions on {\em Gaia}'s impact on M-dwarf multiplicity.
For instance, \cite{Win19a} foresaw that {\em Gaia}'s astrometric measurements over five years would enable the detection of low-mass binaries that remained undetected, providing a more complete picture of the nearby M dwarf population.

This is the ninth paper of the series of publications on the CARMENES input catalogue of M dwarfs. 
CARMENES aims to look for Earth-like planets around the closest, brightest, late-type stars 
with the radial velocities technique.
In this work, we present an updated and systematic study of the multiplicity across all spatial separations in the brightness-spectral type-limited CARMENES input catalogue.
The sample under study is a collection of more than two thousand M dwarfs as described in Sect.~\ref{section:sample}.
Our study exploits the fruitful {\em Gaia} mission (up to the third data release, DR3, \citealt{Gaia23}) to provide an updated revision of the multiplicity of M dwarfs, accounting for the impact of unresolved binaries on the MF.
Since many definitions in this investigation are based on the resolution of the system components by the {\em Gaia} mission, in our analysis (in Sect.~\ref{section:analysis}) we distinguish between systems with resolved and unresolved components in DR3.
Section~\ref{section:results} is centred on results and discussion, which includes the description of the identified systems and their fundamental parameters.
In particular, we differentiate between the canonical MF (Sect.~\ref{ssection:fraction}) and the expected multiplicity fraction, MF+, which accounts for potential unresolved systems that could boost MF by more than 10\,\%.
The conclusions are summarised in Sect.~\ref{section:conclusions}.
Finally, the appendix compiles and organises useful data produced in this work. 
Table~\ref{tab:meta} offers {twelve} tables with a variety of content.
In particular, Table~\ref{tab:mother} includes an abridged version of the full dataset compiled and produced in this work.

\begin{table}[]
\caption{Summary of the tables appended to this work.}
\label{tab:meta}
\centering
\begin{tabular}{ll}
\hline
\hline 
\noalign{\smallskip}    
Table   &       Description     \\
\noalign{\smallskip}
\hline
\noalign{\smallskip}    
Table~\ref{tab:mother}       &       Full sample \\
Table~\ref{tab:outliers}       &       Bona fide binaries with astrometric anomalies        \\
Table~\ref{tab:new}       &       New stellar multiple systems proposed       \\
Table~\ref{tab:spectroscopic}       &       Spectroscopic binaries, triples, and quadruples       \\
Table~\ref{tab:eclipsing}       &       Eclipsing binaries        \\
Table~\ref{tab:description}       &       Description of the full sample table        \\
Table~\ref{tab:widest}       &       Binaries with separations larger than 10$^4$\,au \\
Table~\ref{tab:fgk}       &       M dwarfs + FGK stars      \\
Table~\ref{tab:wd}       &       M dwarfs + white dwarfs   \\
Table~\ref{tab:ucd}      &       M dwarfs + ultra-cool dwarfs \\
Table~\ref{tab:exoplanets}      &       M dwarfs + planets                \\
Table~\ref{tab:references}      &       Bibliographic references and abbreviations    \\
\noalign{\smallskip}
\hline
\end{tabular}
\tablefoot{
{For simplification purposes, we generally abbreviate the name of the objects designated with long catalogue identifiers: Gaia DR3 as G3, Gaia DR2 as G2, and 2MASS as 2M.}
}
\end{table}

\section{Sample}
\label{section:sample}

\begin{figure*}[ht]
    \centering
    \includegraphics[width=0.49\linewidth]{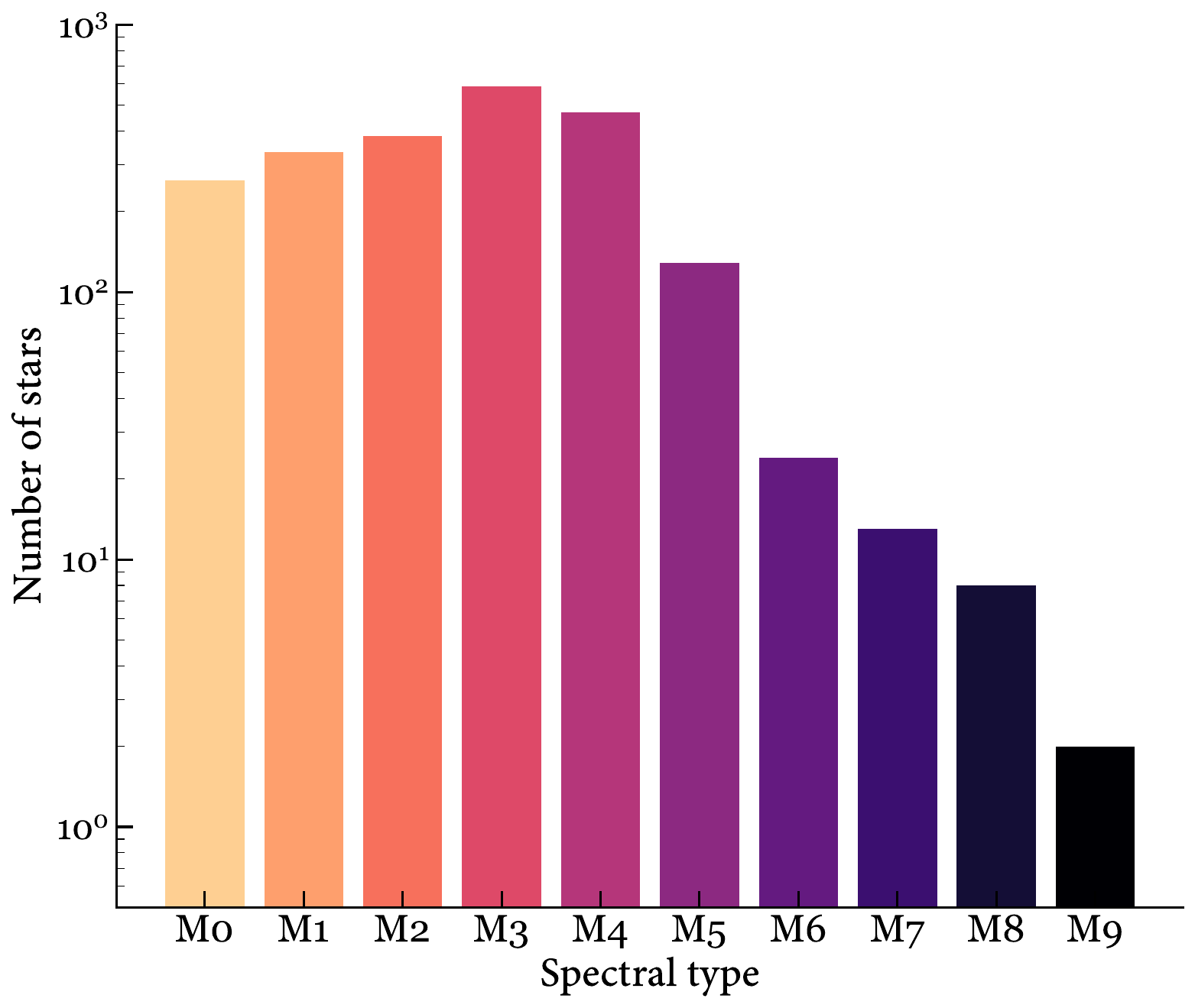}
    \includegraphics[width=0.49\linewidth]{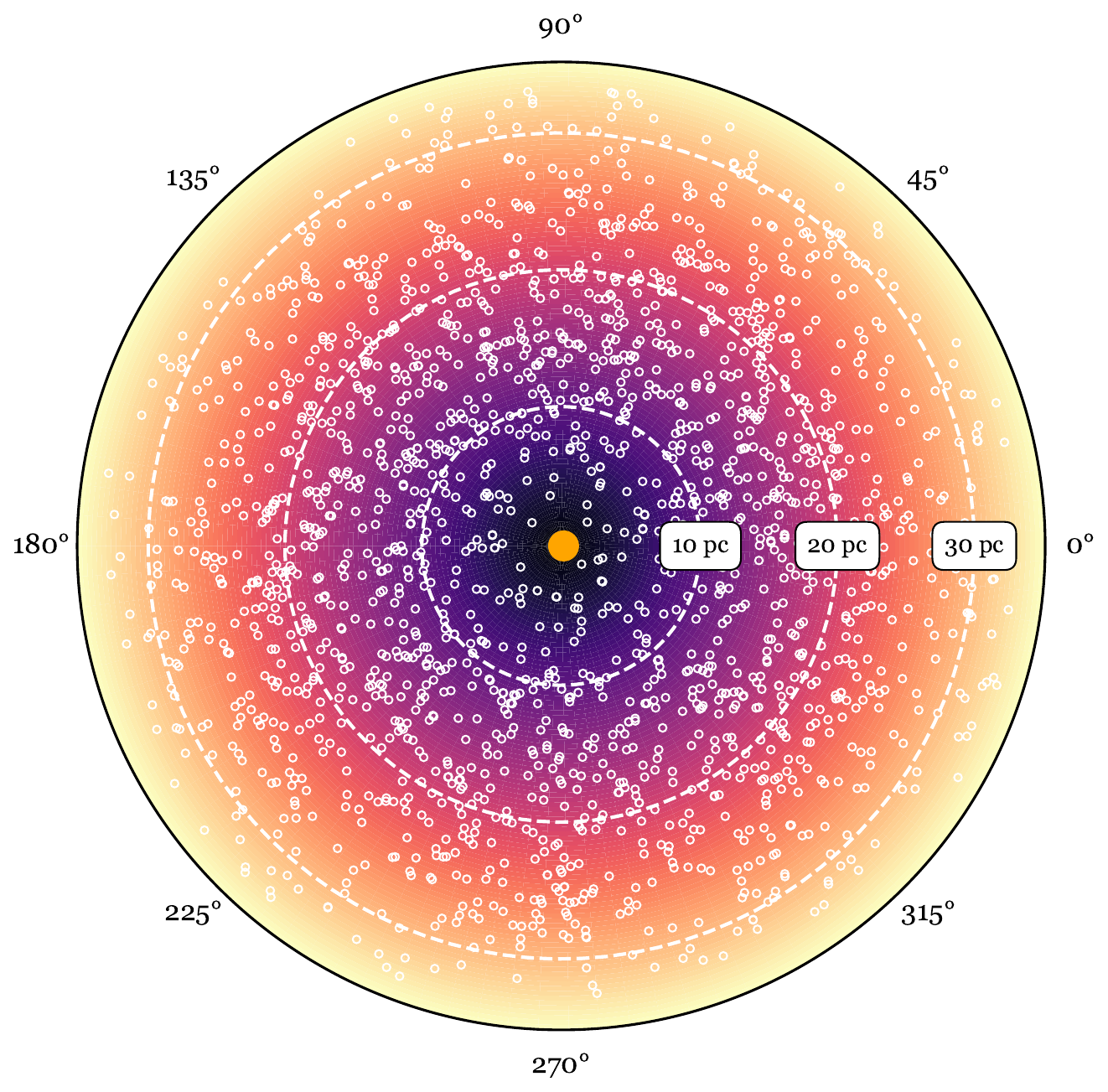}
    \caption{Distribution of spectral types of the stars in the sample ({\em left}) and illustration of the completeness distance of Carmencita ({\em right}).
    The latter panel shows the distance and right ascension $\alpha$ of the Carmencita stars in the solar neighbourhood and colour-codes the spectral subtype as a function of the distance to which our sample is complete.
    The dashed lines depict 10-parsec increments of the distances.
    }
    \label{fig:SpT}
\end{figure*}

The CARMENES project selected 350 M dwarfs as the targets for the main survey, whereby a total of 750 useful nights were reserved as guaranteed time observations, from 1 January 2016 to 31 December 2020.
The observations of the guaranteed time observations sample continue within the CARMENES Legacy+ programme, which aims for 50 measurements for all suitable targets and is expected to run at least until the end of 2025.
The raw data, calibrated spectra, and high-level data products were made publicly available as the CARMENES Data Release 1 \citep{Rib23}.
The continuous update of the catalogue has introduced additional M dwarfs to the original sample, mostly objects of interest (TOI) identified by the Transiting Exoplanet Survey Satellite programme \citep[{\em TESS};][]{Ric14}.
RV follow-up has confirmed many of these transiting exoplanets \citep[e.g.][just to mention a few recent ones]{Gon23b,Gof24,Kuz24,Dai24}.
The left panel of Fig.~\ref{fig:SpT} illustrates the distribution of M dwarfs in our sample according to spectral subtype.
The latest-type star is an M9.5 dwarf, \object{Scholz's~star} (Karmn\footnote{The Karmn nomenclature is Jhhmms$\pm$ddm(N/S/E/W) with J2000 equatorial coordinates. We skip the Karmn prefix from now on.} J07200--087).

The full sample of our study is dubbed Carmencita, the input catalogue for the CARMENES project \citep{Alo15a,Cab16}.
Carmencita contains a total of {2214} M dwarfs, from M0.0\,V to M9.5\,V, including the targets for the main survey.
The stars were intentionally chosen independently of their multiplicity, age, or metallicity.
Only one Carmencita object classified as K7\,V (\object{HD~97101}) remains in the RV-monitored sample, as it is the bound companion of an M2\,V star (\object{HD~97101}\,B).
All these stars satisfy simple selection criteria based on their spectral types, their visibility from the Calar Alto Observatory in Southern Spain ($\delta \gtrsim$ --23\,deg), and on their apparent brightness in the $J$-band magnitude, between 4.2\,mag and 11.5\,mag \citep[this range of magnitudes also depends on spectral type, as detailed by][]{Alo15a}.
The stars in our sample are located at distances ranging between 1.82\,pc (\object{Barnard's star}) and 166.1\,pc (\object{Haro 6-36}), with the majority of them in our immediate vicinity, with a median distance of 22.0\,pc. 

Magnitude-limited samples may be unintentionally overpopulated by intrinsically bright unresolved stellar systems due to the Malmquist bias \citep[see][and references therein]{Duq91}.
In particular, these samples may over-represent spectroscopic binaries because two stars are brighter than one.
Since Carmencita started to be built well ahead of the {\em Gaia} launch, it is not a volume-limited sample.
Furthermore, by construction it is not a magnitude-limited sample (the faintest M0.0--0.5\,V stars are several magnitudes brighter in $J$ than the brightest late-type M dwarfs).
Therefore, we calculated a `completeness distance', $d_{\rm com}$, that depends on the spectral type.
We used the definition of absolute magnitude in the $J$ band, $M_J-J=5-5\log{d_{\rm com}}$, where $J = J({\rm SpT})$ from the construction of Carmencita \citep[Table 1 from][]{Alo15a}, and $M_J = M_J({\rm SpT})$ from an empirical relation \cite[Table 7 in][]{Cif20}.
Therefore, $d_{\rm com}$(SpT) is the radius of the sphere that contains all known M dwarfs with an equal or earlier spectral type.
As a result, Carmencita contains all M dwarfs with spectral types M4.5\,V or earlier up to 30\,pc, and all M dwarfs with spectral types M9.5\,V or earlier up to 10\,pc. 
The latter finding was double-checked against the 10\,pc sample study by \cite{Rey21}.
The right panel of Fig.~\ref{fig:SpT} aims to illustrate this fact.
Of course, this completeness is contingent on the additional condition that the stars must be visible in the northern hemisphere.

\section{Analysis}
\label{section:analysis}

\begin{figure}[]
    \centering
    \includegraphics[width=0.99\linewidth]{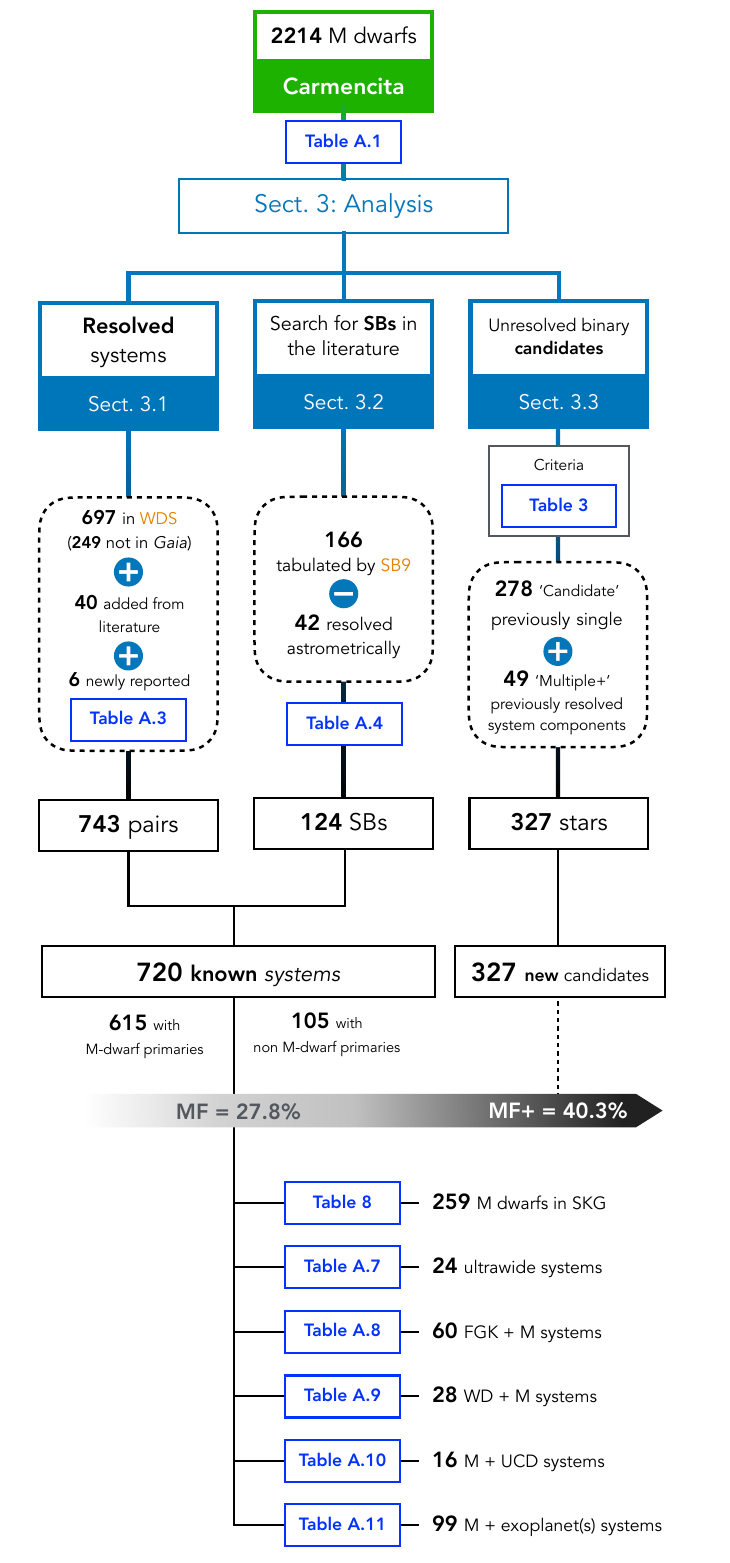}
    \caption{Schematic summary of our analysis (Section~\ref{section:analysis}).
    }
     \label{fig:summary}
\end{figure} 

\subsection{Search for resolved systems}
\label{ssection:resolved}

For each star in our M-dwarf sample, we started by searching for resolved physical companions in two steps.
First, we reviewed the Washington Double Star catalogue\footnote{\url{http://www.astro.gsu.edu/wds/}} \citep[WDS;][]{Mas01}, and then we performed a dedicated, blind search employing the most updated astrometric information available from {\em Gaia} (i.e. DR3). 
In this context, a resolved physical companion (of a given star) is defined as an individual source in an astrometric catalogue such as {\em Gaia} with proper motions and trigonometric parallax that are compatible with physical binding with the primary.

As a preliminary step, we got equatorial coordinates at the 2016.0 {\em Gaia} epoch and apparent magnitudes in $G$ for 100\,\% of our sample using the {\em Gaia} Archive\footnote{\url{https://gea.esac.esa.int/archive/}}.
Next, we ensured that each of our objects had a complete astrometric description.
The full, five-parameter solution from {\em Gaia} DR3 (position, proper motion, and parallax: $\alpha$, $\delta$, $\mu_\alpha \cos{\delta}$, $\mu_\delta$, $\varpi$) is available for {93.0\,\%} of the {2214} M dwarfs.
Among them, {61.2\,\%} also have barycentric radial velocity, $V_r$, from the second or third data releases of {\em Gaia}; we prefer the latter in case of availability in both.
For the sources for which {\em Gaia} does not provide some of these data, we searched in the literature for published measurements \citep[e.g.][]{Gli91,van07,Fah12,Dit14,Fin16}.
Of the remaining 4.9\,\% of stars without full \textit{Gaia} solution, we compiled proper motions and parallaxes from other sources for all except for 34 (see below). 

We performed a cross-match of our \textit{Gaia} stars with WDS utilising the Tool for OPerations on Catalogues And Tables \citep[{\tt TOPCAT};][]{Tay05}.
The WDS is the principal database for astrometric double and multiple star information, collecting {156\,861} systems to submission date.
For many of them, it compiles precise astrometric history and orbital description, making it a valuable resource for our study.
In the WDS catalogue we found {411} systems that contain at least one M dwarf from our sample.
For them, we retrieved the last measured epoch ({\tt sep2}), the corresponding position angle ({\tt pa2}), and separation in the `precise' format (i.e. non-rounded values), and incorporated them in the final table.
We took into account that our M dwarfs could be either the WDS primary or a companion.
At this stage we added one star and eight T-type dwarfs in multiple systems that are not tabulated by \textit{Gaia} because of their brightness (Capella) or their faintness (e.g. GJ~570D, Ross~458C), respectively.

After identifying the resolved pairs already documented by WDS, we also looked for common proper motion and parallax companions within the {\em Gaia} data.
We performed a blind search using the DR3 by dividing the search in two ranges of separations.
First, for the closest systems, we used the automatic positional cross-match tool in {\tt TOPCAT}, {X-match}, limiting to a search radius of 5\,arcsec and setting the `find' option to `all'.
With this configuration we made sure to keep every source found in the vicinity of our sources, regardless of its apparent magnitude, parallax distance, or proper motion, even when some of this information was not available.
Second, we conducted a much wider, separation-limited search using the Astronomical Data Query Language (ADQL) form in the {\em Gaia} Archive.
We limited the search to a physical separation of 10$^5$\,au ($\sim$0.5\,pc), which translates to projected separations of 20\,000--2000\,arcsec for stars located at distances of 5--50\,pc. 
As mentioned in Sect.~\ref{section:introduction}, there is no consensus regarding an upper limit in wide binary separation.
Therefore, we set a safe upper limit of 10$^5$\,au to ensure that actual bound systems with remarkable separations were not missed in the process, but keeping in mind that projected separations are always less than or equal to the true separations \citep[see e.g.][]{Wer06}.
In this search we looked for resolved sources with full five-parameter astrometric solutions compatible with physical binding to our source.
To begin with, we automatically kept all the sources with separations $\rho >$ 5\,arcsec exhibiting a conservative difference in their parallactic distances of 10\,\% with respect to our star, namely, distance ratio:
        
\begin{equation}
\label{eqn:deltad}
        \Delta d = \left|{\frac{d_1-d_2}{d_1}}\right| < 0.10.
\end{equation}

\noindent For these sources, we computed two additional metrics to ensure that they are approximately co-moving.
These are the $\mu\,{\rm ratio}$ and the proper motion position angle difference, $\Delta PA$, defined by \cite{Mon18} and used afterwards by \cite{Cif21} and \cite{GP23}:

\begin{equation}
\label{eqn:muratio}
        \mu\,{\rm ratio} = \left(\frac{(\mu_{\alpha 1} \cos{\delta}_1-\mu_{\alpha 2} \cos{\delta}_2)^2 + (\mu_{\delta 1} - \mu_{\delta 2})^2}{(\mu_{\alpha 1} \cos{\delta}_1)^2 + (\mu_{\delta 1})^2}\right)^{1/2} < 0.15,
\end{equation}

\begin{equation}
\label{eqn:deltapa}
        \Delta PA = |PA_1 - PA_2| < 15\,{\rm deg}.
\end{equation}

\begin{figure}[]
    \centering
    \includegraphics[width=0.99\linewidth]{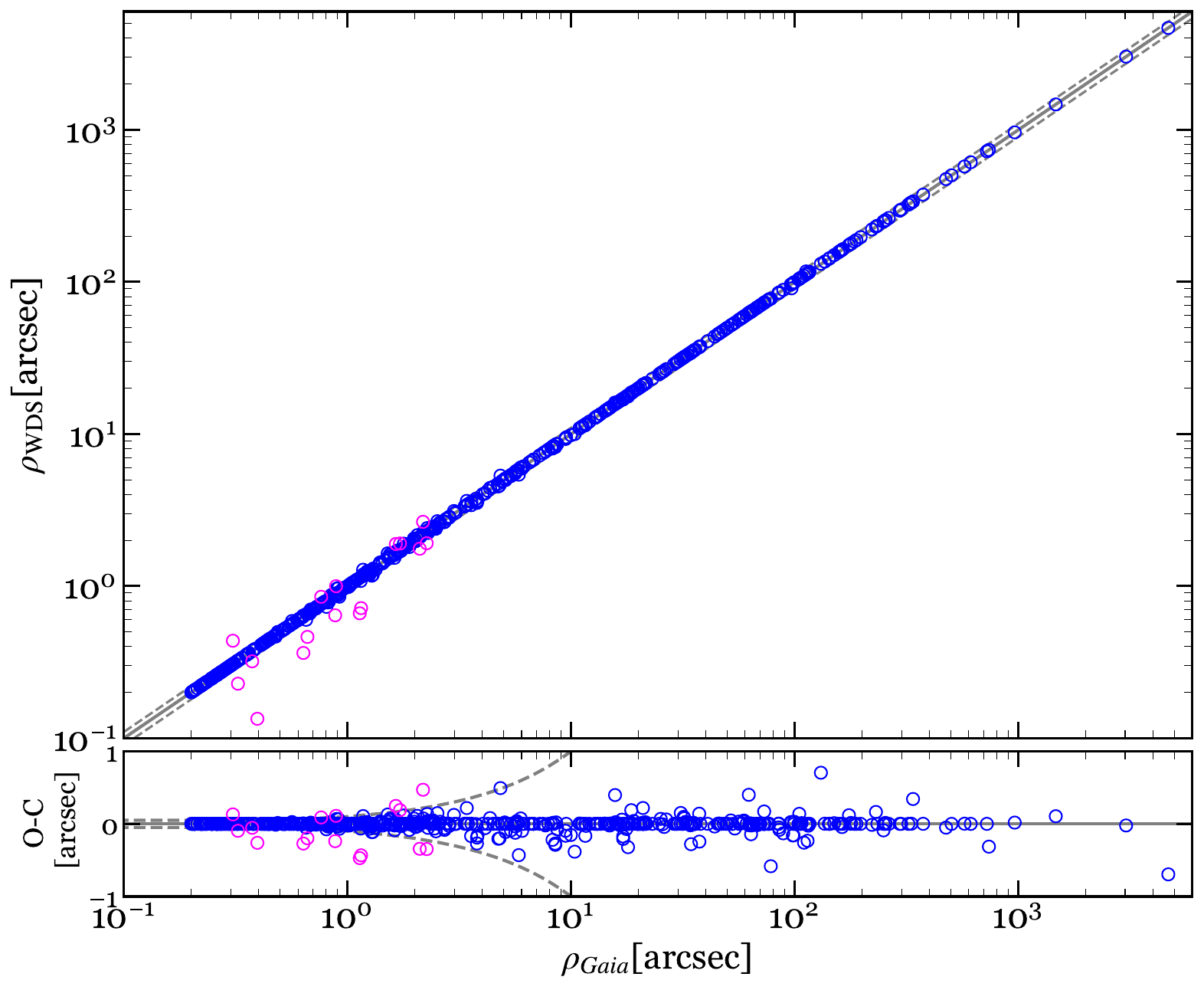}
    \caption{Comparison of projected separations tabulated by the WDS and measured by us using {\em Gaia} astrometry. 
    The solid and dashed grey lines represent the 1:1 relation, and the differences in 10\,\%, respectively.
    The magenta circles are stars beyond this limit.
    The error bars are rather small for almost all cases due to the high precision of {\em Gaia}'s astrometry and they have therefore been omitted.}
     \label{fig:rho_WDS}
\end{figure} 

This search recovered all {568} pairs known in WDS that {\em Gaia} is able to resolve (\ref{sssection:separations}).
In Fig.~\ref{fig:rho_WDS} we compare the values of angular separations tabulated by the WDS with those computed by us using the {\em Gaia} astrometry.
There are only {15} pairs of stars (shown in magenta) with separation differences larger than 10\,\% between WDS and our measurements. 
They correspond to well-documented binaries with very small projected separations ($\rho \lesssim$ 2\,arcsec) that exhibit significant orbital motion over relatively short timescales (of a few years; e.g. \object{BL~Cet} + \object{UV~Cet}, \object{Wolf~424}~AB, or \object{AT~Mic}~AB -- the three of them located at less than 10\,pc from the Sun).
The observed scatter of the 1:1 relation justify a posteriori our 10\,\% distance ratio criterion.
This scatter is due to a mixture of systematic effects on parallax, such as colour-dependent effects, under-estimated uncertainties of faint sources, or impact of unresolved binarity.

\begin{figure}[]
    \centering
    \includegraphics[width=0.99\linewidth]{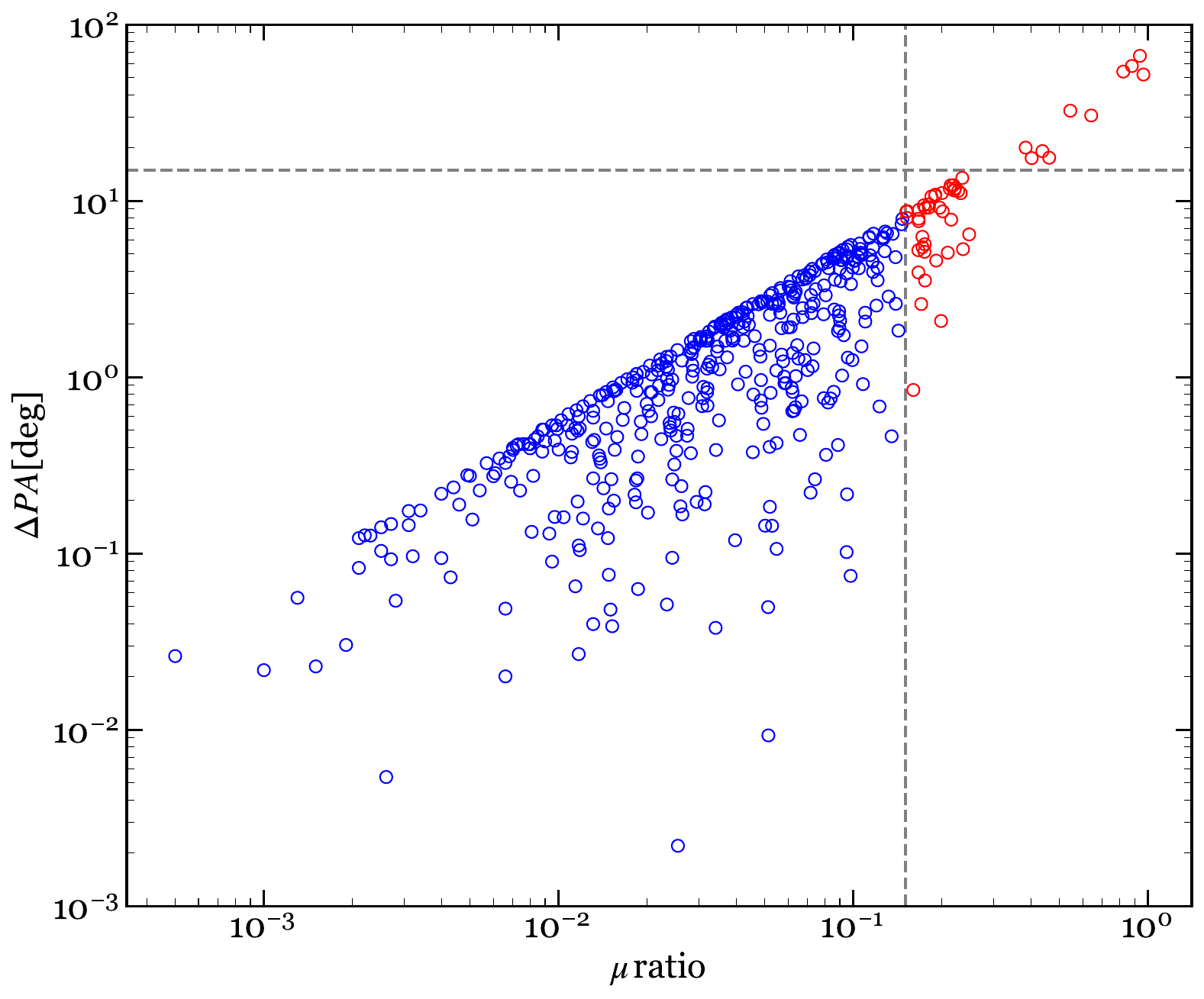}
    \includegraphics[width=0.99\linewidth]{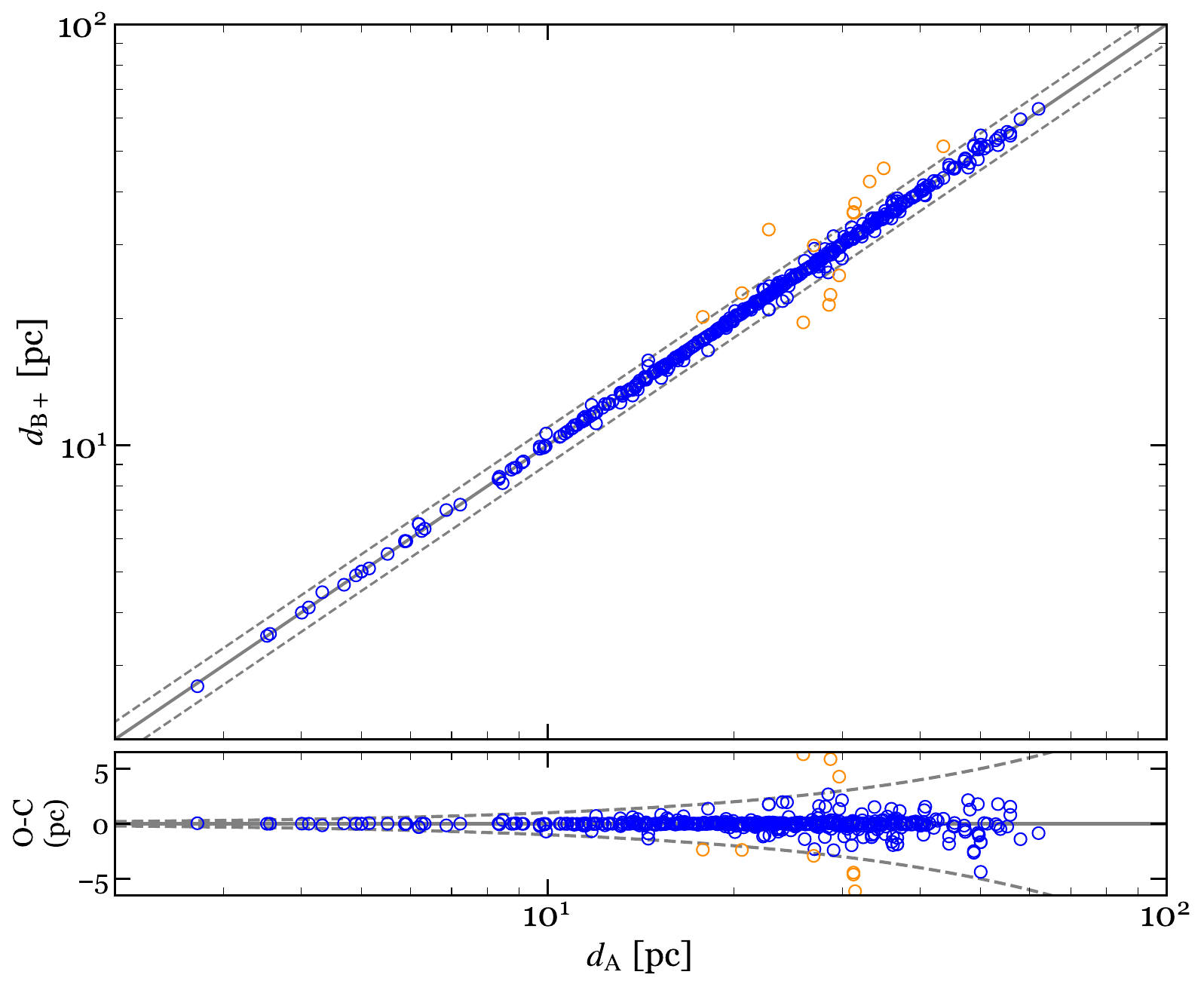}
    \caption{Comparative analysis of the proper motions and distances of the identified pairs. {\em Top}: $\Delta PA$ vs $\mu\,{\rm ratio}$ diagram, where the dashed grey lines set the upper limits of our criteria for physical association (Eqs.~\ref{eqn:muratio} and \ref{eqn:deltapa}). 
     The red {open} circles are pairs that do not comply, or do so partially, with those criteria, respectively.
     {\em Bottom}: Comparison of distances of the companions (denoted `B+') and their corresponding primaries (`A').
    The amber circles are pairs that do not comply with our criterion (Eq.~\ref{eqn:deltad}).
     Our Carmencita M dwarf may be the primary (i.e. the most massive) or a companion in the system.
     }
     \label{fig:criteria}
\end{figure}

Figure~\ref{fig:criteria} shows a comparative analysis of the proper motions ($\Delta PA$ vs $\mu$ ratio; top panel) and distances (companion's $d$ vs primary's $d$; bottom panel) of the identified pairs.
The majority of our pairs comply with the criteria for physical association (shown in blue), although there are {78} sources (shown in red and amber) that exhibit anomalies in their relative positional angles, proper motions, or distances.
Our individual examination of these sources allowed us to propose the most likely reasons for these outliers, and to keep all of them as very interesting instances of binarity, as summarised in Table~\ref{tab:outliers}.
Most of these outliers exhibit proper motion and parallax anomalies \citep{Ker19,Bra21} due to the closeness of the components. 
Their orbital periods are for this reason relatively short, of a few tens of years or less {(Sect.~\ref{sssection:periods})}.
The systems containing some of these pairs are valuable because their dynamical masses could be determined in the near future.

In addition, there are 34 systems in which distances and proper motions are missing for one of the components, and so $\Delta d$, $\mu$ ratio, and $\Delta PA$ are unknown.
For these, our criteria for physical associations are not fully conclusive, but they do all correspond to well-characterised close binaries listed by WDS, which we retained in our analysis nevertheless. 

Even though for most of the cases, the criteria presented here are effective for determining whether a pair of components in a system is physical or optical (unbound), there are some caveats. 
False positives (or chance alignments) can be found in wide pairs (Sect.~\ref{sssection:separations} offers a deeper analysis), while false negatives, attributed to imprecise astrometry, are more commonly found in very close pairs.
The nominal operations of {\em Gaia} up to DR3, spanning 1028 days, may not provide sufficient coverage, leading to ill-defined proper motions in those cases for which orbital periods notably exceed this timespan.
Thus, these may reflect the instantaneous tangential path, including orbital motion, meaning that some bound systems might not qualify as binary.
Given the small probability of finding a source at a similar distance  within a small search radius, the benefit of finding a close companion justifies the effort of checking individually all the potential pairs by accessing to images and catalogues.
For this particular inspection we used the {Aladin} interactive sky atlas \citep{Bon00} and the SIMBAD database \citep{Wen00}.
With Aladin we carefully examined every source at a separation $\rho <$ 5\,arcsec from our targets, discarding those with an astrometric description that identified them as background objects.

Most of the systems discovered in our \textit{Gaia} blind search are tabulated by the WDS, as illustrated by Fig.~\ref{fig:rho_WDS}. 
From a total of {720} systems found in our sample, {697} are tabulated in the WDS, of which {249} are not resolved by {\em Gaia}.
Many of them are very close sub-arcsecond binaries detected with adaptive optics, lucky imaging, or speckle interferometry (Sect.~\ref{sssection:separations}), but there is also room for resolved ultra-cool dwarfs that are fainter than the \textit{Gaia} magnitude limit (Sect.~\ref{sssection:ucds}).

Despite the continued efforts by the WDS team at the United States Naval Observatory to keep this catalogue updated, there can still be missing pairs that appear in the literature.
To address this gap, we also cross-matched our stars with 
the Robo-AO surveys (with $d \lesssim$~30\,pc) of \cite{Lam20} and \cite{Sal22},
the {\em Gaia} catalogue of nearby stars \citep[GCNS,][]{Sma21}, 
the million binaries from {\em Gaia} EDR3 of \cite{EB21},
the 10-parsec sample of \cite{Rey21},
the full-sky 20-pc census of \cite{Kir24},
and the ultra-cool dwarf companion catalogue by \cite{Bai24}.
Some of these compilations (mostly \citealt{EB21} and \citealt{Sma21}) tabulated bound systems that are not yet tabulated in the WDS, but were also identified by us.
For completeness, we also cross-matched the remaining systems with the Washington Double Star Supplemental Catalog\footnote{\url{http://www.astro.gsu.edu/wds/Supplement/wdss.html}}, which besides compiles the input from those and other large faint duplicity surveys such as those by 
{\cite{Dhi15}, \cite{Oh17}, \cite{Tia20}, or \cite{Hart20}}.
Their findings are duly credited in the `Discoverer' column across the various tables in this work.

While most pairs found in our custom \textit{Gaia} search are known, either from WDS or the recent searches mentioned before, we report six physically bound systems for the first time.
All but two are considered to be single stars; the exceptions are the triple systems HD~230017 in the Carina moving group (Sect.~\ref{sssection:young}) and GJ~3261, which were thought to be close binaries.
We show the {six} new systems in Table~\ref{tab:new}.
Apart from {\em Gaia} DR3 astrometric solutions ($\alpha$, $\delta$, $\varpi$, $\mu_{\rm total}$), for each pair, we tabulated their angular separations ($\rho$) and position angles ($\theta$).

\subsection{Search for known spectroscopic binaries}
\label{ssection:unresolved}

After determining the systems in our sample that are astro-photometrically resolved by {\em Gaia}, and those tabulated by WDS and other recent catalogues, we carefully reviewed the literature for references to unresolved systems, namely spectroscopic binaries\footnote{It is customary to use the broad term `spectroscopic binary' to also refer to triples or even quadruples spectroscopically detected, despite the specific implication of two components in the term.} (SBs). 
For a comprehensive list of SBs, the main source of reference used in this work was the 9th Catalogue of Spectroscopic Binary Orbits \citep[SB9,][]{Pou04}.
In this catalogue, we found {166} spectroscopic multiples among the stars in our sample and their companions for which a spectroscopic investigation has been conducted (we refer to Sect.~\ref{ssection:fraction} for an explanation of the detection limits of the Carmencita stars and their physical companions).
Among them, we found {42} cases in which the components have also been astrometrically resolved and compiled in the WDS.
By checking the reported orbital periods and magnitude differences we ensured that both measurements refer in fact to the same pair.
For these cases we removed the spectroscopic binary designation (Aab) and classified them as close resolved, AB or (AB).
In case of doubt, we preferred to stay conservative and avoided losing a potential triple system (thought to be simply double) in a compact configuration.
For instance, the physical pair \object{GJ 3481} and \object{GJ 3482} (COU~91), where the latter is further resolved as a spectroscopic binary \citep[$a<$ 0.66\,au according to][]{Shk10} and as an astrometric visual pair with the FastCam lucky imager \citep[$\rho \approx$ 0.94\,arcsec or $s \sim$ 16.7\,au according to][]{Cor17a}, might potentially constitute a hierarchical quadruple.
Another example is \object{GJ~3522} (\object{LHS~6158}), a 7.6-day double-lined spectroscopic binary \citep{Rei97b,Tok18}, with a third component in a 5.7-year period \citep{Har12} revealed using adaptive optics \citep{Del99b}, with spectral classification assumed roughly similar to the primary's \citep[M3.5\,V,][]{Kir12} and constituting one of the nearest ($d=$ 6.7\,pc) hierarchical triples.
A total of {124} spectroscopic systems found in our sample and are listed in Table~\ref{tab:spectroscopic}, including their published orbital periods, $P_{\rm orb}$, semimajor axes, $a$, and mass ratios, $q = \mathcal{M}_B/\mathcal{M}_A$, together with their references.
Values estimated from mass-luminosity relations, or directly from spectral typing, are not collected.
None of the 123 SBs have been spatially resolved to date.

\begin{table*}[]
\caption[]{Criteria for the detection of unresolved sources based on {\em Gaia} DR3 statistical indicators.}
\label{tab:criteria}
 \centering
\begin{tabularx}{\textwidth}{cXX}
     \hline\hline 
   {Criterion} & {Selection} & {Remarks} \\\hline
     \noalign{\medskip}
1       &        {\tt RUWE} > 2 &       Goodness of fit between the observed astrometric data and a single-star model.  \\
    \noalign{\smallskip}                                        
2       &       {\tt ipd\_gof\_harmonic\_amplitude} > 0.1 \& \newline {\tt RUWE} > 1.4     &       Flag for spurious solutions of resolved doubles, not correctly handled in the {\em Gaia} EDR3 astrometric processing \citep{Fab21}.      \\
    \noalign{\smallskip}
3       &       {\tt ipd\_frac\_multi\_peak} $> 30$     &       
Fraction of windows for which the algorithm has identified a double peak (or double transit), meaning that the detection may be a visually resolved double star \citep[e.g.][]{Tok23,Hol23b,Med23}. \\

    \noalign{\smallskip}                                        
4       &       {\tt rv\_chisq\_pvalue} < 0.01 \& \newline {\tt rv\_renormalised\_gof} > 4 \&  \newline {\tt rv\_nb\_transits} $\geq$ 10       &       Measure of variability in RV among all the {\em Gaia} measurement epochs \citep[][]{Kat23}.        \\
    \noalign{\smallskip}                                        
5       &       {\tt radial\_velocity\_error} $\geq$ 10\,km\,s$^{-1}$   &       Same as above.
For the bright stars ($G \lesssim$ 13\,mag) it is the uncertainty on the median of the epoch radial velocities, to which a constant shift of 0.11\,km\,s$^{-1}$ was added to take into account a calibration floor.     \\
    \noalign{\smallskip}                                        
                                        
6       &        {\tt duplicated\_source} = 1   &       Flag for the existence of a duplicated source during data processing, which may indicate observational, cross-matching, or processing problems, or stellar multiplicity, and probable astrometric or photometric problems in all cases. 
This metric works in support to the others, rather than being a standalone indicator on itself.    
\\
    \noalign{\smallskip}                                        
7       &       {\tt non\_single\_star} &
Flag for a possible non-constant behaviour using binary orbit models.
Three bits indicate whether it has been identified as astrometric (bit 1 is set to one), spectroscopic (bit 2 is set to one), or eclipsing (bit 3 is set to one) binary, respectively \citep{Pou22}.      \\
                                                                                                
\hline
        \noalign{\smallskip}
        \noalign{\smallskip}
\end{tabularx}
\end{table*}

The cases in which the orbital plane of a binary system aligns with our line of sight are rare. 
Eclipsing binaries (EBs) are a unique category and serve as a valuable opportunity for determining empirical masses and radii \citep{Hua56,Pop80,And91}. 
EBs enable the measurement of dynamical masses, especially in detached double-lined SBs, 
and provide accurate radius measurements with precision around 1--2\,\% \citep{Rib03a,Tor10,Schw19}. 
These parameters have the advantage of not relying on models, so they can serve to evaluate the accuracy of theoretical predictions.
\cite{Sha15} demonstrated an elevated occurrence of eclipsing binaries among detached M-dwarf SBs with orbital periods of 1--90\,d, exceeding previous RV-based inferences.
Stellar activity levels, particularly in young, magnetically active, or fast-rotating tidally-locked eclipsing binaries, can also introduce biases, leading to observed inflated radii \citep{Cab10b, Jac18, Kes18, Par18}.
All-sky surveys such as {\em Kepler} and TESS have made possible the identification of eclipsing binaries by the thousands \citep{Kir16,Prs22}.
In our sample there are {five} known eclipsing binaries, which are listed along with their masses and radii in Table~\ref{tab:eclipsing}.
Additionally, the objects 
\object{GJ~3547} (J09193+620), 
\object{GJ~3461} (J07418+050), and
\object{GJ~3793} (J13348+201) are suggestive of eclipsing events from the study of {\em TESS} light curves \citep{Skr21}.
The first two are double-lined spectroscopic binaries (in one case also featured in the variability catalogue of \citealt{Eye23}), while the third is a single star but flagged by us as a likely unresolved binary (`candidate'; see Sect.~\ref{ssection:candidates}).

\subsection{Search for new binary candidates using {\em Gaia}}
\label{ssection:candidates}

Next, we exploited the wealth of {\em Gaia} data in order to identify binary candidates that have not been identified in previous work.
This analysis also extended to companions of known multiple systems found in the previous sections, which added complexity to a number of these systems. 

The spatial resolution of {\em Gaia} was limited to 0.4--0.5\,arcsec in the second data release \citep[DR2,][]{Gaia18bro}, and slightly improved in the early third data release \citep[EDR3,][]{Gaia21a}.
In particular, \cite{Fab21} showed that EDR3 achieves completeness for separations larger than 1.5--2.0\,arcsec, with a severe incompleteness below 0.7\,arcsec.
Objects closer than this limit can be identified with {\em Gaia}, but this identification depends on the magnitude difference, current angular separation, and orientation along the dominating scan directions \citep[e.g.][]{Sma21}.
{\em Gaia} EDR3 data can be affected by spurious signals linked to the time-dependent scan angle of the instrument, leading to false periodic signals in photometry and astrometry. 
Using numerical simulations, \cite{Hol23b} explored how these biases occur and provided statistics to identify and filter affected sources.
They found that these signals often originate from unresolved binaries or other close optical pairs with fixed orientations and separations of less than 0.5\,arcsec (including binaries with orbital periods of several years).
Therefore, {\em Gaia} EDR3 is capable of resolving visual double stars, but unable to resolve the closest pairs, with $\rho \lesssim$ 0.13\,arcsec, although upcoming releases are expected to enhance these resolution capabilities \citep{Bru15}.
Still, {\em Gaia} has been proven highly valuable to date in the detection of binary stars \citep[see][]{EB24}.

The third data release, DR3, maintains the same astrometric data {as EDR3}, but it includes a rich set of new data products that we exploited in this work.
For instance, {\em Gaia} DR3 is the first release that provides an analysis of the RV time-series.
By the time of this release, every source was observed with the Radial Velocity Spectrometer an average of $\sim$70 times, varying from $\sim$30 to $\sim$240, depending on the sky coordinates. 
 
\begin{figure}
   \centering
   \includegraphics[width=0.99\linewidth]{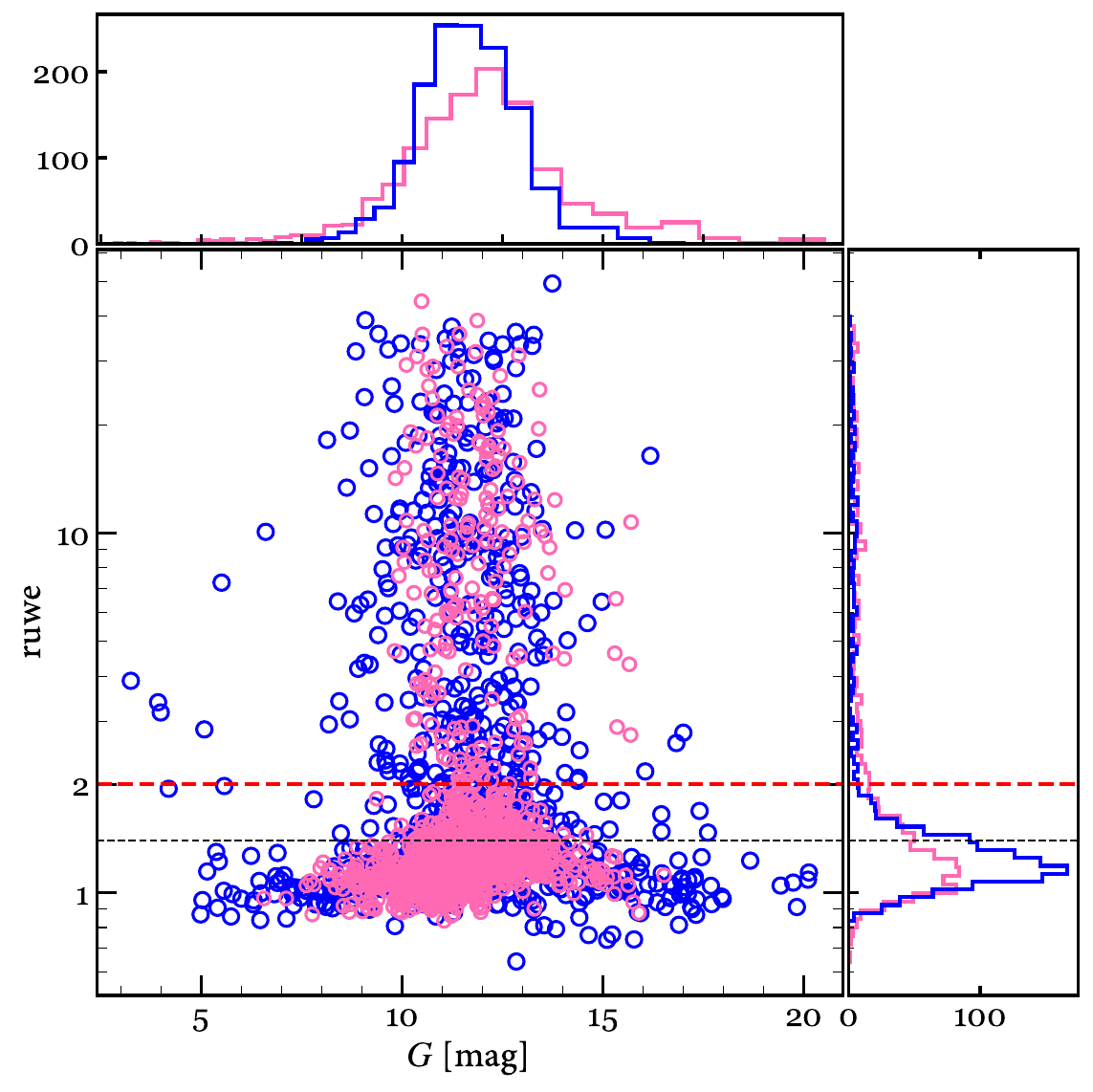}
   \caption{{\tt RUWE} as a function of $G$ magnitude for single stars 
   ({pink open} circles) and stars 
   in multiple systems ({open blue} circles).
   The histograms follow the same colour coding.
   {\tt RUWE} = 2.0 (thick red dashed line), used in this work, leaves behind $\sim$80\,\% of the sample and is more conservative than the traditionally used {\tt RUWE} = 1.4 (thin black  dashed line).}
    \label{fig:G_RUWE}
\end{figure}

\begin{figure}
    \centering
    \includegraphics[width=0.99\columnwidth]{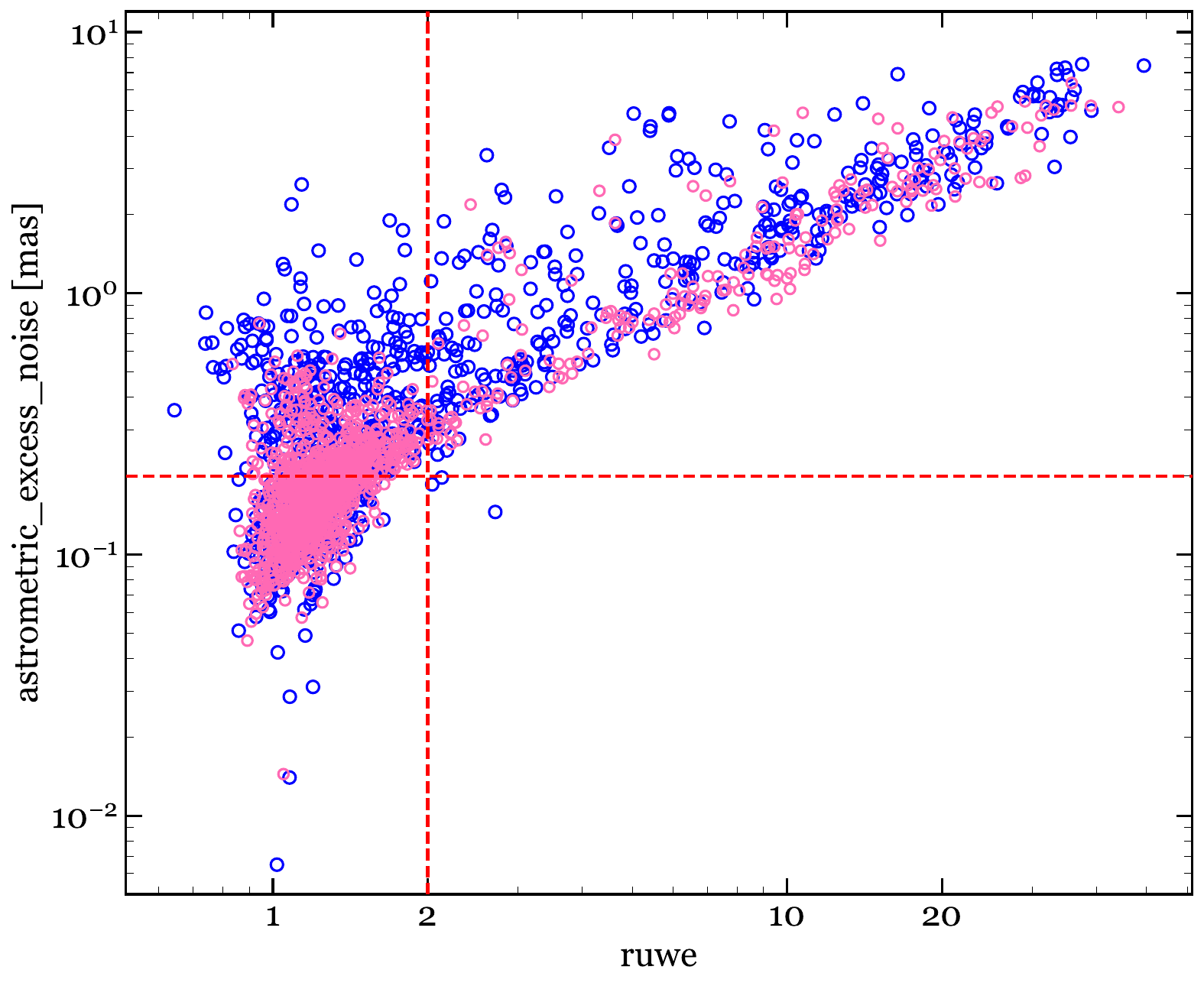}
    \includegraphics[width=0.99\columnwidth]{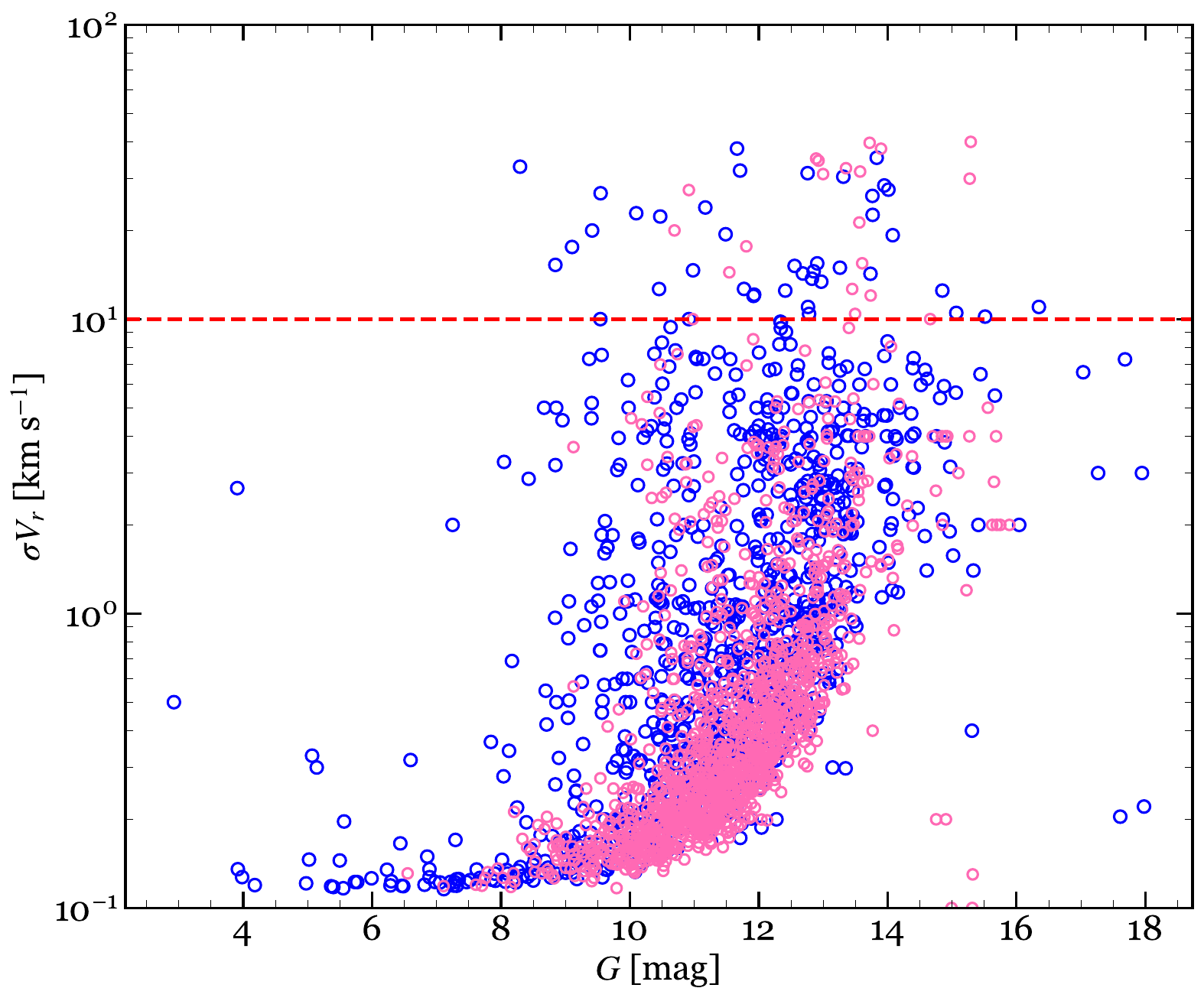}
    \includegraphics[width=0.99\columnwidth]{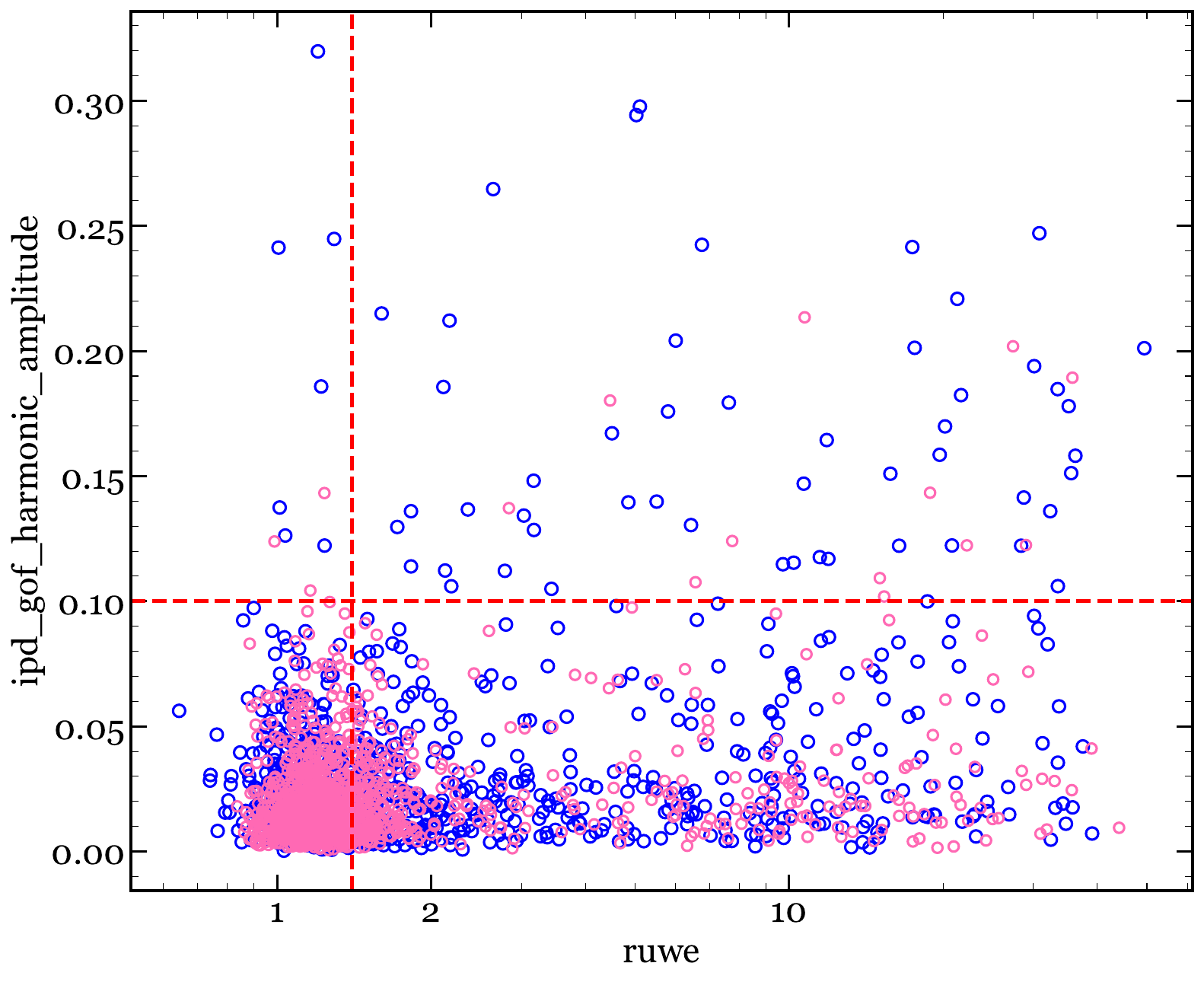}
    \caption{\textit{Gaia} DR3 statistical indicators of very close multiplicity.
    \textit{Top}: {\tt astrometric\_excess\_noise} ($\epsilon$) vs {\tt RUWE}.
    \textit{Middle}: $\sigma({V_r})$ vs $G$.
    \textit{Bottom}: {\tt ipd\_gof\_harmonic\_amplitude} vs {\tt RUWE}.
    In all panels, {pink} and blue {open} circles are for stars that were considered single and part of multiple systems, respectively, and the red thick dashed lines indicate the corresponding close binary selection criterion in Table~\ref{tab:criteria}.
    }
     \label{fig:AllGaia}
\end{figure}

{\em Gaia} DR3 comes with numerous statistical parameters to assess the reliability of the astrometric data.
We summarise in Table~\ref{tab:criteria} some of these quality indicators or combinations of them, which serve to flag stars with astrometric issues.
They have been proven to be sensitive to the presence of potential unresolved companions \citep[e.g.][]{Fab21,Kat23,Pou22,Pen22a,Sha23,Swa23}. 
A detailed description of them can be found in the {\em Gaia} DR3 documentation\footnote{\href{https://gea.esac.esa.int/archive/documentation/GDR3/}{\url{https://gea.esac.esa.int/archive/documentation/GDR3}}}.
The usage of these parameters is explained next (Sects.~\ref{sssection:ruwe}--\ref{sssection:lacking}), as well as complementary methods of assessing unresolved multiplicity exploiting the {\em Gaia} products, which overlap with the criteria of Table~\ref{tab:criteria} (Sects.~\ref{sssection:non-single}--\ref{sssection:excess}).
Figures~\ref{fig:G_RUWE}~and~\ref{fig:AllGaia} illustrate the different unresolved multiplicity criteria.

\subsubsection{{\tt RUWE} (Criteria 1 and 2)}
\label{sssection:ruwe}

Perhaps the most widely used metric in Table~\ref{tab:criteria} is the renormalised unit weight error ({\tt RUWE}), which evaluates the behaviour of the centre of light.
\cite{Bel20} showed that the amplitude of the centroid perturbation correlates with the physical separation between companions and scales with the binary period and mass ratio.
Unresolved binaries with periods similar to or longer than the {\em Gaia} timespan (which increases with each subsequent release) can show elevated {\tt RUWE} due to orbital motion affecting the astrometric fit (see \citealt{Pen22a} for a detailed discussion).
\cite{Cas24} suggested that the majority of binary systems in our Galaxy will remain undetected because the wobble of the centre of mass around the photocentre is largely masked by the astrometric noise from {\em Gaia}, leading to a {\tt RUWE} value below the detection threshold (we refer to their {Fig}.~1 for an illustration).
However, the median distance of our M-dwarf stars is 22.0\,pc, which locates them in our immediate vicinity and allows to detect that wobble in many cases.
\cite{Pen20} noted that assuming an object as a single point mass in astrometry can bias measurements due to unresolved binaries. 
They also concluded that orbital motion with period near one year can mimic parallax, distorting distance estimates.
Additionally, distant objects (further than 100\,pc) can show disproportionate values of {\tt RUWE} \citep{Bai24}, which has been mitigated by an additional renormalisation, namely local unit weight error \citep[{\tt LUWE},][]{Pen22b}, with well-behaved single stars having a value close to 1.0.
Still, {\tt RUWE} remains a powerful and simple metric, especially when complemented with other indicators.
Instead of the generally adopted value of 1.4 \citep[e.g.][]{Are18,Lin18,Cif20}, or even the position-dependent range of values from 1.15 to 1.37 developed by \citet[][their {Fig}.~3]{Cas24}, in our analysis we set a conservative minimum of {\tt RUWE} $>$ 2.0, which leaves behind $\sim$80\,\% of the sample.

\subsubsection{{\tt ipd} metrics (Criteria 2 and 3)}
\label{sssection:ipd}

The broad applications of many of the {\tt ipd} metrics in the targeting of binary stars has been recognised, for instance, by \cite{Vri20}, who  applied them to two decades of astrometric data from the RECONS program along with {\em Gaia} DR2 observations.
\cite{Cla22} demonstrated that {\em Gaia} was unable to resolve a great portion (58.9\,\%) of the close companions that they detected using speckle imaging, which motivates the need for additional high-resolution imaging ($\sim$40\,mas).
They investigated the usefulness of the metrics {\tt RUWE} along with the parameter {\tt ipd\_frac\_multi\_peak} for assessing the likelihood of an unseen stellar companion (IPD stands for `image parameter determination').
For instance, \cite{Bai24} built the ultra-cool dwarf companion catalogue exploring the likelihood of hidden binarity using {\tt LUWE} and IPD.

Likewise, \cite{Gol23} identified sources with spurious astrometric solutions in their Fifth Catalogue of Nearby Stars (CNS5) by applying a simple cut on the {\tt ipd\_gof\_harmonic\_amplitude} (see their Eq.~2). 
We did not include this criterion among ours, but we nevertheless applied it to every star and companion in our sample for which a measure of the parallax is available from {\em Gaia} DR3 (but excluding DR2).
We found that only {nine} objects show spurious solution using this cut, with eight of them being known binaries, and one being single, but categorised as `Candidate' via several other criteria in Table~\ref{tab:criteria}.

\subsubsection{Variability in RV (Criteria 4 and 5)}
\label{sssection:variability}

The standard deviation of RV at several epochs (i.e. $\sigma({V_r})$ = {\tt e\_RV}) is a powerful method of identifying (spectroscopic) binary systems. 
Furthermore, it allows to discriminate true orbital acceleration due to multiplicity from the acceleration due to an effect of perspective. 

Among the stars with the largest $\sigma({V_r})$ measured in {\em Gaia} DR3, greater than 10\,km\,s$^{-1}$, we selected the brightest ones to study their close multiplicity using medium-resolution spectra.
We are carrying several observation campaigns with the high-resolution FIbre-fed Echelle Spectrograph (FIES) mounted on the 2.56\,m Nordic Optical Telescope (NOT), in the medium-resolution mode (${R} = 46\,000$), as well as with the High-Efficiency and high-Resolution Mercator \'Echelle Spectrograph (HERMES) at the 1.2\,m Mercator telescope. 
From this ongoing program, we confirm the multiple nature of several of these candidates. 
Although the detailed results will be published in a forthcoming article, the preliminary results from those observations confirmed that our limit of 10\,km\,s$^{-1}$ is very robust.

\subsubsection{Lacking or poor data (Criteria 6 and 7)}
\label{sssection:lacking}

Sources in DR3 apparently unaffected by the proximity of other objects, but showing excessive uncertainties in parallaxes and proper motions nonetheless, or lacking entries in these fields altogether, could actually be close binary candidates.
For example, \cite{Sma21} noted that spurious astrometric solutions can be due, among a number of reasons, to the presence of more than one object in the astrometric window (close double systems, either real or in projection) or to binary orbital motion that is not accounted for.
The odds for a chance alignment are much lower than for a physical connection, although these odds increase at wider separations and fainter magnitudes \citep[e.g.][]{EB21,Chu22}.
Chance alignments are more likely to occur in regions of high surface density of sources such as open clusters and near the galactic plane \citep{Sma21}.
While DR3 benefits from a larger number of observation epochs with respect to previous releases, some very close binaries not resolved in DR2 were handled as single objects, with blended photometry and occasional spurious astrometric solutions \citep[][]{Are18,Zie18}. 
The {\tt duplicated\_source} flag is usually of help for those sources that also exhibit poor astrometric quality or lack some data (mainly proper motions and parallaxes).
This flag is set to `1' in those instances where the detection system on board {\em Gaia} generates multiple detections for the same source, which results in different data sets for the same target. 
The final DR3 catalogue retained only the solution with the best astrometric quality and flagged it as a {\tt duplicated\_source}, while the poorer ones were discarded.

\subsubsection{Non-single star tables by {\em Gaia} DPAC}
\label{sssection:non-single}

Among the built-in data products in {\em Gaia} DR3 by the Data Processing and Analysis Consortium (DPAC), the `non-single stars' tables \citep{Pou22} enable the identification of unresolved astrometric, spectroscopic, and eclipsing binaries \citep[][]{Are23}. 
Despite their name, these tables are unrelated to the {\tt non\_single\_star} flag, which is primarily a modelling quality flag and not an identifier of multiplicity.
These solutions are distributed in four tables: {\tt nss\_two\_body\_orbit} when the full orbital motion is known, {\tt nss\_acceleration\_astro} and {\tt nss\_non\_linear\_spectro} when a trend is known, and {\tt nss\_vim\_fl} for photometrically variable unresolved binaries. More details on the processing scheme, its validation, and the various types of reached solutions can be found in Chapter~{7} of the {\em Gaia} DR3 documentation.

\subsubsection{Astrometric excess noise}
\label{sssection:excess}

In addition to {\tt RUWE}, another measure of \textit{Gaia}'s astrometric goodness-of-fit is the {\tt astrometric\_excess\_noise} ($\epsilon$).
It quantifies the disagreement between the observations of a source and the best-fitting standard astrometric model.
Both are sensitive to the photocentric motions of unresolved objects, such as astrometric binaries, which are not revealed by the 
IPD statistics, and therefore complement the latter in binary detection \citep{Lin21}.
For well-behaved sources, $\epsilon$ should be zero, but since it accommodates excess noise originating from both the source and the instrument attitude, non-zero values are inevitable \citep{Lin12}. 
For instance, in DR1 nearly all sources show significant excess noise ($\epsilon \sim$ 0.5\,mas), but only unusually large values ($\epsilon\gtrsim$ 1--2\,mas) are indicative of astrometric binarity or other issues \citep{Lin16}.
In DR2 roughly 20\,\% of the sources between $G$ = 12\,mag and 20\,mag have excess noise \citep{Gaia18lin}.
In EDR3, \cite{Lin21} observed an improved homogeneity of $\epsilon$ despite increased noise in crowded regions.
Nevertheless, $\epsilon$ can be regarded as insignificant (i.e. effectively zero) if the significance, {\tt astrometric\_excess\_noise\_sig} ($D$), is less than 2\,mas.

The {\tt RUWE} includes a scaling factor to compensate for calibration errors that correlate with colour and magnitude, but $\epsilon$ does not.
For example, \cite{Gan22} used astrometric excess noise to search for candidate X-ray binaries, selecting sources with $\epsilon \geq$ 0.01\,mas, $D \geq$ 2\,mas, {\tt visibility\_periods\_used} $>$ 10, and $G$ between 13\,mag and 20\,mag. 
They found that systematic effects, such as attitude errors, partially resolved double stars, and source variability, can be sources of contamination, especially when $\epsilon <$ 1\,mas, making interpretation more difficult.

In the stars of our sample, we found a linear correlation between the excess $\epsilon$ and its significance $D$, meaning that larger values of $\epsilon$ are generally associated to larger values of $D$.
In particular, all the instances with notable $\epsilon$ happen to be known binaries in close configurations.  
Therefore, we found $\epsilon$ and $D$ to be redundant and equivalent to {\tt RUWE}, and {stuck} to the latter, as detailed in Sect.~\ref{sssection:ruwe}. 

\subsubsection{Photometric variability}
\label{sssection:photometric}

\begin{figure*}
    \centering
    \includegraphics[width=0.33\linewidth]{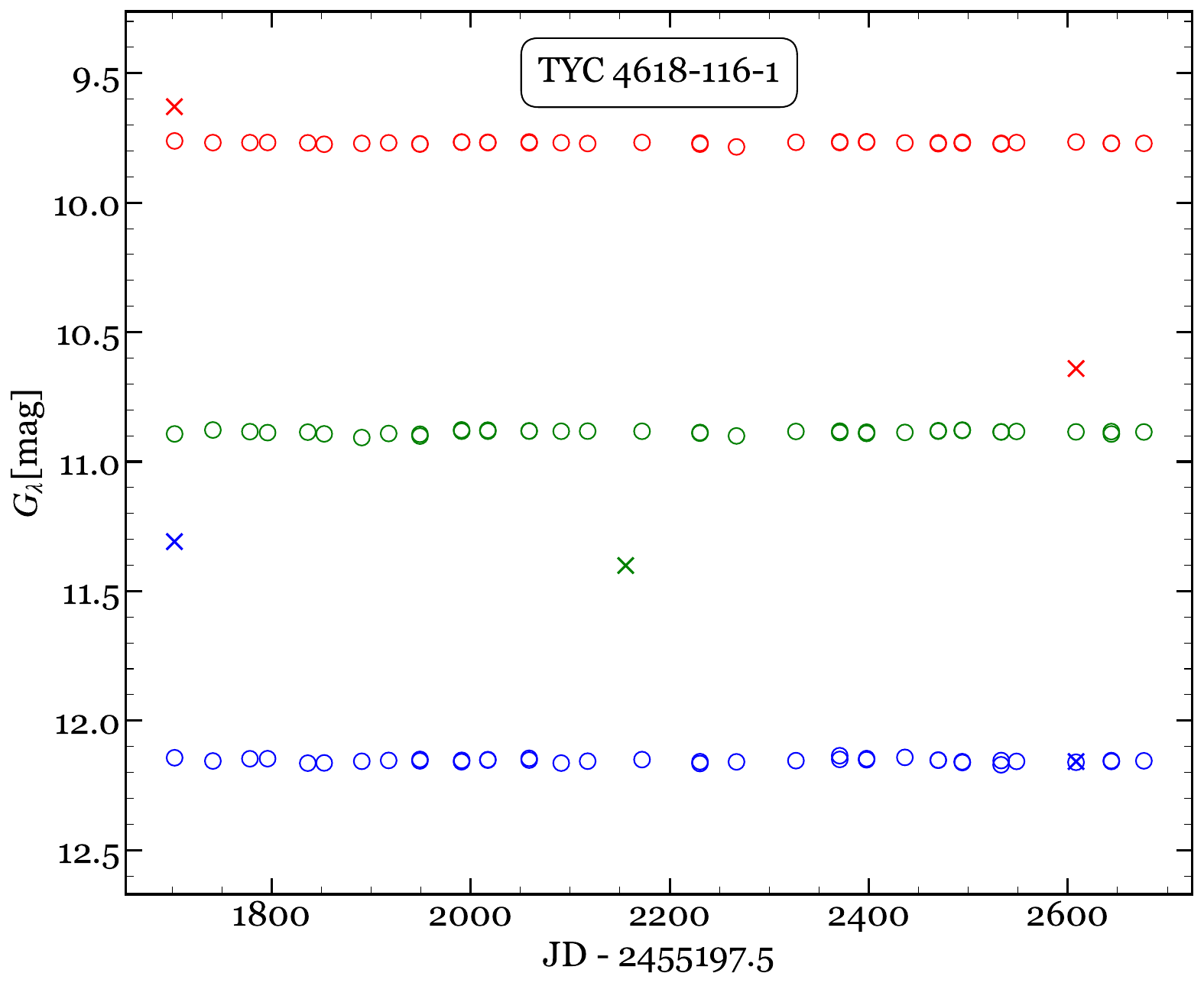}
    \includegraphics[width=0.33\linewidth]{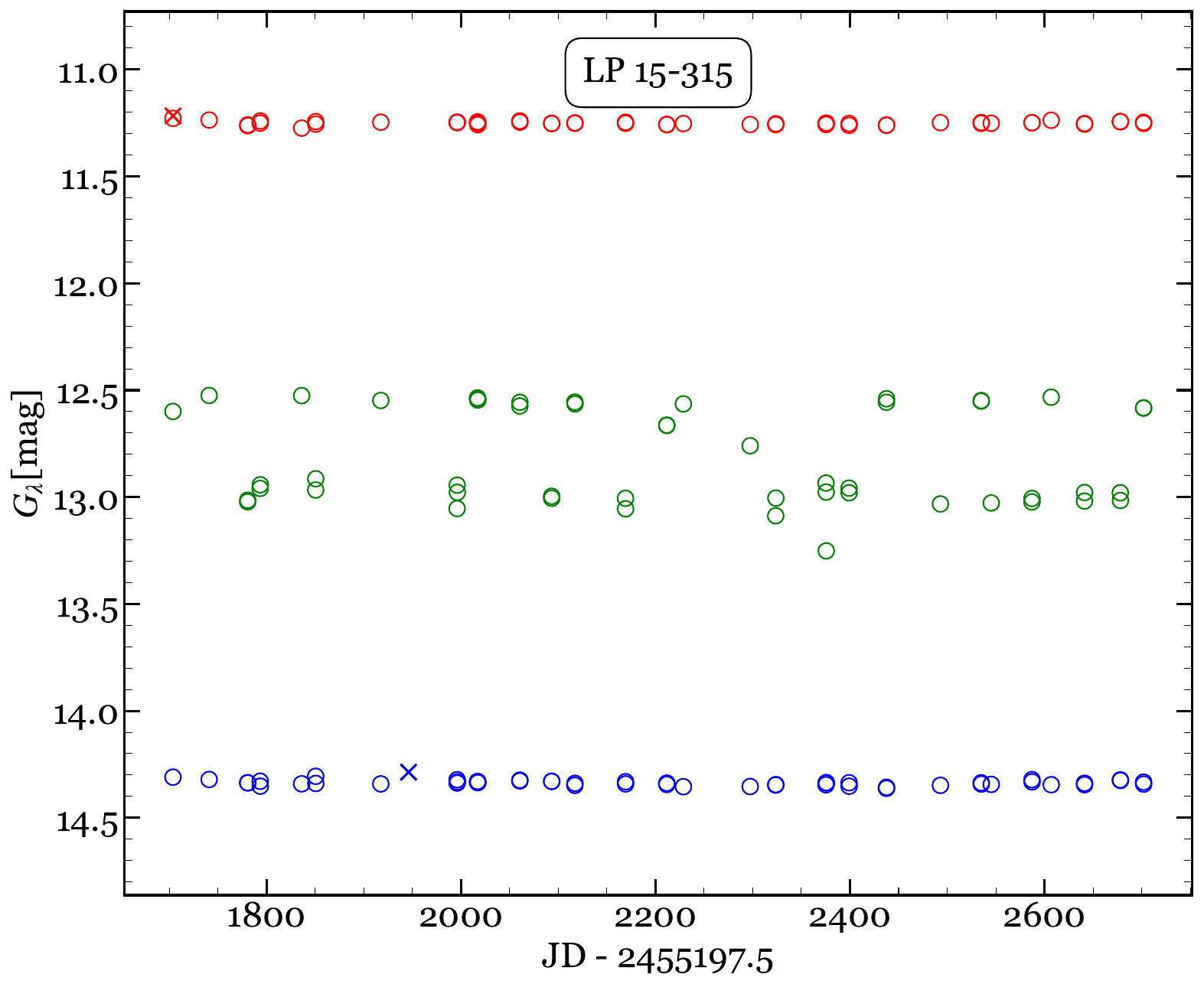}
    \includegraphics[width=0.33\linewidth]{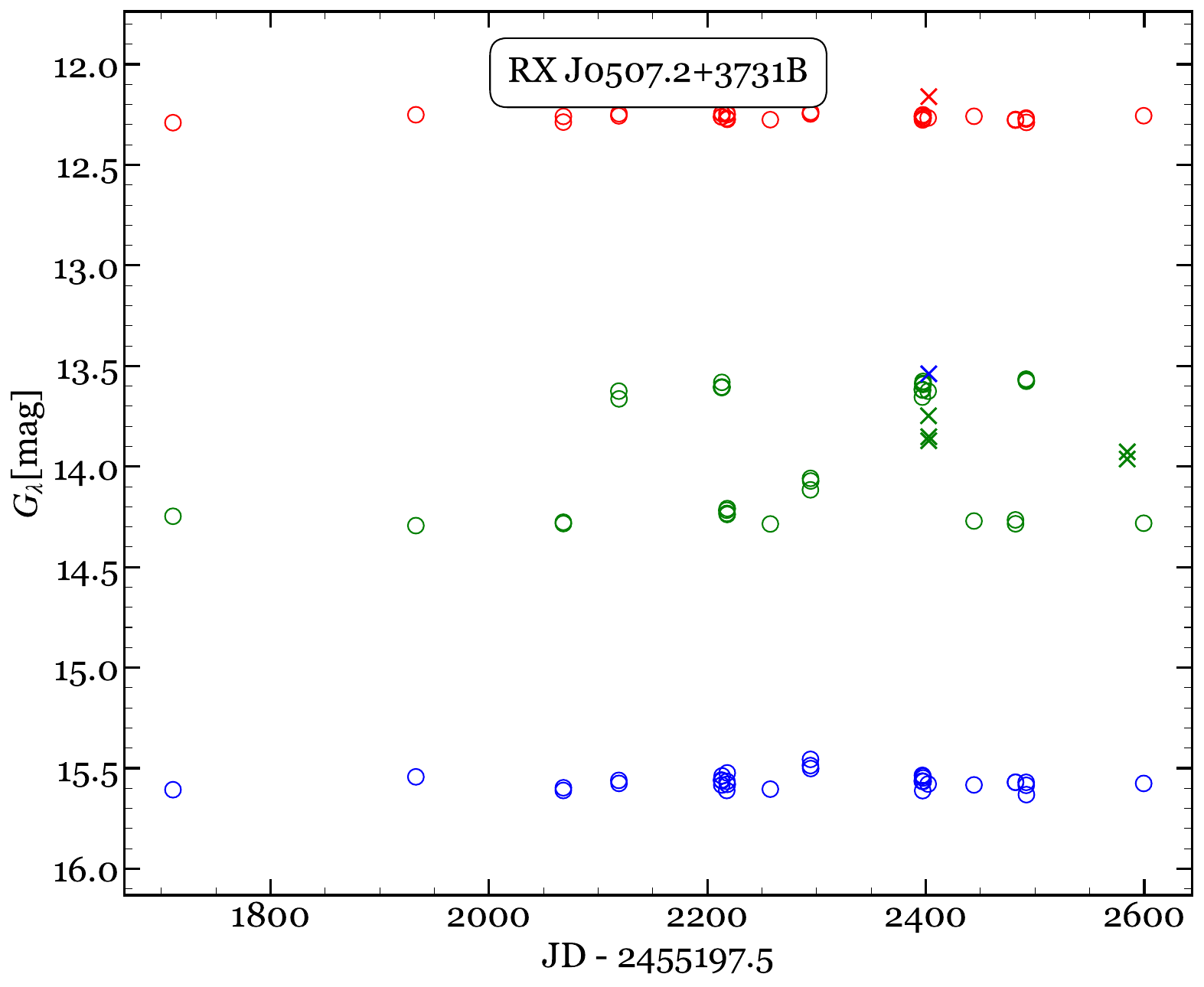}
    \caption{Light curves corresponding to the {\em Gaia} passbands $G_{RP}$ (red), $G$ (green), and $G_{BP}$ (blue) for three cases: a single star ({\em left}), a known close ($\rho=$ 0.1--0.2\,arcsec) binary system (JNN~266, \citealt{Jan14a}; {\em middle}), and a newly reported close ($\rho \approx$ 0.48\,arcsec) binary system ({\em right}).
    Isolated outliers (represented by crosses) are photometric errors, not consistent with flares, automatically rejected by variability processing. 
    The remaining light curves for the unresolved binary candidates with double-sequence patterns are displayed in Fig.~\ref{fig:light_curves_appendix}.}
     \label{fig:light_curve}
\end{figure*}

\begin{figure}
    \centering
    \includegraphics[width=0.99\linewidth]{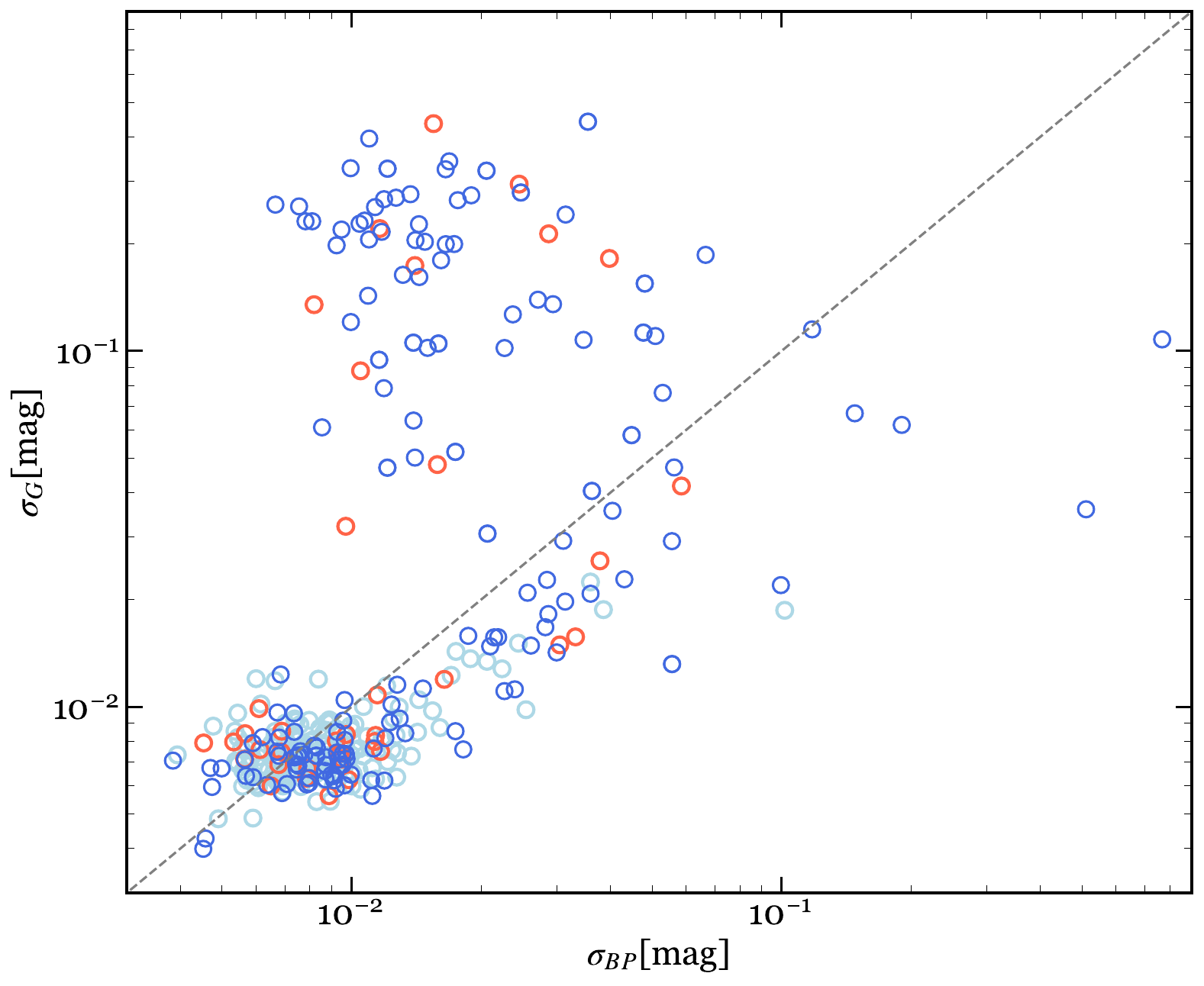}
    \caption{Comparison of the standard deviations of $G_{BP}$  against that of $G$ for single stars (light blue), stars in multiple systems (blue), and new unresolved binary candidates (red).
    The dashed line represent a 1:1 relation.
    The similar plot for $G_{RP}$ is omitted for simplicity.}
     \label{fig:sigma_comparison}
\end{figure}

Measuring the stellar brightness over time can reveal the presence of binaries that are otherwise indistinguishable in static images. 
{\em Gaia} DR3 provides a variability analysis of many objects using data from the 34 months.
The variability processing and analysis was based mostly on time series of field-of-view transit (integrated) photometry in the calibrated $G$, $G_{BP}$, and $G_{RP}$ bands, with additional input data, such as RV time series. 
We refer to \citealt{Eye17} for a more complete description of the models and methods.
In {\em Gaia} DR3, {\tt phot\_variable\_flag} tags with `VARIABLE' those sources identified and processed as variable from the photometric data.
The variables {\tt GrVFlag}, {\tt BPrVFlag}, and {\tt RPrVFlag} accompanying the photometric measurements indicate the photometry rejected by variability processing \citep{Eye23}.
Even so, {346} stars in our sample are identified as variables by {\em Gaia} DR3.

Figure~\ref{fig:light_curve} shows the photometric time series (or light curves) in the three {\em Gaia} passbands for three selected cases: a single star, a known binary, and a binary candidate.
When an object crosses the focal plane of {\em Gaia} \citep[see Fig.~4 of][]{Gaia16pru}, its flux is measured nearly simultaneously in the three passbands, but the $G$-band photometry is more precise and has a better spatial resolution \citep[][additionally, the object is typically detected by nine CCDs in the green photometer, which is larger than the red and the blue]{Eye17}.
Stellar variability of intrinsic nature (i.e. not due to the presence of a companion) is apparent in the three passbands simultaneously and with a common pattern.
However, there are \textit{Gaia} sources with relatively flat $G_{BP}$ and $G_{RP}$ light curves, but much scattered $G$ light curves.
Sometimes they even display a double $G$ light curve separated by up to 0.7\,mag; these sources are actually close binary systems unresolved in $G_{BP}$ and $G_{RP}$ but resolved or partially resolved, depending mostly on the scan angle, in $G$ (\citealt{Vin23}; \citealt{Mai23}; \citealt{GP24}; T.~Prusti, priv. comm.).
In the three examples of Fig.~\ref{fig:light_curve}, only the $G$ light curves of the known binary (LP~15--315, JNN~266) and our new candidate binary (RX~J0507.2+3731B, $\rho \approx$ 0.48\,arcsec) display a ``double-sequence'' pattern.

Figure~\ref{fig:sigma_comparison} compares the standard deviation, $\sigma$, between two {\em Gaia} passbands.
It distinguishes between multiple systems and single stars, and among these, the ones that are new unresolved binary candidates.
Except for a few intrinsically variable young stars, single stars systematically show low standard deviations in the three passbands.
All the sources with significant deviations are either confirmed binaries or close binary candidates.
All the known binaries and new unresolved binary candidates in the agglomeration in top left corner show the $G$ double-sequence pattern that is not detected in the $G_{BP}$ (and $G_{RP}$) light curves.

\subsubsection{Selection of candidates}
\label{sssection:candidates}

We used all the seven criteria of Table~\ref{tab:criteria} to identify new candidate binaries among the M dwarfs of our sample and their {\em Gaia} companions (2634 stars, white dwarfs, and ultra-cool dwarfs in total).
We found that {327} of them meet one or more of the criteria.
Of them, {278} were thought to be singles and {49} formed part of multiple systems with known companions but are separated enough to exert a negligible influence.
When the identified unresolved binary candidate is part of a know multiple system, we flagged them as `Multiple+' in the full version of Table~\ref{tab:mother}.
If the candidate is a single star without known companions, we flagged them as `Candidate'.

Figures~\ref{fig:G_RUWE} and~\ref{fig:AllGaia} illustrate the correlation of four selected statistics of {\em Gaia}: {\tt ipd\_gof\_harmonic\_amplitude}, {\tt ruwe}, {\tt radial\_velocity\_error} (criteria 1, 2, and 5 in Table~\ref{tab:criteria}), and {\tt astrometric\_excess\_noise}.
Stars in multiple systems may only have wide companions and, therefore, experience insignificant (if any) change in {\tt ruwe} or $\sigma_{V_r}$.
Single stars are generally more scarce as we approach to the limits set by the criteria.

We also looked for matches of our sample (and their resolved companions) within non-single star tables provided by {\em Gaia}, and found {49}, {6}, {1}, {0}, and {2} coincidences in the tables enumerated in Sect.~\ref{sssection:non-single}, respectively.
These coincidences translate into {55} individual stars proposed as unresolved pairs by the {\em Gaia} processing scheme.
Of them, {24} are known pairs with very close-in orbits, and {31} are bona fide single stars, or in a few cases component in very wide pairs, presumably  unaffected by their distant companion.
The {2} eclipsing binaries are the known systems of {Castor~C} \citep{Joy26,Giz02} and {GJ~3547} \citep{Shk10,Rein12}.
The former belongs to the sextuplet system $\alpha$~Gem, one of the most complex configurations in our sample.

Additionally, we matched our list of candidates with those identified in the {GCNS} by \cite{Pen22b}.
This yielded 227 objects with both {\tt RUWE} and {\tt LUWE} values pointing to possible binarity.
All of them are also tagged by us as candidates, except in those cases where close (or very close) multiplicity is already known.
Additionally, we cross-matched the table {\tt vari\_eclipsing\_binary} \citep{Mow22,Eye23}, which was the first {\em Gaia} catalogue of eclipsing binaries from the study of variability.
No new EB candidates were found as a result.

\section{Results and discussion}
\label{section:results}

\subsection{Multiplicity fraction}
\label{ssection:fraction}

In the present work, we adopted the traditional definition for stellar multiplicity, requiring that the M dwarf is the primary (i.e. the most massive) component of its system.
{This means that M dwarfs as companions to A-type stars\footnote{{The three A-type stars with M-dwarfs companions are: \object{Castor}, \object{HD~140232}, and \object{HD~29391}.}}, FGK stars (Sect.~\ref{sssection:fgk}) and white dwarfs (Sect.~\ref{sssection:wd}) do not count towards multiplicity statistics}.
In order to quantify the observed multiplicity frequency or multiplicity fraction (MF) of M dwarfs, we followed the convenient notation of \cite{Bat73}, who denoted as $f_n$ the fraction of primaries that have $n$ companions:

\begin{equation}
        {\rm MF} = \dfrac{\sum\limits_{n=1}f_n}{\sum\limits_{n=0}f_n} = \frac{B+T+Q+\ldots}{S+B+T+Q+\ldots},
\end{equation}

\noindent where $S, B, T, Q$ represent the number of single, binary, triple, and quadruple systems, respectively.
In our sample of {2214} M dwarfs, {834} ({37.7\,\%}) belong to a multiple system, and from these almost three out of four ({73.8\,\%} or {615}) are the primaries of their systems, which implies a multiplicity fraction of MF = 27.8$^{+1.9}_{-1.8}$\,\%, 
where the uncertainties correspond to the 95\,\% confidence interval using
the Wilson formula \citep{Wil27}.
The remaining {1377} ({62.2\,\%} of the total) stars did not have suspected companions at any separation until now, with the exception of exoplanet hosts (Table~\ref{tab:exoplanets}).
However, {278} of them (plus {{49}} wide components among those in multiple systems) are proposed as new unresolved binary candidates in this work.
These values imply that the multiplicity fraction of M dwarfs (from M0.0\,V to M9.5\,V) could be increased by 12.5\,\%, potentially reaching 
MF+ = 40.3$^{+2.1}_{-2.0}$\,\%, 
which is higher than the values typically {found} for M dwarfs.
Our canonical MF is in agreement with the values reported in the literature, especially in those cases where a sizeable sample was studied.
However, our expected MF+ notably exceeds previous estimations, with the exceptions of the early studies of \cite{Hen90}, \cite{Fis92}, and \cite{Sim96}.

For a proper comparison with Table~\ref{tab:literature}, besides in the full M-dwarf spectral range, we also calculated the {MF} in the ranges from M0.0 to M4.5\,V and from M5.0\,V to M9.5\,V.
The former is volume-complete up to a distance of $d_{\rm com} =$ 30\,pc, while the later is only up to 10\,pc 
(Sect.~\ref{section:sample}).
Besides, given the uncertainties due to the smaller sample size at the lowest masses, we cannot confirm that the MF+ decreases with decreasing primary mass.
While MF+ = $40.5^{+2.1}_{-2.1}$ in the M0.0--4.5\,V spectral type range, it is $38.1^{+7.4}_{-6.8}$ in the M5.0--M9.5\,V range.

Likewise, the companion star fraction (CSF)\footnote{In order to avoid confusion, we retain here the term `star' even when the companion object may be substellar or a stellar remnant.} is the ratio of the total number of companions to the total number of stars in the sample:

\begin{equation}
        {\rm CSF} = \dfrac{\sum\limits_{i=1}(n-1)f_n}{\sum\limits_{i=0}f_n} = \frac{B+2T+3Q+\ldots}{S+B+T+Q+\ldots}.
\end{equation}

\noindent The CSF is a measure of the average number of companions per system, and can be larger than one. 
In our sample, M dwarfs have a  
CSF $= 0.332^{+0.020}_{-0.019}$ 
(CSF+ $= 0.462^{+0.038}_{-0.079}$ if the new candidates are included).
These values imply that roughly one in three M dwarfs have at least one (less massive) stellar or brown-dwarf companion (one in two M dwarfs if the new candidates are confirmed). 

Regarding the configurations of the multiple systems with M dwarfs as primaries, binary arrangements embody the majority of architectures ({83.1\,\%}), 
followed by triple systems ({14.3\,\%}), 
quadruples (2.1\,\%), 
and quintuples (0.3\,\%).
\object{V1311~Ori} is either a marginally stable hierarchy or a disintegrating mini-cluster \citep{Tok22} and it is the only sextuple system with an M-dwarf primary (J05320--030) in our sample.

The multiplicity fractions provided above (MF, MF+, CSF, and CSF+) are intended to offer results that can be compared with previous investigations.
However, one of the main concerns when claiming multiplicity fractions is selection bias.  
In other words, we need to understand the limitations of the observations available, and also factoring in the potential for undetected companions.
Here, the `detection limits' define the minimum separations at which one can be confident about the absence of companions (above a certain mass).
This definition ensures that the probability of missing a bound companion above that limit is minimal.
These limits are primarily based on the spatial resolution of the applied observational techniques, and the contrast limit between the primary star and potential companions. Following this idea, we assigned one of the following categories to each M dwarf
of our sample, including their confirmed companions:

\begin{itemize}
    \item {Category 1}: Stars observed with extreme precision spectrographs, with ten or more spectra in the CARMENES survey or monitored by other programmes \citep[][and their Table~4]{Rib23}. 
    Some of them may have also been observed with high-resolution imagers (AO, LI), the \textit{Hubble} Space Telescope, or even interferometric instruments (e.g. \citealt{Cab22}).
    \item {Category 2}: Stars observed with high-resolution imagers or those having less than ten high-resolution spectra.
    \item {Category 3}: Stars with only \textit{Gaia} DR3 data.
\end{itemize}

Categories 1--3 refer to the {maximum} precision (in decreasing order) with which each star stands in our study.
Among the 2214 M dwarfs in our sample of study, {447} are category 1, {408} are category 2, and {1359} are category 3 (included in Table~\ref{tab:mother}).
We computed the MF, MF+, CSF, and CSF+ values, ultimately concluding that the expected (MF+, CSF+) fractions are larger than the canonical ones (MF, CSF) in all cases, but only significantly for category 3 stars.
That is to say, \textit{Gaia} DR3 is not enough for imposing restrictive detection limits to close multiplicity.  Therefore more high-resolution spatial and spectral monitoring of stars from the ground are needed.
{Furthermore, it is more difficult to detect faint companions at further distance. 
Thus, the binary fraction measured with the whole sample may also be biased.
We could have measured the binary fractions in more complete subsamples by limiting ourselves to shorter distances.
However, since Carmencita was defined in the pre-\textit{Gaia} era, it is better to get rid of this bias by repeating the analysis in well-defined volume-limited samples, such as those of \citet[][at 10\,pc]{Rey21,Rey22}, \citet[][at 20\,pc]{Kir24}, and \citet[][at 100\,pc]{Sma21}.
This new analysis is part of forthcoming work (e.g. Gonz\'alez-Payo et al., in prep.).}

\subsection{Astrophysical parameters}
\label{ssection:parameters}

The main stellar parameters that we inferred were luminosities ($\mathcal{L}$), radii ($\mathcal{R}$), and masses ($\mathcal{M}$).
The radii and masses can be directly measured but only for a limited number of stars, which usually belong to multiple systems.
Element abundances and surface gravities can be studied from high-resolution spectroscopy, which is not always available.
Nevertheless, broadband multi-wavelength photometric data have almost always been measured for relatively bright, nearby stars.
With these data, the spectral energy distribution (SED) can be built and fitted to theoretical models.
These fits provide a good estimation of the bolometric flux, which results in the luminosities and effective temperatures ($T_{\rm eff}$), provided that the distance to the star is known \citep{Cif20}.
Parallaxes were available in \textit{Gaia} DR3, obviating the usage of photometric distances, subject to much larger uncertainties.

\begin{figure}[]
    \centering
    \includegraphics[width=.99\linewidth]{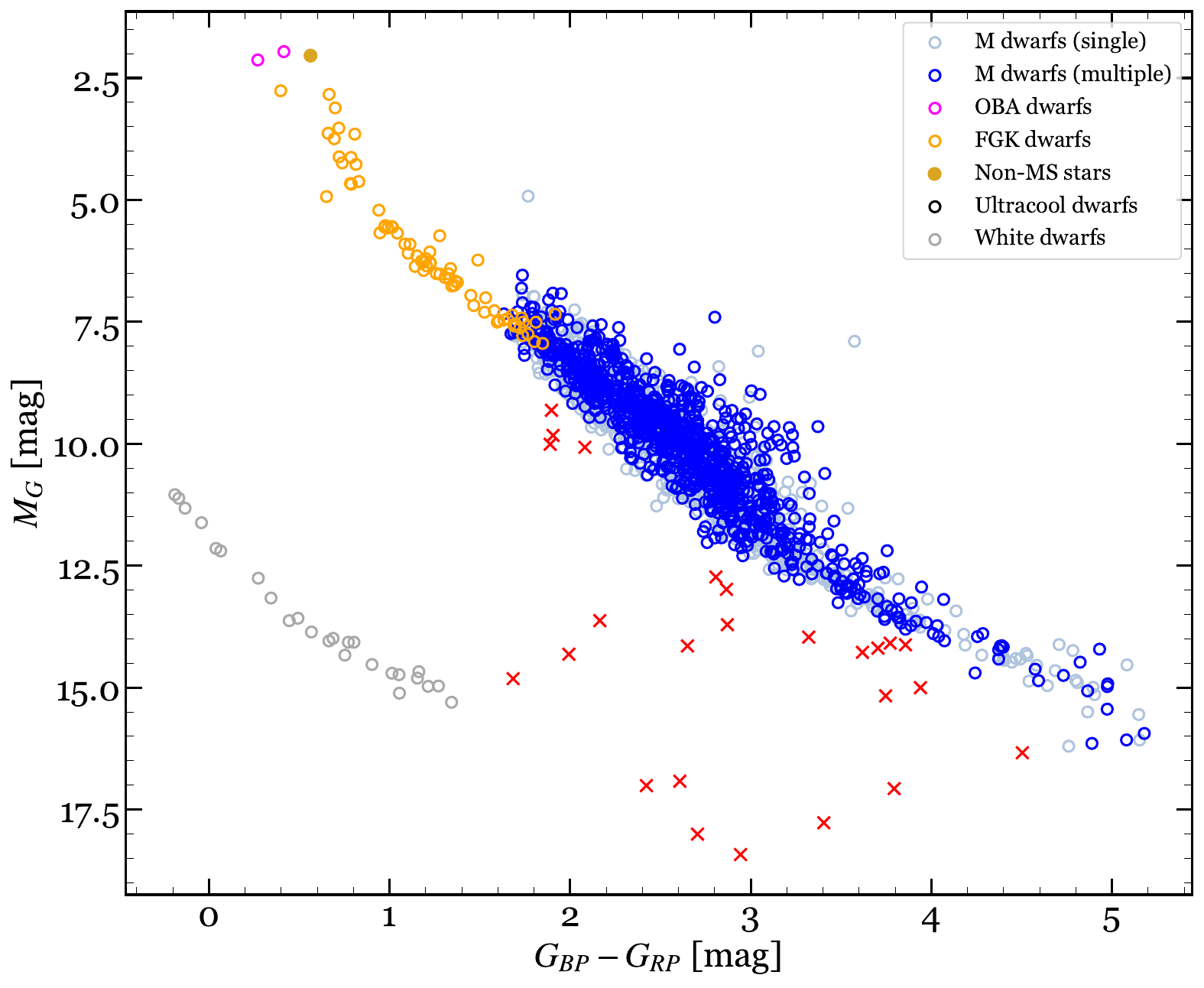}
    \caption{Absolute magnitude $M_G$ against $G_{BP}-G_{RP}$ colour for all the stars in our sample and their resolved companions with full photometry available in {\em Gaia} DR3. 
    Red crosses correspond to components of multiple systems that are very close, very faint, or both, resulting in compromised photometry.
    Very bright stars (e.g. {Capella} or {Castor}) are necessarily excluded for lacking \textit{Gaia} photometric measurements.
    Very compact systems, not resolved by {\em Gaia} (Sect.~\ref{ssection:unresolved}), and resolved young stars (Sect.~\ref{sssection:young}) lie above the main sequence.
    }
     \label{fig:MG}
\end{figure}

For each star in our sample and their resolved companions, first we compiled up to ten different magnitudes from the optical blue to the mid-infrared: three from {\em Gaia} ($G_{BP}$, $G$, $G_{RP}$), three from the Two Micron All Sky Survey 
(2MASS, \citealt{Skr06} -- $J$, $H$, $K_s$),
and four form the Wide-field Infrared Survey Explorer All-Sky Data Release
(AllWISE, \citealt{Cut14} -- $W1$, $W2$, $W3$, $W4$).
Our attempt to automatically include these catalogues by using the {\tt best\_neighbour} automatic cross-match from the {\em Gaia} Archive turned out to be unsuccessful.
Instead, we performed this search manually to ensure a correct discrimination of the components of the systems as \citet{Cif20}. 
This approach is of crucial importance in this work, because the description of each system fundamentally relies on whether 2MASS and {\em Gaia} are able to resolve the system or not.

At least one spectral classification is available in the literature for all but {15} of our {2214} M dwarfs.
For these {15} M dwarfs and for {227} companions of M dwarfs, we photometrically estimated the spectral type using Table~7 from \cite{Cif20} for late-K and M dwarfs (242 cases), or the public table derived from \cite{Pec13}\footnote{``A Modern Mean Dwarf Stellar Color and Effective Temperature Sequence'', \url{http://www.pas.rochester.edu/~emamajek/EEM_dwarf_UBVIJHK_colors_Teff.txt}} for stars other than M dwarfs (three ultra-cool dwarfs and five solar-type stars; see Sects.~\ref{sssection:fgk} and \ref{sssection:ucds}).
{None of these 242 stars have available spectra in the LAMOST DR9 database \citep{Zha12}.}

One object is classified as a white dwarf, as discussed in Sect.~\ref{sssection:wd}.
If both $M_G$ and $M_J$ were missing, we took advantage of the magnitude difference reported by the WDS, given that the spectral type of the primary is known and assuming that the two stars are at equal distance.
There are {six} stars with spectral types ranging from G5\,V to K7\,V as displayed by SIMBAD, but without an assigned bibliographic reference.
Finally, we reclassified all `K6\,V' and `K8\,V' stars as K7\,V, as this is widely accepted in the literature \citep{Mor73,Kir91,Alo15a,Mai24}.

The faintest components of both very close and wide pairs have had their photometry compromised, and this has negatively impacted their photometrically derived parameters.
For the closest ones, the photometry is affected by the brighter nearby source; 
for the wide ones, it is not feasible to obtain a good measure of the flux from {\em Gaia}'s blue filter, $G_{BP}$.

Using the compiled photometry and distances only, we constructed the empirical SEDs and fitted them to  synthetic models.
For the fitting, we used the Virtual Observatory Spectral energy distribution Analyzer \citep[{VOSA};][]{Bay08} and the grid of BT-Settl CIFIST theoretical spectra \citep{Bar98,All12}, as \cite{Cif20}.
Because these models reproduce the stellar photospheres, we did not include magnitudes from passbands with $\lambda_{\rm eff} \lesssim$ 420\,nm (i.e. $u$ and bluer) because they are mostly of chromospheric origin.
VOSA calculates the flux and provides $T_{\rm eff}$ and $\mathcal{L}$ for a given metallicity, which we set to solar ([Fe/H] = 0.0\,dex).
We performed this process exclusively for the objects whose photometric measurements are not compromised by the presence of a companion that is very close, very bright, or both.
As \cite{Cif20}, we imposed that the flux of the secondary does not exceed 1\,\% that of the primary, this is, $\Delta G = |G_A - G_B| >$ 5\,mag. Therefore, we excluded known spectroscopic binaries and resolved, but very close binaries.
In particular, we did not determine $\mathcal{L}$ and $T_{\rm eff}$ for binaries not resolved by both {\em Gaia} and 2MASS.

From $\mathcal{L}$ we derived 
$\mathcal{R}$ using the Stefan-Boltzmann law, $\mathcal{L} = 4\pi \mathcal{R}^2 \sigma T_{\rm eff}^4$, where $\sigma$ is the Stefan-Boltzmann constant.
M-dwarf $\mathcal{M}$ are empirically related to $\mathcal{R}$ via Eq.~6 of \cite{Schw19}.
This relation was derived from the study of detached, double-lined, double-eclipsing, main sequence M-dwarf binaries from the literature, which is valid across a wide range of metallicities for stars older than a few hundred million years. 
For the companions to stars in our sample that are outside the M-dwarf range, we used the mean values of $\mathcal{R}$ and $\mathcal{M}$ provided by \cite{Pec12} and \cite{Pec13}.

\begin{figure*}
    \centering
    \includegraphics[width=.49\linewidth]{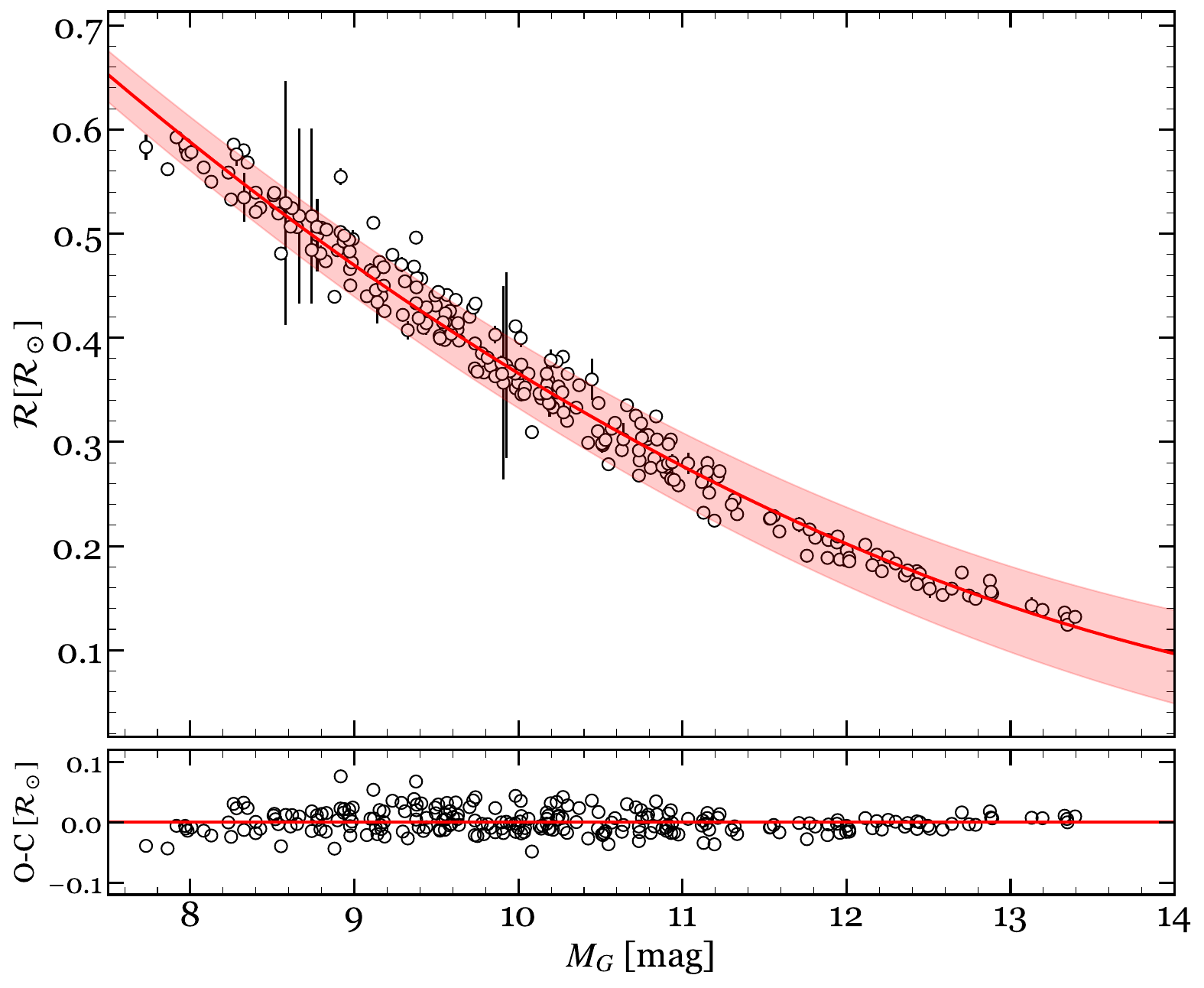}
    \includegraphics[width=.49\linewidth]{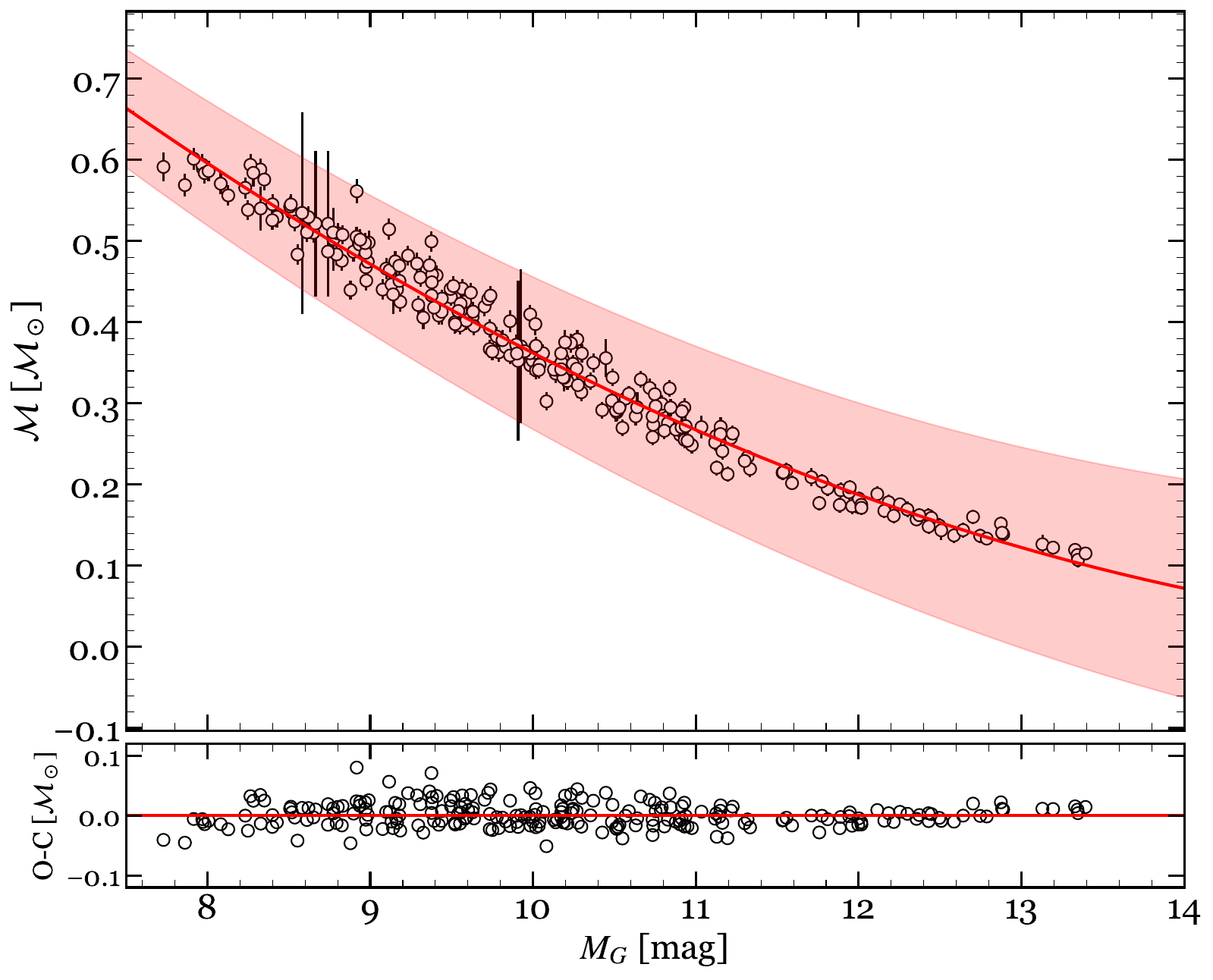}
    \caption{Stellar radius ($\mathcal{R}$, {\em left}) and mass ($\mathcal{M}$, {\em right}) as a function of the absolute magnitude ($M_G$) valid in the M-dwarf domain. 
    The red line represents the polynomial fit described in Table~\ref{tab:coefficients}, and the red shaded area indicates the 1-$\sigma$ level of uncertainty.
    }
     \label{fig:M_MG}
\end{figure*}

\begin{table}
\caption{Coefficients for the polynomial fits of $G$-band absolute magnitudes to masses and radii.}
\label{tab:coefficients}
\centering
\begin{tabular}{lcc}
\hline\hline \noalign{\smallskip}       
Parameter & $\mathcal{M}$ & $\mathcal{R}$ \\
\noalign{\smallskip}\hline\noalign{\smallskip}  
$a$ & 2.124  $\pm$ 0.040 &  2.0585 $\pm$ 0.0026 \\
\noalign{\smallskip}    
$b$ & --0.2503 $\pm$ 0.0075 &  --0.24208 $\pm$ 0.00012 \\
\noalign{\smallskip}
$c$ & 0.00741 $\pm$ 0.00035 &  0.00728 $\pm$ 0.00012 \\
\noalign{\smallskip}
\hline
\end{tabular}
\tablefoot{
The polynomial fit takes the form $Y = a + bX + cX^2$, where $Y=M_G$ (in mag) and $X$ can be either $\mathcal{M}$ (in $\mathcal{M_\odot}$) or $\mathcal{R}$ (in $\mathcal{R_\odot}$).
}
\end{table}

We did not tabulate the bolometric luminosity for {603} stars in our sample for two main reasons: 
Lack of trigonometric distances, or the presence of companion(s) at short angular separations. 
{\em Gaia} data offer the capability to separate sub-arcsecond binaries that have not been resolved by other all-sky surveys (2MASS, WISE).
Therefore, the individual components of stars without computed $\mathcal{L}$ still have  an  $M_G$ value, which we use as a proxy for luminosity \citep[see][]{Cif20}. 
For those stars without luminosities and with components resolved by {\em Gaia} (`AB' or `A+B' in our nomenclature; see Sect.~\ref{ssection:description}), we estimated $\mathcal{M}$ and $\mathcal{R}$ using their $M_G$ absolute magnitudes instead. 

To do so, we fit ex professo $\mathcal{M}$-$M_G$ and $\mathcal{R}$-$M_G$ relations (Fig.~\ref{fig:M_MG}) using the radii and masses derived from a subsample consisting of {240} M dwarfs with several restrictions:
({\em i}) they must be single (i.e. avoiding those in multiple systems, even with companions of large separation); 
({\em ii}) they are not new binary candidates (Sect.~\ref{ssection:candidates}); 
and 
({\em iii}) they are not classified as young stellar objects or members of young kinematic groups 
(Sect.~\ref{sssection:young}), for which the masses have been underestimated.
Both fits are second-grade polynomials (degree determined by the Bayesian information criterion) with the coefficients given in Table~\ref{tab:coefficients} and Pearson's $r$ equals 0.986 for both.
They hold within the range $M_G \in$ [7.5, 14.0]\,mag or K7--M0\,V to M7\,V \citep{Cif20}.
The $\mathcal{M}$-$\mathcal{R}$ relation from \cite{Schw19} links both fits, therefore the uncertainties of $\mathcal{M}$-$M_G$ are larger due to the error propagation. 
They can still be used up to $M_G$ = 16\,mag if necessary, but with extreme caution, being aware that the photometrically derived masses of ultra-cool dwarfs strongly depend on age \citep[][and Fig.~13 of \citealt{Sah20}]{Sod10}.
For the stars outside our range of validity, we used the $M_G$--$\mathcal{M}$ and $M_G$--$\mathcal{R}$ relations by \cite{Pec13}, where we assumed an uncertainty of at least 15\%. Here,
{$M_{K_s}$ is better correlated to $\mathcal{M}$ and less dependent on metallicity and age than $M_G$ \citep{Man15}.
However, we used $M_G$ to maximise the number of stars with homogeneous $\mathcal{M}$ and $\mathcal{R}$ determination, as there is a large number of close pairs resolved by \textit{Gaia} (AB), but unresolved by 2MASS (A+B; Sect.~\ref{ssection:description}).}

For white dwarfs, we retrieved the masses from the literature when possible (see Sect.~\ref{sssection:wd}) or assigned a mean mass of 0.6\,$\mathcal{M}_\odot$ otherwise \citep{Kep07,Bed20}. 
For objects cooler than L2, we did not estimate their masses or radii.

\subsection{Description of the systems}
\label{ssection:description}

\begin{figure*}[]
    \centering
    \includegraphics[width=.24 \linewidth]{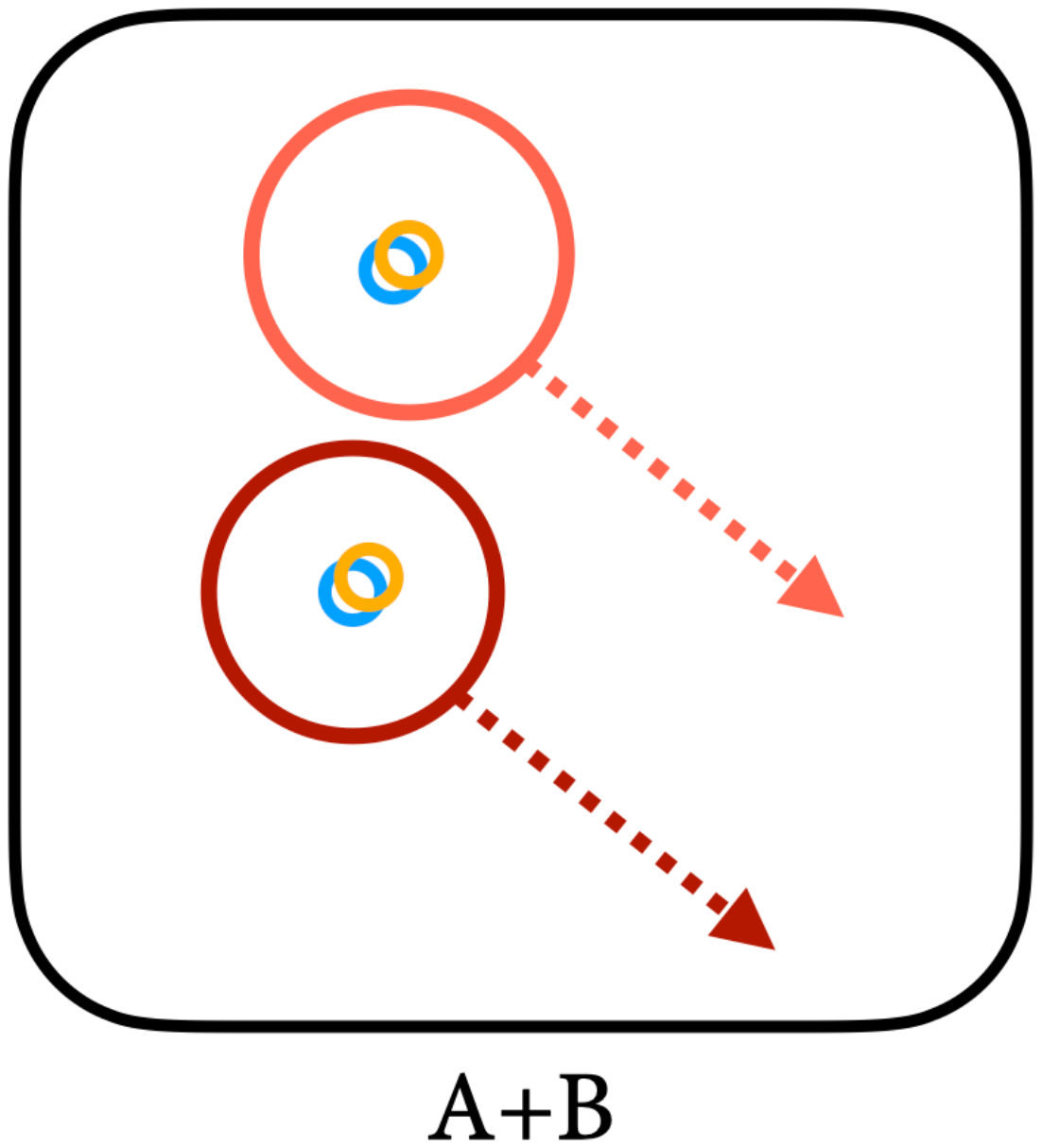}
    \includegraphics[width=.24 \linewidth]{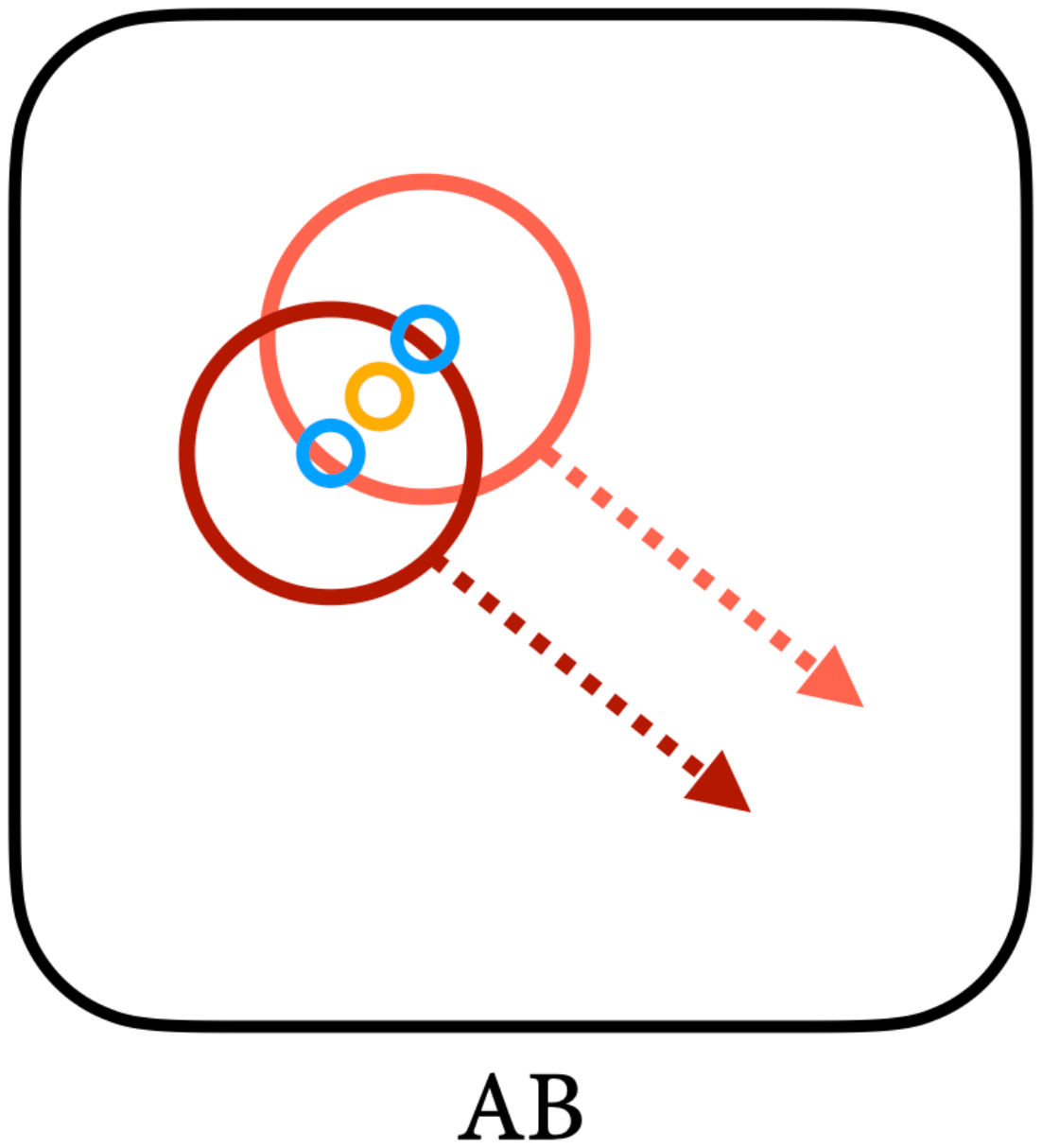}
    \includegraphics[width=.24 \linewidth]{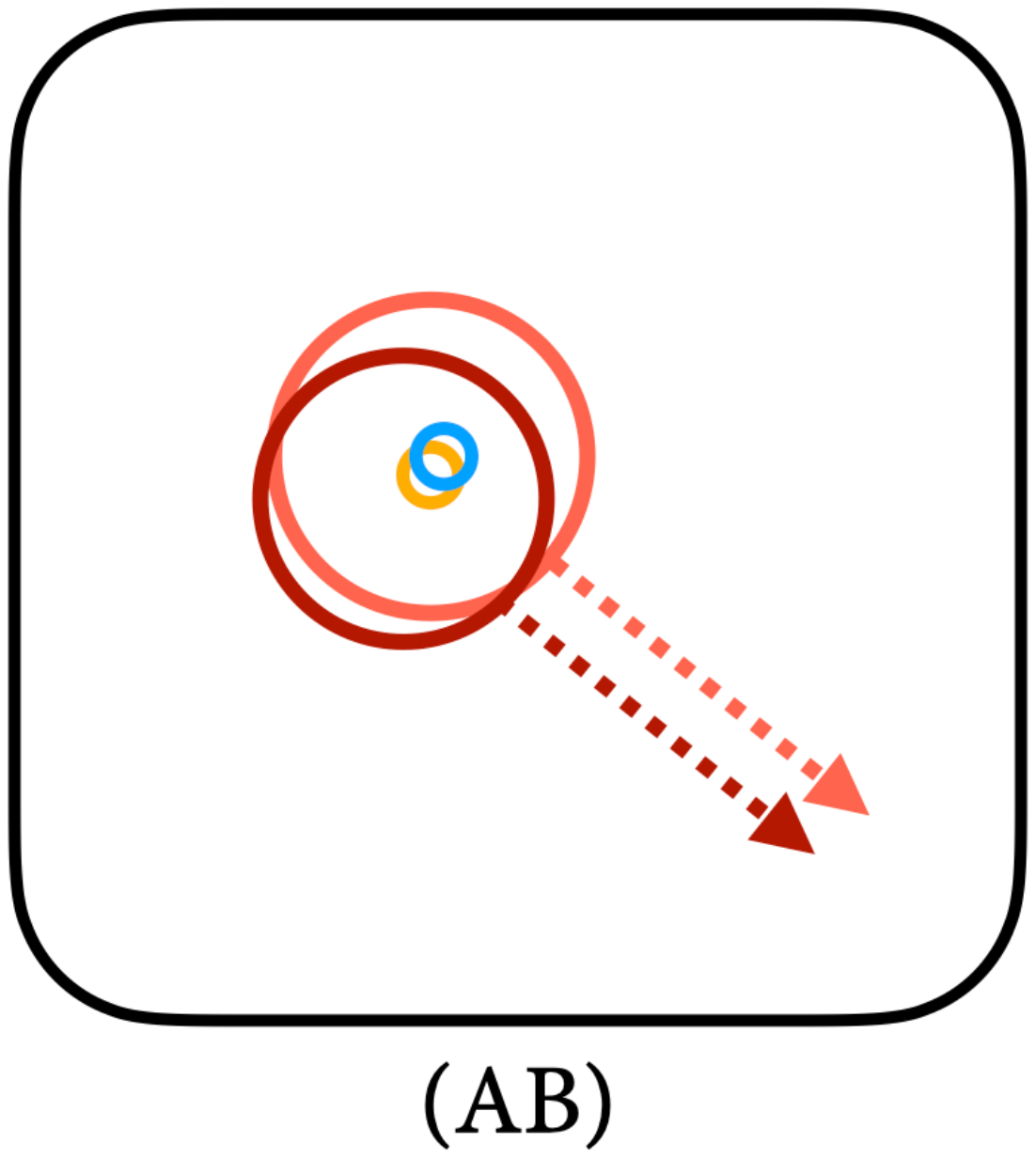}
    \includegraphics[width=.24 \linewidth]{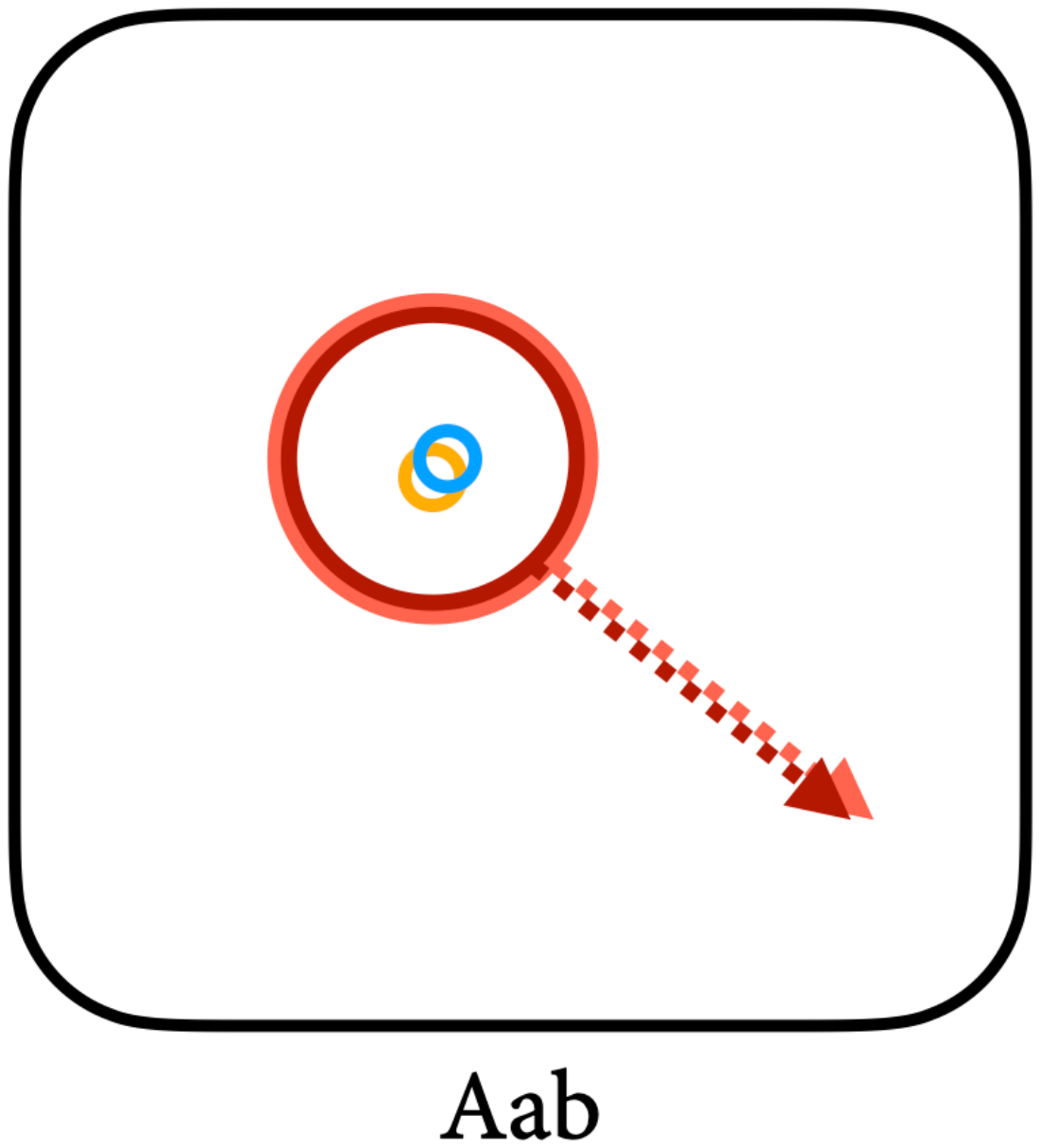}
    \caption{Nomenclature of multiple systems based on the resolution of {\em Gaia} DR3 (blue circles) and 2MASS (orange circles). 
    {\em From left to right}: `A+B' (resolved in both surveys), `AB' (only resolved in {\em Gaia}), `(AB)' (not resolved either in {\em Gaia} or 2MASS, but resolved by adaptive optics, lucky imaging, or space imaging), and `Aab' (spectroscopic binaries). 
    Multiple systems can often be a combination of these cases.
    }
     \label{fig:nomenclature}
\end{figure*}

\begin{figure}[]
    \centering
    \includegraphics[width=0.99 \linewidth]{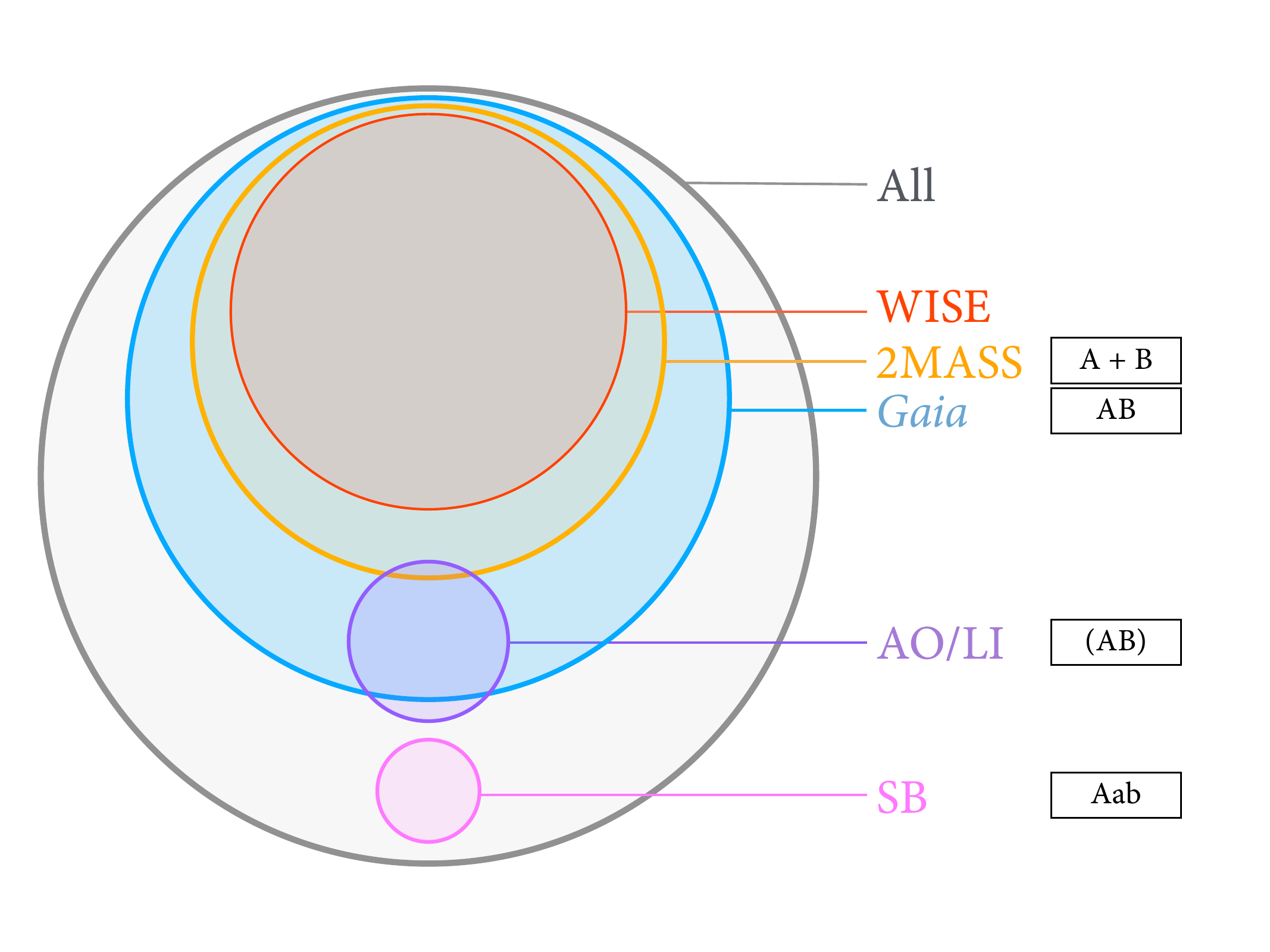}
    \includegraphics[width=0.99\linewidth]{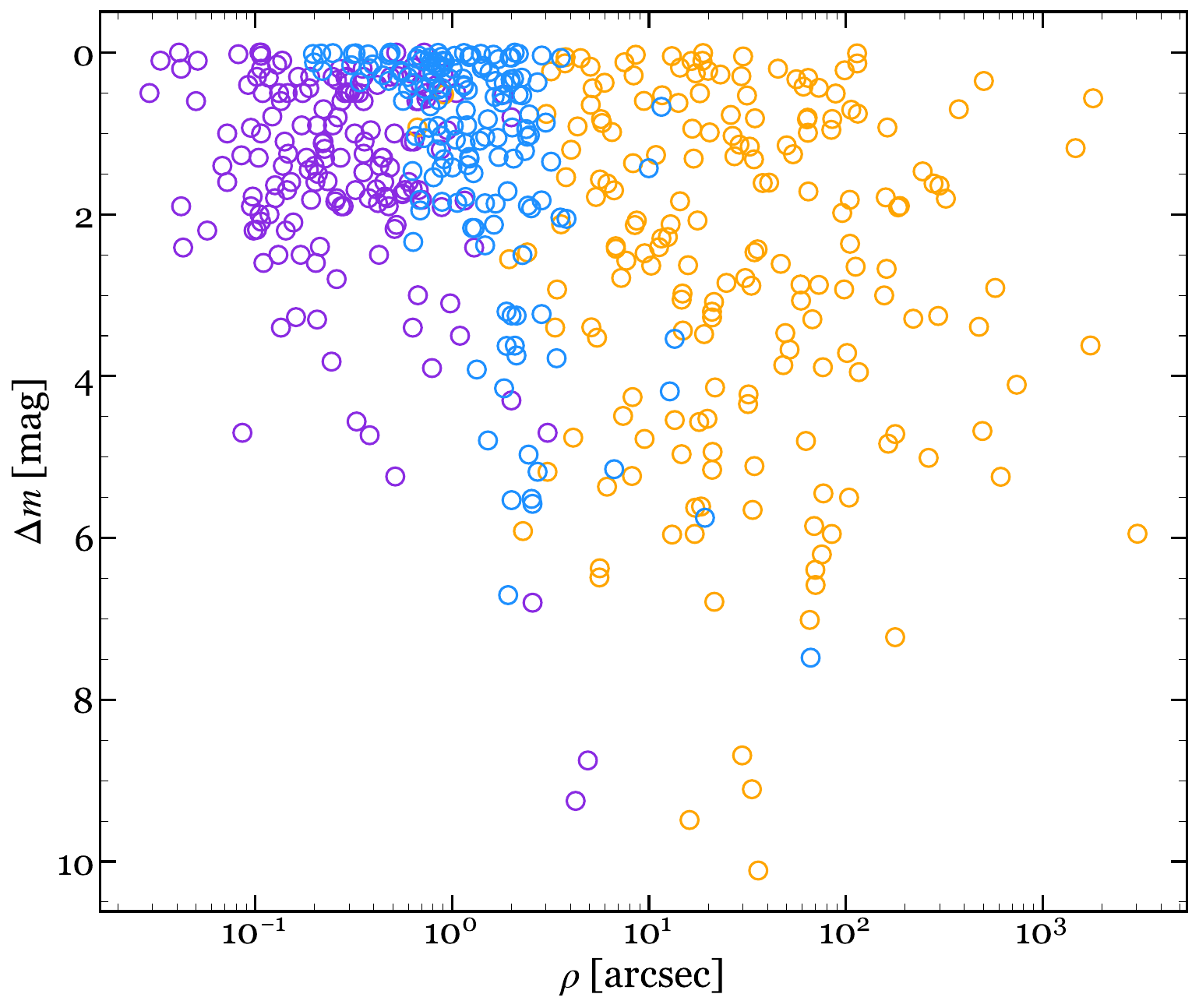}
    \caption{Schematic of the capabilities to identify binaries of the surveys and techniques exploited in this work ({\em top}) and difference in $G$ magnitude as a function of angular separation for pairs that {\em Gaia} resolves (``AB'' or ``A+B'') and those for which WDS tabulated magnitude difference ({\em bottom}).
    Both figures follow the same scheme of colours.
    The nomenclature assigned to each scenario (see again Fig.~\ref{fig:nomenclature}) is also displayed.
    }
     \label{fig:resolution}
\end{figure}

Table~\ref{tab:mother} displays all the M dwarfs in Carmencita plus their companions (in the case of multiple systems) resolved by {\em Gaia}.
It contains a total of {2634} rows ({2214} M dwarfs in the study sample plus {420} resolved companions) and 131 columns.
Its structure is meant to be both human- and machine-readable.
For the former, the systems are displayed with one component of the systems resolved by {\em Gaia} per row. 
A few notable cases lack {\em Gaia} identification, such as the very bright \object{Capella} or the very faint \object{Wolf~1130~B}, but they sill have their rows assigned in the table.
The stars are sorted by right ascension, but ensuring that those that belong to the same system are consecutive, in order of decreasing brightness.

The nomenclature of the system follows the scheme shown in Fig.~\ref{fig:nomenclature}.
The primary components (`A', or their variations) are the most massive ones of the systems, which are typically the brightest ones.
However, white dwarfs are an exception.
Although they are dimmer than M-dwarf companions in most cases (see again Fig.~\ref{fig:MG}), white dwarfs are known to be the remnants of late B to early G main sequence stars, {which were more massive than M dwarfs.
However, for historical reasons, we considered white dwarfs as companions} in all the instances found in this work (see more details in Sect.~\ref{sssection:wd}).
A comparison of the different surveys (WISE, {\em Gaia}, and 2MASS) and techniques (AO, LI, SB), with a focus on their ability to identify resolved binaries, is shown in the upper panel of Fig.~\ref{fig:resolution}; the lower panel shows the difference in magnitude (in general, {\em Gaia}'s $G$) as a function of the angular separation.
The resolving capabilities of \textit{Gaia} DR3 do not only depend on the separation, but also on the flux ratio, with considerable difficulty involved in telling individual sources apart with $\rho \lesssim 0.6$\,arcsec and $\Delta G > 0.1$\,mag.
In addition, it is known that the rectangular pixels of {\em Gaia} induce a dependence on the position angle between the two stars, influenced by the scanning direction \citep{Bru15}; however, this fact is alleviated by the large number of transits at different scanning directions.

{Figure~\ref{fig:resolution} also helps to shed light on the boundary between `close' and `wide' pairs.
These terms are generally defined in a static way, with separations that depend on the context of the study.
However, the distinction between close and wide should be dynamic rather than static, based on the detection limits and spatial resolution of the used facilities.
A practical definition of wide pairs could be those that can be resolved using natural seeing conditions (1--2\,arcsec) without the need for advanced techniques such as AO and LI. 
The 2MASS survey, with a spatial resolution of 2--4\,arcsec, can serve as a useful gauge to define these `wide' pairs. 
In this work, we aim to adhere to this definition. 
However, the terms `wide' and `close' may occasionally be used to define specific separations, particularly when referencing the literal works of others.}

An adapted version of the complete table can be found in the Appendix as Table~\ref{tab:mother}, which provides: numerical ID for each star and system, Karmn, common name, and GJ identifiers, coordinates, spectral type, multiplicity type (single, part of multiple systems, or new unresolved binary candidates), and availability in other tables of this work. We give a column-by-column description in Table~\ref{tab:description}. 
A comma-separated values (csv) file of the complete table is available on a dedicated GitHub repository\footnote{\url{https://github.com/ccifuentesr/CARMENES-IX}}.

\subsubsection{Physical separations}
\label{sssection:separations}

\begin{figure}
    \centering
    \includegraphics[width=0.99\linewidth]{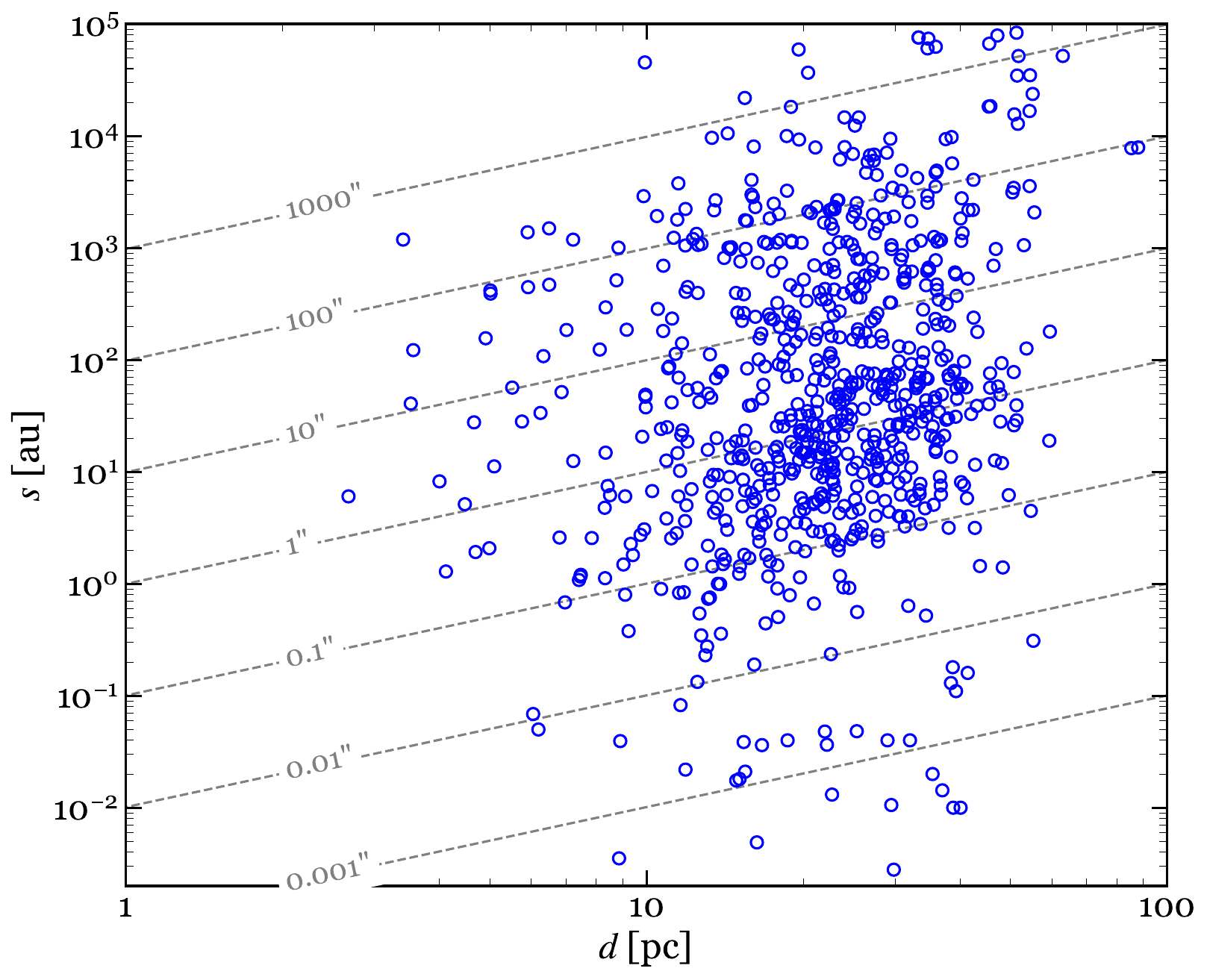}
    \caption{
     Physical {separation as} a function of the distance for all the pairs in our sample.
     The grey dashed lines represent projected angular separations, $\rho = s / d$, in dex.}
     \label{fig:s_d}
\end{figure}

{Projected angular separation, $\rho$, and projected linear separation, $s$, relate as $s = \rho \cdot d$ under the small-angle approximation ($\sin{\rho} \approx \rho$, with an error below 1\,\% for $\rho <$ 0.24\,radians or 14\,deg).}
The coverage in projected linear separation, $s = \rho \cdot d$ (or $a$), of this work extends from 2.8$\cdot$10$^{-3}$\,au to 2.7$\cdot$10$^{5}$\,au, spanning eight orders of magnitude; namely, from close spectroscopic binaries to ultra-wide systems (Fig.~\ref{fig:s_d}).
The smallest orbital separation in our sample corresponds to a double-lined spectroscopic binary (SB2), with a minimum value of $a \sin{i} =$ 0.002797\,au \citep[\object{1RXS~J070005.1--190115},][]{Bar21}.
On the other hand, the largest orbital separation (273\,660\,au) corresponds to the widest pair, which is part of a quadruple system  \citep[\object{GJ~282}~A+B+Cab,][]{Pov09,Bar21} in the Ursa Major stellar kinematic group \citep{Tab17}; therefore, its actual binding may be called into question.
The widest pair not associated to any stellar kinematic group is in the quintuple system \object{$\sigma$~CrB}, which consists of three Sun-like main-sequence stars and two low-mass stars: One of them, the M2.5\,V \object{$\sigma$~CrB~C} star is at 14\,346\,au to the bright central star.

In terms of angular separation in resolved pairs, the range is 0.020--12\,954\,arcsec (or 3.6\,deg).
If we extrapolated to non-resolved systems such as SBs, the minimum separation would be as short as 0.0003\,arcsec.
\object{G~146-35}~B is the known equal-mass pair \citep{Lam20} with the smallest angular separation that {\em Gaia} DR3 resolves, with just $\rho \approx$ 0.196\,arcsec.
Among those reported for the first time, \object{LP~780-23}~B is the star ($\sim$M3.0\,V) with the smallest separation from an M dwarf primary, with $\rho =$ 0.199\,arcsec.
In cases like these, the advantage of being able to resolve the pairs with {\em Gaia} comes at the cost of being significantly limited by problems affecting the  astrometric measurements.
In a few cases we identified members of stellar kinematic groups or associations that were either previously known (therefore confirming their membership) or unknown (therefore assigning them for the first time).
These young stars are \object{V1221~Tau} with $s\simeq$ 52\,000\,au \citep[in the $\beta$ Pictoris moving group,][]{Gag18b},
and \object{PM~J18542+1058} with $s\simeq$ 10\,000\,au \citep[primary in the Carina moving group,][]{Gag18b}, proposed for the first time in this work.

In the study of binaries, we faced the problem of translating projected angular separation, $\rho$, to the actual semimajor axis, $a$ \citep[see e.g.][]{Kui35,Cou60,van68b}.
For circular orbits, $s$ is always equal or smaller than $a$; 
for eccentric orbits, it is the opposite case.
The approximation to this problem varies between authors.
Many incorporate a fixed correction factor, such as 
${\pi}/{2} \simeq 1.57$ \citep{Abt76} or
1.26 \citep[][and later \citealt{Dhi10}]{Fis92}.
In contradiction with previous works \citep[e.g.][]{Hal83}, \cite{Tor99} suggested that this factor varies with eccentricity. \cite{Dup11} incorporated this notion (as well as a discovery bias) to yield a factor of $1.16^{+0.81}_{-0.31}$.
Other investigations have supported the assumption that $s$ serves as a reasonable estimate for $a$ when there is not enough information about the orbital elements of the system. 
For instance, \cite{Kui42} derived empirically, from 62 known orbits, a relation between the semimajor axis and the projected separation on the sky that $E(\log{a})-E(\log{s}) = -0.11$, where $E$ indicates the expectation value.
A similar result was later obtained by \cite{Cou60} using 410 orbits.
When the eccentricity is taken into account, \cite{van68b} calculated the expectation values between 0.0 (for $e=1$) and --0.133 (for $e=0$).
More recently, \cite{Jia10} stated that for a population of binaries at a given $a$, the median projected separation is 0.978$a$.
Given the broad scope intended for this work (we refer to the title) and the need for a proper comparison with previous work, we did not apply a conversion factor between $a$ and $s$, while also assuming random orbital plane orientations.

\begin{figure}[]
    \centering
    \includegraphics[width=0.99\linewidth]{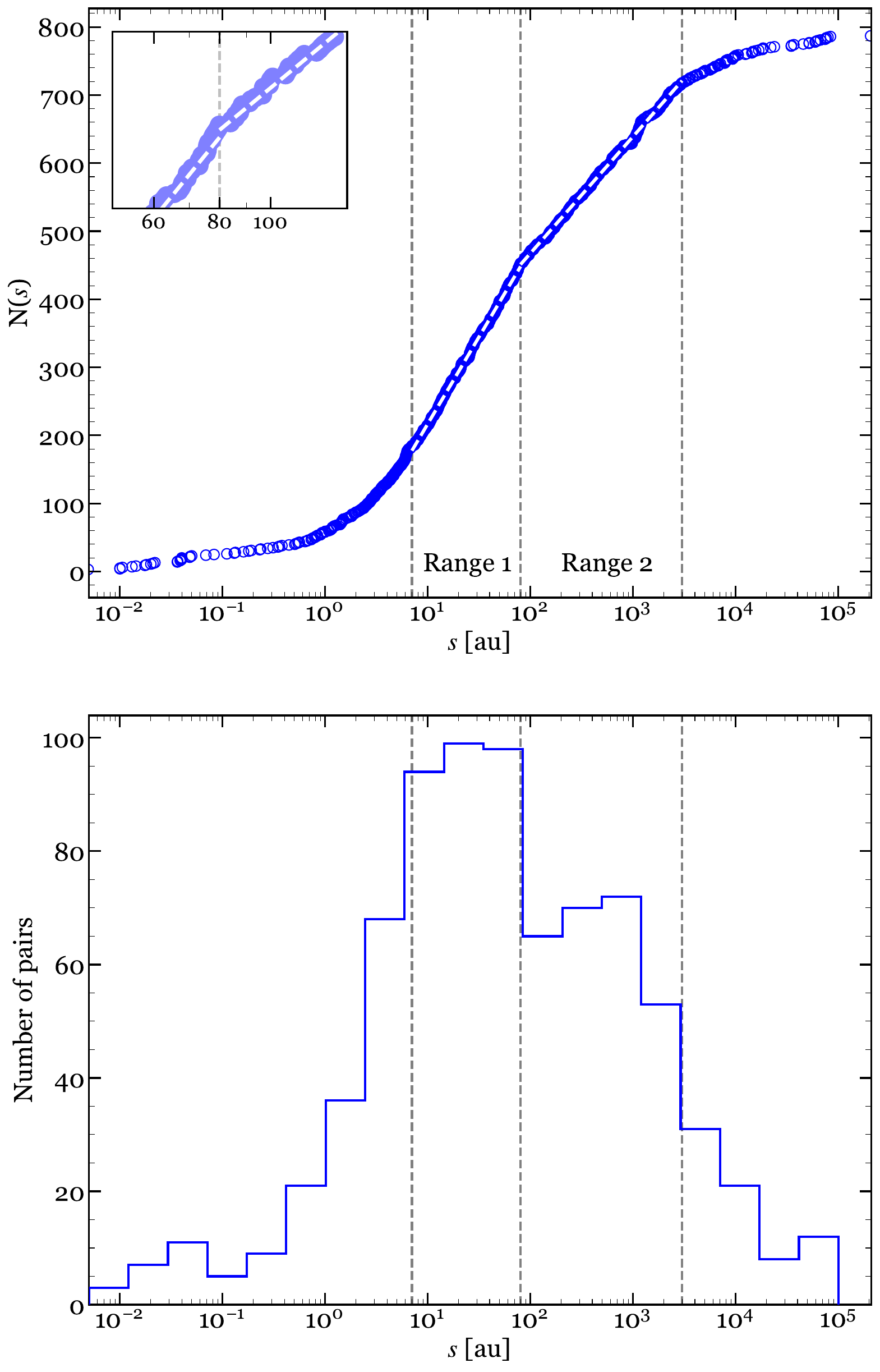}
    \caption{Cumulative ({\em top}) and non-cumulative ({\em bottom}) distribution of physical separations for all the pairs in our sample.
    The white dashed lines in the upper panel are power-law fits for the ranges 7--80\,au and 80--3000\,au, respectively.
    The grey vertical dashed lines delimit these ranges.}
     \label{fig:hist_combined}
\end{figure}

\subsubsection{\"Opik's law}
\label{sssection:Oepik}

It has been observed in a variety of ranges of separation that the number of pairs per unit of logarithmic separation remains roughly constant \citep{Opi24,Ver88,Clo90,All00,Pov04,Pov07,Lep07a}.
This empirical relation, often referred to as \"Opik's law, captures the intuitive idea that the probability of finding a bound companion decreases as we go to larger separations.
It admits different formulations, either through the frequency distribution $f(a) \propto a^{-1}$ or $f(s) \propto s^{-1}$, or via the cumulative distribution, which increases linearly on a logarithmic scale $N(\log{a}) \propto \log{a}$ or $N(\log{s}) \propto \log{s}$, where $a$ and $s$ can be used indistinctly \citep[e.g.][]{All00}.

We show in Fig.~\ref{fig:hist_combined} the cumulative (top panel) and non-cumulative (bottom panel) distributions of projected physical separations between pairs in our sample.
These data include all the possible separations between the elements in each system, namely, $n-1$ measurements for a system of $n$ components.
We fit two well-sampled ranges in the cumulative distribution, delimited by vertical dashed lines, 
in which the number of multiple stars is uniformly distributed, separated by a change in slope at $s \sim$ 80\,au.
We did not make any assumption on the optimal fitting intervals beforehand.
Rather, we chose the optimal limits based on the maximisation of the $\chi^2$ value.
Following \"Opik's law, which states that there are roughly equal numbers of binaries ($n_0$) in each logarithmic interval of $s$, we expressed the parametrisation as:

\begin{equation}
\label{eqn:oepik}
 N({s}) = C + n_0(\log{s} - \log{s_0}),
\end{equation}

\noindent where $s_0$ corresponds to the smallest $s$ value in each range, and $C$ and $n_0$ are free parameters. 
Obtaining very discrepant values for these parameters between both ranges points to a break in \"Opik's law.
This behaviour had been already described \citep[e.g.][]{Lep07a}, with variations observed across studies due to differences in population characteristics, sample size, and data completeness.
For instance, \cite{All00} found that `wide binaries' ($a>$ 25\,au, as defined by the authors) follow \"Opik's law all the way up to 10\,000\,au (and up to 20\,000\,au for those with the most halo-like orbits). Meanwhile, \cite{Pov07} found binaries in the field following \"Opik's law in the separation range 100\,au $\leq a \leq$ 3000\,au, which corresponds to the second range in our plot. 

The cumulative distribution shown in Fig.~\ref{fig:hist_combined} departs from \"Opik's at very close ($s \lesssim$ 1\,au) and very wide ($s \gtrsim$ 10$^4$\,au) orbital separations, a fact repeatedly observed in empirical studies \citep[e.g.][but see \citealt{Pov07} for the exception of very young  populations of binaries]{Pov04}.
The upper limit of \"Opik's law decreases with increasing system age as older systems tend to have fewer wide binaries, simply because they have had more time to be disrupted \citep[see e.g. ][]{Wei87}. More recently,
\cite{Kin12b}, who studied the separation distributions in young star-forming regions, found an excess of binaries with $s<100$\,au when compared to the field population.
As \cite{Lep07a} pointed out that even if a population initially follows the canonical form of \"Opik's law, it will eventually exhibit a truncated distribution with a steeper power law, $f(s) \propto s^{-\lambda}$, where $\lambda > 1$.

It has been argued \citep[e.g.][]{Alle97} that it might be preferable to describe this distribution function in accordance with the two main mechanisms of binary formation: disk fragmentation for the close systems, and first-collapse (or turbulent) fragmentation or the dynamical unfolding of compact triples for wider systems.
Some authors \citep[e.g.][]{EB18,Tia20} have adopted a `broken' power law instead.
The existence of two formation mechanisms dominating in different separation regimes was also suggested by \cite{Tok20b} and \cite{Hwa22} when studying the distribution of eccentricities. 
This observational result alleviates the problem that a single theory (namely, disk fragmentation) faces when explaining the existence of binaries with very wide separations.
This behaviour has been observed both in pre-main and in main sequence stars \citep{Pov88,Pov82,Pov94,Pov04}.
All in all, it is accepted that multiple systems are formed through different processes and that the statistical properties of these systems are influenced by the environment in which they form.

The broad peak around 20\,au (Fig.~\ref{fig:hist_combined}, bottom panel) was identified by \cite{Win19a}, who noted that many stars could potentially harbour close companions, eventually altering the multiplicity rate and shifting the most common separations to smaller orbital sizes, once their orbital semimajor axes are determined.
For 89 unresolved systems they adopted fixed projected separations of 1\,arsec, which substantially underestimate the number of very close companions (at sub-Solar System scales, or less than 1\,au).
The authors noted that their distribution figures differed from that of \cite{Jan14b}, but highlighted the methodological differences between the two; this  specifically includes the primary mass (\citealt{Jan14b} focussed {on M5--9\,V}).
Our study cannot be compared directly with those above, since our distribution cannot be described with a single Gaussian given the broad range of separations that we cover.

Regarding the observed flattening of the distribution at very short and very wide separations, we assert that the former is caused by a fundamental limitation of the main high-resolution techniques (AO, LI, SI), while
the latter is a consequence of the difficulty of the formation and eventual survival of the most separated binary systems, rather than observational constraints \citep[see e.g.][]{Pov04,Lep07a,Cab09,GP23}.
Indeed, {\em Gaia} DR3 is complete for separations above 1.5--2.0\,arcsec \citep{Fab21} and so, we did not expect to miss the most separated binaries in our search.
The evidence shows that in binaries of $a \lesssim$ 50\,au, disruption by dyamical influences is negligible. They evolve essentially unaffected, either in the field or inside clusters \citep[see e.g.][]{Moe10}.
\cite{Par09} referred to these close-in pairs as `hard binaries', while `intermediate binaries' was used to refer to those with 50\,au $< a <$ 1000\,au; the latter are  severely affected by dynamical processes, especially when formed in dense star clusters. 
Unevolved `wide binaries' with $s>$ 1000\,au can only originate in low-density star-forming regions with less than a few stars per cubic parsec \citep[`isolated star formation mode',][]{Goo10} or, in the words of \cite{Dea20}, `wide binaries are rare in open clusters'.
Early investigations, such as those by \cite{van68a}, suggested that binaries observed at separations greater than about 1000\,au must be either `escaping' double stars (or run-away stars; see also \citealt{Pov67}) or one of the components is actually a  double star (yielding to a triple system) that supplies more mass (and therefore more gravitational support) to the system \citep{Reip12,Cif21}.

The non-cumulative distribution (Fig.~\ref{fig:hist_combined}, bottom panel) exhibits a rather bimodal shape, peaking at $\sim$\,10--100\,au and around $\sim\,10^{3}$\,au, respectively, which roughly correspond to the changes in slope in the cumulative distribution, as marked by the range boundaries.
It also shows that a large percentage of M dwarf companions are found to be as close as $s \lesssim$ 50\,au.
Studies such as \cite{Hwa23} found that the fraction of wide (10$^{3}$--10$^{4}$\,au) triple systems increases with decreasing orbital periods of the inner binaries, and the authors also found a tentative excess at $\sim$10$^{4}$\,au for tertiaries of eclipsing binaries.
The bimodality in this distribution of separations has been reported before in the literature for wide ($s > 10^3$\,au) binaries \citep[e.g.][]{Dhi10,Oel17,Jim19} and also for close (<10$^3$\,au) binaries \cite[][see their {Fig}.~8]{Kou10}.
\cite{Dhi10} suggested that this bimodality most likely reveals two distinct populations of wide binaries, possibly representing systems that form or dissipate through differing mechanisms: 
(1) systems of stars that formed with sufficient binding energy to survive for the age of the Galaxy, and 
(2) relatively young systems that formed within the past 1--2\,Ga but that are not likely to survive much longer.
The authors noted their high confidence that the bimodal structure is not due to a large contamination of very wide chance pairs, but likely that this bimodality reveals these two distinct populations of wide binaries.
\cite{Oel17} noted that the bimodality is split near separations of 10$^{4.7}$\,au, also suggesting two separate binary populations, which follow different formation and evolutionary paths.
Using {\em Gaia} DR2 data, \cite{Jim19} correlated the number of false positives with the separation of the system, estimating that it is $\sim$1--5\,\% of the candidate pairs with $s<$ 50\,000\,au, $\sim$10--18\,\% for $s$ between 50\,000 and 100\,000\,au, and increasing to $\sim$40--51\,\% for the largest separations, therefore resulting in most of their comoving candidate binaries with $s \lesssim$10$^4$\,au being real.

Before the {\em Gaia} mission, only relatively small samples of nearby and bright stars had useful parallaxes.
Foundational works such as \cite{Kui42} only had limited data available, but recognised the limitations due to the `incomplete discoveries of spectroscopic and close visual binaries' [sic].
More recently, \cite{Lep07a} suggested that, using CCD imaging and AO, it should be possible to perform a complete census of common proper motion companions of Hipparcos stars down to angular separations of a fraction of an arcsecond, underlining the usefulness in extending the census to pairs with smaller orbital separations ($s<$ 1000\,au) and verify whether this regime can be consistently modelled with \"Opik’s law.

Pre-{\em Gaia} investigations were fundamentally limited by the astrometric accuracy (in distance measurements and proper motion data), which made the characterisation of binary stars much more challenging.
For example, the SLoWPoKES catalogues I and II \citep{Dhi10,Dhi15} were based on Sloan Digital Sky Survey (SDSS) data limited by their use of distance–colour relations rather than direct distance measurements.
The SLoWPoKES extension catalogue, GAMBLES \citep{Oel17}, used the published distances and proper motions from the Tycho-Gaia Astrometric Solution (TGAS), which was based on the first data release of {\em Gaia} and on the Tycho-2 catalogue, although also relied on SDSS data.
TGAS uncertainties are typically around 0.3\,mas\,a$^{-1}$, with positional uncertainties around 1--2\,mas.
{\em Gaia} DR3 significantly improved these measurements, with typical uncertainties for parallaxes around 0.02--0.03\,mas, and 0.01--0.02\,mas for positional measurements. 
Recent catalogues have used {\em Gaia} DR1 \citep{Oh17}, DR2 \citep{Tia20,Har20}, and EDR3 \citep{EB21}, which compile hundreds of thousands of pairs with different levels of contamination by chance alignments, sometimes using Bayesian formulation in the process. 

Also using TGAS astrometry, \cite{And17} recognised that the excess of the distribution at $s \sim$ 10$^4$\,au corresponds to the transition from genuine binaries to random alignments, implying that the reported bimodality within separations of 1\,pc in samples of wide binaries is most likely due to contamination from random alignments.
\cite{EB21} showed that the contamination rate from chance alignments increases rapidly with separation.
For instance, they distinguished between `initial candidate pairs' and `high bound probability pairs' (see their {Fig}.~2), suggesting that most initial candidates are chance alignments, which dominate at $s \gtrsim$ 3000\,au. 
They concluded that the separation distribution of binaries with high probability of being bound falls off steeply at wide separations.
In particular, they suggested that at wide separations (meaning $10^{4.5}$\,au) most candidates are chance alignments, an observation already made by \cite{Tia20} relying on data from {\em Gaia} DR2 (see their {Fig}.~4).
Prior to that study, the results from detailed works on the probabilities of chance alignments by \cite{Lep07a} displayed an agreement with these results. 
By using astrometric data with unprecedented accuracy, the evidence presented here (see Sect.~\ref{sssection:binding} for the study of binding energies and see again Figure~\ref{fig:hist_combined}) also suggests that some of the ultra-wide binaries ($s \gtrsim$ 10$^{4}$\,au) claimed to be genuine are, in fact, random alignments.

\subsubsection{Orbital periods}
\label{sssection:periods}

Orbital periods ($P_{\rm orb}$) relate to the semimajor axis, $a$, and the total mass of the system, $\mathcal{M}_T$, via Kepler's third law:

\begin{equation}
\label{eqn:kepler}
        P_{\rm orb} = 2\pi \sqrt{\frac{a^3}{G\mathcal{M}_T}}.
\end{equation}

\noindent As stated above, the impact of the correction factor between $a$ and $s$ can be neglected.
As mentioned in Sect.~\ref{section:introduction}, close compact arrangements of stars can be adequately treated as a single star (with the aggregated masses) when considered from a distance much larger than the separation between its components.
In this sense, triple, quadruple, or even higher order systems can be treated as binaries in the majority of cases.
Therefore, we estimated orbital periods of the pairs in our sample only for the systems with available masses for all the components, either measured dynamically in the past or estimated by us (see Sect.~\ref{ssection:parameters}), and with the suitable hierarchical arrangement.

Almost all the estimated $P_{\rm orb}$ lie between 10\,a and 10$^5$\,a, with 40\,\% of the total corresponding to periods of less than 10$^3$\,a (given the dependence of $P_{\rm orb}$ on $s$, we omit the corresponding histogram, but the bottom panel of Fig.~\ref{fig:hist_combined} may serve as a representative plot).
Without the information about their orbital eccentricity and orientation, these values could be over- as well as under-estimated.
For the vast majority of pairs in our sample, the orbital periods are measured by millennia, so the prospect of following them during one single orbit is unrealistic in practice. 
As an example, we refer again to Proxima Centauri, which has an estimated orbital period $P_{\rm orb} \sim 5.5 \cdot 10^5$\,a \citep{Ker17}.
One consequence is that, in most cases, it will not be possible to discriminate among actual bound multiple systems and disintegrating clusters.

The largest orbital periods estimated correspond to \object{GJ~1284} and \object{GJ~282~C}, with extreme values of $\sim$10$^8$\,a.
On the other hand, several systems have estimated periods of less than half a century.
Table~\ref{tab:periods} lists these systems, along with their masses and separations, both at the epoch of {\em Gaia} and at the time of the first satisfactory observation as listed by the WDS.
In all cases these systems are binaries resolved by {\em Gaia} with angular separations $\rho \lesssim$ 2\,arcsec.
One exception could be the system \object{GJ~400}, which might be a triple system made of HO~532 (shown in Table~\ref{tab:periods}) and the innermost CHR~191 ($\rho \sim$ 0.37\,arcsec), but that has been poorly investigated \citep{McA87}.
If the system were triple, the orbital period would be of about 20\,a.

In addition to the periods estimated by us, we also collected those reported in the literature, mainly derived from the study of spectroscopic binaries.
Most of these are measured by days, or even by hours, such as that of \object{LP~86-173} (\object{LHS~1817}), an M4.5-dwarf in a 7\,h orbit with a possible white dwarf \citep{New16,Win20}.
The periods computed by us, as well as those collected from the literature, are included in the full version of Table~\ref{tab:mother} (see Sect.~\ref{ssection:description}).

\begin{table*}[]
\caption[]{Systems with estimated minimum orbital periods of less than 50 years.}
\label{tab:periods}
\centering
\begin{tabular}{lllcccc}
\hline\hline 
\noalign{\smallskip}
Name    &       Comp.   &       Spectral        &       $\rho$ (J2016.0)        &         $\rho$ (epoch)  &       $\mathcal{M}$                   &       $P_{\rm orb}^{\rm \,min}$                       \\
        &               &       Type    &       {[}arcsec{]}    &       {[}arcsec{]}    &         {[}$\mathcal{M}_\odot${]}                       &       {[}a{]}                 \\
   \noalign{\smallskip}\hline \noalign{\smallskip}
        BL Cet  &       A       &       M5.0 V  &               &               &       0.129    $\pm$   0.044   &                               \\
        UV Cet  &       B       &       M6.0 V  &       2.259   &       1.4 (1935)  &       0.118    $\pm$  0.045   &       29.9     $\pm$  3.8     \\
        \noalign{\smallskip}        \noalign{\smallskip}                                                                                                                                        
        HD 32450 A      &       A       &       {M0.0 V}        &               &               &       0.611    $\pm$   0.027   &                               \\
        HD 32450 B      &       B       &       M3.0 V  &       0.888   &       0.5 (1929)  &       0.372    $\pm$  0.032   &       20.60    $\pm$  0.44    \\
        \noalign{\smallskip}        \noalign{\smallskip}                                                                                                                                        
        V577 Mon        &       A       &       M4.5 V  &               &               &       0.230    $\pm$   0.037   &                               \\
        Ross 614 B      &       B       &       M4.5 V  &       0.313   &       1.3 (1939)  &       0.228    $\pm$  0.037   &       2.16     $\pm$  0.12    \\
        \noalign{\smallskip}        \noalign{\smallskip}                                                                                                                                        
        LP 780--32      &       A       &       M4.0 V  &               &               &       0.326    $\pm$   0.013   &                               \\
        Gaia DR3 2926756741750933120    &       B       &       M4.0 V  &       0.556   &       0.6 (2016)  &       0.266    $\pm$  0.036   &       32.30    $\pm$  1.0     \\
        \noalign{\smallskip}        \noalign{\smallskip}                                                                                                                                        
        GJ 400 A        &       A       &       M0.5 V  &               &               &       0.642    $\pm$   0.027   &                               \\
        Gaia DR3 776067093937332992     &       B       &       M2.5 V  &       0.684   &       1.2 (1896)  &       0.393    $\pm$  0.032   &       28.22    $\pm$  0.57    \\
        \noalign{\smallskip}        \noalign{\smallskip}                                                                                                                                        
        GJ 3673 &       A       &       M3.5 V  &               &               &       0.386    $\pm$   0.033   &                               \\
        G 122--34 B     &       B       &       M3.0 V  &       0.339   &       0.2 (2012)  &       0.346    $\pm$  0.034   &       42.2     $\pm$  1.4     \\
        \noalign{\smallskip}        \noalign{\smallskip}                                                                                                                                        
        Wolf 424 A      &       A       &       M5.0 V  &               &               &       0.140    $\pm$   0.043   &                               \\
        Wolf 424 B      &       B       &       M5.5 V  &       1.149   &       0.7 (1938)  &       0.143    $\pm$  0.042   &       21.9     $\pm$  2.3     \\
        \noalign{\smallskip}        \noalign{\smallskip}                                                                                                                                        
        GJ 3760 A       &       A       &       M2.5 V  &               &               &       0.464    $\pm$   0.030   &                               \\
        GJ 3760 B       &       B       &       M2.0 V  &       0.477   &       0.4 (1991)  &       0.463    $\pm$  0.030   &       49.9     $\pm$  1.2     \\
        \noalign{\smallskip}        \noalign{\smallskip}                                                                                                                                        
        GJ 3775 &       A       &       M3.5 V  &               &               &       0.218    $\pm$   0.038   &                               \\
        Gaia DR3 3688439268658769408    &       B       &       M4.5 V  &       0.701   &       0.6 (2012)  &       0.211    $\pm$  0.038   &       44.2     $\pm$  2.8     \\
        \noalign{\smallskip}        \noalign{\smallskip}                                                                                                                                        
        Ross 52 A       &       A       &       M3.5 V  &               &               &       0.327    $\pm$   0.034   &                               \\
        Ross 52 B       &       B       &       M4.5 V  &       0.880   &       0.9 (1940)  &       0.238    $\pm$  0.037   &       43.3     $\pm$  1.9     \\
        \noalign{\smallskip}        \noalign{\smallskip}                                                                                                                                        
        HD 239960 A     &       A       &       M3.0 V  &               &               &       0.306    $\pm$   0.034   &                               \\
        DO Cep  &       B       &       M4.0 V  &       2.052   &       2.3 (1890)  &       0.199    $\pm$  0.039   &       33.2     $\pm$  1.7     \\
        \noalign{\smallskip}        \noalign{\smallskip}                                                                        
        
\hline
\end{tabular}
\tablefoot{
{Values for the orbital periods are published for some of these stars, namely: 
BL~Cet + UV~Cet: 26.380 $\pm$ 0.002\,a \citep{Gra24};
HD~32450\,AB: 43.55 $\pm$ 0.27\,a \citep{Har01};
V577~Mon + Ross~614: 16.5777 $\pm$ 0.0027\,a \citep{Man19};
\object{GJ~400}\,AB: 160.67\,a \citep{Man00};
\object{Wolf~424}\,AB: 15.826 $\pm$ 0.017\,a \citep{Man19};
\object{Ross~52}\,AB: 31.45 $\pm$ 0.42\,a \citep{Mas18};
HD~239960\,A + DO~Cep: 44.5814 $\pm$ 0.0345\,a \citep{Izm19}.
}
}
\end{table*}

\subsubsection{Binding energies}
\label{sssection:binding}

\begin{figure}[]
    \centering
    \includegraphics[width=0.99\linewidth]{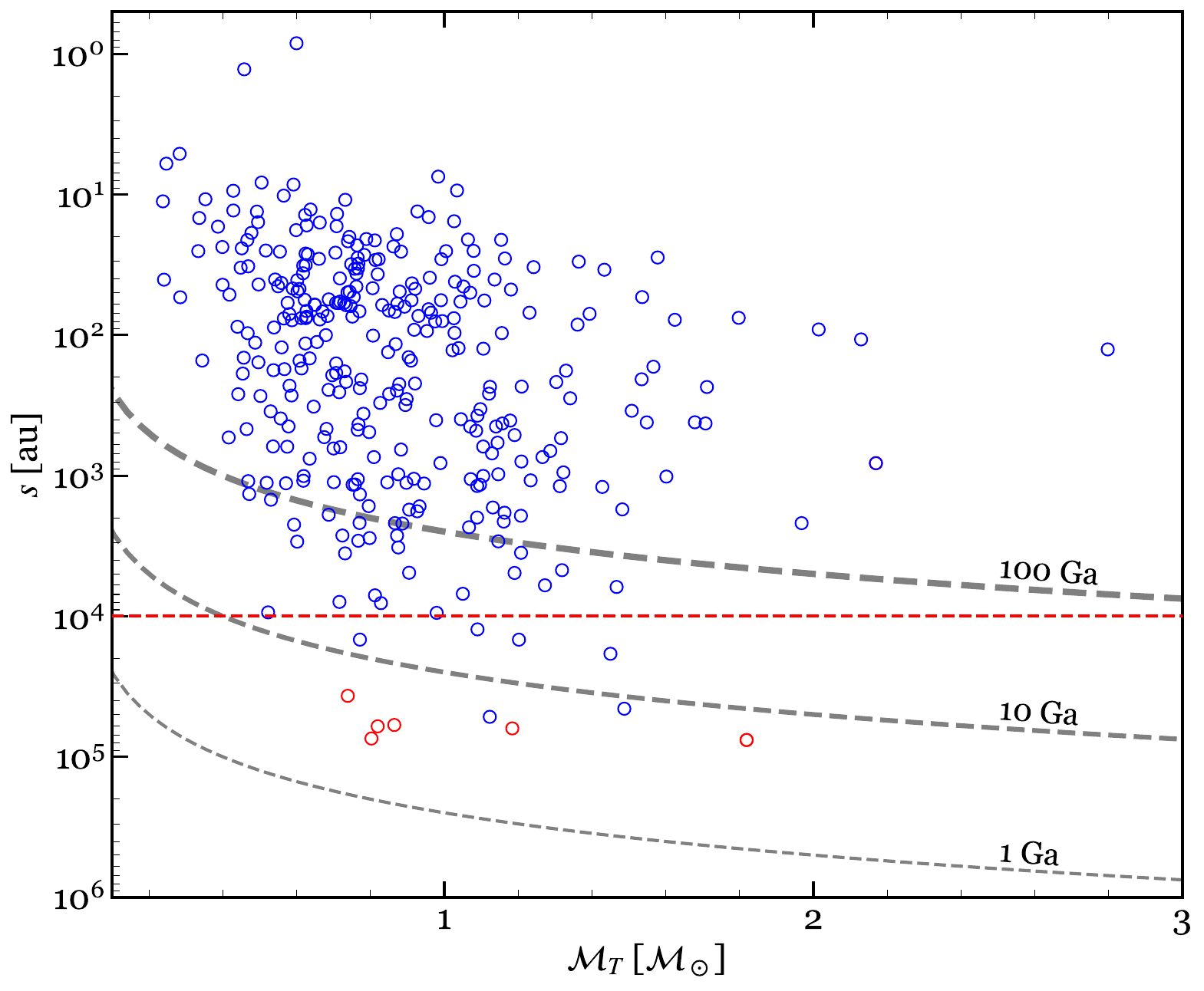}
    \includegraphics[width=0.99\linewidth]{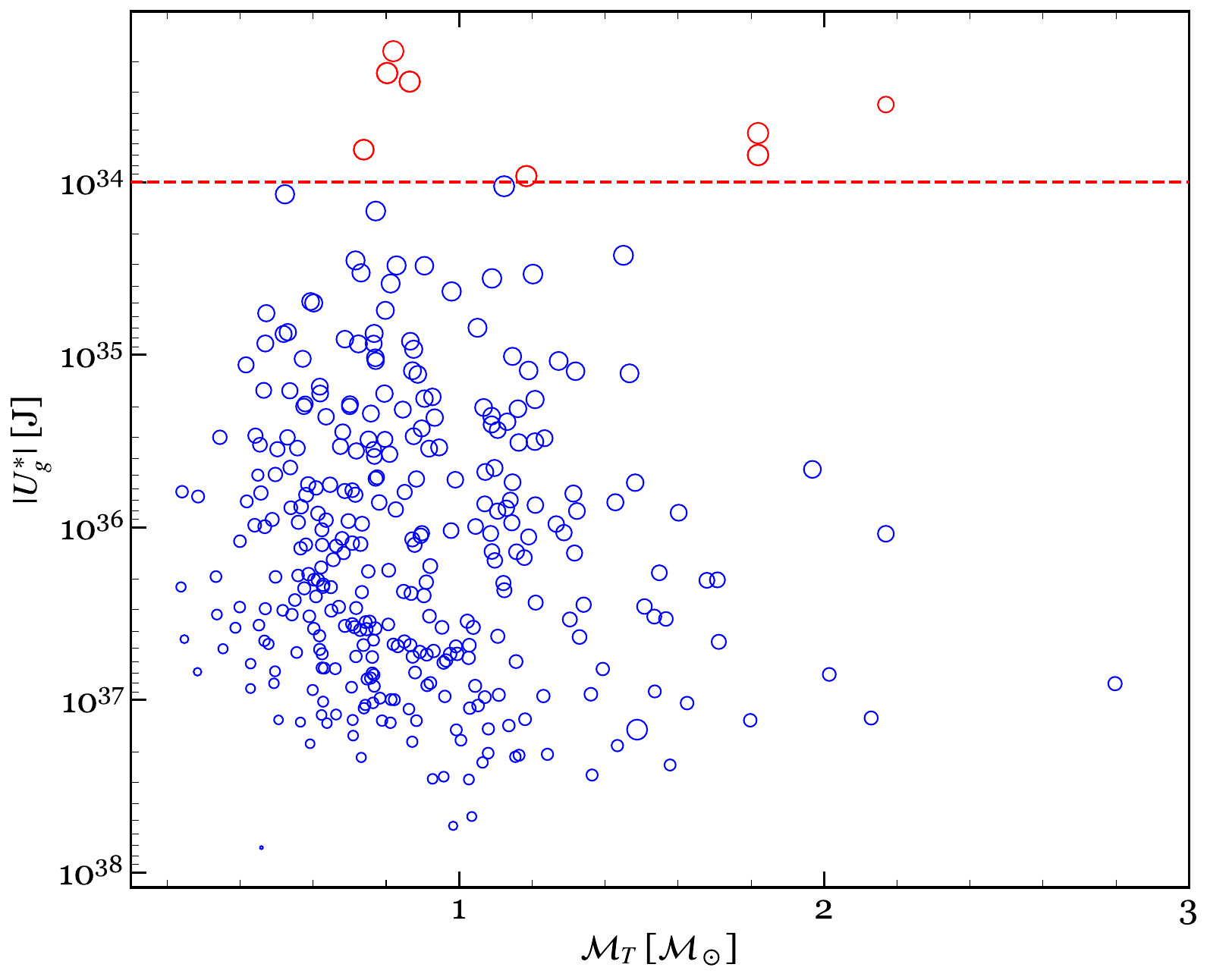}
    \caption{Physical separation ({\em top}) and binding energy ({\em bottom}) as a function of system total mass.
    In the bottom panel, sizes are proportional to the physical separation from the primary.
    Red {open} circles highlight in both panels the most fragile pairs ($|U^\ast_g| <$ 10$^{34}$\,J -- In the upper panel two of the datapoints overlap, as they share a common $\mathcal{M}_T$ and roughly equal separation.)
    The grey dashed lines mark the maximum separation, $s$, for expected survival in the case of 100, 10, and 1\,Ga (from top to bottom). 
    The red horizontal dashed lines mark the representative limits of 10$^4$\,au ({\em top}) and 10$^{34}$\,J ({\em bottom}).
    }
     \label{fig:Mt_Ug}
\end{figure}

The gravitational binding energy of a binary system is a measure of the strength of their attachment.
In this sense, \cite{Heg75} introduced the categories `soft' and `hard' binaries as a broad classification for the class of attachment that stellar pairs can experience:
The former are fragile and easy to break, while the latter are highly resilient to encounters.
These definitions resemble those of `intermediate' and `hard' binaries as used by \citet{Par09} and summarised in Sect.~\ref{sssection:Oepik}.
The reduced binding energy of a binary system of masses $\mathcal{M}_1$ and $\mathcal{M}_2$, with a projected physical separation $s$, can be written as \citet{Cab09}: 

\begin{equation}
\label{eq:binding}
        U_g^\ast = -G\frac{\mathcal{M}_1\mathcal{M}_2}{s},
\end{equation}

\noindent where the asterisk denotes the use of projected separation instead of the actual one, $r$, as the latter is unknown in most cases.
Given that always $r \geq s$, the computed $U_g^\ast$ values represent upper limits to the actual binding energies.
Eq.~\ref{eq:binding} also applies for triple and higher order systems, with $\mathcal{M}_1$ and $\mathcal{M}_2$ corresponding to aggregated masses.
This is, binding energies of high-order systems can be computed as long as they dynamically behave like a binary (as we proceeded when estimating the orbital periods).
Figure~\ref{fig:Mt_Ug} shows $s$ and $|U^{\ast}_g|$ as a function of the total stellar mass, $\mathcal{M}_{T} = \mathcal{M}_1 + \mathcal{M}_2$, in the top and bottom panels, respectively.

The dynamical evolution of the stars over time, determined by their environment, sets a time limit for the stability of multiple systems, especially in the case of the most separated pairs.
For example, \cite{EB18} suggested that most wide binaries with separations exceeding a few thousand {au} may become unbound during post-main sequence evolution.
However, a binary never decays without neighbours.
While catastrophic encounters such as collisions are not common, there is a myriad of subtle chances to disrupt their balance.
The gravitational interaction with nearby stars can be disruptive in the long term, as shown by \cite{Wei87}.
Considering the encounters with stars, the estimated average lifetime of a binary (Eq.~28 of their work) is:  

\begin{equation}
\begin{aligned}
\label{eqn:time}
        t_\ast (a) = 1.8 \times 10^4 \,{\rm Ma}\, \left(\frac{n_\ast}{0.05\,{\rm pc}^{-3}}\right)^{-1} \left(\frac{\mathcal{M}_{T}}{\mathcal{M}_\odot}\right)\left(\frac{\mathcal{M}_\ast}{\mathcal{M}_\odot}\right)^{-2} \\
 \times \left(\frac{\langle 1/V_{\rm rel}\rangle^{-1}}{20\,{\rm km\,s}^{-1}}\right)\left(\frac{a}{0.1\,{\rm pc}}\right)^{-1} \ln^{-1}{\Lambda},  
\end{aligned}
\end{equation}

\noindent where $n_\ast$ and $\mathcal{M}_\ast$ account for number density and mass of stellar perturbers, $V_{\rm rel}$ is the relative velocity between the binary and the perturber, $\mathcal{M}_T$ and $a$ refer to the total mass and semimajor axis of the binary, and $\Lambda$ is the Coulomb logarithm.
The expression accounts for the stochastic gravitational perturbations and the encounters with passing stars.
Assuming an average perturber mass of 0.7\,$\mathcal{M}_\odot$, $V_{\rm rel}=20$\,km\,s$^{-1}$, $\Lambda=1$, and
using the average Galactic disk mass density of 0.11 $\mathcal{M}_\odot$\,pc$^{-3}$ \citep{Clo07,Dhi10}, Eq.~\ref{eqn:time} can be greatly simplified as:

\begin{equation}
        s_{\rm max} \simeq 1.212 \frac{\mathcal{M_{T}}}{t_\star}\cdot206\,265 \,\mathrm{au}.
\end{equation}

\noindent Here, $s_{\rm max}$ represents the maximum separation for a total mass $\mathcal{M}_{T}$ in solar masses, to survive for a given age $t_\star$ in gigayears.
The separations that correspond to ages of 1, 10, and 100\,Ga are plotted 
in Fig.~\ref{fig:Mt_Ug}.

\cite{Par09} suggested that in high- to moderate-density environments, most binaries with separations ranging from a few hundred to a few thousand {au} are disrupted within a few million years.
Conversely, in lower density regions, galactic tides and weak interactions with passing stars gradually separate binaries with separations of around 10$^4$\,au over approximately 10\,Ga \citep{Heg75,Wei87}.
The majority of pairs, separated some dozens or hundreds of au, are expected to be stable for hundreds of gigayears, assuming a negligible occurrence of catastrophic encounters.
However, there are {eight} pairs in our sample with expected survival periods ranging from 1\,Ga to 10\,Ga.
Among them there are the most weakly bound systems, with $|U^{\ast}_g| < $10$^{34}$\,J, highlighted in red in both panels of Fig.~\ref{fig:Mt_Ug}.
The smallest binding energy, of $|U^\ast_g| = 1.74 \pm 0.21\,10^{33}$\,J, corresponds to \object{LP~404--54}, a resolved component of the M3.5 dwarf primary \object{GJ~3022} at an orbital separation of 60\,680\,au in a hierarchical triple system.
These configurations are remarkably fragile and, therefore, more susceptible to be unbound by dynamical encounters with a plethora of sources of perturbation, such as stars, molecular clouds, clusters, and even stellar-mass black holes \citep{Hil75,Ret82,Wei87,Jia10,Par11,Dea20,Ryu22}.

During our search for physical pairs resolved with {\em Gaia} up to 10$^5$\,au, we found some physically bound components with $s > 10^4$\,au.
There are {24} systems with at least one companion of this kind in our sample.
Among them there are the {eight} pairs with the most fragile bindings ($|U^{\ast}_g| < $10$^{34}$\,J), as defined above.
They are collected in Table~\ref{tab:widest}, along with the most relevant parameters.
{Two} of them (\object{V1221~Tau} and \object{PMJ18542+1058}) are identified as comoving pairs in this work for the first time (see also Table~\ref{tab:new}). 
Among them, \object{PM~J18542+1058} is assigned by us as a new member of the Carina association.

\subsubsection{Mass ratios}
\label{sssection:massratios}

\begin{figure}[]
    \centering
    \includegraphics[width=0.99\linewidth]{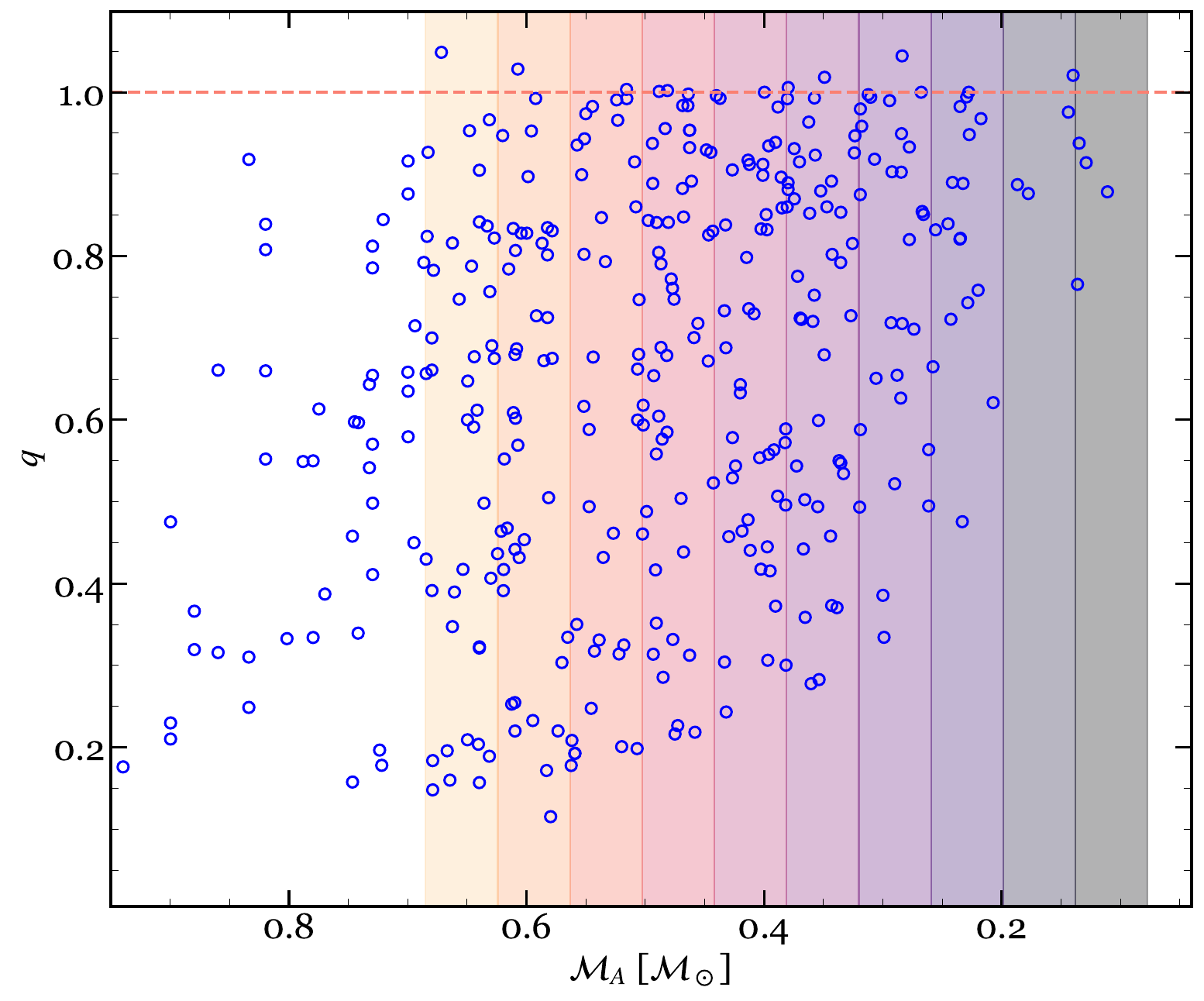}
    \includegraphics[width=0.99\linewidth]{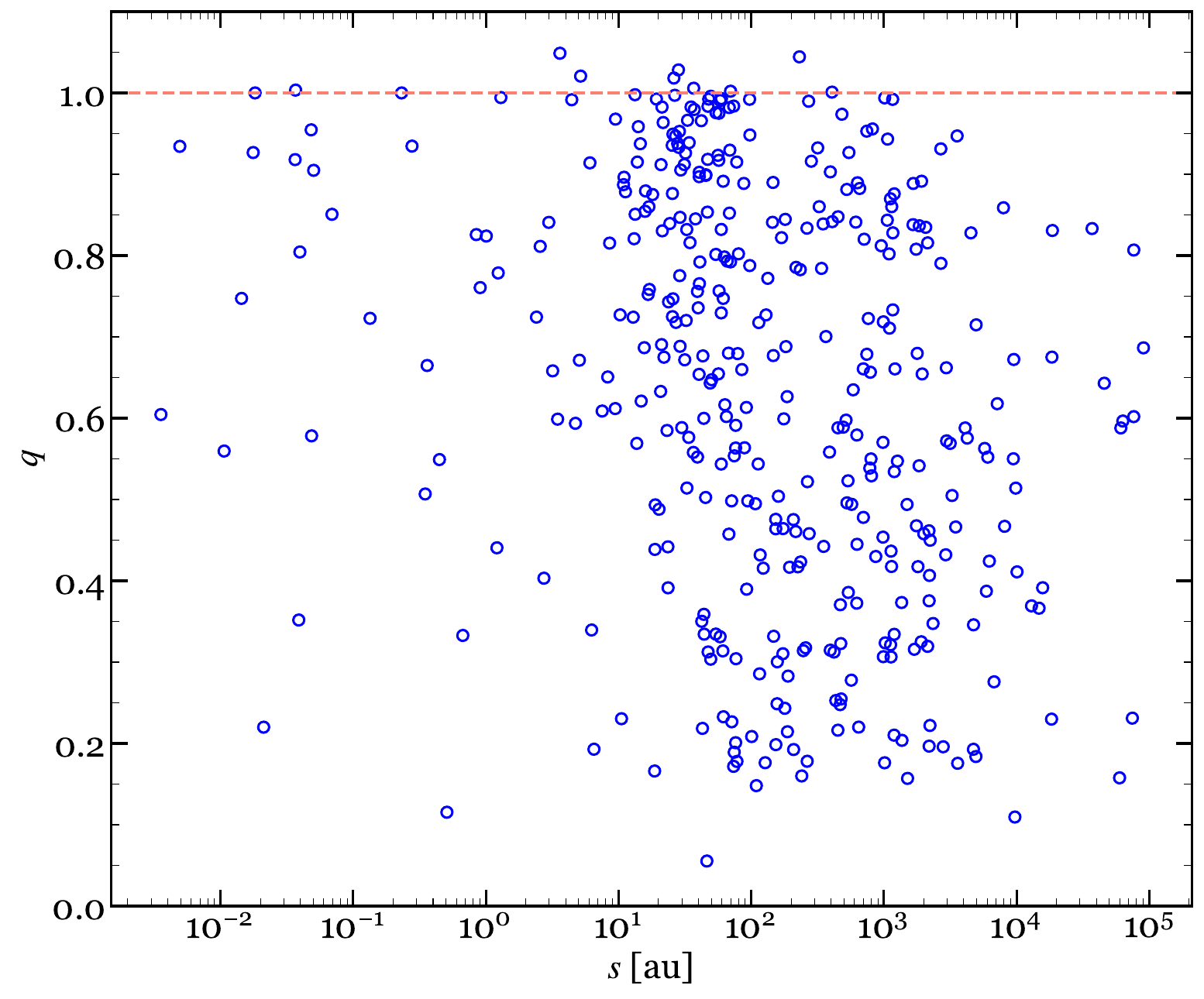}
    \caption{Mass ratios as a function of primary star mass ({\em top}) and projected physical separation ({\em bottom}), only for M-dwarf primaries.
    The red dashed line marks $q=1$.
    The gradient map in the background represents the spectral types from M0.0\,V to M9.0\,V \citep{Cif20} following the same scheme as in Fig.~\ref{fig:SpT}.
    }
     \label{fig:M_q}
\end{figure}

The mass ratio is defined as the fraction $q = \mathcal{M}_B/\mathcal{M}_A$, where `A' refers to the primary component and `B' refers generically to the remaining components.
Given that a large number of the masses of the components of multiple systems are known or, in most cases, estimated when possible (see Sect.~\ref{ssection:parameters}), we were able to calculate $q$ for a considerable fraction of the pairs.
We did not compute mass ratios for the M dwarf-white dwarf pairs (Sect.~\ref{sssection:wd}), given the special nature of these configurations, in which one of the components has lost much of its mass during the post-main-sequence evolution.
Moreover, many $q$ values involving spectroscopic binaries in one of the components may be absent given that only lower limits of the mass of the SB are often available.

The top panel of Fig.~\ref{fig:M_q} displays the mass ratios of {424} pairs as a function of the physical separation and the primary mass\footnote{\cite{Goo13} proposed that the total mass of the system ($\mathcal{M}_T$) should replace the primary masses in the study of binary systems. 
They suggested that the mass ratio, $q$, aligns with a universal, flat distribution of $\mathcal{M}_T$.
For the sake of comparison, we used the primary mass, $\mathcal{M}_A$.}.
An absence of small $q$ for small primary masses (bottom right corner in the panel) has been observed in previous studies \citep[e.g.][]{Ber10,Jan12,Cor17a}.
This is arguably an observational bias, either because most surveys systematically miss either the least massive (faintest) companions ($\mathcal{M} \lesssim 0.1\,\mathcal{M}_\odot$) or the most compact binaries ($a \lesssim 10$\,au).

In the bottom panel of Fig.~\ref{fig:M_q}, the latter observational selection effect is visible, with a remarkable scarcity of data for the closest pairs.
This lack of information prevents from testing if near equal-mass pairs ($q \gtrsim$ 0.8) are typically found at smaller separations \citep{Jod13,Jan14a}, or if on the contrary mass ratio distributions are independent of separation \citep{Duq91,Fis92,Reg11}.
Nevertheless, based on the analysis of high-resolution images of 201 nearby systems collected with the \textit{Hubble} Space Telescope, \cite{Die12} showed that the mass ratios of low-mass binaries tend to approach to unity as masses approach the hydrogen burning limit at about 0.07\,$\mathcal{M}_\odot$, which means that brown dwarf companions are scarce.
Studies involving the mass-ratio distribution in M dwarf binaries suggest that the formation mechanisms might differ from those of more massive stars, potentially involving more complex interactions and accretion processes (we refer to several of the studies in Table~\ref{tab:literature}). 
In this sense, we are unable to reproduce the results by \cite{Gol03}, who analysed 162 SBs and observed two distinct populations ($q \sim$ 0.2 and $q \sim$ 0.8).
But similarly to \cite{Win19a}, we found no preference for equal-mass systems. 
When complemented with other orbital parameters, $q$ also conveys important information regarding the long-term dynamical stability of these systems.

In both panels of Fig.~\ref{fig:M_q}, some values of $q$ are slightly larger than unity, which at first glance seems contradictory.
The most plausible reason has to do with the definition of primary (component `A') as the brightest in the {\em Gaia} $G$ magnitude that we followed in this work.
This definition can disagree with the original nomenclature of some systems, which based this naming convention typically in Johnson $V$ filter, a narrower passband in comparison with {\em Gaia}'s.
Another reason could involve a secondary that is actually a binary system of two equal-mass stars, which together could indeed be more massive than the primary, but a more detailed investigation of these secondaries is needed to clarify this hypothesis.

\subsection{Companions to M dwarfs}
\label{ssection:companions}

M dwarfs can be found with a variety of stellar and substellar companions, ranging from massive OBA stars to ultra-cool dwarfs and planets. 
In particular, in our sample, only two M dwarfs are companions to A-type stars: Castor, an A1\,V spectroscopic binary in a hierarchical sextuple system, and $\tau^{07}$~Ser, an A8\,V star in a triple system.
There are no earlier OB-type primaries.
This lack of OBA+M systems is a direct consequence of the initial mass function and the short life of massive stars in the main sequence.
The vast majority of M dwarfs in our sample (around 93\,\%) are found to be paired with other M dwarfs.
Table~\ref{tab:companions} displays a summary of the different types of companions to our M dwarfs.
A given system can harbour more than one type of companion.
We describe below the systems with FGK-type primaries, white and ultra-cool dwarf companions, and, for completeness, exoplanets.

\begin{table}[]
\caption{Companions to M dwarfs found in our sample.}
\label{tab:companions}
\centering
\begin{tabular}{lcc}
\hline\hline \noalign{\smallskip}       
Type    & Number of systems & Tables    \\
\noalign{\smallskip}\hline\noalign{\smallskip}  
M dwarfs        & 720 & \ldots  \\
OBA stars       & 3 & \ldots    \\
FGK stars       & 60 & \ref{tab:fgk}      \\
White dwarfs    & 28 & \ref{tab:wd}      \\
Ultra-cool dwarfs        & 16 & \ref{tab:ucd}     \\
Exoplanet hosts & {99}  & \ref{tab:exoplanets}\\
\noalign{\smallskip}
\hline
\end{tabular}
\end{table}

\subsubsection{FGK primaries}
\label{sssection:fgk}

Modelling the atmospheres of a star is a challenging task, and the cooler the star, the harder it becomes \citep[see e.g.][]{Val98,Woo06,One12}.
Consequently, advances in new grids of theoretical models (MARCS: \citealt{Gus75,Gus08} -- PHOENIX: \citealt{Hau97,Hus13} -- NextGen: \citealt{Hau99} -- BT-Settl: \citealt{Bar98,All12}), lists of atomic and molecular lines and opacities, computational efficiency, and high-resolution spectroscopy are needed to determine effective temperatures, surface gravities, and, especially, element abundances.
Gaining precision in the determination of stellar abundances is important for various reasons, including better constrains on planet formation, mantle composition, and relative core size \citep{Dor17}.
As a result, the CARMENES Consortium has devoted efforts in the precise determination of abundances in M dwarfs \citep{Abi20,Marf21,Pas22,Tab24}.

In this effort of calibrating the metallicity of the coolest stars, solar-type physical companions can provide a shortcut because their abundances are much easier to determine {\citep{Nor04,Adi12,Ben14}}.
A few investigations have estimated M-dwarf metallicities using wide binary pairs with FGK primaries \citep[e.g.][]{Roj12,Mon18,Bir20,Ish22,Lim24}.
Among them, \cite{Mon18} presented a relatively large sample of 192 wide visual binaries covering a reasonably broad range in metallicity and spectral type, with atmospheric parameters of the primaries homogeneously derived with the \texttt{StePar} code \citep{Tab19}, and chemical abundances for 13 atomic species. 
We expect the list of investigated systems and species to increase in the near future, partly because of this work.

In Table~\ref{tab:fgk} we list 60 systems that contain 64 stars with spectral types between {F5} and K5, and 70 M-dwarf companions.
Earlier F and later K stars are either too hot or too cool for precise abundance determinations.
Besides, the FGK stars are either dwarfs (60) or giants and subgiants (7).
We tabulate the systems, their main astrometric properties, and their iron abundance ([Fe/H] as a proxy of metallicity) from the literature.
The abundances of some FGK+M systems in Table~\ref{tab:fgk} were investigated for the first time by \cite{Duq23}.

\subsubsection{White dwarfs}
\label{sssection:wd}

White dwarfs (WDs) are remnants of stars less massive than 8--10\,$\mathcal{M}_\odot$ and represent the final evolutionary stage for the majority of the stars in the Milky Way \citep{Fon01,Cat08,Doh15}.
When physically paired with M dwarfs, WDs are a valuable source of information because their modelling is easier \citep{Ber95,Ren10}.
For instance, WD companions can serve as effective chronometers for their hosts \citep{Mon06,Fou19,Qiu21,Kim21} because the age estimation for WDs is based on their cooling, which is governed by well-understood physics, but one has to add the main-sequence life of the progenitor to get the total age \citep[see][]{Sod10}.

The white dwarfs observed today are probably the remnants of F to late-B main sequence stars ($\sim$1.2--8\,$\mathcal{M}_\odot$), with the number of cool unevolved dwarf companions peaking at mid-M type \citep{Far05}.
M dwarfs are the most common type of star, yet they do not seem to often pair with white dwarfs. 
According to \cite{Fer12}, this might indicate that the way binary star systems form tends to avoid pairing them with stars that were originally F to late B-type stars.
Regarding this observed deficit, \cite{Wil04} suggested that a large portion of the missing white dwarfs might be explained if these are part of unresolved systems \citep[e.g.][]{MR05,Too17}. 
\cite{Fer12} estimated that there must be an additional $\sim$30\,\% of as yet undiscovered white dwarfs hidden within some kind of binaries.
Also, studying wide binary systems from the Sloan Digital Sky Survey, \cite{Fer12} noted that if both M dwarfs and white dwarfs were drawn from the same initial mass function, the number of such systems would be significantly higher than observed, suggesting a different distribution of masses for M dwarf companions.
Due to the limited number of WD+M systems, this specific aspect is not addressed in the present work (refer to Sect.~\ref{sssection:separations}).

Table~\ref{tab:wd} displays the {28} systems in our sample that have at least one WD component, namely {21} binaries, {6} triples, and {1} quadruple.
We tabulate $\rho$ and $\theta$ with respect to the WD, as the progenitor was originally the most massive and, thus, brightest, star in the system.
The quadruple system contains two white dwarfs (\object{G~107--70}~A and \object{G~107--70}~B, \citealt{Lim15}) and a resolved ($\rho \sim$105\,arcsec) equal-mass M4\,V binary companion \citep[\object{GJ~275.2}~AB,][]{Gic61}.
Among the triple systems is \object{$o^{02}$~Eri~B}, arguably the first white dwarf discovered \citep[][and refer to the previous foundational work by Russell and Hertzprung]{Ada14,Hol05}.
It orbits the solar-type star \object{$o^{02}$~Eri~A} in close binarity with the M4.5 dwarf \object{DY~Eri} (or \object{$o^{02}$~Eri~C}; \citealt{Her85}).

The object {{\em Gaia}~DR3~2005884249925303168}, a physical companion of the M0.0\,V star \object{LF~4+54~152}, is several magnitudes fainter than a main-sequence star of the same colour based on its position in the Hertzsprung-Russell diagram (Fig.~\ref{fig:MG}).
Although it has not been spectroscopically confirmed, \cite{Jim18} and \cite{Gen19} classified it as an astro-photometric WD candidate, as we also do.
All the remaining WDs have been spectroscopically investigated, and many of them are tabulated by the Montreal White Dwarf Database\footnote{\url{https://www.montrealwhitedwarfdatabase.org}} (MWDD). 
Masses, luminosities, and effective temperatures from evolutionary models are available for several of them in the MWDD, which we retrieved and incorporated in our study.

There have been several observational claims on the projected physical separation distribution of stellar systems with WDs.
For example, \cite{EB18} suggested that the distribution of $s$ of systems of WDs and main-sequence stars breaks at $\sim$3000\,au, twice the break of WD+WD pairs.
Unfortunately, the population of binaries in our sample is not sufficiently large to perform an analogous study (there are only five systems with $s >$ 3000\,au), but will allow to determine ages complementary to the kinematic study of \citet{Cor24}.

\subsubsection{Ultra-cool dwarfs}
\label{sssection:ucds}

Ultra-cool dwarfs (UCDs) are a mixture of very low-mass stars and brown dwarfs with spectral types M7.0\,V and later.
UCDs can provide important insights into the formation and evolution of low-mass stars and brown dwarfs, and can also help quantifying properties such as masses and ages, eventually testing the accuracy of ultra-cool evolutionary models.

In our sample, there are 16 systems with M dwarfs and UCDs, shown in Table~\ref{tab:ucd}.
The spectral types of the UCDs were taken from the literature, except for five, which we derived from $G$-band absolute magnitudes and the relationships of \citet{Cif20} and \citet{Pec13}.
The list of UCDs contains 1 late-M-, 7 L-, and 8 T-type UCDs, including the cornerstone late T8.0--8.5 dwarfs Ross~458C (Wolf~462~C), GJ~570~D, Wolf~1130~B, and Wolf~940~B, which were not detected in our \textit{Gaia} search. 
Projected physical separations vary from roughly 25\,au of the \object{GJ~9492} system to over 3000\,au of the {\object{Wolf~1130}\,AB} system.

We provide astrometric measurements of two pairs for the first time: an $\sim$L4 dwarf at just $\sim$0.53\,arcsec to the X-ray M2.0\,V star 1RXS~J190405.9+211030, and an $\sim$L2 dwarf at $\sim$11\,arcsec to the the M3.0+M3.5: double star LP~395--8~AB. 
The $\sim$L2 dwarf is reported here for the first time, while the $\sim$L4 dwarf would need an independent astrometric confirmation given its close separation to the brighter primary and lack of five-parameter \textit{Gaia} solution.

Our collection of systems is smaller than the recent and comprehensive work by \citet{Bai24}, who compiled 278 systems with UCDs at $d <$ 100\,pc, or the encyclopedic work by \citet{Kir24}, who tabulated all known UCDs (either single or in multiple systems) at $d <$ 20\,pc.
However, all primaries in Table~\ref{tab:ucd} (except for the system containing GJ~570~D, whose primary is a K4\,V star -- Table~\ref{tab:fgk}) are nearby, bright, Carmencita M dwarfs, some of them with CARMENES monitoring and, therefore, high signal-to-noise ratio, high-resolution spectroscopy.
As a result, our collection of systems with well-investigated M dwarfs and UCDs can be of help for forthcoming ultra-cool studies (metallicity, kinematics, age dating, dynamical mass determination, etc.).

\subsubsection{Exoplanets}
\label{sssection:planets}

\begin{table*}[]
\caption{System components with new assigments to SKGs.}
\label{tab:skg_new}
\centering
\begin{tabular}{llccll}
\hline\hline 
\noalign{\smallskip}
Star    & Comp. &       $\alpha$        &       $\delta$         &      SKG      & Ref.$^a$\\
\noalign{\smallskip}
\hline
\noalign{\smallskip}
1RXS J011549.5+470159   &       AB      &       01:15:50.51     &       +47:02:02.1     &       (Hya)   &       This work    \\
LP 151-21       &       CD      &       01:15:49.20     &       +47:02:25.7     &       Hya     &       Lod19, Fre20   \\
\noalign{\bigskip}                                                                                      
LP 296-57       &       A       &       01:56:45.99     &       +30:33:28.6     &       (Hya)   &       This work    \\
LP 296-56       &       B       &       01:56:41.74     &       +30:28:34.6     &       Hya     &       Ros11   \\
\noalign{\bigskip}                                                                                      
PM J06102+2234  &       A       &       06:10:17.81     &       +22:34:17.2     &       Hya     &       Ros11   \\
LP 362-121      &       B       &       06:10:22.52     &       +22:34:18.1     &       (Hya)   &       This work    \\
Gaia DR3 3425067888342287616    &       C       &       06:10:22.50     &       +22:34:17.9     &       (Hya)   &       This work    \\
\noalign{\bigskip}                                                                                      
HD 230017A      &       A       &       18:54:53.67     &       +10:58:42.4     &       (Car)   &       This work    \\
HD 230017B      &       B       &       18:54:53.86     &       +10:58:45.1     &       Car     &       Gag18b  \\
PM J18542+1058  &       C       &       18:54:17.14     &       +10:58:11.0     &       (Car)   &       This work    \\
\noalign{\bigskip}                                                                                      
GJ 9721 &       AB      &       21:08:45.41     &       --04:25:36.7    &       UMa     &       Mon01   \\
DENIS J210844.8-042517  &       C       &       21:08:44.75     &       --04:25:18.3    &       (UMa)   &       This work    \\
\noalign{\smallskip}\hline
\end{tabular}
\tablefoot{
\tablefoottext{a}{
Fre20: \cite{Fre20};
Gag18b: \cite{Gag18b};
Lod19: \cite{Lod19};
Mon01: \cite{Mon01};
Ros11: \cite{Ros11}.}
}
\end{table*}

On one hand, the multiplicity of stars with exoplanets comes with an inherent observational bias \citep{Lil12,Cia15,Fon21,Mul21,GP24}, although close stellar companions can also modify the formation, migration, or dynamical evolution of planets \citep{Hol99,Mat00,Des07,Kai13,Kra16}.
On the other hand, the discovery of exoplanets around M dwarfs is a relatively new topic in astrophysics, since at the end of the first decade of the 21st century only a bunch such systems were known 
(e.g. GJ~876\,bcd -- {\citealt{Marc98,Mar01,Riv05}}, 
GJ~436\,b-- \citealt{Butl04}, 
GJ~581\,bce -- \citealt{Bon05,Udr07,May09}).

It was only after the advent of new dedicated extreme precision spectrographs (e.g. CARMENES) and the TESS space mission that the `M-dwarf Opportunity' was eventually made possible.
Nowadays, there are hundreds of M dwarfs with known exoplanets 
(e.g. TRAPPIST-1 a-g -- \citealt{Gil16},
LTT~1445~A\,bc -- \citealt{Win19b}, 
GJ~3929\,b -- \citealt{Kem22}, 
GJ~486\,b -- \citealt{Tri21},
Barnard\,b -- \citealt{Gon24}).
However, the multiplicity of M dwarfs with exoplanets has not yet been investigated in detail.

Here, we pave the way for future analyses on this topic by compiling the {{99}} stars in our initial sample that have exoplanets, 91 of which are M dwarfs, 8 are FGK stars (3 of them non-MS stars), and one is a WD (\object{LP~141--14}). 
We list the planet host stars together with the multiplicity status, number of planets, and corresponding references in Table~\ref{tab:exoplanets}.
For building it, we removed those that have been challenged by CARMENES \citep[see Table 5 of][]{Rib23} or by other authors \citep{Bra20,Sub23}.
We identified 73 single- and 28 multiple-star planetary systems. 
The {eight} FGK stars and one white dwarf with planets but without a Karmn identifier are companions to Carmencita stars.
Not by chance, a significant fraction of the exoplanets in multiple systems in Table~\ref{tab:exoplanets} were discovered by CARMENES \citep{Rei18a,Kam18,Per19,Gon20,Gon21,Kem20,Now20,Sto20b,Kos21,Esp22}.
This observational bias is due to the careful preparation of the input catalogue, by which only systems with companions at $\rho <$ 5\,arcsec that could contaminate the target's optical fibre aperture were discarded from the RV survey \citep{Cor17a}.

\subsection{Young systems}
\label{sssection:young}

\begin{table}[]
\caption{Number of kinematically young stars in multiple systems.}
\label{tab:skg}
\centering
\begin{tabular}{llcc}
\hline\hline 
\noalign{\smallskip}
SKG or assoc.   &       Descriptor      &       $N_{\rm input}$  &      $N_{\rm system}$        \\
\noalign{\smallskip}
\hline
\noalign{\smallskip}
Taurus-Auriga   &       Tau     &       7       &       5       \\
Upper Scorpius  &       USco    &       3       &       1       \\
Argus   &       Arg     &       10      &       5       \\
IC 2391 Supercluster    &       IC 2391 &       19      &       4       \\
Pleiades Supercluster   &       Ple     &       6       &       4       \\
TW Hydrae       &       TWA     &       2       &       2       \\
$\beta$ Pictoris        &       $\beta$ Pic     &       64      &       44      \\
Columba &       Col     &       11      &       8       \\
Carina  &       Car     &       13      &       9       \\
Tucana-Horologium       &       Tuc-Hor &       6       &       5       \\
AB Doradus      &       AB Dor  &       39      &       28      \\
Hercules-Lyrae  &       Her-Lyr &       3       &       3       \\
Ursa Major      &       UMa     &       35      &       18      \\
Castor  &       Cas     &       34      &       16 \\
Hyades (Supercluster)   &       Hya     &       100     &       49      \\
\noalign{\smallskip}\hline
\end{tabular}
\tablefoot{
Young SKG, stellar association, supercluster, or open cluster sorted by increasing age, descriptor in Table~\ref{tab:mother}, and number of stars in the input sample ($N_{\rm input}$) and in multiple systems (in $N_{\rm system}$).
}
\end{table}

Stellar associations and moving groups are loose unbound star agglomerations that contain from dozens to hundreds of stars with a common origin in space \citep{Egg58,Egg75,Mon01,Zuc04,Gag24}.
Since young stars in stellar kinematic groups (SKGs) represent a transitional stage between the original birthplace and the field, so do young multiple stellar systems in SKGs.
Likewise, since wide multiple systems are disrupted by the galactic gravitational potential \citep{Bah80,Wei87}, youth is a common characteristic of the widest pairs \citep[e.g.][]{Cab10a,Sha11,GP24}.

Recently, \cite{Cor24} characterised the kinematic properties of the CARMENES input catalogue, identified M-dwarf members in the different galactic populations (thin, transition, and thick discs and halo) and in young SKGs, stellar associations, superclusters, and open clusters with ages from 1\,Ma to 800\,Ma.
They used the public codes {\tt SteParKin}\footnote{\url{https://github.com/dmontesg/SteParKin}} \citep{Mon01}, {\tt LACEwING} \citep{Rie17a}, and {\tt BANYAN} $\Sigma$ \citep{Mal13,Gag18a} for the identification.
We assigned the same kinematic properties as \cite{Cor24}, and extrapolated them to their physical companions not listed in Carmencita.
After cross-matching with the literature, we found 7 new SKG members 
in 5 systems from this extrapolation. 
We list them in Table~\ref{tab:skg_new} (see \citealt{Alo15a} for an extrapolation example in the $\beta$~Pictoris moving group).

Table~\ref{tab:skg} summarises the number of stars classified as kinematically young by \cite{Cor24} that are part of our input sample (Carmencita M dwarfs plus confirmed resolved companions, $N_{\rm input}$) and of multiple systems ($N_{\rm system}$).
Of the 352 young stars in our input sample ($\Sigma N_{{\rm input},i}$), flagged in Table~\ref{tab:mother} with the corresponding SKG descriptor, a total of 201 are part of multiple systems ($\Sigma N_{{\rm system},i}$).
The large corresponding proportion of young multiple systems is due to the our wide angular search radius, which corresponds to projected physical separations of $10^5$\,au ($\sim$0.5\,pc).
At these separations, it is difficult to delimit the boundary between relatively close, but unbound, SKG members and actual very wide binaries, especially in the core of the Hyades supercluster, which is the Hyades cluster itself (see the series of papers started by \citealt{Cab09}).
As a result, many of the systems in Table~\ref{tab:skg} should be taken with a grain a salt.
However, as discussed in Sect.~\ref{sssection:binding} and by \citet{GP24}, binding energies above a certain limit (e.g. $10^{33}$--$10^{34}$\,J) could discriminate among actually bound systems and coincidental common proper motion pairs in SKGs.
Delimiting such a boundary is beyond the scope of this work.

\section{Conclusions}
\label{section:conclusions}

In this work, we explore the prevalence of M dwarfs as part of stellar systems based on a comprehensive photometric and astrometric characterisation.
We examined individually each of the 2214 stars in the CARMENES input catalogue in search for known and potential physical companions.
We identified a total of 835 M dwarfs as part of 720 multiple systems, predominantly binaries. 
Additionally, we take advantage of the statistical products of {\em Gaia} to propose another {327} candidates to binaries.
They are found to exhibit astrometric and photometric anomalies compatible with the presence of an unresolved companion.
They should be discarded from input target lists of exoplanet searches with forthcoming space missions \citep{Tuc24,Har24,Men24}.

Our findings translate into a minimum multiplicity fraction of 27.8\,\%, which is comparable to the binarity rates reported in previous studies of M dwarfs, although differences in methodology and sample make this comparison somewhat nuanced.
Despite this, we suggest that the actual multiplicity fraction of M dwarfs may be significantly higher, potentially reaching 40.3\,\%, if all newly proposed binary candidates are confirmed.
This revised estimate brings M dwarf multiplicity more in line with the observed rates among Sun-like stars.
Further high-resolution spectral and spatial follow-up studies are mandatory to capture the complete population of close companions.
All these analyses will also be complemented with {\em Gaia} DR4 (second half of 2026) and DR5 (end of 2030).

A number of other conclusions can be derived from our analyses.
In our sample a two-slope distribution of projected physical separations may support the idea that distinct formation mechanisms are at play for different separation regimes, as previous investigations have proposed:
the wide-separation pairs suggest a population of systems formed through dynamical interactions, while closer binaries may arise from more traditional fragmentation processes. 
Pre-{\em Gaia} investigations of ultra-wide physical pairs heavily relied on photometric distances due to the lack of precise astrometric data. 
Previous studies, while valuable, have often suggested the existence of a large number of very wide real binaries, but were dominated by chance alignments at $s \gtrsim$10$^4$\,au.
The mass ratio distribution observed in our sample is slightly skewed towards equal-mass systems, compared to binaries of more massive stars; nonetheless, selection biases are likely to make us miss part of the picture.
Given the significant binding energies computed for M-dwarf systems that have not been found to be young, they are likely to remain bound over tens of gigayears, enduring through the majority of the Galaxy’s lifetime. 

Finally, we identified remarkable systems with M dwarfs that have FGK star, white dwarf, ultra-cool dwarf, and exoplanet companions, and which belong to young stellar kinematic groups or associations. 
Fort these systems, we provide a homogeneous compilation of fundamental parameters and observables, and a detailed examination of architectures.
These diverse pairings provide unique opportunities for deriving characteristics of M dwarfs that would be otherwise inaccessible or difficult to determine in M dwarfs that are single (e.g. ages, element abundances).

While our results are consistent with previous studies in terms of observed binary frequency, separation distributions, and mass ratio trends, they also highlight the need for continued investigation into close companions and the impact of observational biases. 
We expect that the data gathered in this work will serve as a valuable asset for testing and informing theoretical predictions, and for providing a foundation for future investigations on multiplicity.

\section*{Data availability}

Tables \href{https://doi.org/10.5281/zenodo.14248958}{A.1}--\href{https://doi.org/10.5281/zenodo.14251637}{A.12} are only available in electronic form at Zenodo.
The full version of Table~\ref{tab:mother} is available at the CDS via anonymous ftp to \url{cdsarc.cds.unistra.fr} (\url{130.79.128.5})
or via \url{http://cdsweb.u-strasbg.fr/cgi-bin/qcat?J/A+A/}, and at the GitHub repository \url{https://github.com/ccifuentesr/CARMENES-IX}. 

\begin{acknowledgements}

We are grateful to X.~Delfosse, J.~Gagn\'e, C.~Reyl\'e, and, especially, A.~Tokovinin {and the reviewer Z.~Zhang} for their valuable suggestions, which have greatly enriched our work.
We acknowledge financial support from the Agencia Estatal de Investigaci\'on of the Ministerio de Ciencia, Innovaci\'on y Universidades and the ERDF through projects 
{2023AT003 (PIE, CSIC) associated to RYC2021-031640-I,}
{CNS2023-144309,}
PID2023-150468NB-I00, 
PID2022-137241NB-C4[1:4], 
PID2021-125627OB-C31, 
{PID2019-107061GB-C61,}
BES-2017-080769, 
and the Centre of Excellence ``Severo Ochoa'' and ``Mar\'ia de Maeztu'' awards to the 
Instituto de Astrof\'isica de Canarias (CEX2019-000920-S), 
Instituto de Astrof\'isica de Andaluc\'ia (CEX2021-001131-S), 
Institut de Ci\`encies de l'Espai (CEX2020-001058-M), and 
Centro de Astrobiolog\'ia (MDM-2017-0737).

CARMENES is an instrument at the Centro Astron\'omico Hispano en Andaluc\'ia (CAHA) at Calar Alto (Almer\'{\i}a, Spain), operated jointly by the Junta de Andaluc\'ia and the Instituto de Astrof\'isica de Andaluc\'ia (CSIC).
  
CARMENES was funded by the Max-Planck-Gesellschaft (MPG), 
  the Consejo Superior de Investigaciones Cient\'{\i}ficas (CSIC),
  the Ministerio de Econom\'ia y Competitividad (MINECO) and the European Regional Development Fund (ERDF) through projects FICTS-2011-02, ICTS-2017-07-CAHA-4, and CAHA16-CE-3978, 
  and the members of the CARMENES Consortium 
  (Max-Planck-Institut f\"ur Astronomie,
  Instituto de Astrof\'{\i}sica de Andaluc\'{\i}a,
  Landessternwarte K\"onigstuhl,
  Institut de Ci\`encies de l'Espai,
  Institut f\"ur Astrophysik G\"ottingen,
  Universidad Complutense de Madrid,
  Th\"uringer Landessternwarte Tautenburg,
  Instituto de Astrof\'{\i}sica de Canarias,
  Hamburger Sternwarte,
  Centro de Astrobiolog\'{\i}a and
  Centro Astron\'omico Hispano-Alem\'an), 
  with additional contributions by the MINECO, 
  the Deutsche Forschungsgemeinschaft through the Major Research Instrumentation Programme and Research Unit FOR2544 ``Blue Planets around Red Stars'', 
  the Klaus Tschira Stiftung, 
  the states of Baden-W\"urttemberg and Niedersachsen, 
  and by the Junta de Andaluc\'{\i}a.

This work has made use of data from the European Space Agency (ESA) mission {\it Gaia} (\url{https://www.cosmos.esa.int/gaia}), processed by the {\it Gaia} Data Processing and Analysis Consortium (DPAC,
\url{https://www.cosmos.esa.int/web/gaia/dpac/consortium}). 
Funding for the DPAC has been provided by national institutions, in particular the institutions participating in the {\it Gaia} Multilateral Agreement.

This publication made use of the Washington Double Star Catalog maintained at the U.S. Naval Observatory,
VOSA and the Filter Profile Service developed and maintained by the Spanish Virtual Observatory {through grant AYA2017-84089},
the SIMBAD database, the Aladin sky atlas, and the VizieR catalogue access tool developed at CDS, Strasbourg Observatory, France, 
the NASA Exoplanet Archive, which is operated by the California Institute of Technology, under contract with the National Aeronautics and Space Administration under the Exoplanet Exploration Program,
and the Python libraries {Matplotlib, NumPy, SciPy, Pandas} and collection of software packages {AstroPy}.

\end{acknowledgements}

\bibliographystyle{aa}
\bibliography{biblio}

\begin{appendix}
\section{Online material}
\label{appsection:tables}

\begin{figure*}[]
    \begin{subfigure}{\linewidth}
	\includegraphics[width=0.49\linewidth]{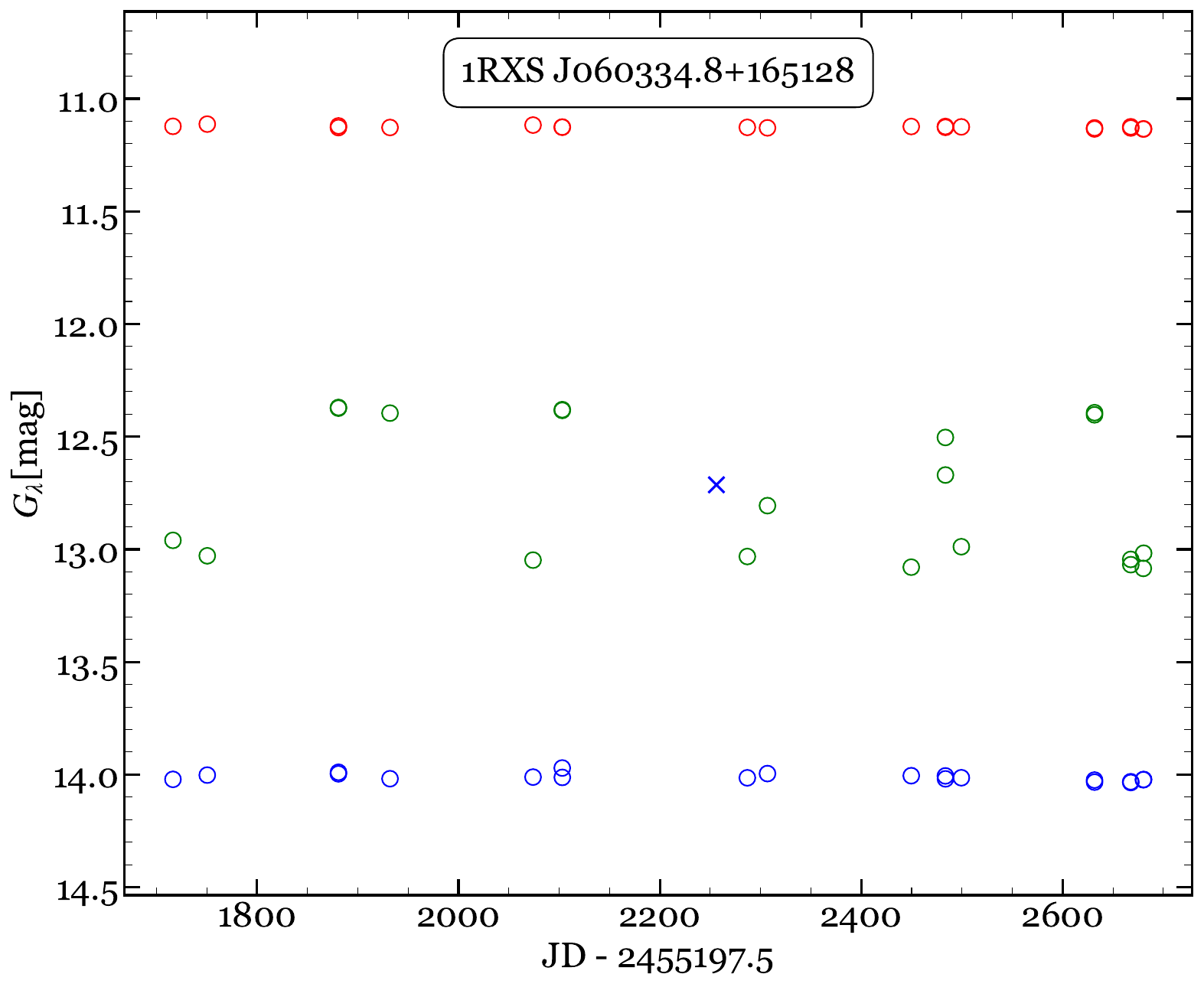}
	\includegraphics[width=0.49\linewidth]{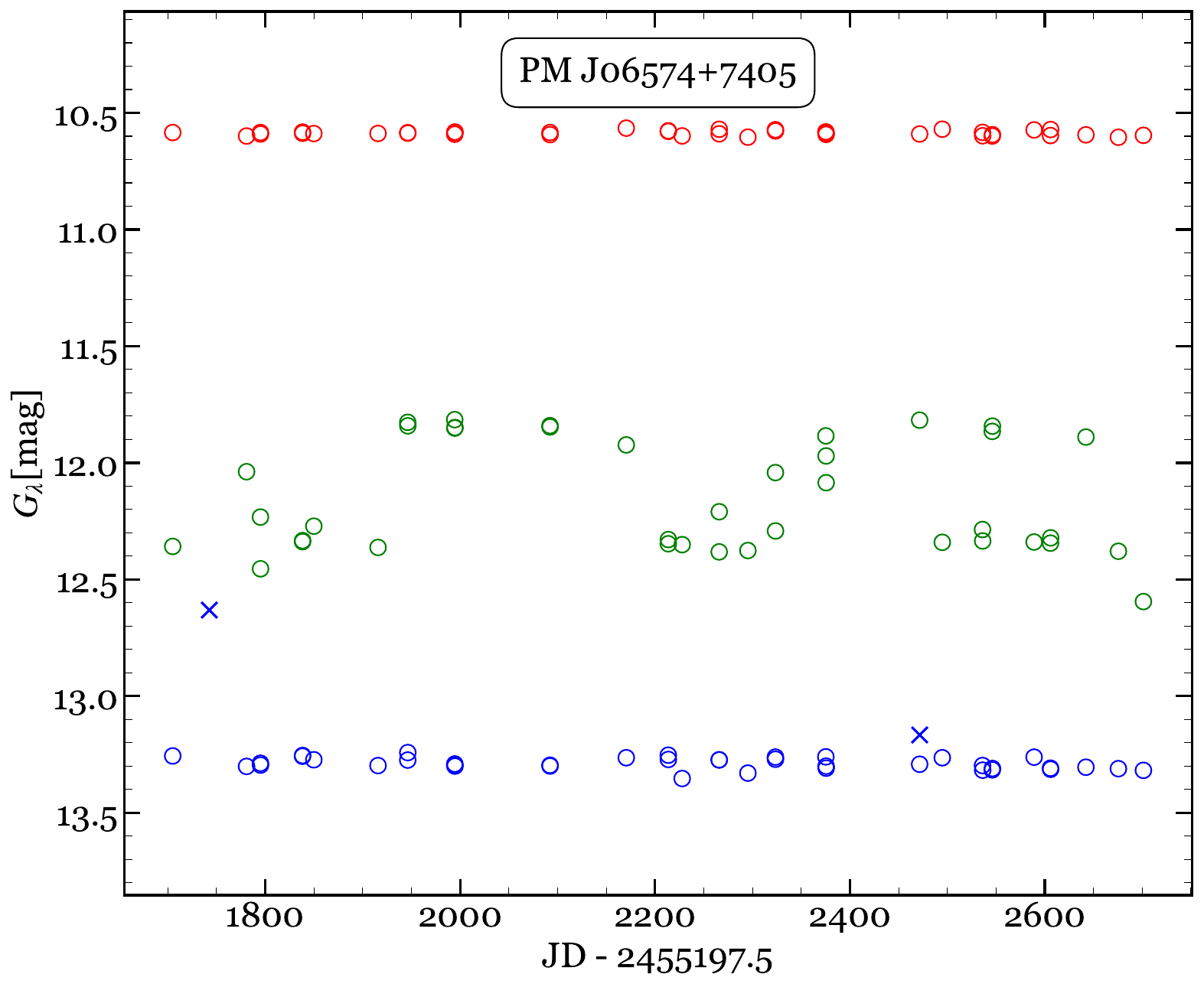}
    \end{subfigure}
    \begin{subfigure}{\linewidth}
    \includegraphics[width=0.49\linewidth]{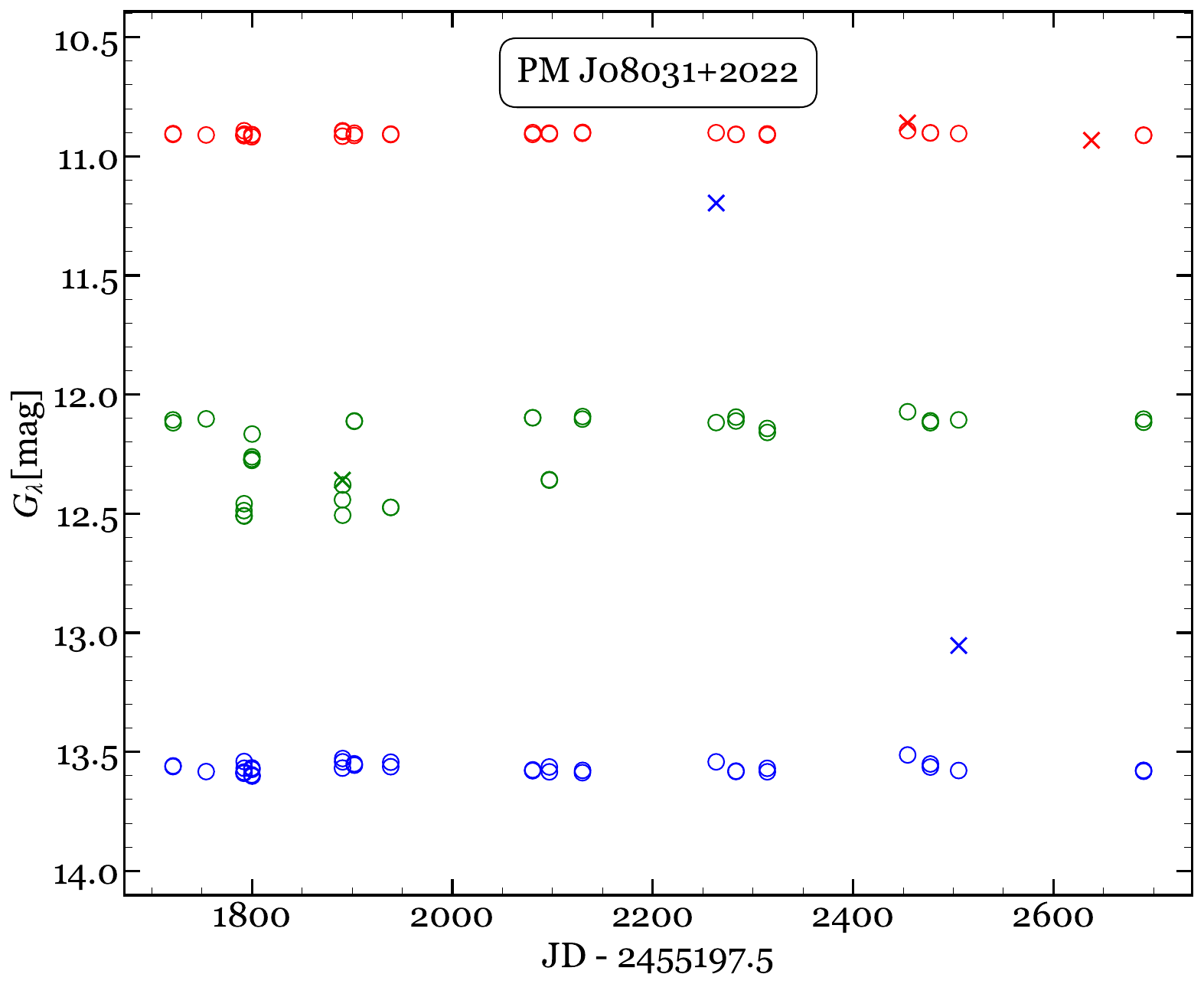}
    \includegraphics[width=0.49\linewidth]{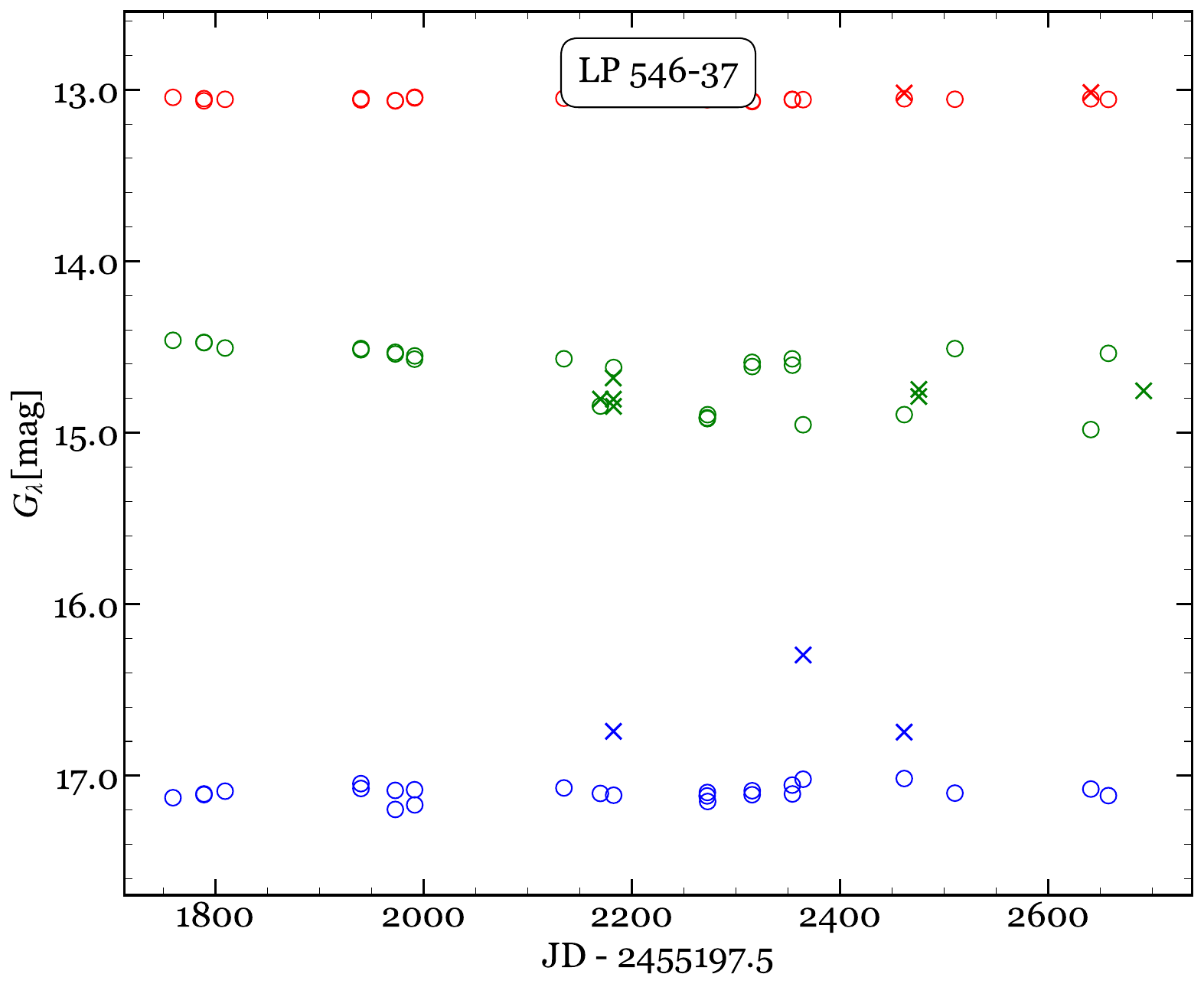}  
    \end{subfigure}
	\begin{subfigure}{\linewidth}
    \includegraphics[width=0.49\linewidth]{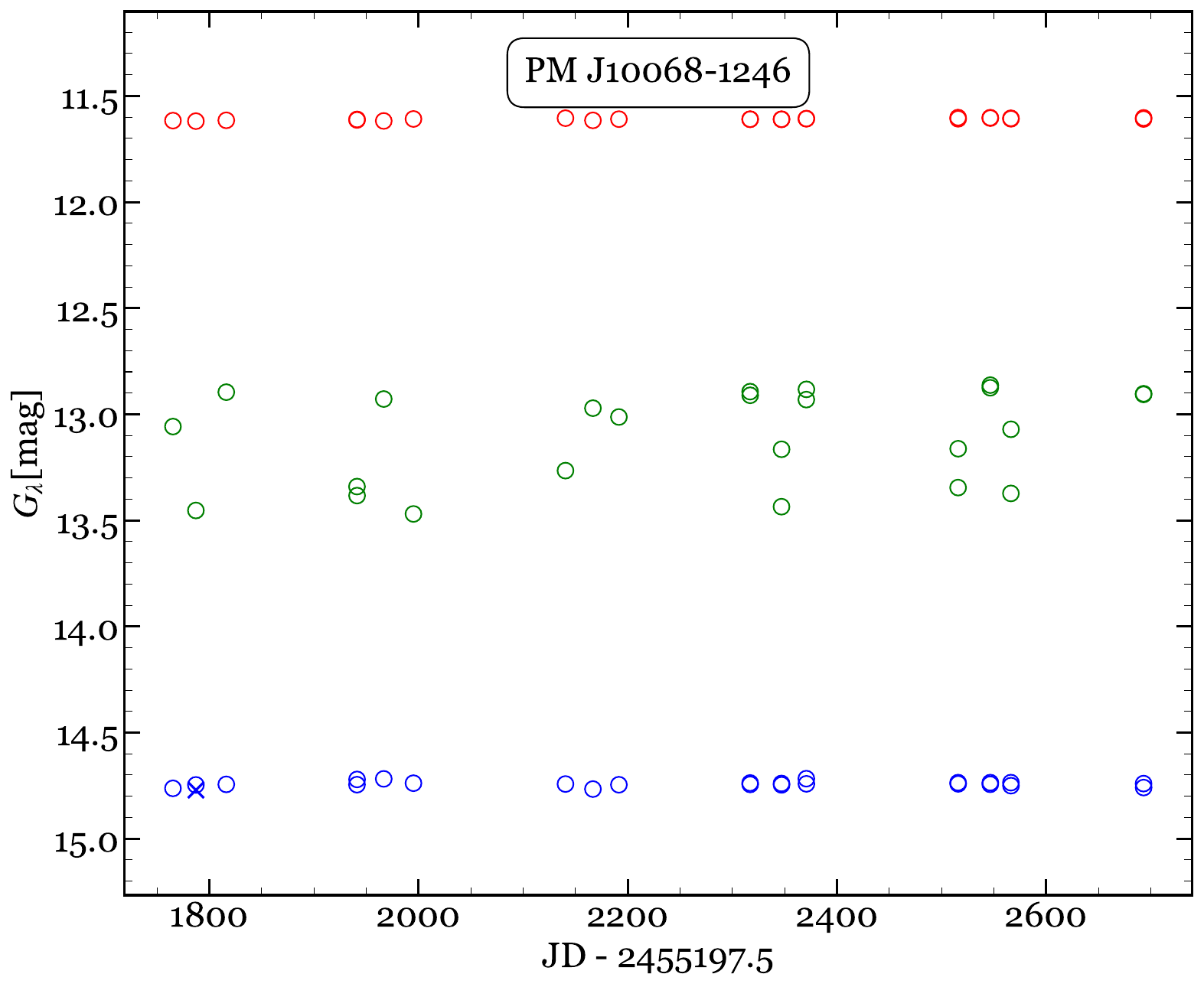}
    \includegraphics[width=0.49\linewidth]{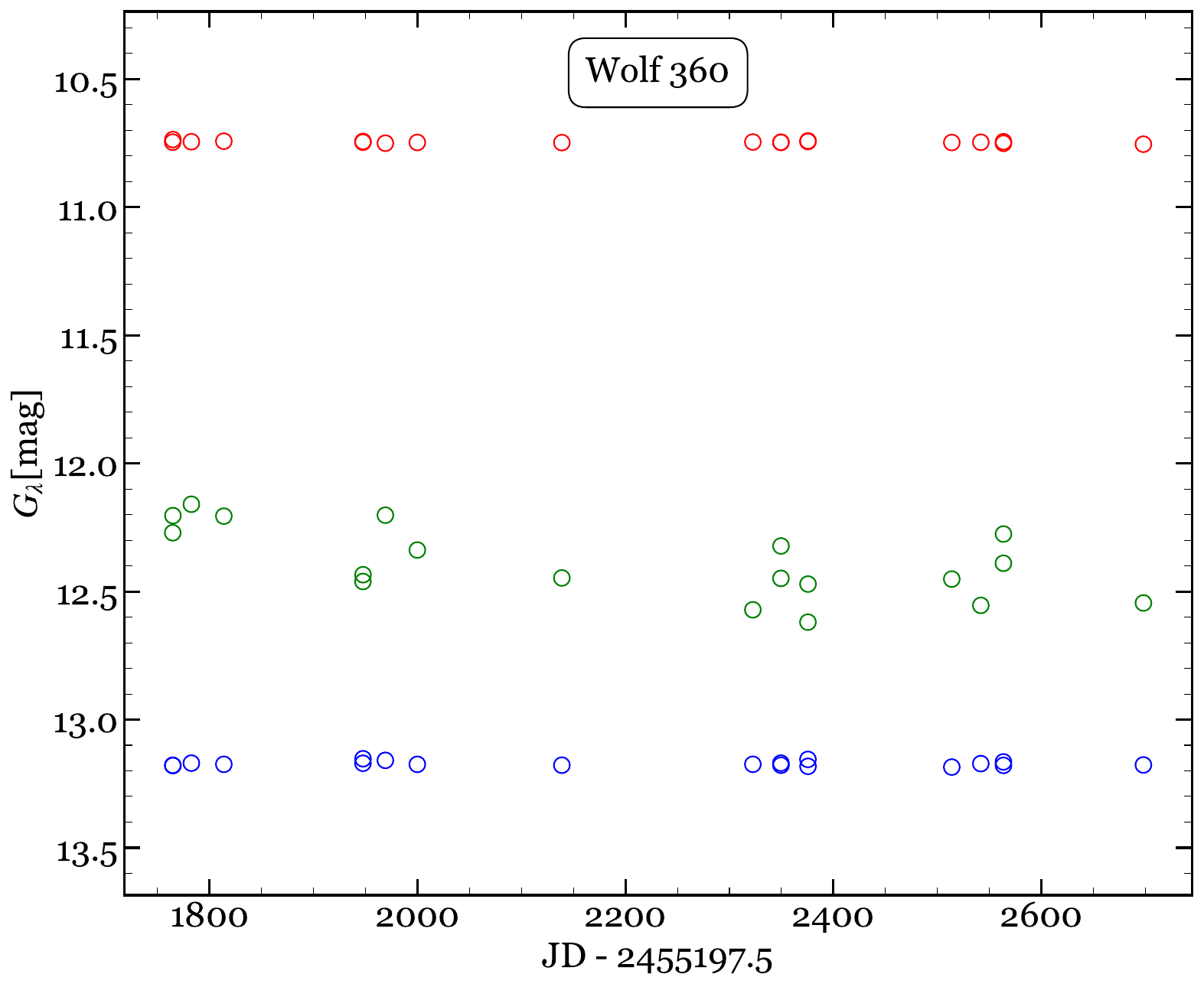}
    \end{subfigure}
    \caption{Same as Fig.~\ref{fig:light_curve} for single stars reported in this work as unresolved binary candidates.}
     \label{fig:light_curves_appendix}
\end{figure*}
\clearpage
\begin{figure*}[]
    \ContinuedFloat
    \begin{subfigure}{\linewidth}
	\includegraphics[width=0.49\linewidth]{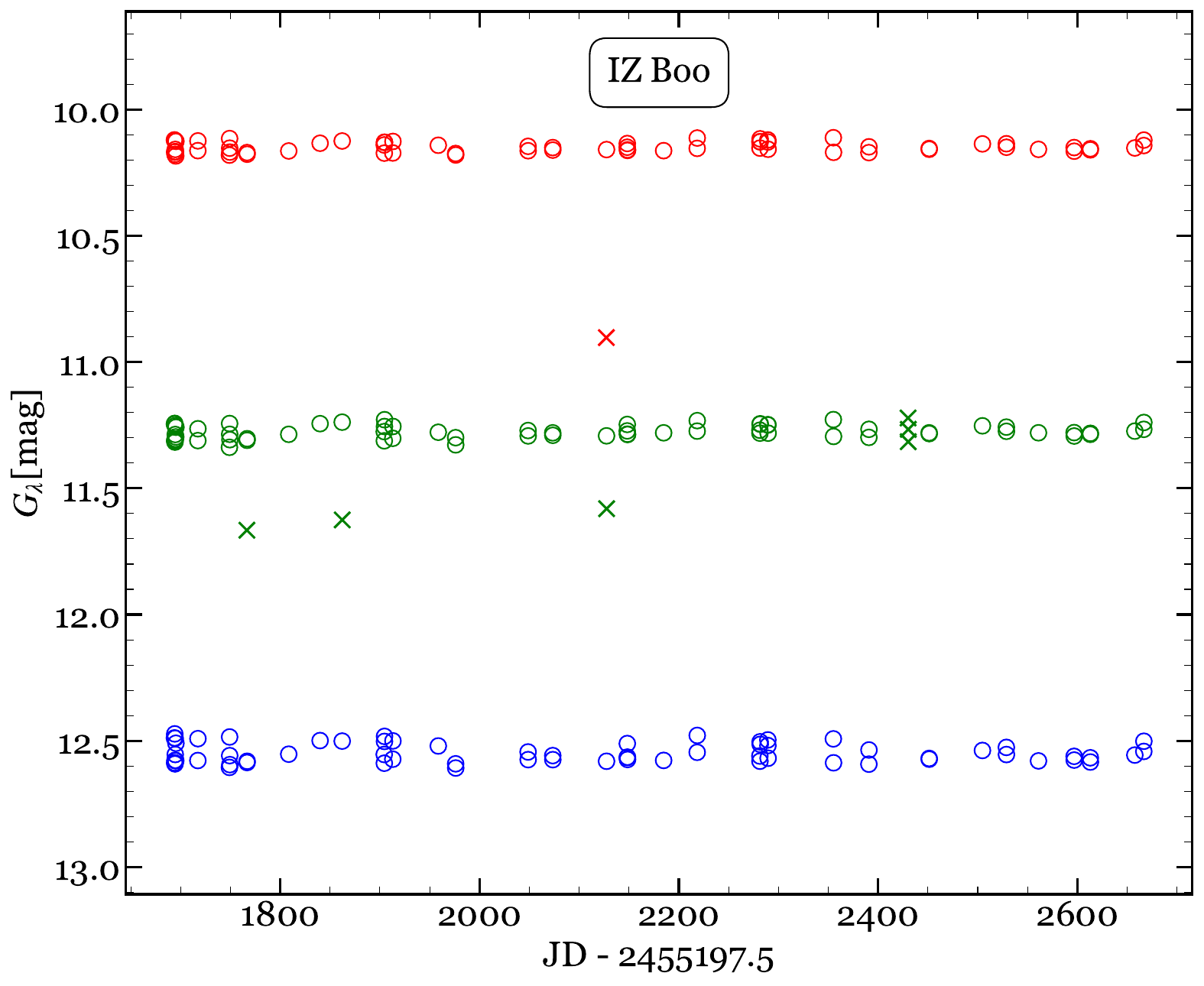}
	\includegraphics[width=0.49\linewidth]{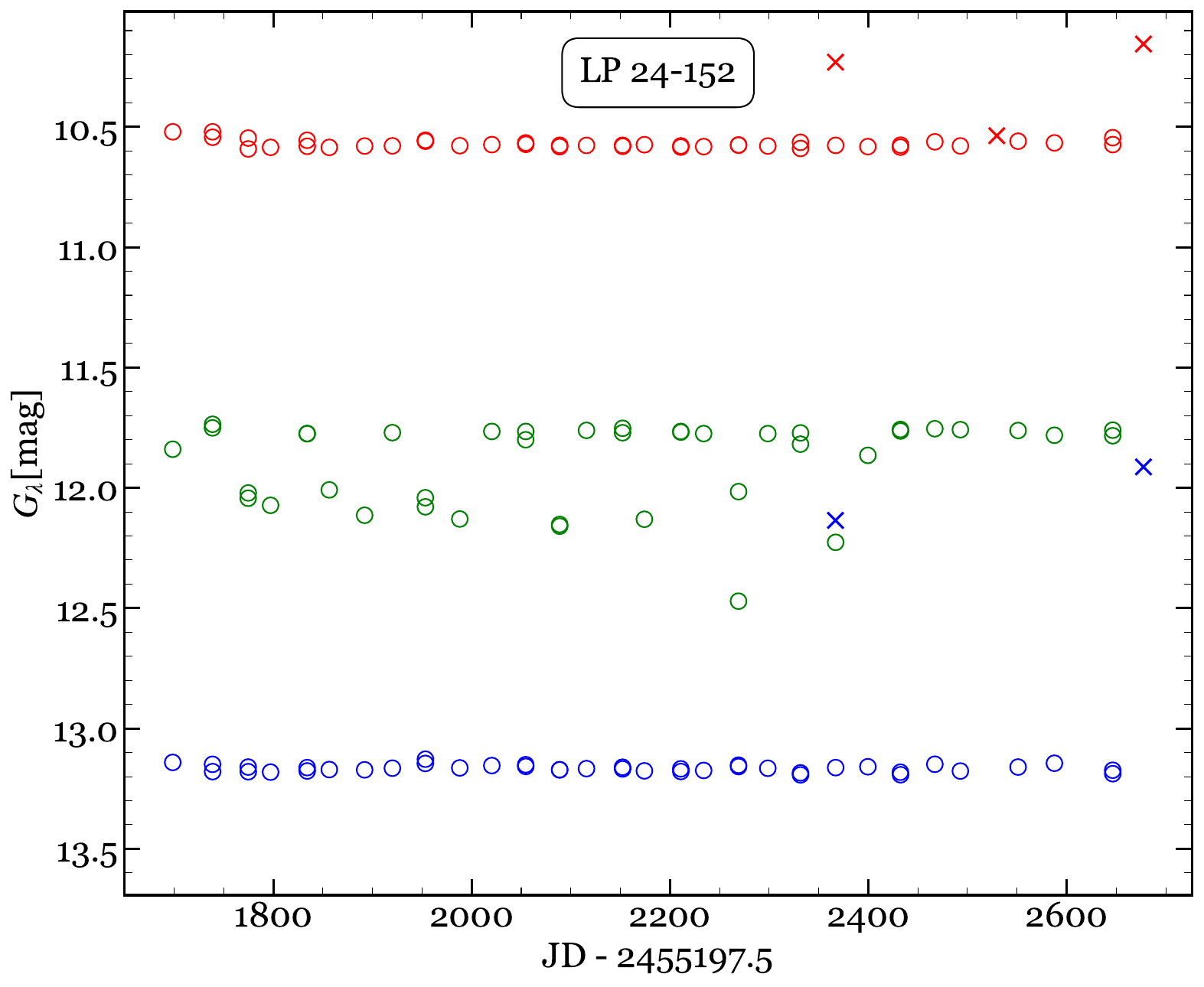}
    \end{subfigure}
    \begin{subfigure}{\linewidth}
    \includegraphics[width=0.49\linewidth]{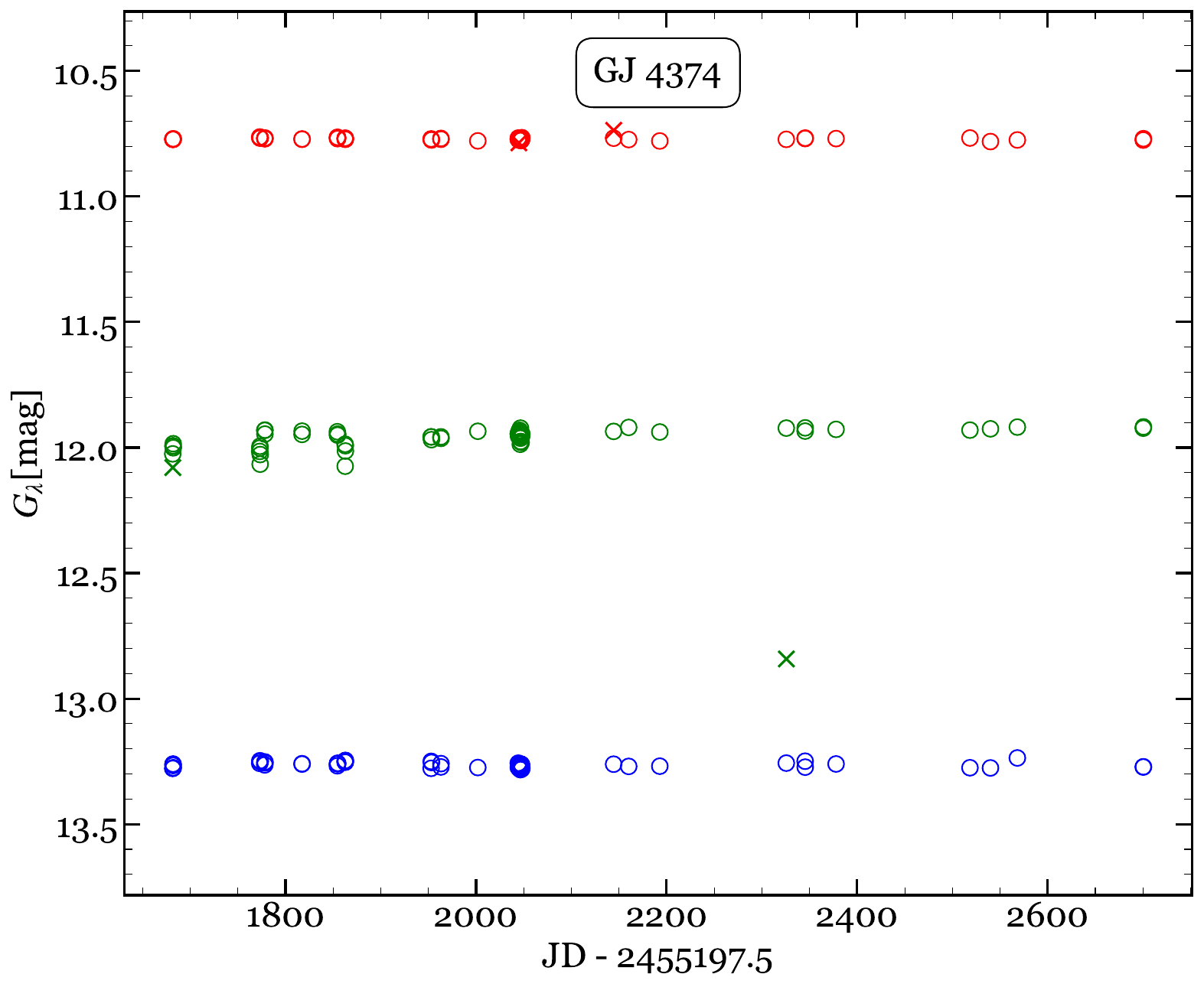}
    \includegraphics[width=0.49\linewidth]{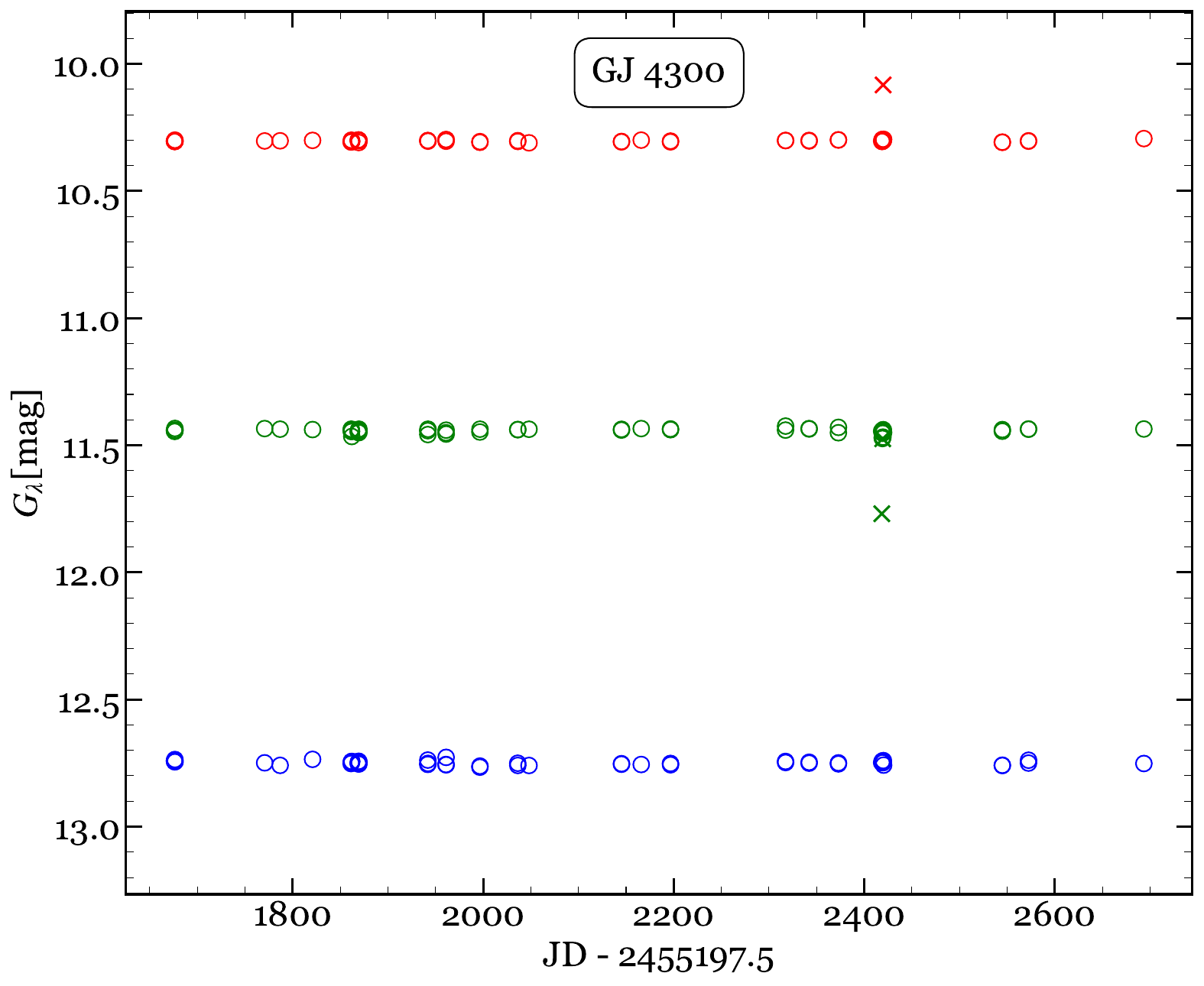}  
    \end{subfigure}
	\begin{subfigure}{\linewidth}
    \includegraphics[width=0.49\linewidth]{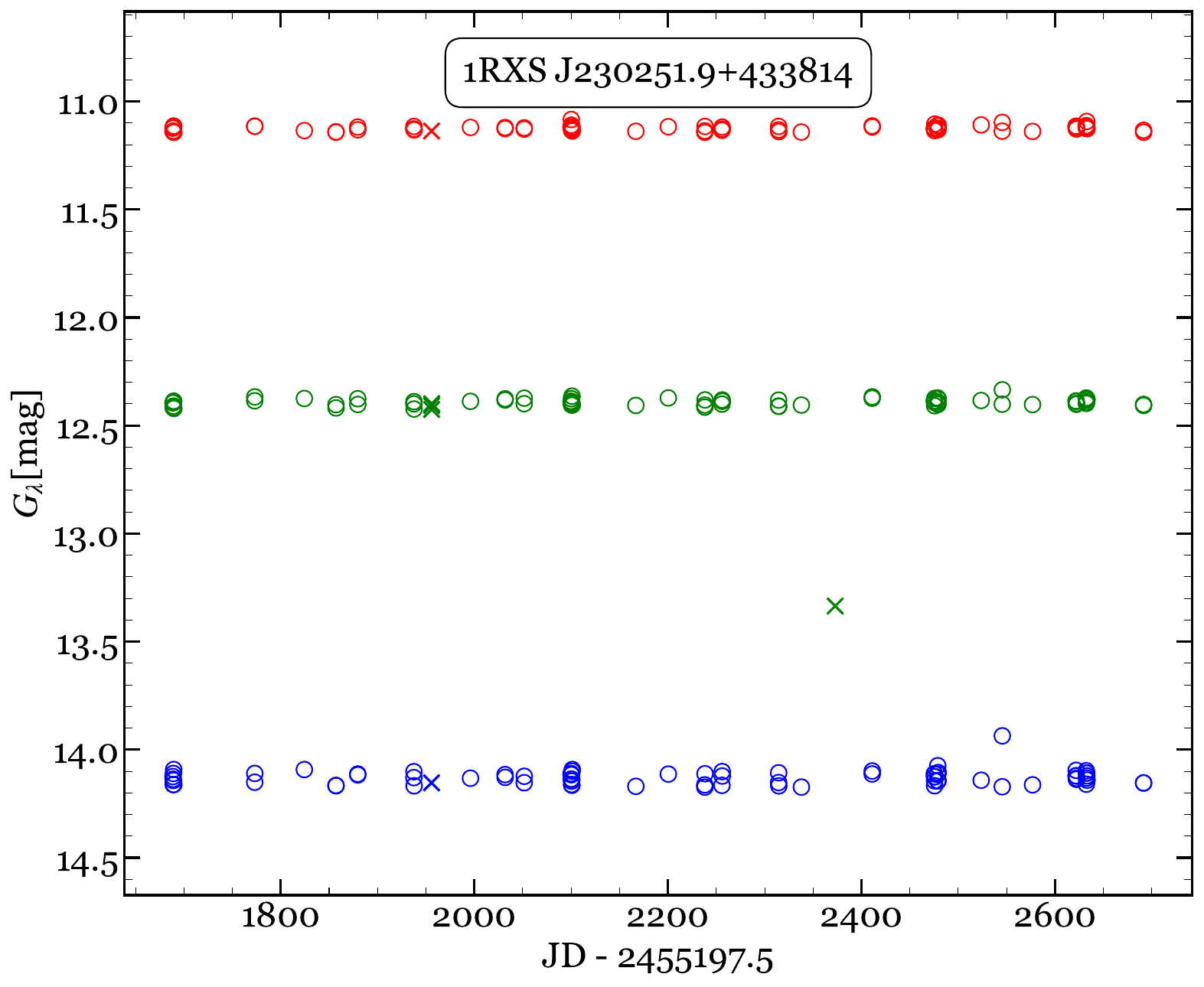}
    \includegraphics[width=0.49\linewidth]{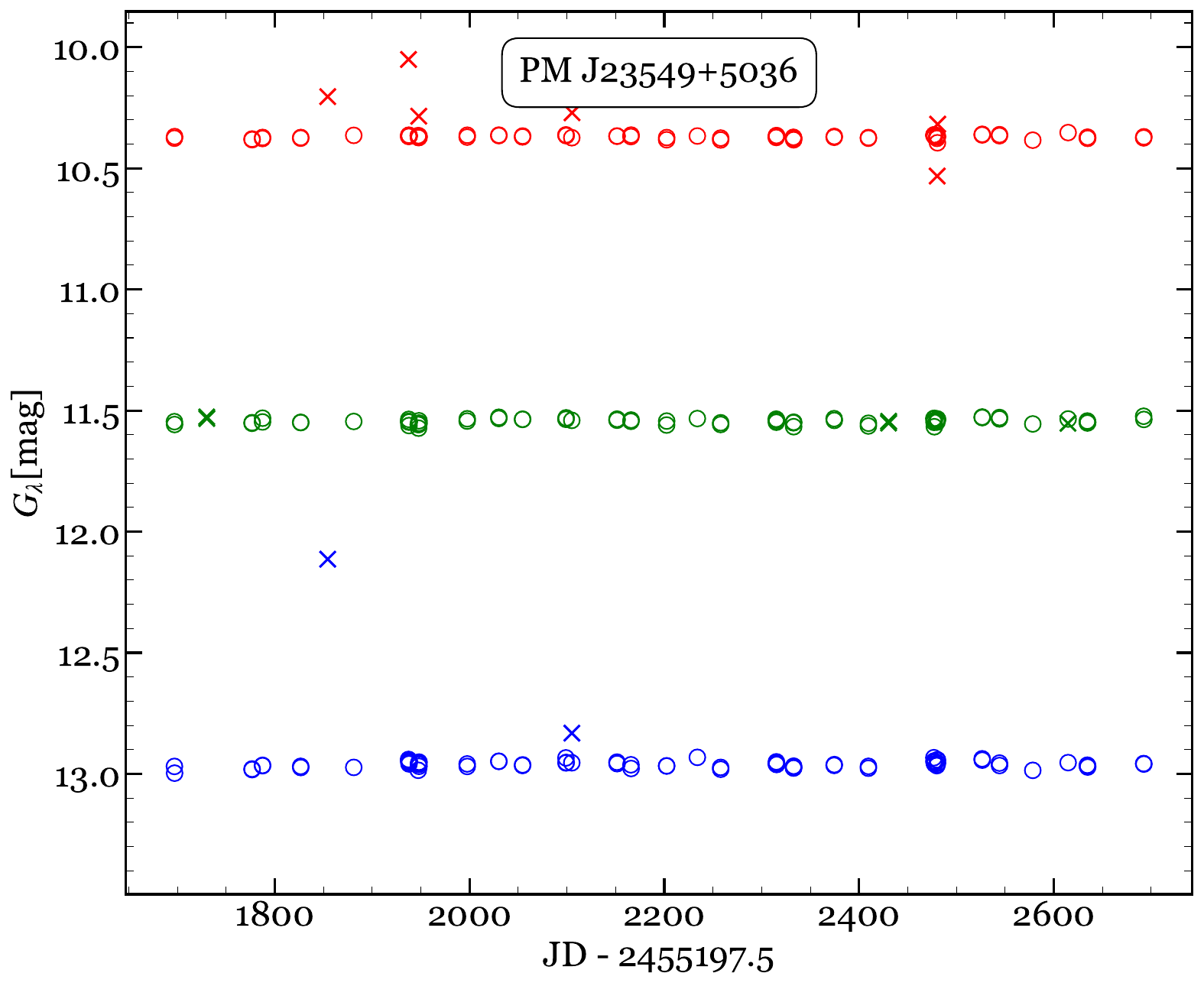}
    \end{subfigure}
    \caption{Same as Fig.~\ref{fig:light_curve} for the single stars reported in this work as candidates to unresolved binaries (cont.).}
\end{figure*}

{\scriptsize
\begin{longtable}{ccllcccclcl}
\caption{\label{tab:mother} Complete sample studied in this work$^a$.}
\\
\noalign{\smallskip}
\noalign{\smallskip}
\noalign{\smallskip}
\hline
\hline
\noalign{\smallskip}
Star	&	System	&	Karmn	&	Name	&	GJ	&	$\alpha$	&	$\delta$	&	Spectral	&	Multiplicity	&	In tables	\\
id.	&	id.	&		&		&		&	(J2016.0)	&	(J2016.0)	&	type	&	type	&		\\
\noalign{\smallskip}
\hline
\noalign{\smallskip}
\endfirsthead
\caption{Complete sample studied in this work$^a$ (cont.).}\\
\hline\hline
        \noalign{\smallskip}
Star	&	System	&	Karmn	&	Name	&	GJ	&	$\alpha$	&	$\delta$	&	Spectral	&	Multiplicity	&	In tables	\\
id.	&	id.	&		&		&		&	(J2016.0)	&	(J2016.0)	&	type	&	type	&		\\
\hline
\noalign{\smallskip}
\endhead
\hline
\endfoot
\noalign{\smallskip}
1	&	1	&	J00012+139N	&	BD+13 5195A	&		&	00:01:13.21	&	+13:58:32.7	&	M0.5 V	&	Multiple	&		\\
2	&	1	&	J00012+139S	&	BD+13 5195B	&		&	00:01:12.89	&	+13:58:22.0	&	M0.0 V	&	Multiple	&		\\
3	&	2	&	J00026+383	&	PM J00026+3821A	&		&	00:02:40.00	&	+38:21:44.1	&	M4.0 V	&	Multiple	&		\\
4	&	2	&		&	PM J00026+3821B	&		&	00:02:40.05	&	+38:21:45.3	&	M3.5 V	&	Multiple	&	\ref{tab:outliers}	\\
5	&		&	J00033+046	&	StKM 1--2199	&		&	00:03:18.97	&	+04:41:11.6	&	M1.5 V	&	Single	&		\\
6	&	3	&		&	HD 38A	&	4 A	&	00:05:42.38	&	+45:48:40.9	&	K7 V	&	Multiple	&	\ref{tab:spectroscopic}	\\
7	&	3	&	J00056+458	&	HD 38B	&	4 B	&	00:05:42.30	&	+45:48:35.0	&	M0.0 V	&	Multiple	&		\\
8	&	3	&	J00051+457	&	GJ 2	&	2	&	00:05:12.22	&	+45:47:09.2	&	M1.0 V	&	Multiple	&		\\
\end{longtable}
\tablefoot{
\tablefoottext{a}{
The full version of this table ({2634} rows, {132} columns) is available in electronic form at the CDS and on a dedicated GitHub repository (\url{https://github.com/ccifuentesr/CARMENES-IX}).}
}
}

\footnotesize{
\begin{longtable}{lcccccccc}
\caption{\label{tab:outliers} Components of multiple system that do not comply with one or more of the criteria for physical association.}
\\
        \noalign{\smallskip}
        \noalign{\smallskip}
        \noalign{\smallskip}
\hline\hline
        \noalign{\smallskip}
Name	& 	Comp.$^a$	& 	$\alpha$	&	$\delta$	&	$\mu$ ratio	& 	$\Delta PA$	& 	$\Delta d$	& 	$\rho^b$	& 	Cause$^c$	\\
	& 		& 	(J2016.0)	&	(J2016.0)	&		& 		& 		& 	{[}arcsec{]}	& 		\\
        \noalign{\smallskip}
\hline
        \noalign{\smallskip}
\endfirsthead
\caption{Components of multiple system that do not comply with one or more of the criteria for physical association (cont.).}\\
\hline\hline
        \noalign{\smallskip}
Name	& 	Comp.$^a$	& 	$\alpha$	&	$\delta$	&	$\mu$ ratio	& 	$\Delta PA$	& 	$\Delta d$	& 	$\rho^b$	& 	Cause$^c$	\\
	& 		& 	(J2016.0)	&	(J2016.0)	&		& 		& 		& 	{[}arcsec{]}	& 		\\
\hline
        \noalign{\smallskip}
\endhead
\hline
\endfoot
        \noalign{\smallskip}
PM J00026+3821 B	&	B	&	00:02:40.05	&	+38:21:45.3	&	0.214	&	12.254	&	0.031	&	1.415	&	1	\\
G 217--32 B	&	B	&	00:07:43.40	&	+60:22:53.8	&	0.196	&	9.161	&	0.0032	&	0.848	&	1	\\
LP 296--56	&	B	&	01:56:41.74	&	+30:28:34.6	&	0.012	&	0.027	&	0.163	&	299.1	&	3	\\
PM J02024+1034 A	&	B	&	02:02:28.18	&	+10:34:52.7	&	0.545	&	32.513	&	0.012	&	0.911	&	1	\\
LP 198--637 B	&	B	&	03:20:45.35	&	+39:42:59.6	&	0.152	&	8.738	&	0.0051	&	0.795	&	1	\\
PM J03247+4447 B	&	B	&	03:24:42.23	&	+44:47:39.7	&	0.191	&	4.590	&	0.0048	&	1.912	&	1	\\
Gaia DR3 3296932486866670720	&	B	&	04:05:38.89	&	+05:44:40.1	&	0.331	&	1.614	&	\ldots	&	0.817	&	1, 2	\\
GJ 2033 B	&	B	&	04:16:41.80	&	--12:33:19.8	&	0.334	&	11.137	&	0.0069	&	2.990	&	1	\\
\end{longtable}
\tablefoot{
\tablefoottext{a}	{Component in system.}
\tablefoottext{b}	{Angular separation of the closest component.}
\tablefoottext{c}	{Cause of the deviation: 
1:	Presence of a very close companion (either resolved or not). This is also applicable to the primary;
2:	Distances and proper motions from a source different than {\em Gaia} DR3, with larger uncertainties, or missing values;
3: New unresolved binary candidates as described in Sect.~\ref{ssection:candidates}. This is applicable to the primary.
}
\tablefoottext{$\dagger$}	{
\object{Castor~C}: Detached eclipsing double-lined spectroscopic binary (DESB2) with $a\lesssim$ 0.01\,au \citep{Tor02};  
\object{FP~Cnc~B}: SB2 \citep{Shk10} in a quadruple system;
\object{GJ~569~B}: Ultra-cool companion resolved ($\rho \lesssim$ 0.1\,arcsec) with the Keck Adaptive Optics Facility \citep{Mar00,Lan01}; 
\object{GJ~897}: Parallax from Hipparcos removed because of extremely large relative error.
}
}
}

\begin{landscape} 
\begin{longtable}{lllccccccc}
\caption{\label{tab:new} New stellar multiple systems not reported in the literature.}
\\
        \noalign{\smallskip}
        \noalign{\smallskip}
        \noalign{\smallskip}
\hline\hline
        \noalign{\smallskip}
Name	&	Karmn	&	Spectral	&	Component$^b$	&	$\alpha$	&	$\delta$	&	$\varpi$$^c$			&	$\mu_{\rm total}$			&	$\theta$	&	$\rho$	\\
	&		&	type$^a$	&		&		&		&	{[}mas{]}			&	{[}mas\,a$^{-1}${]}			&	{[}deg{]}	&	{[}arcsec{]}	\\
        \noalign{\smallskip}
\hline
        \noalign{\smallskip}
\endfirsthead
\caption{New stellar multiple systems not reported in the literature (continued).}\\
\hline\hline
        \noalign{\smallskip}
Name	&	Karmn	&	Spectral	&	Component$^b$	&	$\alpha$	&	$\delta$	&	$\varpi$$^c$			&	$\mu_{\rm total}$			&	$\theta$	&	$\rho$	\\
	&		&	type$^a$	&		&		&		&	{[}mas{]}			&	{[}mas\,a$^{-1}${]}			&	{[}deg{]}	&	{[}arcsec{]}	\\
\hline
        \noalign{\smallskip}
\endhead
\hline
\endfoot
        \noalign{\smallskip}
GJ 3261	&	J04056+057	&	M3.5 V	&	A	&	04:05:38.94	&	+05:44:40.4	&	16.094	$ \pm $	0.103	&	47.98	$ \pm $	0.15	&		&		\\
Gaia DR3 3296932486866670720	&		&	M4.0: V	&	B	&	04:05:38.89	&	+05:44:40.1	&				&				&	251.9	&	0.817	\\
V1221 Tau	&		&	M3.0: V	&	(C)	&	04:05:53.46	&	+05:31:24.6	&	15.876	$ \pm $	0.026	&	49.85	$ \pm $	0.03	&	164.8	&	824.8	\\
\noalign{\smallskip}        \noalign{\smallskip}																							
RX J0507.2+3731 A	&	J05072+375	&	M5.0 V	&	A	&	05:07:14.33	&	+37:30:42.1	&	43.800	$ \pm $	4.000	&	102.11	$ \pm $	4.90	&		&		\\
RX J0507.2+3731 B	&		&	M5.0: V 	&	(B)	&	05:07:14.37	&	+37:30:42.1	&	44.022	$ \pm $	0.461	&	96.09	$ \pm $	0.82	&	93.2	&	0.476	\\
\noalign{\smallskip}        \noalign{\smallskip}																							
LP 780--23	&	J06401--164	&	M2.5 V	&	A	&	06:40:08.72	&	--16:27:21.5	&		\ldots		&	319.03	$ \pm $	11.31	&		&		\\
LP 780--23 B	&		&	M2.5: V 	&	(B)	&	06:40:08.72	&	--16:27:21.7	&		\ldots		&		\ldots		&	187.9	&	0.199	\\
\noalign{\smallskip}        \noalign{\smallskip}																							
GJ 375.2	&	J10004+272	&	M0.5 V	&	A	&	10:00:26.69	&	+27:16:03.5	&	27.767	$ \pm $	0.016	&	118.42	$ \pm $	0.02	&		&		\\
2MASS J10003572+2717054	&		&	M6.5 V	&	B	&	10:00:35.70	&	+27:17:07.5	&	27.723	$ \pm $	0.103	&	118.34	$ \pm $	0.13	&	61.9	&	136.0	\\
Gaia DR3 740041664172917376	&		&	M7.5: V 	&	(C)	&	10:00:26.56	&	+27:16:05.7	&	26.561	$ \pm $	0.349	&	117.20	$ \pm $	0.34	&	320.9	&	2.882	\\
\noalign{\smallskip}        \noalign{\smallskip}																							
HD 230017 A$\dagger$	&		&	M0.0 V	&	A	&	18:54:53.67	&	+10:58:42.4	&	53.423	$ \pm $	0.108	&	133.27	$ \pm $	0.16	&		&		\\
HD 230017 B	&	J18548+109	&	M3.5 V	&	B	&	18:54:53.86	&	+10:58:45.1	&	53.644	$ \pm $	0.043	&	89.29	$ \pm $	0.06	&	45.1	&	3.792	\\
PM J18542+1058	&		&	M4.0 V	&	(C)	&	18:54:17.14	&	+10:58:11.0	&	53.838	$ \pm $	0.024	&	114.61	$ \pm $	0.03	&	266.7	&	538.9	\\
\noalign{\smallskip}        \noalign{\smallskip}																							
G 28--46 &	J23113+085	&	M3.0 V	&	A	&	23:11:23.45	&	+08:30:56.4	&	49.800	$ \pm $	3.300	&	458.92	$ \pm $	11.31	&		&		\\
G 28--46 B$\dagger$	&		&	M3.0: V 	&	(B)	&	23:11:23.47	&	+08:30:56.4	&		\ldots		&		\ldots		&	77.7	&	0.248	\\
\noalign{\smallskip}        \noalign{\smallskip}	
\end{longtable}
\tablefoot{
\tablefoottext{a}	{A colon (:) indicates an estimated spectral type from photometry.}
\tablefoottext{b}	{The new component found in each system is written in parentheses.}
\tablefoottext{$\dagger$} {
\object{HD~230017}\,ABC: Members of the Carina association \citep{Gag18b};
\object{G~28--46}\,AB: System identified as a close binary during the CARMENES exoplanet search \citep{Rib23}. A minimum period of 2225\,d was tabulated by \cite{Sab21} from spectroscopic observations.}
}
\end{landscape}

{\fontsize{8}{10}\selectfont 
\begin{longtable}{llccccl}
\caption{Spectroscopic binaries, triples, and quadruples in our sample.}
\label{tab:spectroscopic}
\\

    \hline\hline 
    \noalign{\smallskip}
    Name	&	Karmn	&	Type$^a$	&	$P_{\rm orb}$ [d]			&	$a \sin{i}$ [au]	&	$q = \mathcal{M}_B/\mathcal{M}_A$&	Reference$^b$	\\
    \noalign{\smallskip}
    \hline
    \noalign{\smallskip}
\endfirsthead
\caption{Spectroscopic binaries, triples, and quadruples in our sample (cont.).}\\
    \hline\hline 
    Name	&	Karmn	&	Type$^a$	&	$P_{\rm orb}$ [d]			& $a \sin{i}$ [au]	&	$q = \mathcal{M}_B/\mathcal{M}_A$&	Reference$^b$	\\
    \noalign{\smallskip}
    \hline
    \noalign{\smallskip}
\endhead
\endfoot
HD 38A	&		&	SB	&		\ldots		&	\ldots	&	\ldots	&	Abt70, Str93	\\
1RXS J000806.3+475659	&	J00081+479	&	SB2	&	<4.4			&	0.04	&	\ldots	&	Shk10, Fou18	\\
EZ Psc	&	J00162+198W	&	SB2	&	3.95652	 $\pm$ 	0.00008	&	0.038623	&	0.35	&	Bar18	\\
FF And	&	J00428+355	&	SB2	&		\ldots		&	\ldots	&	\ldots	&	Bop77, Giz02	\\
RX J0050.2+0837	&	J00502+086	&	SB2	&		\ldots		&	\ldots	&	\ldots	&	Jef18	\\
GJ 1029	&	J01056+284	&	SB2	&	95.76	 $\pm$ 	0.18	&	0.1331	&	0.72	&	Bar18, Win20	\\
PM J01437--0602	&	J01437--060	&	SB2	&		\ldots		&	\ldots	&	\ldots	&	Fou18	\\
G 173--18	&	J01453+465	&	SB2	&	<157.9			&	0.56	&	\ldots	&	Giz02, Shk10	\\
BD--21 332	&	J01531--210	&	SB2	&	<2.9			&	0.04	&	\ldots	&	Shk10	\\
GJ 1041 B	&	J01592+035W	&	SB2	&	<356.4			&	0.93	&	\ldots	&	Shk10	\\
GJ 3129	&	J02027+135	&	SB2	&	<27.8			&	0.13	&	\ldots	&	Jen09, Shk10	\\
GJ 3131	&	J02033--212	&	SB2	&		\ldots		&	\ldots	&	\ldots	&	Jef18	\\
V374 And	&	J02069+451	&	SB2	&	897.0	 $\pm$ 	2.3	&	\ldots	&	0.70	&	Spe19	\\
GJ 3160	&	J02289+120	&	SB2	&		\ldots		&	\ldots	&	\ldots	&	Jef18	\\
GJ 3235	&	J03346--048	&	ST3	&		\ldots		&	\ldots	&	\ldots	&	Jef18	\\
GJ 3236	&	J03372+691	&	DESB2	&	0.7712600	 $\pm$ 	0.0000023	&	0.0143	&	0.75	&	Irw09, Shk10	\\
GJ 3239	&	J03375+178N	&	SB2	&	<33.2			&	0.18	&	\ldots	&	Shk10, Fou18	\\
GJ 3240	&	J03375+178S	&	SB2	&	<0.4			&	0.01	&	\ldots	&	Shk10	\\
HD 278874	&		&	SB2	&		\ldots		&	\ldots	&	\ldots	&	Mon18	\\
Wolf 227	&	J03526+170	&	SB2	&		\ldots		&	\ldots	&	\ldots	&	Bon13a, Jef18	\\
HD 24916 B	&	J03574--011	&	SB	&		\ldots		&	\ldots	&	\ldots	&	Zak79	\\
LP 414--117	&	J04123+162	&	SB2	&	128.114			&	\ldots	&	0.60	&	Ben08	\\
1RXS J042441.9--064725	&	J04247--067	&	ST3	&	<1.9			&	0.02	&	\ldots	&	Shk10	\\
GJ 3282	&	J04252+080S	&	SB2	&		\ldots		&	\ldots	&	\ldots	&	Jef18	\\
LP 775--31	&	J04352--161	&	SB2	&		\ldots		&	\ldots	&	\ldots	&	Rei09	\\
TYC 694--1183--1	&	J04414+132	&	SB	&		\ldots		&	\ldots	&	\ldots	&	Mer09	\\
LP 415--345	&	J04425+204	&	SB2	&		\ldots		&	\ldots	&	\ldots	&	Sta97	\\
LP 416--43	&	J04480+170	&	SB	&	8.49474	 $\pm$ 	0.00007	&	6.2	&	\ldots	&	Gri85	\\
1RXS J044847.6+100302	&	J04488+100	&	SB2	&		\ldots		&	\ldots	&	\ldots	&	Jef18	\\
GJ 3322 A	&	J05019+099	&	SB2	&		\ldots		&	\ldots	&	0.57	&	Del98, Del99b	\\
HD 285190	&	J05032+213	&	SB2	&		\ldots		&	\ldots	&	\ldots	&	Jef18, Fou18	\\
Capella	&		&	SB2	&	104.02173	 $\pm$ 	0.00022	&	0.7357	&	0.99	&	She85, Tor09	\\
GJ 1080	&	J05282+029	&	SB	&		\ldots		&	\ldots	&	\ldots	&	Jen09	\\
HD 35956	&		&	SB1	&		\ldots		&	\ldots	&	\ldots	&	Kat13	\\
Ross 42	&	J05322+098	&	SB2	&	<61.1			&	0.23	&	1.00	&	Mar87, Shk10	\\
V371 Ori	&	J05337+019	&	SB1	&	0.60417356	 $\pm$ 	0.00000025	&	\ldots	&	\ldots	&	Rein12, Bar21	\\
Ross 45 B	&	J05342+103S	&	SB	&		\ldots		&	\ldots	&	\ldots	&	Rein12	\\
V1402 Ori	&	J05402+126	&	SB2	&	<138.9			&	0.52	&	\ldots	&	Shk10	\\
Wolf 237	&	J05466+441	&	SB2	&		\ldots		&	\ldots	&	\ldots	&	Jef18	\\
Ross 59	&	J05532+242	&	SB2	&	721	 $\pm$ 	2	&	0.6656	&	0.33	&	Bar18	\\
LP 86--173	&	J06054+608	&	SB1	&	0.30992678	 $\pm$ 	0.00000048	&	0.0049	&	0.93	&	New16, Win20	\\
TYC 4525--194--1	&	J06171+751	&	ST3?	&		\ldots		&	\ldots	&	\ldots	&	Fou18	\\
G 108--4	&	J06298--027	&	SB2	&	<46.9			&	0.16	&	\ldots	&	Shk10	\\
1RXS J070005.1--190115	&	J07001--190	&	SB2	&	6.56025	 $\pm$ 	0.00030	&	0.002797	&	\ldots	&	Bar21	\\
QY Aur	&	J07100+385	&	SB2	&	10.42673	 $\pm$ 	0.00010	&	0.0687	&	0.85	&	TP86, Win20	\\
TYC 4530--1414--1	&	J07119+773	&	SB1	&		\ldots		&	\ldots	&	\ldots	&	Jef18	\\
1RXS J073138.4+455718	&		&	SB2?	&		\ldots		&	\ldots	&	\ldots	&	Fou18	\\
GJ 3447	&	J07320+173E	&	SB	&		\ldots		&	\ldots	&	\ldots	&	Gaia22a	\\
Castor	&		&	SB2 + SB2	&	167570	 $\pm$ 	840	&	83.9052	&	0.79	&	Bel97, Cur06, Tor22	\\
Castor C	&	J07346+318	&	DESB2	&	0.8142822	 $\pm$ 	0.0000010	&	0.01809	&	1.00	&	Joy26, Giz02, Tor02, Tor22	\\
GJ 282 C	&	J07361--031	&	SB1	&	6591	 $\pm$ 	177	&	6.224	&	0.34	&	Bar21	\\
GJ 3461	&	J07418+050	&	EB?/SB2	&		\ldots		&	\ldots	&	\ldots	&	Jef18	\\
1RXS J075434.3+083213	&	J07545+085	&	SB1	&		\ldots		&	\ldots	&	\ldots	&	Jef18	\\
FP Cnc B	&	J08089+328	&	SB2	&	<395.5			&	0.92	&	\ldots	&	Shk10	\\
CU Cnc	&	J08316+193S	&	DESB2	&	2.771468	 $\pm$ 	0.000004	&	0.03624	&	0.92	&	Del99a, Del99b, Tor10	\\
LSPM J0835+1408	&	J08353+141	&	ST3	&		\ldots		&	\ldots	&	\ldots	&	Ski18	\\
GJ 319 A	&	J08427+095	&	SB1	&	20.9491	 $\pm$ 	0.0019	&	0.021	&	0.22	&	Duq88	\\
GJ 3522	&	J08589+084	&	SB2	&	7.5555	 $\pm$ 	0.0002	&	\ldots	&	0.82	&	Rei97b, Del99b	\\
GJ 3540	&	J09120+279	&	SB2	&		\ldots		&	\ldots	&	\ldots	&	Jef18	\\
LP 427--16	&	J09140+196	&	SB1	&		\ldots		&	\ldots	&	\ldots	&	Bar21	\\
HD 79210	&	J09143+526	&	SB1	&		\ldots		&	\ldots	&	\ldots	&	Jef18	\\
GJ 9303	&	J09362+375	&	SB2	&		\ldots		&	\ldots	&	\ldots	&	Mal14a	\\
LP 728--70	&	J09506--138	&	SB2	&		\ldots		&	\ldots	&	\ldots	&	Jef18	\\
GJ 372	&	J09531--036	&	SB2	&	47.709	 $\pm$ 	0.053	&	39.06	&	0.76	&	Har96, Jef18, Bar18	\\
GJ 373	&	J09561+627	&	SB?	&		\ldots		&	\ldots	&	\ldots	&	Rein12	\\
LP 790--2	&	J10182--204	&	SB2	&	5.922845	 $\pm$ 	0.000061	&	0.048242	&	0.58	&	Bar18	\\
GJ 3612	&	J10354+694	&	SB2	&	119.41	 $\pm$ 	0.04	&	0.3464	&	0.51	&	Bar18	\\
LP 127--502	&	J10368+509	&	SB2?	&		\ldots		&	\ldots	&	\ldots	&	Fou18	\\
GJ 3626	&	J10504+331	&	SB1	&	2996	 $\pm$ 	31	&	0.2355	&	\ldots	&	Bar21	\\
GJ 3630	&	J10520+005	&	ST3/SQ4	&	<28.1			&	0.11	&	\ldots	&	Shk10	\\
LP 491--51	&	J11036+136	&	SB1	&		\ldots		&	\ldots	&	\ldots	&	Jef18	\\
GJ 426.1 A	&		&	SB	&		\ldots		&	\ldots	&	\ldots	&	Abt76	\\
GJ 455	&	J12023+285	&	SB2	&		\ldots		&	\ldots	&	\ldots	&	Giz02	\\
LP 734--34	&	J12104--131	&	SB2	&	33.6551	 $\pm$ 	0.0046	&	0.0479	&	0.95	&	Win20	\\
GJ 3719	&	J12169+311	&	SB2	&		\ldots		&	\ldots	&	\ldots	&	Fou18	\\
G 148--43	&	J12191+318	&	SB2	&		\ldots		&	\ldots	&	\ldots	&	Jef18	\\
GJ 3731	&	J12299--054W	&	SB2	&		\ldots		&	\ldots	&	\ldots	&	Jef18	\\
G 123--45	&	J12364+352	&	SB1	&	34.7557	 $\pm$ 	0.0041	&	0.0219	&	\ldots	&	Win20	\\
DP Dra	&	J12490+661	&	ST3	&	54.075	 $\pm$ 	0.006	&	\ldots	&	0.95	&	Del99b	\\
GJ 507 B	&	J13195+351E	&	SB1	&		\ldots		&	\ldots	&	\ldots	&	Jef18	\\
GQ Vir	&	J14130--120	&	SB2	&		\ldots		&	\ldots	&	\ldots	&	Jef18	\\
GJ 1182	&	J14155+046	&	SB2	&	154.2	 $\pm$ 	0.1	&	0.35835	&	0.66	&	Bar18	\\
PM J14171+0851	&	J14171+088	&	SB2	&		\ldots		&	\ldots	&	\ldots	&	Jef18	\\
GJ 3861	&	J14368+583	&	SB2	&		\ldots		&	\ldots	&	\ldots	&	Giz02, Jef18	\\
GJ 3875	&	J14549+411	&	SB	&		\ldots		&	\ldots	&	\ldots	&	Ski18	\\
G 136--35	&	J14564+168	&	SB	&		\ldots		&	\ldots	&	\ldots	&	Ski18	\\
StKM 1--1240	&	J15238+561	&	SB2	&		\ldots		&	\ldots	&	\ldots	&	Fou18	\\
GJ 3910	&	J15319+288	&	SB1	&		\ldots		&	\ldots	&	\ldots	&	Jen09	\\
UU UMi	&	J15412+759	&	SB2	&	5240	 $\pm$ 	410	&	4.7	&	0.59	&	Bar18, Bar21	\\
GJ 595	&	J15421--194	&	SB1	&		\ldots		&	\ldots	&	\ldots	&	Nid02	\\
GJ 3916	&	J15474--108	&	ST3	&	3028	 $\pm$ 	23	&	3.59	&	1.05	&	Bar21	\\
LP 177--102	&	J15474+451	&	DESB2	&	3.5500184	 $\pm$ 	0.0000018	&	0.036535	&	1.00	&	Moc02, Har11, Bir12	\\
$\sigma$ CrB A	&		&	SB2	&	1.139791423	 $\pm$ 	0.00000008	&	0.013106	&	0.96	&	Str03, Rag09	\\
CM Dra	&	J16343+571	&	DESB2	&	1.268389985	 $\pm$ 	0.000000005	&	0.017502	&	0.93	&	Mor09, Tor10, Scha19	\\
V1054 Oph	&	J16554--083S	&	ST3	&	2.965522	 $\pm$ 	0.000014	&	0.05	&	0.90	&	Joy47, Pet84, Del98	\\
LP 331--57 A	&	J17038+321	&	SB2?	&		\ldots		&	\ldots	&	\ldots	&	Shk09, Fou18	\\
GJ 3991	&	J17095+436	&	SB1	&		\ldots		&	\ldots	&	\ldots	&	Rei97b, Del99b	\\
Wolf 1473	&	J17464--087	&	SB2	&	83.926	 $\pm$ 	0.032	&	0.27559	&	0.93	&	Mal14a, Win20	\\
GJ 1230 A	&	J18411+247S	&	SB2	&	5.0688	 $\pm$ 	0.00005	&	\ldots	&	0.95	&	GR96, Del99b  	\\
GJ 735	&	J18554+084	&	SB2	&		\ldots		&	\ldots	&	\ldots	&	Mar87, Giz02, Kar04	\\
GJ 4091 B	&	J18563+544	&	SB2	&		\ldots		&	\ldots	&	\ldots	&	Fou18	\\
G 125--15	&	J19312+361	&	SB2	&	<0.7			&	0.01	&	\ldots	&	Shk10	\\
RX J1935.4+3746	&	J19354+377	&	SB1	&		\ldots		&	\ldots	&	\ldots	&	Shk09, Jef18	\\
LP 869--19	&	J19420--210	&	SB2	&		\ldots		&	\ldots	&	\ldots	&	Mal14a	\\
{V1513 Cyg} & {J20050+544} & {SB1} & {0.49670418 $\pm$ 0.00000005} & {0.010537 $\pm$ 0.000005} & \ldots & {Giz97b, Mac18} \\
LP 574--21	&	J20105+065	&	SB2	&	<40.1			&	0.19	&	\ldots	&	Shk10	\\
LP 395--8 A	&	J20198+229	&	SB2	&	1.1293392	 $\pm$ 	0.0000067	&	0.01057	&	0.56	&	Bar18	\\
1RXS J203011.0+795040	&	J20301+798	&	SB2	&		\ldots		&	190.469	&	\ldots	&	Jef18	\\
GJ 4155	&	J20409--101	&	SB	&		\ldots		&	\ldots	&	\ldots	&	Rei12	\\
GJ 4161	&	J20445+089N	&	SB2	&		\ldots		&	\ldots	&	\ldots	&	Jef18	\\
GJ 810 A	&	J20556--140N	&	SB2	&	812	 $\pm$ 	51	&	0.841	&	0.83	&	Bar18	\\
FR Aqr	&	J20568--048	&	SB2	&		\ldots		&	\ldots	&	\ldots	&	Jef18	\\
Ross 775	&	J21296+176	&	SB2	&	53.221	 $\pm$ 	0.004	&	\ldots	&	1.00	&	Mar87, Del99b	\\
GJ 4213	&	J21442+066	&	SB1	&		\ldots		&	\ldots	&	\ldots	&	Jef18	\\
EZ Aqr	&	J22385--152	&	ST3	&	822.6	 $\pm$ 	0.6	&	1189	&	0.53	&	Del99c	\\
FK Aqr	&	J22387--206S	&	SB2	&	4.08322	 $\pm$ 	0.00004	&	0.039372	&	0.80	&	Her65, Del99b	\\
FL Aqr	&	J22387--206N	&	SB1	&	1.795	 $\pm$ 	0.017	&	0.00353	&	0.60	&	Dav14	\\
LP 521--79	&	J23063+126	&	SB2(3?)	&	<58.9			&	0.31	&	\ldots	&	Shk10	\\
GJ 4314	&	J23096--019	&	SB2	&		\ldots		&	\ldots	&	\ldots	&	Jef18	\\
GJ 4327	&	J23174+382	&	SB2	&		\ldots		&	\ldots	&	\ldots	&	Jef18	\\
GJ 1284	&	J23302--203	&	SB2	&	11.838033	 $\pm$ 	0.000076	&	6.48275	&	0.19	&	Giz02, Jef18, Car21	\\
Ross 676	&	J23439+647	&	SB2	&	<3			&	0.04	&	\ldots	&	Shk10	\\
GJ 912	&	J23556--061	&	SB1	&	5188	 $\pm$ 	58	&	0.505	&	0.12	&	Bar21	\\
GJ 4379	&	J23573--129W	&	SB2	&		\ldots		&	\ldots	&	\ldots	&	Giz02, Jef18	\\
Wolf 1051	&	J23585+076	&	ST3	&	4634	 $\pm$ 	17	&	0.4436	&	0.55	&	Bar21	\\\hline
\noalign{\smallskip}

\end{longtable}
\tablefoot{
\tablefoottext{a}{ 
A spectroscopic binary (SB) that displays one or two lines in the spectrum is a single- or double-lined spectroscopic binary, respectively, and abbreviated as SB1 or SB2.
A spectroscopic triple (ST) showing two or three lines is a double- or triple-lined spectroscopic triple, and abbreviated as ST2 or ST3.
Spectroscopic quadruples (SQ) are much more rare, and typically are three- or four-lined systems, denominated SQ3 and SQ4, respectively. 
In our sample, only one star is classified as SQ4 (\object{GJ~3630}, J10520+005).
}
\tablefoottext{b}{While recognising the contributions of contemporary investigations 
(e.g. \citealt{Shk10}, \citealt{Jef18}, \citealt{Bar18,Bar21}, and \citealt{Win20} are the source of many of the parameters in this table), references are often limited for simplicity to the original sources that first identified the systems. 
However, exceptions are made for prominent systems such as \object{Castor~C} or nearby stars with a long history of observations, such as \object{CM~Dra} and \object{CU~Cnc}. References --
Abt70: \cite{Abt70};
Bar18: \cite{Bar18};
Bar21: \cite{Bar21};
Bel97: \cite{Bel97};
Ben08: \cite{Ben08};
Bir12: \cite{Bir12};
Bon13a: \cite{Bon13a};
Bop77: \cite{Bop77};
Cur06: \cite{Cur06};
Dav14: \cite{Dav14};
Del98: \cite{Del98};
Del99a: \cite{Del99a};
Del99b: \cite{Del99b};
Del99c: \cite{Del99c};
Fou18: \cite{Fou18};
Gaia22a: \cite{Gaia22a};
Giz02: \cite{Giz02};
{Giz97b: \cite{Giz97b};}
GR96: \cite{GR96};
Har11: \cite{Har11};
Har96: \cite{Har96};
Her65: \cite{Her65};
Irw09: \cite{Irw09};
Jef18: \cite{Jef18};
Jen09: \cite{Jen09};
Joy26: \cite{Joy26};
Joy47: \cite{Joy47};
Kar04: \cite{Kar04};
Kat13: \cite{Kat13};
Mac02: \cite{Moc02};
{Mac18: \cite{Mac18};}
Mal14a: \cite{Mal14a};
Mar87: \cite{Mar87};
Mer09: \cite{Mer09};
Moc02: \cite{Moc02};
Mon18: \cite{Mon18};
Mor09: \cite{Mor09};
New16: \cite{New16};
Nid02: \cite{Nid02};
Pet84: \cite{Pet84};
Rag09: \cite{Rag09};
Rei09: \cite{Rei09};
Rein12: \cite{Rein12};
Rei97b: \cite{Rei97b};
Scha19: \cite{Scha19};
She85: \cite{She85};
Shk09: \cite{Shk09};
Shk10: \cite{Shk10};
Shk12: \cite{Shk12};
Ski18: \cite{Ski18};
Spe19: \cite{Spe19};
Sta97: \cite{Sta97};
Ste14: \cite{Ste14};
Str03: \cite{Str03};
Str93: \cite{Str93};
TP86: \cite{TP86};
Tor09: \cite{Tor09};
Tor22: \cite{Tor22};
Win20: \cite{Win20};
Zak79: \cite{Zak79}.
}
}
}

\newpage
\begin{table}
\footnotesize
\caption{Eclipsing binaries in our sample$^a$.}
\label{tab:eclipsing} 
\centering
\begin{tabular}{lccccc}
    \hline\hline 
    \noalign{\smallskip}  
&	GJ 3236			&	Castor C			&	CU Cnc			&	LP 177--102			&	CM Dra			\\
\hline
\noalign{\smallskip}
\noalign{\smallskip}																		
Karmn	&	J03372+691			&	J07346+318			&	J08316+193S			&	J15474+451			&	J16343+571			\\
\noalign{\smallskip}																					
$\mathcal{M}_1$ {[}$\mathcal{M}_\odot${]}	&	0.376	 $\pm$ 	0.017	&	0.5992	 $\pm$ 	0.0047	&	0.4349	 $\pm$ 	0.0012	&	0.2576	 $\pm$ 	0.0085	&	0.23102	 $\pm$ 	0.00089	\\
\noalign{\smallskip}																					
$\mathcal{M}_2$ {[}$\mathcal{M}_\odot${]}	&	0.281	 $\pm$ 	0.015	&	0.5992	 $\pm$ 	0.0047	&	0.39922	 $\pm$ 	0.00089	&	0.2585	 $\pm$ 	0.0080	&	0.21409	 $\pm$ 	0.00083	\\
\noalign{\smallskip}																					
$\mathcal{R}_1$ {[}$\mathcal{R}_\odot${]}	&	0.3828	 $\pm$ 	0.0070	&	0.6191	 $\pm$ 	0.0057	&	0.4323	 $\pm$ 	0.0055	&	0.2895	 $\pm$ 	0.0068	&	0.2534	 $\pm$ 	0.0019	\\
\noalign{\smallskip}																					
$\mathcal{R}_2$ {[}$\mathcal{R}_\odot${]}	&	0.2992	 $\pm$ 	0.0075	&	0.6191	 $\pm$ 	0.0057	&	0.3916	 $\pm$ 	0.0094	&	0.2895	 $\pm$ 	0.0068	&	0.2398	 $\pm$ 	0.0018	\\
\noalign{\smallskip}																					
$P_{\rm orb}$ {[}d{]}	&	0.7712600	 $\pm$ 	0.0000023	&	0.8142822	 $\pm$ 	0.0000010	&	2.771468	 $\pm$ 	0.000004	&	3.5500184	 $\pm$ 	0.0000018	&	1.268389985	 $\pm$ 	0.000000005	\\
\noalign{\smallskip}																					
$a$ {[}au{]}	&	0.01430			&	0.01819			&	0.03624			&	0.03654			&	0.01750			\\
\noalign{\smallskip}																					
Reference	&	{Irw09, Shk10}			&	{Joy26, Giz02, Tor22}			&	{Del99a, Del99b, Tor10}			&	{Moc02, Har11, Bir12}			&	{Mor09, Tor10, Scha19}			\\
\noalign{\smallskip}
\hline
\end{tabular}
\tablefoot{
\tablefoottext{a}{
\,\,Bir12	: \cite{Bir12};
Del99a: \cite{Del99a};
Del99b: \cite{Del99b};
Giz02: \cite{Giz02};
Har11: \cite{Har11};
Irw09: \cite{Irw09};
Joy26: \cite{Joy26};
Moc02: \cite{Moc02};
Mor09: \cite{Mor09};
Scha19: \cite{Scha19};
Shk10: \cite{Shk10};
Tor10: \cite{Tor10};
Tor22: \cite{Tor22}.}   
}
\end{table}

{\fontsize{8}{10}\selectfont 
\begin{longtable}{lccl}
\caption{\label{tab:description} Description of the online table.}
\\
    \hline
    \hline 
    \noalign{\smallskip}
    Parameter & Units & Column(s) & Description \\
    \noalign{\smallskip}
    \hline
    \noalign{\smallskip}
\endfirsthead
\caption{Description of the online table (cont.).}\\
\hline\hline
\noalign{\smallskip}
    Parameter & Units & Column(s) & Description \\
\hline
\noalign{\smallskip}
\endhead
\hline
\endfoot
\multicolumn{4}{l}{\em Identification} \\		\hline			
{\tt ID\_star, ID\_system}	&	\ldots	&	1, 2	&	Star and system identifiers$^a$	\\
{\tt Name}	&	\ldots	&	3	&	Discovery name or most common name$^b$	\\
{\tt GJ}	&	\ldots	&	4	&	Gliese-Jahreiss catalogue number$^c$	\\
{\tt Karmn}	&	\ldots	&	5	&	Carmencita star identifier (JHHMMm+DDd)$^d$	\\
{\tt RA\_J2016, DE\_J2016}	&	hms, dms	&	6, 7	&	Right ascension and declination in the epoch J2016.0	\\
{\tt SpT, SpTnum, SpT\_ref}	&	\ldots	&	8--10	&	Spectral type, its numerical format, and the reference$^e$	\\
\hline		\multicolumn{4}{l}{\em Multiplicity} \\		\hline			
{\tt Type, Class}	&	\ldots	&	11, 12	&	Type of system and multiplicity class$^f$	\\
{\tt Component, System}	&	\ldots	&	13, 14	&	Component designation and their resolution$^g$	\\
{\tt SB, SB\_ref}	&	\ldots	&	15, 16	&	Type of spectroscopic system and reference$^h$	\\
{\tt Discoverer}	&	\ldots	&	17	&	Discoverer code$^i$	\\
{\tt Group, Group\_ref}	&	\ldots	&	18, 19	&	Stellar group (SKG, cluster or association) and reference$^j$	\\
{\tt Nplanet, Planet\_ref}	&	\ldots	&	20, 21	&	Number of confirmed planets and reference	\\
{\tt Category}	&	\ldots	&	22	&	Category 1--3 as described in Sect.~\ref{ssection:fraction}	\\
{\tt Remarks}	&	\ldots	&	23	&	Comments and remarks on multiplicity	\\
{\tt WDS\_...}	&	\ldots	&	24--32	&	Washington double star catalogue data$^k$	\\
{\tt s01, s02}	&	au	&	33, 37	&	Physical separation between the components in multiple systems$^l$	\\
{\tt s01\_ref}	&	\ldots	&	36	&	Reference for the closest component$^m$	\\
{\tt rho01, rho02}	&	arcsec	&	34, 38	&	Projected angular separation ($s = \rho d$)	\\
{\tt theta01, theta02}	&	deg	&	35, 39	&	Positional angle$^n$	\\
{\tt muratio, deltaPA, deltad}	&	\ldots	&	40--42	&	$\mu{\rm ratio}$, $\Delta PA$, $\Delta d$ criteria for physical association$^o$	\\
\hline		\multicolumn{4}{l}{\em Stellar parameters} \\		\hline			
{\tt M\_Msol, eM\_Msol}	&	$\mathcal{M}_\odot$	&	43, 44	&	Stellar mass and its error 	\\
{\tt R\_Rsol, eR\_Msol}	&	$\mathcal{R}_\odot$	&	45, 46	&	Stellar radius and its error 	\\
{\tt MR\_ref}	&	\ldots	&	47	&	Reference for mass and radius	\\
{\tt L\_Lsol, eL\_Lsol, L\_ref}	&	$\mathcal{L}_\odot$	&	48--50	&	Luminosity, error, and reference	\\
{\tt MG\_mag, eMG\_mag}	&	mag	&	51, 52	&	Absolute magnitude in $G$ and its error	\\
{\tt Teff\_K, eTeff\_K, Teff\_K\_ref}	&	K	&	53--55	&	Effective temperature and its error	\\
{\tt logg, elogg}	&	dex	&	56, 57	&	Surface gravity and its error (reference shared with {\tt L\_Lsol})	\\
{\tt Fe\_H, Fe\_H\_ref }	&	\ldots	&	58, 59	&	Iron abundance (metallicity) and its error	\\
\hline		\multicolumn{4}{l}{\em System parameters} \\		\hline			
{\tt q}	&	\ldots	&	60	&	Mass ratio	\\
{\tt P01, eP01, P01\_ref, P02, eP02}	&	d, a	&	61--65	&	Orbital periods, errors and reference	\\
{\tt Ug\_J, eUg\_J}	&	J	&	66, 67	&	Binding energy and its error	\\
{\tt crit\_...}	&	Boolean	&	68--75	&	Criteria for unresolved companions$^p$	\\
{\tt Candidate}	&	Boolean	&	76	&	Candidate to unresolved companion$^q$	\\
\hline		\multicolumn{4}{l}{\em Astrometry} \\		\hline			
{\tt \_id}	&	\ldots	&	77--79	&	Catalogue identifiers	\\
{\tt ra, ra\_error}	&	deg	&	80, 81	&	Barycentric right ascension and its error in the epoch {J2016.0}	\\
{\tt dec, dec\_error}	&	deg	&	82, 83	&	Barycentric declination and its error in the epoch {J2016.0} 	\\
{\tt parallax, parallax\_error, parallax\_ref}	&	mas	&	84--86	&	Parallax, error and reference	\\
{\tt pmra, pmra\_error, pmdec, pmdec\_error, pm\_ref}	&	mas\,a$^{-1}$	&	87--91	&	Proper motion in right ascension and declination, errors, and reference	\\
{\tt rv, rv\_error, rv\_ref}	&	km\,s$^{-1}$	&	92--94	&	Radial velocity, error, and reference	\\
{\tt ruwe}	&	\ldots	&	95	&	Renormalised unit weight error from {\em Gaia} DR3	\\
{\tt {[}Statistics DR3{]}}	&	\ldots	&	96--110	&	Additional statistical indicators in {\em Gaia} DR3$^r$	\\
\hline		\multicolumn{4}{l}{\em Photometry} \\		\hline

{\tt NN\_mag, eNN\_mag, Qf\_MM}	&	mag	&	111--132	&	Photometric magnitudes and quality flags in up to 10 passbands$^s$	\\

\end{longtable}
\tablefoot{
\tablefoottext{a}{\texttt{ID\_star} is a unique identifier that sorts the table by right ascension but prioritising that components of the same system (equal \texttt{ID\_system}) are together and sorted by decreasing brightness.}\\
\tablefoottext{b}{Name of the star, Simbad-searchable, obeying the following priority ($n$ designates a natural number): 
Proper name,
variable in constellation (V* V$n$ Con; but not suspected variables, SV*),
Henry Draper (HD~$n$, with $n \leq$ 225300),
Gliese-Jahreiss (GJ~$n$, only if $n<$ 4000),
Bonner Durchmusterung (BD$\pm n$~$n$),
Luyten (LP~$n$-$n$),
Giclas (G~$n$-$n$, only if unique Giclas designation),
Luyten (LHS~$n$),
other designations in chronological order (Haro, StKM/StM, 1RXS/RXS, HIP, LSPM, PM, NLTT, GSC, TYC, MCC, R78b, I81, 2MUCD),
catalog identifier ({\em Gaia}~DR3, 2MASS, UCAC4).}\\
\tablefoottext{c}{Gliese-Jahreiss (GJ) designation entry, when available. This includes the first catalogue by \cite{Gli57} and its update \citep{Gli69}, the supplement by \cite{Woo70}, and the succeeding editions \citep{GJ79}.}\\
\tablefoottext{d}{``HHMMm$\pm$DDd'' are the truncated equatorial coordinates.
For close stars with separations less than 5\,arcsec, a position of the star in the system is added as ``N'', ``S'', ``E'' or ``W''.}\\
\tablefoottext{e}{{\tt SpTnum} = 10.0 for O0.0\,V, 20.0 for B0.0\,V, 30.0 for A0.0\,V, 40.0 for F0.0\,V, 50.0 for G0.0\,V, 60.0 for K0.0\,V, 70.0 for M0.0\,V, 70.5 for M0.5\,V, 80.0 for L0. The values 0.0 and 999 are reserved for white dwarfs and stars with a luminosity class other than main sequence (V), respectively.
Spectral types were rounded when necessary (e.g. M0.0 instead of M0.1, or M4.0 instead of M3.8).
`This work' refers to spectral type photometric estimation from absolute magnitudes as \cite{Cif20}.}\\
\tablefoottext{f}{`Candidate' and `Multiple+' must be read as `new unresolved binary candidate'.}\\
\tablefoottext{g}{The system nomenclature follows the scheme in Fig.~\ref{fig:nomenclature}.}\\
\tablefoottext{h}{Definitions of SB1, SB2, ST2, ST3, and SQ are shown in Table~\ref{tab:spectroscopic}.}\\
\tablefoottext{i}{WDS discoverer code or literature reference.}\\
\tablefoottext{j}{Abbreviations follow the scheme outlined in Table~\ref{tab:skg}.}\\
\tablefoottext{k}{{\tt id}: WDS identifier (see also column {\tt Discoverer}); 
{\tt comp}: Component designations;
{\tt obs1, obs2}: First and last observation years;
{\tt pa2}: Positional angle in the most recent measurement;
{\tt sep2}: Separation in the most recent measurement;
{\tt mag1, mag2}: Magnitudes of the two components.
More details can be found in the WDS website: \url{http://www.astro.gsu.edu/wds/}.}\\
\tablefoottext{l}{In systems of three or more components, {\tt 01} denotes the separation between the closest components, i.e., {\tt s01} $<$ {\tt s02}.}\\
\tablefoottext{m}{The reference is displayed when the value is not calculated in this work using {\em Gaia} data. The reference for {\tt s02} is {\em Gaia} in all cases. This also applies to the orbital periods.}\\
\tablefoottext{n}{Measured eastward from the north in the epoch 2016.0. In the programmatic computation of this parameter caution must be taken regarding the ``quadrant ambiguity'', i.e., a 180-deg erroneous difference.}\\
\tablefoottext{o}{For triple or higher-order systems, this association refers to the primary component.}\\
\tablefoottext{p}{Criteria for physical association ({\tt crit\_association}) and statistical metrics from {\em Gaia} as described in Table~\ref{tab:criteria} ({\tt crit\_ruwe, crit\_ipd\_ruwe, crit\_rv, crit\_rv\_error, crit\_ipd\_fmp, crit\_dupl\_source, crit\_non\_single, crit\_DR3\_non\_single}).
Orbital: Orbital model for an astrometric binary; OrbitalTargetedSearch: Orbital model for a priori known systems, with a subset containing suffix `Validated'; SB1: Single-lined spectroscopic binary; SB2: Double-lined spectroscopic binary; SB2C: Double-lined spectroscopic binary with circular orbit; AstroSpectroSB1: Combined astrometric and single lined spectroscopic orbital model.}\\
\tablefoottext{q}{It refers to unknown unresolved binaries. That is, if an unresolved companion already exists (e.g. an spectroscopic binary), the value is `false’ regardless of the {\em Gaia} metrics (Table~\ref{tab:criteria}).}\\
\tablefoottext{r}{Statistics from DR3 related to the criteria for unresolved binarity:
{\tt 
astrometric\_excess\_noise, 
astrometric\_excess\_noise\_sig, 
phot\_bp\_rp\_excess\_factor, 
phot\_bp\_n\_blended\_transits, 
phot\_rp\_n\_blended\_transits,
phot\_variable\_flag,
rv\_chisq\_pvalue, 
rv\_amplitude\_robust,
rv\_nb\_transits,  
renormalised\_gof,
astrometric\_n\_obs\_al, 
astrometric\_n\_good\_obs\_al,
ipd\_gof\_harmonic\_amplitude,
duplicated\_source
} (see Sect.~\ref{ssection:unresolved}).}\\
\tablefoottext{s}{
{\tt BP}, {\tt G}, {\tt RP}: $G_{B_P}$, $G$, and $G_{R_P}$ from {\em Gaia} DR3 \citep{Gaia23}; 
{\tt J}, {\tt H}, {\tt Ks}: $J$, $H$, and $Ks$ from 2MASS \citep{Skr06}; 
{\tt W1}, {\tt W2}, {\tt W3}, {\tt W4}: $W1$, $W2$, $W3$, and $W4$ from AllWISE \citep{Cut14}.
The photometric uncertainties in the {\em Gaia} passbands have been calculated by us $\Delta \lambda = |-2.5/ \ln{10} \times \Delta F_\lambda/F_\lambda|$, where $F_\lambda$ and $\Delta F_\lambda$ are the flux and its error in the $\lambda$ passband, using the errors in the corresponding fluxes, and the zero points as provided by VizieR.}\\
}}

{\fontsize{7}{10}\selectfont 
\begin{landscape}
\begin{longtable}{lllllcccccll}
\caption{Multiple systems with components at separations larger than 10$^4$\,au.}
\label{tab:widest} 
\\
\hline\hline \noalign{\smallskip}
Name	& 	Karmn	& 	Spectral	& 	Comp.	& 	Discoverer$^a$	& 	$s$ {[}au{]}		& 	$\pi$			& 	$\mu_{\rm total}$			& 	$|U^\ast_g|$			& 	SKG	& 	SKG	\\
	& 		& 	type	& 		& 		& 			& 	{[}mas{]}			& 	{[}mas\,a$^{-1}${]}			& 	{[}$10^{33}$\,J{]}			& 		& 	Ref.$^b$	\\
\noalign{\smallskip}
\hline \noalign{\smallskip}
\endfirsthead
\caption{Multiple systems with components at separations larger than 10$^4$\,au. (cont.).}\\
\hline\hline 
\noalign{\smallskip}
Name	& 	Karmn	& 	Spectral	& 	Comp.	& 	Discoverer$^a$	& 	$s$ {[}au{]}		& 	$\pi$			& 	$\mu_{\rm total}$			& 	$|U^\ast_g|$			& 	SKG	& 	SKG	\\
	& 		& 	type	& 		& 		& 			& 	{[}mas{]}			& 	{[}mas\,a$^{-1}${]}			& 	{[}$10^{33}$\,J{]}			& 		& 	Ref.$^b$	\\
\hline \noalign{\smallskip}
\endhead
\hline
\endfoot
\noalign{\smallskip}
GJ 3022	&	J00169+200	&	M3.5 V	&	A	&		&			&	29.310	 $\pm$ 	0.097	&	239.537	 $\pm$ 	0.209	&				&		&		\\
G 131--47 B	&		&	M3.5 V	&	B	&	CRC 43	&	36.96		&	29.139	 $\pm$ 	0.071	&	233.362	 $\pm$ 	0.318	&	4764	 $\pm$ 	727	&		&		\\
LP 404--54	&		&	M5.0 V	&	C	&	Sma21	&	60679.5		&	28.873	 $\pm$ 	0.034	&	233.800	 $\pm$ 	0.048	&	1.742	 $\pm$ 	0.211	&		&		\\
\noalign{\smallskip}        \noalign{\smallskip}																												
V493 And A	&	J00341+253	&	M0.0 V	&	A	&		&			&	20.101	 $\pm$ 	0.039	&	127.914	 $\pm$ 	0.054	&				&	AB Dor	&	Jan17	\\
V493 And B	&		&	K7 V	&	B	&	SKF 220	&	77.91		&	19.720	 $\pm$ 	0.037	&	129.463	 $\pm$ 	0.056	&	10445	 $\pm$ 	567	&	AB Dor	&	$\star$	\\
UCAC4 578--001365	&		&	M4.0 V	&	C	&	Sma21	&	15617.5		&	19.672	 $\pm$ 	0.234	&	127.043	 $\pm$ 	0.353	&	20.44	 $\pm$ 	1.29	&	AB Dor	&	$\star$	\\
\noalign{\smallskip}        \noalign{\smallskip}																												
EX Cet	&		&	K0.5 V	&	A	&		&			&	41.564	 $\pm$ 	0.024	&	197.856	 $\pm$ 	0.032	&				&	Ple/beta Pic	&	$\star$	\\
LP 648--20$^c$	&	J01369--067	&	M3.5 V	&	B	&	CAB 3	&	14678.4		&	41.699	 $\pm$ 	0.045	&	200.547	 $\pm$ 	0.052	&	34.09	 $\pm$ 	5.33	&	Ple/beta Pic	&	Alo15b	\\
\noalign{\smallskip}        \noalign{\smallskip}																												
G 221--21	&	J03454+729	&	M1.5 V	&	A	&		&			&	39.088	 $\pm$ 	0.013	&	487.290	 $\pm$ 	0.016	&				&		&		\\
LP 31--200	&		&	M3.5 V	&	B	&	WIS  99	&	14667.0		&	39.146	 $\pm$ 	0.015	&	488.277	 $\pm$ 	0.018	&	14.70	 $\pm$ 	1.08	&		&		\\
\noalign{\smallskip}        \noalign{\smallskip}																												
LSPM J0401+5131	&		&	DC8	&	A	&		&			&	39.836	 $\pm$ 	0.077	&	883.245	 $\pm$ 	0.117	&				&		&		\\
Ross 25	&	J04011+513	&	M3.5 V	&	B	&	Sma21	&	12401.7		&	39.816	 $\pm$ 	0.021	&	883.463	 $\pm$ 	0.028	&	23.97	 $\pm$ 	4.90	&		&		\\
\noalign{\smallskip}        \noalign{\smallskip}																												
GJ 3261	&	J04056+057	&	M3.5 V	&	A	&		&			&	16.094	 $\pm$ 	0.103	&	47.984	 $\pm$ 	0.155	&				&	beta Pic	&	Gag18b	\\
G3 3296932486866670720	&		&	M4.0 V	&	B	&	MCT   3	&			&		\ldots		&		\ldots		&				&	beta Pic	&	$\star$	\\
V1221 Tau	&		&	M3.0 V	&	C	&	This work	&	51950.3		&	15.876	 $\pm$ 	0.026	&	49.854	 $\pm$ 	0.034	&	10.56	 $\pm$ 	0.62	&	beta Pic	&	Gag18b	\\
\noalign{\smallskip}        \noalign{\smallskip}																												
LP 414--117	&	J04123+162	&	M4.0 V	&	Aab	&	Ben08	&			&	28.240	 $\pm$ 	0.091	&	156.844	 $\pm$ 	0.155	&				&	Hya	&	Ros11	\\
LSPM J0409+1622	&		&	M5.5 V	&	B	&	Sma21	&	74092.6		&	28.769	 $\pm$ 	0.051	&	161.152	 $\pm$ 	0.075	&	2.339	 $\pm$ 	0.672	&	Hya	&	Ros11	\\
\noalign{\smallskip}        \noalign{\smallskip}																												
TYC 78--257--1	&		&	K3.0 V	&	A	&		&			&	26.974	 $\pm$ 	0.022	&	139.559	 $\pm$ 	0.028	&				&	Hya	&	Ros11	\\
RX J0422.4+0337	&	J04224+036	&	M3.5 V	&	B	&	Sma21	&	62796.5		&	27.846	 $\pm$ 	0.026	&	143.007	 $\pm$ 	0.039	&	9.22	 $\pm$ 	0.46	&	Hya	&	Ros11	\\
\noalign{\smallskip}        \noalign{\smallskip}																												
HD 27848$^d$	&		&	F5 V	&	AB	&	OCC 615	&			&	19.959	 $\pm$ 	0.026	&	103.388	 $\pm$ 	0.039	&				&	Hya	&	Ros11	\\
V991 Tau	&		&	K4 V	&	C	&		&	34929.6		&	18.356	 $\pm$ 	0.019	&	94.411	 $\pm$ 	0.025	&				&	Hya	&	Ros11	\\
V805 Tau	&	J04252+172	&	M3.5 V	&	DE	&	AST 4	&	51890.1		&	19.313	 $\pm$ 	0.193	&	110.653	 $\pm$ 	0.259	&				&	Hya	&	Ros11	\\
LP  415--881	&		&	M7.0 V	&	F	&		&	79117.5		&	21.204	 $\pm$ 	0.061	&	105.698	 $\pm$ 	0.088	&				&	Hya	&	Lod19	\\
\noalign{\smallskip}        \noalign{\smallskip}																												
LP 415--345	&	J04425+204	&	M3.0 V	&	Aab	&	Sta97	&			&	20.476	 $\pm$ 	0.022	&	97.452	 $\pm$ 	0.034	&				&	Hya	&	Gag18b	\\
LP 415--3051	&		&	M3.0 V	&	B	&	Sma21	&	34713.1		&	19.426	 $\pm$ 	0.020	&	91.262	 $\pm$ 	0.028	&				&	Hya	&	Ros11	\\
G2 3411054848866601472	&		&	M6.0 V	&	C	&	Sma21	&	83728.3		&	19.484	 $\pm$ 	0.118	&	93.825	 $\pm$ 	0.180	&				&	Hya	&	Lod19	\\
\noalign{\smallskip}        \noalign{\smallskip}																												
PM J05334+4809	&		&	M0.0 V	&	A	&		&			&	30.270	 $\pm$ 	0.018	&	66.932	 $\pm$ 	0.022	&				&		&		\\
PM J05341+4732 A	&	J05341+475	&	M2.5 V	&	B	&	EB21	&	75945.4		&	30.050	 $\pm$ 	0.027	&	69.013	 $\pm$ 	0.033	&	6.96	 $\pm$ 	0.52	&		&		\\
PM J05341+4732 B$^e$	&		&	M3.0 V	&	C	&	EB21	&	75899.1		&	30.039	 $\pm$ 	0.023	&	61.853	 $\pm$ 	0.028	&	5.20	 $\pm$ 	0.52	&		&		\\
UPM J0533+4809	&		&	M3.5 V	&	D	&	Sma21	&	4219.5		&	30.229	 $\pm$ 	0.235	&	65.011	 $\pm$ 	0.311	&				&		&		\\
\noalign{\smallskip}        \noalign{\smallskip}																												
1RXS J073138.4+455718	&		&	M3.0 V	&	Aab	&	Fou18	&			&	17.880	 $\pm$ 	0.416	&	93.776	 $\pm$ 	0.481	&				&		&		\\
1RXS J073101.9+460030	&	J07310+460	&	M4.0 V	&	B	&	Cif21	&	23780.3		&	18.141	 $\pm$ 	0.052	&	101.748	 $\pm$ 	0.059	&				&		&		\\
G3 975312928903090560	&		&	M4.5 V	&	C	&	Sma21	&	16738.6		&	18.388	 $\pm$ 	0.034	&	100.736	 $\pm$ 	0.042	&				&		&		\\
\noalign{\smallskip}        \noalign{\smallskip}																												
V869 Mon	&		&	K3 V	&	A	&		&			&	71.032	 $\pm$ 	0.024	&	286.810	 $\pm$ 	0.030	&				&	UMa	&	Tab17	\\
HD 61606 B	&		&	K7 V	&	B	&	BGH 3	&	815.63		&	70.992	 $\pm$ 	0.025	&	294.206	 $\pm$ 	0.031	&	1090	 $\pm$ 	231	&	UMa	&	Tab17	\\
GJ 282 C	&	J07361--031	&	M1.0 V	&	Cab	&	Pov09	&	6.2	273662.8018	&	70.275	 $\pm$ 	0.131	&	302.347	 $\pm$ 	0.173	&	3.766	 $\pm$ 	0.638	&	UMa	&	Tab17	\\
\noalign{\smallskip}        \noalign{\smallskip}																												
PM J13255+2738	&		&	M1.0 V	&	A	&		&			&	21.996	 $\pm$ 	0.019	&	71.210	 $\pm$ 	0.029	&				&		&		\\
PM J13260+2735 A	&	J13260+275	&	M3.0 V	&	B	&	Sma21	&	18493.6		&	21.868	 $\pm$ 	0.070	&	72.891	 $\pm$ 	0.109	&	26.54	 $\pm$ 	2.08	&		&		\\
PM J13260+2735 B	&		&	M2.5 V	&	C	&	KPP 3896	&	18343.6		&	22.030	 $\pm$ 	0.057	&	63.569	 $\pm$ 	0.092	&	21.74	 $\pm$ 	2.06	&		&		\\
\noalign{\smallskip}        \noalign{\smallskip}																												
HD 140232	&		&	A8 V	&	A	&		&			&	18.719	 $\pm$ 	0.041	&	82.169	 $\pm$ 	0.044	&				&		&		\\
G3 1197801408884577408	&		&	M3.5 V	&	B	&	DRS 17	&	126.52		&	18.642	 $\pm$ 	0.113	&	82.285	 $\pm$ 	0.167	&	8053	 $\pm$ 	1482	&		&		\\
StKM 1--1264	&	J15416+184	&	M1.5 V	&	C	&	TOK 302	&	12869.9		&	19.367	 $\pm$ 	0.139	&	91.885	 $\pm$ 	0.173	&	165.8	 $\pm$ 	25.7	&		&		\\
\noalign{\smallskip}        \noalign{\smallskip}																												
$\sigma$ CrB A	&		&	F6 V	&	Aab	&	Str03, Rag09	&	0.01		&	44.057	 $\pm$ 	0.046	&	282.061	 $\pm$ 	0.072	&				&		&		\\
$\sigma$ CrB B	&		&	G1 V	&	B	&	STF 2032	&	163.85		&	44.134	 $\pm$ 	0.018	&	301.273	 $\pm$ 	0.026	&	2567	 $\pm$ 	385	&		&		\\
$\sigma$ CrB C	&	J16139+337	&	M2.5 V	&	CD	&	STF 2032	&	11.5	14345.9663	&	44.267	 $\pm$ 	0.159	&	300.693	 $\pm$ 	0.241	&	12.28	 $\pm$ 	0.88	&		&		\\
\noalign{\smallskip}        \noalign{\smallskip}																												
HD 160269 A	&		&	G0 IV/V	&	AB	&	BU  962	&	9.02		&	69.283	 $\pm$ 	0.200	&	522.563	 $\pm$ 	0.334	&				&		&		\\
GJ 685	&	J17355+616	&	M0.5 V	&	C	&	LDS 2736	&	10561.5		&	69.892	 $\pm$ 	0.015	&	577.333	 $\pm$ 	0.023	&				&		&		\\
\noalign{\smallskip}        \noalign{\smallskip}																												
HD 230017 A	&		&	M0.0 V	&	A	&		&			&	53.423	 $\pm$ 	0.108	&	133.273	 $\pm$ 	0.161	&				&	Car	&	This work	\\
HD 230017 B	&	J18548+109	&	M3.5 V	&	B	&	VYS   8	&	70.68		&	53.644	 $\pm$ 	0.043	&	89.289	 $\pm$ 	0.063	&	6625	 $\pm$ 	637	&	Car	&	Gag18b	\\
PM J18542+1058	&		&	M4.0 V	&	C	&	This work	&	10009.5		&	53.838	 $\pm$ 	0.024	&	114.612	 $\pm$ 	0.034	&	38.59	 $\pm$ 	2.21	&	Car	&	This work	\\
\noalign{\smallskip}        \noalign{\smallskip}																												
AU Mic	&	J20451--313	&	M0.5 V	&	A	&		&			&	102.943	 $\pm$ 	0.023	&	456.998	 $\pm$ 	0.029	&				&	beta Pic	&	Cab09	\\
 AT Mic A	&	J20418--324	&	M4.5 V	&	B	&	LDS 720	&	45473.4		&	100.792	 $\pm$ 	0.073	&	484.794	 $\pm$ 	0.094	&				&	beta Pic	&	Ell14	\\
 AT Mic B	&		&	M4.5 V	&	C	&	LDS 720	&	20.62		&	101.972	 $\pm$ 	0.077	&	423.100	 $\pm$ 	0.097	&	14879	 $\pm$ 	1575	&	beta Pic	&	$\star$	\\
\noalign{\smallskip}        \noalign{\smallskip}																												
Wolf 1548	&	J22058--119	&	M0.0 V	&	A	&		&			&	38.616	 $\pm$ 	0.533	&	319.261	 $\pm$ 	1.369	&				&	Cas	&	Cab10	\\
LP 759--25	&		&	M6.0 V	&	B	&	WNO 57	&	59342.7		&	51.070	 $\pm$ 	0.054	&	322.495	 $\pm$ 	0.074	&	2.615	 $\pm$ 	0.227	&	Cas	&	$\star$	\\
\noalign{\smallskip}        \noalign{\smallskip}																												
G 29--19	&	J23175+063	&	M3.0 V	&	A	&		&			&	48.918	 $\pm$ 	0.027	&	302.148	 $\pm$ 	0.042	&				&		&		\\
G 28--50	&	J23161+067	&	M3.5 V	&	B	&	Sma21	&	36823.1		&	48.976	 $\pm$ 	0.026	&	304.894	 $\pm$ 	0.037	&	6.48	 $\pm$ 	0.38	&		&		\\
\noalign{\smallskip}        \noalign{\smallskip}																												
V368 Cep	&		&	G9 V	&	A	&		&			&	52.784	 $\pm$ 	0.014	&	215.918	 $\pm$ 	0.022	&				&	Col	&	$\star$	\\
HD 220140 B	&	J23194+790	&	M3.5 V	&	B	&	LDS 2035	&	206.38		&	52.840	 $\pm$ 	0.021	&	218.080	 $\pm$ 	0.036	&	3291	 $\pm$ 	548	&	Col	&	Gag18b	\\
LP 12--90	&	J23228+787	&	M5.0 V	&	C	&	MKR 1	&	18221.8		&	52.834	 $\pm$ 	0.030	&	217.441	 $\pm$ 	0.051	&	18.03	 $\pm$ 	2.89	&	Col	&	$\star$	\\
\noalign{\smallskip}        \noalign{\smallskip}																												
AF Psc	&	J23317--027	&	M4.5 V	&	A	&		&			&	28.631	 $\pm$ 	0.035	&	118.451	 $\pm$ 	0.051	&				&	Tuc-Hor/beta Pic	&	Kra14b	\\
2M J23301129--0237227	&	J23301--026	&	M6.0 V	&	B	&	CAB24	&	66833.9		&	21.982	 $\pm$ 	0.070	&	122.009	 $\pm$ 	0.110	&				&	Tuc-Hor/beta Pic	&	Alo15b	\\
\noalign{\smallskip}        \noalign{\smallskip}																												
HD 221503	&		&	K7 V	&	A	&		&			&	68.736	 $\pm$ 	0.027	&	405.450	 $\pm$ 	0.036	&				&		&		\\
GJ 1284	&	J23302--203	&	M3.0 V	&	Bab	&	Giz02, SHY 110	&	6.5	206051.1082	&	62.868	 $\pm$ 	0.076	&	375.234	 $\pm$ 	0.092	&				&		&		\\
GJ 897$^f$	&	J23327--167	&	M2.0 V	&	C	&	LDS 816	&	21894.8		&	64.8	 $\pm$ 	4.1	&	413.04	 $\pm$ 	3.75	&				&		&		\\
G3 2395220664463236992	&		&	M3.0 V	&	D	&	VOU 28	&			&		\ldots		&		\ldots		&				&		&		\\
\noalign{\smallskip}        \noalign{\smallskip}											
\end{longtable}
\tablefoot{
\tablefoottext{a}	{References for discovery --
Ben08: \cite{Ben08};
EB21: \cite{EB21};
Fou18: \cite{Fou18};
Giz02: \cite{Giz02};
Gri85: \cite{Gri85};
Pov09: \cite{Pov09};
Rag09: \cite{Rag09};
Sma21: \cite{Sma21};
Sta97: \cite{Sta97};
Str03: \cite{Str03};}
\tablefoottext{b}	{References for first assignation --
Alo15b: \cite{Alo15b};
Cab09: \cite{Cab09};
Cab10a: \cite{Cab10a};
Ell14: \cite{Ell14};
Gag18b: \cite{Gag18b};
Jan17: \cite{Jan17};
Kra14b: \cite{Kra14b};
Lod19: \cite{Lod19};
Ros11: \cite{Ros11};
Tab17: \cite{Tab17}.
The star symbol ($\star$) means the same SKG assignation as its physical companion, therefore sharing bibliographic reference. The list of abbreviations for the SKG can be found in Sect.~\ref{sssection:young}.}
\tablefoottext{c}	{LP~648--20: The $\sim$L0 companion 
reported by \cite{Ber10} based on lucky imaging  at 5.587\,arcsec is a background star ($\pi =$ 1.63\,mas).}
\tablefoottext{d}	{HD~27848: System dissected by \cite{Ben21}. Bona fide single star according to \cite{Bra23}.}
\tablefoottext{e}	{PM~J05341+4732\,B: Absent in the WDS but see \cite{Ans15} ($\rho=$ 2.33\,arcsec).}
\tablefoottext{f}	{GJ~897: Parallax and proper motions from Hipparcos \citep{van07}.}
}
\end{landscape}
}

{\fontsize{7}{10}\selectfont 
\begin{landscape}
\begin{longtable}{lllllccclcc}
\caption{\label{tab:fgk} Multiple systems containing M dwarfs and FGK primaries.}
\\
\noalign{\smallskip}
\noalign{\smallskip}
\noalign{\smallskip}
\hline\hline
\noalign{\smallskip}
Name	&	Karmn	&	Spectral	&	Component	&	Discoverer$^a$	&	$\pi$			&	$\mu_{\rm total}$			&	[Fe/H]	&	[Fe/H]	&	$\rho$	&	$\theta$	\\
	&		&	type	&		&		&	{[}mas{]}			&	{[}mas\,a$^{-1}${]}			&	{[}dex{]}	&	Ref.$^b$	&	{[}arcsec{]}	&	{[}deg{]}	\\
\noalign{\smallskip}
\hline
\noalign{\smallskip}
\endfirsthead
\caption{Multiple systems containing M dwarfs and FGK primaries (cont.).}\\
\hline\hline
        \noalign{\smallskip}
Name	&	Karmn	&	Spectral	&	Component	&	Discoverer$^a$	&	$\pi$			&	$\mu_{\rm total}$			&	[Fe/H]	&	[Fe/H]	&	$\rho$	&	$\theta$	\\
	&		&	type	&		&		&	{[}mas{]}			&	{[}mas\,a$^{-1}${]}			&	{[}dex{]}	&	Ref.$^b$	&	{[}arcsec{]}	&	{[}deg{]}	\\
\hline
\noalign{\smallskip}
\endhead
\hline
\endfoot
\noalign{\smallskip}
HD 4967	&		&	K5 V	&	A	&		&	65.090	$ \pm $	0.025	&	672.691	$ \pm $	0.047	&	--0.33	&	Hou16	&		&		\\
HD 4967 B	&	J00515--229	&	M5.5 V	&	B	&	LDS 1082	&	65.073	$ \pm $	0.050	&	680.566	$ \pm $	0.095	&	\ldots	&		&	17.07	&	72.1	\\
\noalign{\smallskip}        \noalign{\smallskip}																									
EX Cet	&		&	K0.5 V	&	A	&		&	41.564	$ \pm $	0.024	&	197.856	$ \pm $	0.032	&	--0.06	&	Zak22	&		&		\\
LP 648--20	&	J01369--067	&	M3.5 V	&	B	&	CAB 3	&	41.699	$ \pm $	0.045	&	200.547	$ \pm $	0.052	&	\ldots	&		&	612.08	&	258.6	\\
\noalign{\smallskip}        \noalign{\smallskip}																									
Wolf 109	&		&	K5 V	&	A	&		&	27.827	$ \pm $	0.021	&	466.135	$ \pm $	0.031	&	\ldots	&		&		&		\\
Wolf 109 B	&	J02020+039	&	M2.0 V	&	B	&	OSV 1	&	28.922	$ \pm $	0.353	&	465.771	$ \pm $	0.482	&	\ldots	&		&	17.96	&	14.4	\\
G3 2517912315149114240	&		&	M5.0 V	&	C	&		&		\ldots		&		\ldots		&		&		&	17.74	&	16.8	\\
\noalign{\smallskip}        \noalign{\smallskip}																									
HD 15468	&		&	K4.5 V	&	AB	&	RST 2280	&	51.606	$ \pm $	0.136	&	646.402	$ \pm $	0.166	&	--0.31	&	Mon18	&	0.18	&	145.3	\\
\end{longtable}
\tablefoot{
\tablefoottext{a}{References for discovery --
Kat13: \cite{Kat13};
Mal14a: \cite{Mal14a};
Mar87: \cite{Mar87};
Pov09: \cite{Pov09};
Rag09: \cite{Rag09};
Ros21: \cite{Ros21};
She85: \cite{She85};
Sma21: \cite{Sma21};
Str03: \cite{Str03};
Tor09: \cite{Tor09};
Zak79: \cite{Zak79}.
}
\tablefoottext{b}{References for metallicities --
Boc18: \cite{Boc18};
Bud21: \cite{Bud21};
Bur13: \cite{Bur13};
Gai14b: \cite{Gai14b};
Gui09: \cite{Gui09};
Hir21: \cite{Hir21};
Hou16: \cite{Hou16};
Jon20: \cite{Jon20};
Kuz19: \cite{Kuz19};
Luc17: \cite{Luc17};
Luc18: \cite{Luc18};
Man15: \cite{Man15};
Mal20: \cite{Mal20};
Man15: \cite{Man15};
Marf21: \cite{Marf21};
Mas08: \cite{Mas08};
Mon18: \cite{Mon18};
Pas18: \cite{Pas18};
Raj18a: \cite{Raj18a};
Ric20: \cite{Ric20};
Ros21: \cite{Ros21};
Stei20: \cite{Stei20};
Ston20: \cite{Ston20};
Tau20: \cite{Tau20};
Ter15a: \cite{Ter15a};
Tin19: \cite{Tin19};
Zak22: \cite{Zak22}.}
}
\end{landscape}
}

{\fontsize{7}{10}\selectfont 
\begin{longtable}{llllcccccccc}
\caption{\label{tab:wd} Multiple systems containing M dwarfs and white dwarfs.}
\\
\noalign{\smallskip}
\noalign{\smallskip}
\noalign{\smallskip}
\hline
\hline
\noalign{\smallskip}
Name	&	Karmn	&	Spectral	&	Discoverer$^b$	&	$\pi$			&	$\mu_{\rm total}$			&	$\rho^c$	&	$\theta$	\\
	&		&	type$^a$	&		&	{[}mas{]}			&	{[}mas\,a$^{-1}${]}			&	{[}arcsec{]}	&	{[}deg{]}	\\
\noalign{\smallskip}
\hline
\noalign{\smallskip}
\endfirsthead
\caption{Multiple systems containing M dwarfs and white dwarfs (cont.).}\\
\hline\hline
        \noalign{\smallskip}
Name	&	Karmn	&	Spectral	&	Discoverer$^b$	&	$\pi$			&	$\mu_{\rm total}$			&	$\rho^c$	&	$\theta$	\\
	&		&	type$^a$	&		&	{[}mas{]}			&	{[}mas\,a$^{-1}${]}			&	{[}arcsec{]}	&	{[}deg{]}	\\
\hline
\noalign{\smallskip}
\endhead
\hline
\endfoot
GJ 1015 B	&		&	DBQ5	&		&	43.659	$ \pm $	0.022	&	324.163	$ \pm $	0.025	&		&		\\
GJ 1015 A	&	J00413+558	&	M4.0 V	&	GIC  13	&	43.735	$ \pm $	0.022	&	332.499	$ \pm $	0.024	&	10.854	&	248.4	\\
\noalign{\smallskip}        \noalign{\smallskip}																			
GJ 3118	&		&	DA5.6	&		&	57.799	$ \pm $	0.016	&	301.944	$ \pm $	0.018	&		&		\\
GJ 3117	&	J01518+644	&	M2.5 V	&	GIC  27	&	57.825	$ \pm $	0.016	&	306.369	$ \pm $	0.018	&	13.503	&	0.7	\\
\noalign{\smallskip}        \noalign{\smallskip}																			
GJ 3151$^d$	&	J02204+377	&	M2.5 V + DA	&	LDS 3370	&	40.867	$ \pm $	0.033	&	349.298	$ \pm $	0.050	&	2.0	&	89.0	\\
\noalign{\smallskip}        \noalign{\smallskip}																			
LSPM J0401+5131	&		&	DC8	&		&	39.836	$ \pm $	0.077	&	883.245	$ \pm $	0.117	&		&		\\
Ross 25	&	J04011+513	&	M3.5 V	&	Sma21	&	39.816	$ \pm $	0.021	&	883.463	$ \pm $	0.028	&	494.0	&	173.4	\\
\noalign{\smallskip}        \noalign{\smallskip}																			
o$^{02}$ Eri B	&		&	DA2.9	&		&	199.691	$ \pm $	0.051	&	4018.591	$ \pm $	0.061	&		&		\\
o$^{02}$ Eri	&		&	K0 V	&	STF 518	&	199.608	$ \pm $	0.121	&	4089.836	$ \pm $	0.165	&	83.337	&	282.2	\\
o$^{02}$ Eri C	&	J04153--076	&	M4.5 V	&	STF 518	&	199.452	$ \pm $	0.069	&	4083.715	$ \pm $	0.084	&	78.097	&	277.5	\\
\noalign{\smallskip}        \noalign{\smallskip}																			
GJ 169.1 B	&		&	DC5	&		&	181.273	$ \pm $	0.020	&	2361.129	$ \pm $	0.028	&		&		\\
GJ 169.1 A	&	J04311+589	&	M4.0 V	&	STI 2051	&	181.244	$ \pm $	0.050	&	2424.355	$ \pm $	0.073	&	10.263	&	238.2	\\
\noalign{\smallskip}        \noalign{\smallskip}																			
GJ 3431	&		&	DQ8	&		&	41.029	$ \pm $	0.036	&	357.871	$ \pm $	0.053	&		&		\\
GJ 3430	&	J07102+376	&	M4.0 V	&	GIC  69	&	44.601	$ \pm $	0.437	&	350.218	$ \pm $	0.608	&	12.472	&	223.1	\\
\noalign{\smallskip}        \noalign{\smallskip}																			
G 107--70 A	&		&	DA	&		&	83.484			&	1281.725	$ \pm $	3.606	&		&		\\
G 107--70 B	&		&	DA	&	GIC  75	&		\ldots		&		\ldots		&	0.667	&	318.1	\\
GJ 275.2 A$^e$	&	J07307+481	&	M4.0 V	&	WNO 49	&	88.723	$ \pm $	0.030	&	1287.988	$ \pm $	0.034	&	103.1	&	333.9	\\
\noalign{\smallskip}        \noalign{\smallskip}																			
GJ 283 A	&		&	DZQA6	&		&	109.344	$ \pm $	0.018	&	1261.341	$ \pm $	0.024	&		&		\\
GJ 283 B	&	J07403--174	&	M6.0 V	&	LUY 5693	&	109.254	$ \pm $	0.039	&	1270.984	$ \pm $	0.052	&	20.338	&	279.5	\\
\noalign{\smallskip}        \noalign{\smallskip}																			
GJ 401 B	&		&	DQ	&		&	53.190	$ \pm $	0.034	&	1962.558	$ \pm $	0.057	&		&		\\
GJ 401	&	J10456--191	&	M0.5 V	&	LDS 4013	&	53.172	$ \pm $	0.022	&	1963.772	$ \pm $	0.034	&	6.655	&	174.5	\\
\noalign{\smallskip}        \noalign{\smallskip}																			
GJ 1142 B	&		&	DA3	&		&	40.293	$ \pm $	0.032	&	446.106	$ \pm $	0.042	&		&		\\
GJ 1142 A	&	J11081--052	&	M3.0 V	&	LDS 852	&	40.180	$ \pm $	0.026	&	443.166	$ \pm $	0.034	&	279.0	&	159.3	\\
\noalign{\smallskip}        \noalign{\smallskip}																			
GJ 1155 B	&		&	DA	&		&	42.773	$ \pm $	0.043	&	711.074	$ \pm $	0.064	&		&		\\
GJ 1155 A	&	J12168+029	&	M3.0 V	&	LDS 935	&	42.820	$ \pm $	0.035	&	700.554	$ \pm $	0.054	&	2.126	&	357.9	\\
\noalign{\smallskip}        \noalign{\smallskip}																			
Wolf 485	&		&	DA3.5	&		&	62.148	$ \pm $	0.044	&	1207.505	$ \pm $	0.056	&		&		\\
Ross 476	&	J13300--087	&	M4.0 V	&	LDS 448	&	62.281	$ \pm $	0.042	&	1210.608	$ \pm $	0.059	&	502.4	&	198.7	\\
\noalign{\smallskip}        \noalign{\smallskip}																			
GJ 1179 B	&		&	DC9	&		&	84.311	$ \pm $	0.029	&	1490.906	$ \pm $	0.041	&		&		\\
GJ 1179 A	&	J13482+236	&	M5.5 V	&	LDS 4410	&	84.225	$ \pm $	0.027	&	1487.630	$ \pm $	0.037	&	188.1	&	49.5	\\
\noalign{\smallskip}        \noalign{\smallskip}																			
GJ 630.1 B	&		&	DQ8	&	LDS 1436	&	67.354	$ \pm $	0.021	&	1634.847	$ \pm $	0.035	&		&		\\
CM Dra	&	J16343+571	&	M4.5 V	&	Mor09, Tor10	&	67.288	$ \pm $	0.034	&	1623.346	$ \pm $	0.061	&	26.726	&	201.9	\\
\noalign{\smallskip}        \noalign{\smallskip}																			
LP 387--36	&		&	DC7	&		&	28.841	$ \pm $	0.054	&	285.375	$ \pm $	0.072	&		&		\\
LP 387--37	&	J17058+260	&	M1.5 V	&	LDS 4721	&	28.746	$ \pm $	0.019	&	291.032	$ \pm $	0.025	&	19.244	&	179.6	\\
\noalign{\smallskip}        \noalign{\smallskip}																			
LSPM J1826+1120S	&		&	DA	&		&	37.055	$ \pm $	0.066	&	281.066	$ \pm $	0.092	&		&		\\
GJ 4059	&	J18264+113	&	M3.5 V	&	NI   38	&	37.056	$ \pm $	0.024	&	276.066	$ \pm $	0.033	&	8.196	&	16.8	\\
\noalign{\smallskip}        \noalign{\smallskip}																			
LP 141--14	&		&	DC	&		&	40.393	$ \pm $	0.051	&	246.401	$ \pm $	0.095	&		&		\\
G 229--20 A	&	J18576+535	&	M3.5 V	&	LDS 4802	&	40.348	$ \pm $	0.015	&	261.445	$ \pm $	0.028	&	43.653	&	343.9	\\
LP 141--13	&		&	M3.0 V	&	LDS 4802	&	40.365	$ \pm $	0.015	&	245.563	$ \pm $	0.027	&	2.290	&	17.0	\\
\noalign{\smallskip}        \noalign{\smallskip}																			
GJ 754.1 A	&		&	DBQA5	&		&	95.176	$ \pm $	0.029	&	173.039	$ \pm $	0.038	&		&		\\
GJ 754.1 B	&	J19205--076	&	M2.5 V	&	LDS 678	&	95.178	$ \pm $	0.031	&	191.462	$ \pm $	0.037	&	27.168	&	306.2	\\
\noalign{\smallskip}        \noalign{\smallskip}																			
V1513 Cyg$^f$	&		&	sdM1.5 + WD 	&	Giz97b	&	60.296	$ \pm $	0.027	&	1470.281	$ \pm $	0.050	&	0.010537 {[}au{]}	&		\\
Wolf 1130 C	&	J20050+544	&	T8p	&	MGN1	&	62.9	$ \pm $	3.3	&		\ldots		&	188.54	&	115.00	\\
\noalign{\smallskip}        \noalign{\smallskip}																			
V1412 Aql	&		&	DC7	&		&	43.574	$ \pm $	0.038	&	635.705	$ \pm $	0.051	&		&		\\
GJ 784.2 A	&	J20139+066	&	M3.5 V	&	GIC 164	&	43.570	$ \pm $	0.022	&	634.751	$ \pm $	0.029	&	101.5	&	150.9	\\
\noalign{\smallskip}        \noalign{\smallskip}																			
Ross 193 B	&		&	DC10	&	LDS 6420	&	61.760	$ \pm $	0.052	&	807.720	$ \pm $	0.068	&		&		\\
FR Aqr	&	J20568--048	&	M4.0 V	&	Jef18	&	61.824	$ \pm $	0.074	&	826.321	$ \pm $	0.091	&	15.006	&	129.4	\\
\noalign{\smallskip}        \noalign{\smallskip}																			
UCAC4 747--070768	&		&	DAH	&		&	118.155	$ \pm $	0.016	&	88.850	$ \pm $	0.026	&		&		\\
TYC 3980--1081--1	&	J21516+592	&	M4.0 V	&	KPP 4252	&	123.057	$ \pm $	0.594	&	79.885	$ \pm $	1.668	&	14.642	&	291.9	\\
\noalign{\smallskip}        \noalign{\smallskip}																			
Gaia DR3 2005884249925303168$^g$	&		&	D?	&		&	18.550	$ \pm $	0.079	&	108.971	$ \pm $	0.123	&		&		\\
LF 4 +54 152	&	J22129+550	&	M0.0 V	&	Sma21	&	18.385	$ \pm $	0.153	&	109.131	$ \pm $	0.248	&	66.265	&	209.8	\\
\noalign{\smallskip}        \noalign{\smallskip}																			
GJ 4305	&		&	DA8.1	&		&	40.325	$ \pm $	0.072	&	439.472	$ \pm $	0.074	&		&		\\
GJ 4304	&	J22559+057	&	M1.0 V	&	LDS 5021	&	40.347	$ \pm $	0.024	&	446.045	$ \pm $	0.029	&	17.168	&	104.6	\\
\noalign{\smallskip}        \noalign{\smallskip}																			
LSPM J2309+5506E	&		&	DA	&		&	60.895	$ \pm $	0.030	&	410.185	$ \pm $	0.039	&		&		\\
G 233--42	&	J23089+551	&	M5.0 V	&	NSN  11	&	60.921	$ \pm $	0.027	&	411.267	$ \pm $	0.035	&	6.150	&	249.6	\\
\noalign{\smallskip}        \noalign{\smallskip}																			
GJ 4357	&		&	DA	&		&	25.367	$ \pm $	0.100	&	325.253	$ \pm $	0.126	&		&		\\
GJ 4356	&	J23389+210	&	M3.5 V	&	LDS 5108	&	25.402	$ \pm $	0.028	&	328.855	$ \pm $	0.036	&	9.495	&	290.7	\\
\noalign{\smallskip}        \noalign{\smallskip}																			
GJ 905.2 B	&		&	DA3.8	&	LDS 1070	&	53.762	$ \pm $	0.027	&	224.052	$ \pm $	0.039	&	174.7	&	10.0	\\
GJ 905.2 A	&	J23438+325	&	M3.0 V	&	JOD 26	&		\ldots		&	223.766	$ \pm $	13.688	&	0.109	&	128.0	\\
\noalign{\smallskip}        \noalign{\smallskip}																													
\end{longtable}
\tablefoot{
\tablefoottext{a}{In Simbad's taxonomy of white dwarfs, `D’ stands for degenerate, followed by an abbreviation of the most significant spectral features: 
`A’ for a hydrogen-rich atmosphere,
`B’ for a helium-rich atmosphere,
`C’ for no strong spectral lines,
`Q’ for carbon lines, and
`Z’ for metal lines.}
\tablefoottext{b}{References —
Mor09: \cite{Mor09};
Tor10: \cite{Tor10};
Jef18: \cite{Jef18};
Sma21: \cite{Sma21}.}
\tablefoottext{c}{{Angular separation $\rho$ and position angle $\theta$ measured from the white dwarf.}}
\tablefoottext{d}{GJ~3151: White dwarf in a compact configuration ($s \simeq 49$\,au) determined spectroscopically \citep{Schi12} but not resolved by {\em Gaia}.}
\tablefoottext{e}{\object{GJ~275.2~A}: it has a closer companion according to \cite{Har81}, who determined a 0.94-year period and individual masses of 0.17 and 0.08\,$\mathcal{M}_\odot$.
They also estimated a mean angular separation of $\sim$ 0.05\,arcsec, which is compatible with the application of Third Kepler's law.}
\tablefoottext{f}{{\object{Wolf~1130} is a 0.4967-day binary composed of a subdwarf (sdM1.5; \citealt{Giz97b}) and a white dwarf, likely bound to evolve into a cataclysmic variable \citep{Mac18}.}}
\tablefoottext{g}{Gaia~DR3 2005884249925303168:
Unconfirmed spectroscopically but proposed candidate to WD in agreement with \cite{Jim18} based on position in HR diagram.
}
}
}

\begin{longtable}{lllllcccc}
\caption{\label{tab:ucd} Multiple systems containing M dwarfs and ultra-cool dwarfs.}
\\
\noalign{\smallskip}
\noalign{\smallskip}
\hline
\hline
\noalign{\smallskip}
Name	&	Karmn	&	Spectral	&	Comp.	&	Discoverer$^b$	&	$\pi$			&	$\mu_{\rm total}$			&	$\rho$	&	$\theta$	\\
	&		&	type$^a$	&		&		&	{[}mas{]}			&	{[}mas\,a$^{-1}${]}			&	{[}arcsec{]}	&	{[}deg{]}	\\
\noalign{\smallskip}
\hline
\noalign{\smallskip}
\endfirsthead
\caption{Multiple systems containing M dwarfs and ultra-cool dwarfs (continued).}\\
\hline\hline
        \noalign{\smallskip}
Name	&	Karmn	&	Spectral	&	Comp.	&	Discoverer$^b$	&	$\pi$			&	$\mu_{\rm total}$			&	$\rho$	&	$\theta$	\\
	&		&	type$^a$	&		&		&	{[}mas{]}			&	{[}mas\,a$^{-1}${]}			&	{[}arcsec{]}	&	{[}deg{]}	\\
\hline
\noalign{\smallskip}
\endhead
\hline
\endfoot
\noalign{\smallskip}
GJ 3131	&	J02033--212	&	M2.5 V	&	Aab	&	Jef18	&	46.672	$ \pm $	0.041	&	471.895	$ \pm $	0.051	&	\ldots	&	\ldots	\\
GJ 3131 B	&		&	L0:	&	B	&	Sma21	&	46.747	$ \pm $	0.307	&	450.045	$ \pm $	0.474	&	4.008	&	252.0	\\
\noalign{\smallskip}        \noalign{\smallskip}																					
G 196--3	&	J10043+503	&	M2.5 V	&	A	&		&	45.854	$ \pm $	0.019	&	246.664	$ \pm $	0.022	&		&		\\
G 196--3 B	&		&	L3$\beta$	&	B	&	REB 1	&	46.195	$ \pm $	0.545	&	246.291	$ \pm $	0.673	&	16.070	&	209.2	\\
\noalign{\smallskip}        \noalign{\smallskip}																					
GJ 3657	&	J11231+258	&	M5.0 V	&	A	&		&	61.652	$ \pm $	0.064	&	1061.305	$ \pm $	0.100	&		&		\\
2M J11225550+2550250	&		&	T6	&	B	&	WIS 183	&	62.9	$ \pm $	2.8	&	1062.3	$ \pm $	4.6	&	254.985	&	221.300	\\
\noalign{\smallskip}        \noalign{\smallskip}																					
Wolf 462	&	J13007+123	&	M1.5 V	&	AB	&	BEU 16	&	86.901	$ \pm $	0.117	&	629.606	$ \pm $	0.227	&	0.525	&	60.0	\\
Ross 458 C	&		&	T8.5p	&	C	&	GDM 1	&	85.540	$ \pm $	1.530	&	639.451	$ \pm $	13.454	&	12.060	&	220.2	\\
\noalign{\smallskip}        \noalign{\smallskip}																					
LP 738--14	&	J13481--137	&	M4.5 V	&	A	&		&	55.028	$ \pm $	0.024	&	858.148	$ \pm $	0.037	&		&		\\
LP 738--14 B	&		&	T5.5	&	B	&	DEA 1	&	59.500	$ \pm $	3.600	&	857.920	$ \pm $	5.371	&	67.340	&	291.50	\\
\noalign{\smallskip}        \noalign{\smallskip}																					
GJ 9492	&	J14423+660	&	M2.0 V	&	A	&		&	91.479	$ \pm $	0.016	&	301.514	$ \pm $	0.028	&		&		\\
GJ 9492 B	&		&	L0	&	B	&	GKI 4	&	91.357	$ \pm $	0.091	&	337.307	$ \pm $	0.177	&	2.290	&	93.6	\\
\noalign{\smallskip}        \noalign{\smallskip}																					
KX Lib	&		&	K4 V	&	A	&		&	169.884	$ \pm $	0.065	&	2008.680	$ \pm $	0.087	&		&		\\
HD 131976	&	J14574--214	&	M1.5 V	&	BC	&	Mar87, HN 28	&	168.770	$ \pm $	21.540	&	1933.943	$ \pm $	26.196	&	25.762	&	306.6	\\
GJ 570 D	&		&	T8	&	D	&	BUG 4	&	169.300	$ \pm $	1.700	&	1972.880	$ \pm $	5.620	&	234.0	&	317.0	\\
\noalign{\smallskip}        \noalign{\smallskip}																					
GJ 618.1 A	&	J16204--042	&	M0.0 V	&	A	&		&	34.076	$ \pm $	0.019	&	416.694	$ \pm $	0.032	&		&		\\
GJ 618.1 B	&		&	L2.5	&	B	&	WIL 3	&	33.872	$ \pm $	0.759	&	415.281	$ \pm $	1.118	&	35.907	&	144.8	\\
\noalign{\smallskip}        \noalign{\smallskip}																					
G 259--20	&	J17431+854	&	M2.0 V	&	A	&		&	45.255	$ \pm $	0.314	&	291.245	$ \pm $	0.527	&		&		\\
2M J17430860+8526594	&		&	L5	&	B	&	LUH 12	&	45.139	$ \pm $	0.280	&	290.996	$ \pm $	0.501	&	29.739	&	358.4	\\
\noalign{\smallskip}        \noalign{\smallskip}																					
GJ 4040	&	J17578+465	&	M2.5 V	&	A	&		&	71.495	$ \pm $	0.015	&	578.336	$ \pm $	0.028	&		&		\\
G 204--39 B	&		&	T7	&	B	&	BDK 9	&	73.750	$ \pm $	1.840	&	594.569	$ \pm $	21.932	&	197.0	&	130.6	\\
\noalign{\smallskip}        \noalign{\smallskip}																					
V1513 Cyg	&	J20050+544	&	M1.0 V	&	A	&		&	60.296	$ \pm $	0.027	&	1470.281	$ \pm $	0.050	&		&		\\
Wolf 1130 B	&		&	T8p	&	B	&	MGN 1	&		\ldots		&		\ldots		&	188.5	&	115.00	\\
\noalign{\smallskip}        \noalign{\smallskip}																					
LP 395--8 A	&	J20198+229	&	M3.0 V	&	Aab	&	Bar18	&	33.897	$ \pm $	0.026	&	135.400	$ \pm $	0.027	&	\ldots	&	\ldots	\\
LP 395--8 B	&		&	M3.5: V	&	B	&	KPP 4191	&	33.896	$ \pm $	0.053	&	137.726	$ \pm $	0.057	&	1.918	&	355.5	\\
G3 1829571684884360832	&		&	L2:	&	C	&	This work	&	33.938	$ \pm $	0.342	&	142.755	$ \pm $	0.359	&	11.019	&	307.4	\\
\noalign{\smallskip}        \noalign{\smallskip}																					
G 126--32 A	&		&	M1.0: V	&	A	&		&	27.661	$ \pm $	0.127	&	219.436	$ \pm $	0.143	&		&		\\
G 126--32 B	&	J21450+198	&	M1.5 V	&	B	&	RAO 466	&	26.818	$ \pm $	0.296	&	240.736	$ \pm $	0.412	&	0.564	&	253.3	\\
G 126--32 C	&		&	L1	&	C	&	Sma21	&	26.252	$ \pm $	0.803	&	232.177	$ \pm $	1.024	&	5.317	&	16.4	\\
\noalign{\smallskip}        \noalign{\smallskip}																					
Wolf 940	&	J21466--001	&	M4.0 V	&	A	&		&	80.738	$ \pm $	0.052	&	920.673	$ \pm $	0.060	&		&		\\
Wolf 940 B	&		&	T8	&	B	&	BNG 2	&	79.800	$ \pm $	4.500	&	919.0			&	31.640	&	250.5	\\
\noalign{\smallskip}        \noalign{\smallskip}																					
Wolf 1154 A	&	J22012+323	&	M1.5 V	&	A	&		&	32.344	$ \pm $	0.024	&	133.621	$ \pm $	0.029	&		&		\\
Wolf 1154 B	&		&	M3.5: V	&	B	&	GRV 1283	&	32.418	$ \pm $	0.083	&	119.082	$ \pm $	0.085	&	1.302	&	235.5	\\
2M J22011701+3222062	&		&	T2.5	&	C	&		&	31.400	$ \pm $	5.400	&	125.765	$ \pm $	2.702	&	80.750	&	147.6	\\
\noalign{\smallskip}        \noalign{\smallskip}																					
GJ 4287	&	J22374+395	&	M0.0 V	&	AB	&	HDS 3211	&	46.887	$ \pm $	0.564	&	365.030	$ \pm $	0.836	&	0.261	&	11.0	\\
G 216--7 B	&		&	M9.5 V	&	C	&	KIR 5	&	47.644	$ \pm $	0.137	&	351.145	$ \pm $	0.206	&	33.332	&	111.5	\\
\noalign{\smallskip}        \noalign{\smallskip}														
\end{longtable}
\tablefoot{
\tablefoottext{a}{When spectroscopic classification is not available, a semicolon (:) denotes an estimated spectral type from absolute magnitude $M_G$.}
\tablefoottext{b}{References:
Jef18: \cite{Jef18};
Mar87: \cite{Mar87};
Bar18: \cite{Bar18};
Sma21: \cite{Sma21}.
Other alphanumeric correspond to the WDS discoverer code. 
}
}

\begin{longtable}{llllcl}
\caption{\label{tab:exoplanets} Stars hosting planets in our sample.}\\
\noalign{\smallskip}
\hline\hline
\noalign{\smallskip} 
Name	&	Karmn	&	Spectral	&	System	&	Number of	&	Discovery$^b$	\\
	&		&	type	&	class	&	planets$^a$	&	\\
\noalign{\smallskip}
\hline
\noalign{\smallskip}
\endfirsthead
\caption{Stars hosting planets in our sample (cont.).}\\
\hline\hline
\noalign{\smallskip}        
Name	&	Karmn	&	Spectral	&	System	&	Number of	&	Discovery$^b$	\\
	&		&	type	&	class	&	planets$^a$	&		\\
\hline
\noalign{\smallskip}
\endhead
\hline
\endfoot
GJ 1002	&	J00067--075	&	M5.5 V	&	Single	&	2	&	Sua23	\\
GJ 12	&	J00158+135	&	M3.0 V	&	Single	&	1	&	Kuz24, Dho24	\\
GX And	&	J00183+440	&	M1.0 V	&	Multiple	&	2	&	How14, Pin18	\\
TOI--1470	&	J00403+612	&	M2.0 V	&	Single	&	2	&	Gon23b	\\
GJ 3053	&	J00449--152	&	M4.5 V	&	Single	&	2	&	Dit17, Men19	\\
GJ 3072	&	J01023--104	&	M0.0 V	&	Single	&	1	&	Fen20	\\
Wolf 46	&	J01026+623	&	M1.5 V	&	Multiple	&	1	&	Per19	\\
LSPM J0106+1913	&	J01066+192	&	M3.0 V	&	Single	&	2	&	Cha22	\\
YZ Cet	&	J01125--169	&	M4.5 V	&	Single	&	2	&	Ast17b, Sto20a	\\
Ross 19	&	J02190+353	&	M3.5 V	&	Single	&	1	&	Sch21	\\
PM J02489--1432W	&	J02489--145W	&	M2.0 V	&	Multiple	&	1	&	Kos21	\\
Teegarden's Star	&	J02530+168	&	M7.0 V	&	Single	&	2	&	Zec19, Dre24	\\
HD 18143 A	&		&	K2 IV	&	Multiple	&	2	&	Fen22	\\
LP 14--53	&	J02573+765	&	M4.0 V	&	Single	&	1	&	Sot21	\\
GJ 3193	&	J03018--165S	&	M3.5 V	&	Multiple	&	2	&	Win19b, Win22	\\
CD Cet	&	J03133+047	&	M5.0 V	&	Single	&	1	&	Bau20	\\
o$^{02}$ Eri	&		&	K0 V	&	Multiple	&	1	&	Dia18	\\
LP 714--47	&	J04167--120	&	M0.0 V	&	Single	&	1	&	Dre20	\\
PM J04343+4302	&	J04343+430	&	M2.5 V	&	Single	&	1	&	Blu21	\\
HD 29391	&		&	F0 V	&	Multiple	&	1	&	Mac15	\\
HD 285968	&	J04429+189	&	M2.0 V	&	Single	&	1	&	For09	\\
Wolf 1539	&	J04520+064	&	M3.5 V	&	Single	&	1	&	How10	\\
GJ 180	&	J04538--177	&	M2.0 V	&	Single	&	3	&	Tuo14	\\
GJ 3323	&	J05019--069	&	M4.0 V	&	Single	&	2	&	Ast17a	\\
GJ 229	&	J06105--218	&	M0.5 V	&	Multiple	&	2	&	Tuo14, Fen20	\\
HD 260655	&	J06371+175	&	M0.0 V	&	Single	&	2	&	Luq22	\\
HD 265866	&	J06548+332	&	M3.0 V	&	Single	&	1	&	Sto20b	\\
Luyten's Star	&	J07274+052	&	M3.5 V	&	Single	&	2	&	Ast17a, Poz20	\\
GJ 3470	&	J07590+153	&	M1.5 V	&	Single	&	1	&	Bon12	\\
GJ 3473	&	J08023+033	&	M4.0 V	&	Multiple	&	2	&	Kem20	\\
GJ 317	&	J08409--234	&	M3.5 V	&	Single	&	2	&	Joh07	\\
GJ 3512	&	J08413+594	&	M5.5 V	&	Single	&	2	&	Mor19	\\
$\rho^{01}$ Cnc	&		&	K0 IV/V	&	Multiple	&	5	&	But97, Mar02, Mc04, Fis08	\\
Ross 623	&	J08551+015	&	M0.0 V	&	Single	&	1	&	Rob13	\\
G 41--13	&	J08588+210	&	M2.0 V	&	Single	&	1	&	Stef20	\\
HD 79211	&	J09144+526	&	M0.0 V	&	Multiple	&	1	&	Gon20	\\
LP 727--31	&	J09286--121	&	M2.5 V	&	Single	&	1	&	Ree22	\\
GJ 357	&	J09360--216	&	M2.5 V	&	Single	&	3	&	Luq19	\\
GJ 378	&	J10023+480	&	M1.0 V	&	Single	&	1	&	Hob19	\\
TYC 4384--1735--1	&	J10088+692	&	M0.5 V	&	Single	&	1	&	Blu20, Clo20a	\\
LP 729--54	&	J10185--117	&	M4.0 V	&	Multiple	&	2	&	Now20, Clo20b	\\
Ross 446	&	J10289+008	&	M2.0 V	&	Single	&	1	&	Ama21	\\
Lalande 21185 	&	J11033+359	&	M1.5 V	&	Single	&	2	&	Dia19, Ros21	\\
HD 97101	&	J11110+304E	&	K7 V	&	Multiple	&	2	&	Ded21	\\
K2--18	&	J11302+076	&	M3.0 V	&	Single	&	1	&	FM15, Clo17, Sar18	\\
Ross 1003	&	J11417+427	&	M4.0 V	&	Single	&	2	&	Hag10, Tri18	\\
Ross 905	&	J11421+267	&	M2.5 V	&	Single	&	1	&	Butl04	\\
LP 375--23	&	J11423+230	&	M0.5 V	&	Multiple	&	1	&	Sta19	\\
FI Vir	&	J11477+008	&	M4.0 V	&	Single	&	1	&	Bon18a	\\
GJ 1151	&	J11509+483	&	M4.5 V	&	Single	&	1	&	Bla23	\\
HD 238090	&	J12123+544S	&	M0.0 V	&	Multiple	&	1	&	Sto20b	\\
Ross 690	&	J12230+640	&	M3.0 V	&	Single	&	1	&	End22	\\
Wolf 433	&	J12388+116	&	M3.0 V	&	Single	&	1	&	Fen20	\\
Wolf 437	&	J12479+097	&	M3.5 V	&	Single	&	1	&	Tri21	\\
PM J13119+6550	&	J13119+658	&	M3.0 V	&	Single	&	2	&	Dem20	\\
HD 115404	&		&	K2 V	&	Multiple	&	2	&	Fen22	\\
Ross 1020	&	J13229+244	&	M4.0 V	&	Single	&	1	&	Luq18	\\
TOI--1238	&	J13255+688	&	M0.0 V	&	Single	&	2	&	Gon22	\\
Ross 490	&	J13299+102	&	M0.5 V	&	Single	&	1	&	Dam22	\\
HD 122303	&	J14010--026	&	M1.0 V	&	Single	&	1	&	Sua17a	\\
HN Lib	&	J14342--125	&	M4.0 V	&	Single	&	1	&	Gon23a	\\
HO Lib	&	J15194--077	&	M3.0 V	&	Single	&	3	&	Bon05, May09, Udr07	\\
Ross 508	&	J15238+174	&	M4.5 V	&	Single	&	1	&	Har22	\\
GJ 3929	&	J15583+354	&	M3.5 V	&	Single	&	2	&	Kem22, Bea22	\\
GJ 3942	&	J16090+529	&	M0.0 V	&	Single	&	1	&	Per17	\\
K2--33	&	J16102--193	&	M3.0 V	&	Single	&	1	&	Dav16	\\
LP 804--27	&	J16126--188	&	M3.0 V	&	Single	&	1	&	App10	\\
HD 147379	&	J16167+672S	&	M0.0 V	&	Multiple	&	1	&	Rei18a	\\
GJ 625	&	J16254+543	&	M1.5 V	&	Single	&	1	&	Sua17b	\\
V2306 Oph	&	J16303--126	&	M3.5 V	&	Single	&	2	&	Wri16	\\
V1090 Her	&		&	K3 V	&	Multiple	&	3	&	Fen22	\\
Ross 860	&	J16581+257	&	M1.0 V	&	Single	&	1	&	Joh10	\\
GJ 1214	&	J17153+049	&	M4.5 V	&	Single	&	1	&	Cha09	\\
GJ 3998	&	J17160+110	&	M1.0 V	&	Single	&	2	&	Aff16	\\
GJ 685	&	J17355+616	&	M0.5 V	&	Multiple	&	1	&	Pin19	\\
GJ 687	&	J17364+683	&	M3.0 V	&	Single	&	2	&	Burt14	\\
GJ 686	&	J17378+185	&	M1.0 V	&	Single	&	1	&	Aff19	\\
GJ 720 A	&	J18353+457	&	M0.5 V	&	Multiple	&	1	&	Gon21	\\
LSPM J1835+3259	&	J18356+329	&	M8.5 V	&	Single	&	1	&	Ber17	\\
LP 141--14	&		&	DC	&	Multiple	&	1	&	Van20	\\
HD 176029	&	J18580+059	&	M0.5 V	&	Single	&	1	&	Tol21	\\
V1428 Aql	&	J19169+051N	&	M2.5 V	&	Multiple	&	1	&	Kam18	\\
2MASS J19204172+7311434	&	J19206+731S	&	M4.0 V	&	Multiple	&	1	&	Cad22	\\
HD 190360	&		&	G7 IV/V	&	Multiple	&	2	&	Nae03, Vog05	\\
Ross 754	&	J20138+133	&	M1.0 V	&	Single	&	1	&	Mal21	\\
Wolf 1069	&	J20260+585	&	M5.0 V	&	Single	&	1	&	Kos23	\\
AU Mic	&	J20451--313	&	M0.5 V	&	Multiple	&	2	&	Pla20, Mart21	\\
GJ 806	&	J20450+444	&	M1.5 V	&	Single	&	2	&	Pal23	\\
LSPM J2116+0234	&	J21164+025	&	M3.0 V	&	Single	&	1	&	Lal19	\\
TYC 2187--512--1	&	J21221+229	&	M1.0 V	&	Single	&	1	&	Qui22	\\
G 264--12	&	J21466+668	&	M4.0 V	&	Single	&	2	&	Ama21	\\
TYC 4266--736--1	&	J21474+627	&	M0.0 V	&	Multiple	&	1	&	Esp22	\\
Wolf 1329	&	J22096--046	&	M3.5 V	&	Single	&	2	&	Butl06, Mone14	\\
UCAC4 744--073158	&	J22102+587	&	M2.0 V	&	Single	&	1	&	Fuk22	\\
GJ 1265	&	J22137--176	&	M4.5 V	&	Single	&	1	&	Luq18	\\
GJ 4276	&	J22252+594	&	M4.0 V	&	Single	&	1	&	Nag19	\\
IL Aqr	&	J22532--142	&	M4.0 V	&	Single	&	4	&	Marc98, Mar01, Riv05, Riv10	\\
2MUCD 12171	&	J23064--050	&	M7.5 V	&	Single	&	7	&	Gil16, Gil17	\\
EQ Peg A	&	J23318+199E	&	M3.5 V	&	Multiple	&	1	&	Cur22	\\
\end{longtable}
\tablefoot{
\tablefoottext{a}{In the majority of cases these planets are classified as `Confirmed' (to the date of publication of this work) by the NASA Exoplanet Archive. A few controversial detections have been excluded (e.g. {Barnard's star} or CN Leo). Numbers in parenthesis denote that at least one planet has been challenged \citep{Rib23}.}
\tablefoottext{b}{References for the discovery -- 
Aff16: \cite{Aff16};
Aff19: \cite{Aff19};
Ama21: \cite{Ama21};
App10: \cite{App10};
Ast17a: \cite{Ast17a};
Ast17b: \cite{Ast17b};
Bau20: \cite{Bau20};
Bea22: \cite{Bea22};
Ber17: \cite{Ber17};
Bla23: \cite{Bla23};
Blu20: \cite{Blu20};
Blu21: \cite{Blu21};
Bon05: \cite{Bon05};
Bon12: \cite{Bon12};
Bon18a: \cite{Bon18a};
Burt14: \cite{Burt14};
But97: \cite{But97};
Butl04: \cite{Butl04};
Butl06: \cite{Butl06};
Cad22: \cite{Cad22};
Cha09: \cite{Cha09};
Cha22: \cite{Cha22};
Clo17: \cite{Clo17};
Clo20a: \cite{Clo20a};
Clo20b: \cite{Clo20b};
Cur22: \cite{Cur22};
Dam22: \cite{Dam22};
Dav16: \cite{Dav16};
Ded21: \cite{Ded21};
Dem20: \cite{Dem20};
Dho24: \cite{Dho24};
Dia18: \cite{Dia18};
Dia19: \cite{Dia19};
Dit17: \cite{Dit17};
Dre20: \cite{Dre20};
Dre24: \cite{Dre24};
End22: \cite{End22};
Esp22: \cite{Esp22};
FM15: \cite{FM15};
Fen20: \cite{Fen20};
Fen22: \cite{Fen22};
Fis08: \cite{Fis08};
For09: \cite{For09};
Fuk22: \cite{Fuk22};
Gil16: \cite{Gil16};
Gil17: \cite{Gil17};
Gon20: \cite{Gon20};
Gon21: \cite{Gon21};
Gon22: \cite{Gon22};
Gon23a: \cite{Gon23a};
Gon23b: \cite{Gon23b};
Hag10: \cite{Hag10};
Har22: \cite{Har22};
Hob19: \cite{Hob19};
How10: \cite{How10};
How14: \cite{How14};
Joh07: \cite{Joh07};
Joh10: \cite{Joh10};
Kam18: \cite{Kam18};
Kem20: \cite{Kem20};
Kem22: \cite{Kem22};
Kos21: \cite{Kos21};
Kos23: \cite{Kos23};
Kuz24: \cite{Kuz24};
Lal19: \cite{Lal19};
Luq18: \cite{Luq18};
Luq19: \cite{Luq19};
Luq22: \cite{Luq22};
Mac15: \cite{Mac15};
Mal21: \cite{Mal21};
Mar01: \cite{Mar01};
Mar02: \cite{Mar02};
Marc98: \cite{Marc98};
Mart21: \cite{Mart21};
May09: \cite{May09};
Mc04: \cite{Mc04};
Men19: \cite{Men19};
Mon14: \cite{Mon14};
Mor19: \cite{Mor19};
Nae03: \cite{Nae03};
Nag19: \cite{Nag19};
Now20: \cite{Now20};
Pal23: \cite{Pal23};
Per17: \cite{Per17};
Per19: \cite{Per19};
Pin18: \cite{Pin18};
Pin19: \cite{Pin19};
Pla20: \cite{Pla20};
Poz20: \cite{Poz20};
Qui22: \cite{Qui22};
Ree22: \cite{Ree22};
Rei18a: \cite{Rei18a};
Riv05: \cite{Riv05};
Riv10: \cite{Riv10};
Rob13: \cite{Rob13};
Ros21: \cite{Ros21};
Sar18: \cite{Sar18};
Sch21: \cite{Sch21};
Sot21: \cite{Sot21};
Sta19: \cite{Sta19};
Stef20: \cite{Stef20};
Sto20a: \cite{Sto20a};
Sto20b: \cite{Sto20b};
Sua17a: \cite{Sua17a};
Sua17b: \cite{Sua17b};
Sua23: \cite{Sua23};
Tol21: \cite{Tol21};
Tri18: \cite{Tri18};
Tri21: \cite{Tri21};
Tuo14: \cite{Tuo14};
Udr07: \cite{Udr07};
Van20: \cite{Van20};
Vog05: \cite{Vog05};
Win19b: \cite{Win19b};
Win22: \cite{Win22};
Wri16: \cite{Wri16};
Zec19: \cite{Zec19}.}
}

{\fontsize{7}{10}\selectfont 
\begin{longtable}{llllll}
\caption{\label{tab:references} Bibliographic references for the abbreviations used in the main table.}\\
\noalign{\smallskip}
\hline\hline
\noalign{\smallskip} 
Abbreviation	& 	Reference	& 	Abbreviation	& 	Reference	& 	Abbreviation	& 	Reference	\\
\noalign{\smallskip}
\hline
\noalign{\smallskip}
\endfirsthead
\caption{Bibliographic references for the abbreviations used in the main table (cont.).}\\
\hline\hline
\noalign{\smallskip}        
Abbreviation 	& 	Reference	& 	Abbreviation	& 	Reference	& 	Abbreviation	& 	Reference	\\
\hline
\noalign{\smallskip}
\endhead
\hline
\endfoot
\noalign{\smallskip}
Abt08	&	\cite{Abt08}	&	Can93	&	\cite{Can93}	&	Fen20	&	\cite{Fen20}	\\
Abt70	&	\cite{Abt70}	&	Car21	&	\cite{Car21}	&	Fen22	&	\cite{Fen22}	\\
Abt76	&	\cite{Abt76}	&	Cas07	&	\cite{Cas07}	&	Fin18	&	\cite{Fin18}	\\
Abt81	&	\cite{Abt81}	&	Cat06	&	\cite{Cat06}	&	Fis08	&	\cite{Fis08}	\\
Ada35	&	\cite{Ada35}	&	Cha09	&	\cite{Cha09}	&	Fle88	&	\cite{Fle88}	\\
Aff16	&	\cite{Aff16}	&	Cha22	&	\cite{Cha22}	&	FM15	&	\cite{FM15}	\\
Aff19	&	\cite{Aff19}	&	Che20	&	\cite{Che20}	&	For04	&	\cite{For04}	\\
Alo15a	&	\cite{Alo15a}	&	Che23	&	\cite{Che23}	&	For09	&	\cite{For09}	\\
Alo15b	&	\cite{Alo15b}	&	Chr38	&	\cite{Chr38}	&	Fou18	&	\cite{Fou18}	\\
Ama21	&	\cite{Ama21}	&	Chr78	&	\cite{Chr78}	&	Fre20	&	\cite{Fre20}	\\
Ans15	&	\cite{Ans15}	&	Cif20	&	\cite{Cif20}	&	Fri13	&	\cite{Fri13}	\\
App10	&	\cite{App10}	&	Clo03	&	\cite{Clo03}	&	Fuk22	&	\cite{Fuk22}	\\
Are19	&	\cite{Are19}	&	Clo17	&	\cite{Clo17}	&	Gag15a	&	\cite{Gag15a}	\\
Ast17a	&	\cite{Ast17a}	&	Clo20a	&	\cite{Clo20a}	&	Gag18a	&	\cite{Gag18a}	\\
Ast17b	&	\cite{Ast17b}	&	Clo20b	&	\cite{Clo20b}	&	Gag18b	&	\cite{Gag18b}	\\
Bar14	&	\cite{Bar14}	&	Cor17a	&	\cite{Cor17a}	&	Gai14b	&	\cite{Gai14b}	\\
Bar18	&	\cite{Bar18}	&	Cor24	&	\cite{Cor24}	&	Gaia22a	&	\cite{Gaia22a}	\\
Bar21	&	\cite{Bar21}	&	Cru02	&	\cite{Cru02}	&	Gal22	&	\cite{Gal22}	\\
Bar70	&	\cite{Bar70}	&	Cru03	&	\cite{Cru03}	&	Geb02	&	\cite{Geb02}	\\
Bau20	&	\cite{Bau20}	&	Cru09	&	\cite{Cru09}	&	Geb16	&	\cite{Geb16}	\\
Bea22	&	\cite{Bea22}	&	Cur06	&	\cite{Cur06}	&	Gen19	&	\cite{Gen19}	\\
Bed20	&	\cite{Bed20}	&	Cur22	&	\cite{Cur22}	&	Gia11	&	\cite{Gia11}	\\
Bel17	&	\cite{Bel17}	&	Cve12	&	\cite{Cve12}	&	Gil16	&	\cite{Gil16}	\\
Bel97	&	\cite{Bel97}	&	Dae07	&	\cite{Dae07}	&	Gil17	&	\cite{Gil17}	\\
Ben00	&	\cite{Ben00}	&	Dam22	&	\cite{Dam22}	&	Gil18	&	\cite{Gil18}	\\
Ben08	&	\cite{Ben08}	&	Dav14	&	\cite{Dav14}	&	Giz00b	&	\cite{Giz00b}	\\
Ben21	&	\cite{Ben21}	&	Dav15	&	\cite{Dav15}	&	Giz02	&	\cite{Giz02}	\\
Ber06	&	\cite{Ber06}	&	Dav16	&	\cite{Dav16}	&	Giz97a	&	\cite{Giz97a}	\\
Ber10	&	\cite{Ber10}	&	Daw05	&	\cite{Daw05}	&	Giz97b	&	\cite{Giz97b}	\\
Ber17	&	\cite{Ber17}	&	Dea12	&	\cite{Dea12}	&	Gli91	&	\cite{Gli91}	\\
Bes15	&	\cite{Bes15}	&	Dea14	&	\cite{Dea14}	&	Gol10	&	\cite{Gol10}	\\
Bes20	&	\cite{Bes20}	&	Ded21	&	\cite{Ded21}	&	Gom15	&	\cite{Gom15}	\\
Bes91	&	\cite{Bes91}	&	Dek18	&	\cite{Dek18}	&	Gon06	&	\cite{Gon06}	\\
Bid85	&	\cite{Bid85}	&	Del10	&	\cite{Del10}	&	Gon20	&	\cite{Gon20}	\\
Bil13	&	\cite{Bil13}	&	Del98	&	\cite{Del98}	&	Gon21	&	\cite{Gon21}	\\
Bin16	&	\cite{Bin16}	&	Del99a	&	\cite{Del99b}	&	Gon22	&	\cite{Gon22}	\\
Bir12	&	\cite{Bir12}	&	Del99b	&	\cite{Del99a}	&	Gon23a	&	\cite{Gon23a}	\\
Bir20	&	\cite{Bir20}	&	Del99c	&	\cite{Del99c}	&	Gon23b	&	\cite{Gon23b}	\\
Bla23	&	\cite{Bla23}	&	Dem20	&	\cite{Dem20}	&	Gon24	&	\cite{Gon24}	\\
Blu20	&	\cite{Blu20}	&	Des12	&	\cite{Des12}	&	GR96	&	\cite{GR96}	\\
Blu21	&	\cite{Blu21}	&	Dho24	&	\cite{Dho24}	&	Gra01	&	\cite{Gra01}	\\
Boc05	&	\cite{Boc05}	&	Dia18	&	\cite{Dia18}	&	Gra03	&	\cite{Gra03}	\\
Boc18	&	\cite{Boc18}	&	Dia19	&	\cite{Dia19}	&	Gra06	&	\cite{Gra06}	\\
Bon05	&	\cite{Bon05}	&	Die14	&	\cite{Die14}	&	Gre84	&	\cite{Gre84}	\\
Bon12	&	\cite{Bon12}	&	Dit14	&	\cite{Dit14}	&	Gre90	&	\cite{Gre90}	\\
Bon13a	&	\cite{Bon13a}	&	Dit17	&	\cite{Dit17}	&	Gri85	&	\cite{Gri85}	\\
Bon18a	&	\cite{Bon18a}	&	DR2	&	\cite{Gaia18bro}	&	Gui09	&	\cite{Gui09}	\\
Bop77	&	\cite{Bop77}	&	DR3	&	\cite{Gaia23}	&	Hag10	&	\cite{Hag10}	\\
Bow15a	&	\cite{Bow15a}	&	Dre19	&	\cite{Dre19}	&	Hal18	&	\cite{Hal18}	\\
Bud19	&	\cite{Bud19}	&	Dre20	&	\cite{Dre20}	&	Har11	&	\cite{Har11}	\\
Bud21	&	\cite{Bud21}	&	Dre24	&	\cite{Dre24}	&	Har12	&	\cite{Har12}	\\
Bur00	&	\cite{Bur00}	&	Dup12	&	\cite{Dup12}	&	Har22	&	\cite{Har22}	\\
Bur11	&	\cite{Bur11}	&	Dup17	&	\cite{Dup17}	&	Har70	&	\cite{Har70}	\\
Bur13	&	\cite{Bur13}	&	Dup19	&	\cite{Dup19}	&	Har81	&	\cite{Har81}	\\
Bur15a	&	\cite{Bur15a}	&	Dup94	&	\cite{Dup94}	&	Har96	&	\cite{Har96}	\\
Bur15b	&	\cite{Bur15b}	&	Duq88	&	\cite{Duq88}	&	Haw96	&	\cite{Haw96}	\\
Burt14	&	\cite{Burt14}	&	Duq91	&	\cite{Duq91}	&	Hen02	&	\cite{Hen02}	\\
But97	&	\cite{But97}	&	Edw76	&	\cite{Edw76}	&	Hen18	&	\cite{Hen18}	\\
Butl04	&	\cite{Butl04}	&	Ell14	&	\cite{Ell14}	&	Her03	&	\cite{Her03}	\\
Butl06	&	\cite{Butl06}	&	End22	&	\cite{End22}	&	Her14	&	\cite{Her14}	\\
Cab07	&	\cite{Cab07}	&	Esp22	&	\cite{Esp22}	&	Her65	&	\cite{Her65}	\\
Cab09	&	\cite{Cab09}	&	Eva67	&	\cite{Eva67}	&	Hir21	&	\cite{Hir21}	\\
Cab10a	&	\cite{Cab10a}	&	Fab02	&	\cite{Fab02}	&	Hob19	&	\cite{Hob19}	\\
Cab10b	&	\cite{Cab10b}	&	Fah10	&	\cite{Fah10}	&	Hog00	&	\cite{Hog00}	\\
Cad22	&	\cite{Cad22}	&	Fah12	&	\cite{Fah12}	&	Hoj19	&	\cite{Hoj19}	\\
										
Hou16	&	\cite{Hou16}	&	Luc17	&	\cite{Luc17}	&	Pha06a	&	\cite{Pha06a}	\\
Hou88	&	\cite{Hou88}	&	Luc18	&	\cite{Luc18}	&	Pha06b	&	\cite{Pha06b}	\\
Hou99	&	\cite{Hou99}	&	Luh12	&	\cite{Luh12}	&	Pin18	&	\cite{Pin18}	\\
How10	&	\cite{How10}	&	Luq18	&	\cite{Luq18}	&	Pin19	&	\cite{Pin19}	\\
How14	&	\cite{How14}	&	Luq19	&	\cite{Luq19}	&	Pla20	&	\cite{Pla20}	\\
Hub22	&	\cite{Hub22}	&	Luq22	&	\cite{Luq22}	&	Pou04	&	\cite{Pou04}	\\
Ire08	&	\cite{Ire08}	&	Mac13	&	\cite{Mac13}	&	Poz20	&	\cite{Poz20}	\\
Irw09	&	\cite{Irw09}	&	Mac15	&	\cite{Mac15}	&	Pre02	&	\cite{Pre02}	\\
Jah08	&	\cite{Jah08}	&	Mac18	&	\cite{Mac18}	&	Put97	&	\cite{Put97}	\\
Jam08	&	\cite{Jam08}	&	Mah21	&	\cite{Mah21}	&	Qui22	&	\cite{Qui22}	\\
Jan12	&	\cite{Jan12}	&	Mal06	&	\cite{Mal06}	&	Rag09	&	\cite{Rag09}	\\
Jan14a	&	\cite{Jan14a}	&	Mal12	&	\cite{Mal12}	&	Raj18a	&	\cite{Raj18a}	\\
Jan17	&	\cite{Jan17}	&	Mal13	&	\cite{Mal13}	&	Reb98	&	\cite{Reb98}	\\
Jef18	&	\cite{Jef18}	&	Mal14a	&	\cite{Mal14a}	&	Ree22	&	\cite{Ree22}	\\
Jen09	&	\cite{Jen09}	&	Mal14b	&	\cite{Mal14b}	&	Rei02a	&	\cite{Rei02a}	\\
Jim18	&	\cite{Jim18}	&	Mal20	&	\cite{Mal20}	&	Rei03	&	\cite{Rei03}	\\
Jod13	&	\cite{Jod13}	&	Mal21	&	\cite{Mal21}	&	Rei04	&	\cite{Rei04}	\\
Joh07	&	\cite{Joh07}	&	Mal24	&	\cite{Mal24}	&	Rei05	&	\cite{Rei05}	\\
Joh10	&	\cite{Joh10}	&	Man14	&	\cite{Man14}	&	Rei07	&	\cite{Rei07}	\\
Jon20	&	\cite{Jon20}	&	Man15	&	\cite{Man15}	&	Rei09	&	\cite{Rei09}	\\
Joy26	&	\cite{Joy26}	&	Mar01	&	\cite{Mar01}	&	Rei18a	&	\cite{Rei18a}	\\
Joy49	&	\cite{Joy49}	&	Mar02	&	\cite{Mar02}	&	Rei18b	&	\cite{Rei18b}	\\
Joy74	&	\cite{Joy74}	&	Mar07	&	\cite{Mar07}	&	Rei95	&	\cite{Rei95}	\\
Kam18	&	\cite{Kam18}	&	Mar87	&	\cite{Mar87}	&	Rei97b	&	\cite{Rei97b}	\\
Kar04	&	\cite{Kar04}	&	Marc98	&	\cite{Marc98}	&	Rein12	&	\cite{Rein12}	\\
Kat13	&	\cite{Kat13}	&	Marf21	&	\cite{Marf21}	&	Ria06	&	\cite{Ria06}	\\
Kee89	&	\cite{Kee89}	&	Mart21	&	\cite{Mart21}	&	Rib23	&	\cite{Rib23}	\\
Kem20	&	\cite{Kem20}	&	Mas08	&	\cite{Mas08}	&	Ric20	&	\cite{Ric20}	\\
Kem22	&	\cite{Kem22}	&	May09	&	\cite{May09}	&	Rie17	&	\cite{Rie17}	\\
Ker19	&	\cite{Ker19}	&	Maz01	&	\cite{Maz01}	&	Riv05	&	\cite{Riv05}	\\
Khr10	&	\cite{Khr10}	&	Mc04	&	\cite{Mc04}	&	Riv10	&	\cite{Riv10}	\\
Kir11	&	\cite{Kir11}	&	McC02	&	\cite{McC02}	&	Riv12	&	\cite{Riv12}	\\
Kir12	&	\cite{Kir12}	&	Men19	&	\cite{Men19}	&	Rob12	&	\cite{Rob12}	\\
Kir19	&	\cite{Kir19}	&	Mer09	&	\cite{Mer09}	&	Rob13	&	\cite{Rob13}	\\
Kir24	&	\cite{Kir24}	&	Mir20	&	\cite{Mir20}	&	Rod74	&	\cite{Rod74}	\\
Kir91	&	\cite{Kir91}	&	Mis15	&	\cite{Mis15}	&	Roj12	&	\cite{Roj12}	\\
Kir94	&	\cite{Kir94}	&	Moc02	&	\cite{Moc02}	&	Ros11	&	\cite{Ros11}	\\
Kiy20	&	\cite{Kiy20}	&	Mon01	&	\cite{Mon01}	&	Ros21	&	\cite{Ros21}	\\
Klu20	&	\cite{Klu20}	&	Mon03	&	\cite{Mon03}	&	Sab21	&	\cite{Sab21}	\\
Koe10	&	\cite{Koe10}	&	Mon14	&	\cite{Mon14}	&	Sal22	&	\cite{Sal22}	\\
Kop16	&	\cite{Kop16}	&	Mon18	&	\cite{Mon18}	&	Sam17	&	\cite{Sam17}	\\
Kos21	&	\cite{Kos21}	&	Mor09	&	\cite{Mor09}	&	Sar18	&	\cite{Sar18}	\\
Kos23	&	\cite{Kos23}	&	Mor10	&	\cite{Mor10}	&	Sch05	&	\cite{Sch05}	\\
Kou19	&	\cite{Kou19}	&	Mor19	&	\cite{Mor19}	&	Sch07	&	\cite{Sch07}	\\
Kra14a	&	\cite{Kra14a}	&	Mor65	&	\cite{Mor65}	&	Sch12	&	\cite{Sch12}	\\
Kra14b	&	\cite{Kra14b}	&	Mut10	&	\cite{Mut10}	&	Sch19	&	\cite{Sch19}	\\
Kra17	&	\cite{Kra17}	&	MWDD	&	Montreal WD Database	&	Sch21	&	\cite{Sch21}	\\
Kuz19	&	\cite{Kuz19}	&	Nae03	&	\cite{Nae03}	&	Scha19	&	\cite{Scha19}	\\
Kuz24	&	\cite{Kuz24}	&	Nag19	&	\cite{Nag19}	&	Schw19	&	\cite{Schw19}	\\
Lafr07	&	\cite{Lafr07}	&	Nap20	&	\cite{Nap20}	&	She85	&	\cite{She85}	\\
Lal19	&	\cite{Lal19}	&	Nes95	&	\cite{Nes95}	&	Shk09	&	\cite{Shk09}	\\
Law06	&	\cite{Law06}	&	New14	&	\cite{New14}	&	Shk10	&	\cite{Shk10}	\\
Law08	&	\cite{Law08}	&	New16	&	\cite{New16}	&	Shk12	&	\cite{Shk12}	\\
Lee84	&	\cite{Lee84}	&	Nid02	&	\cite{Nid02}	&	Shk17	&	\cite{Shk17}	\\
Lep02	&	\cite{Lep02}	&	Now20	&	\cite{Now20}	&	Sim11	&	\cite{Sim11}	\\
Lep03	&	\cite{Lep03}	&	Osw88	&	\cite{Osw88}	&	Sio09	&	\cite{Sio09}	\\
Lep05	&	\cite{Lep05}	&	Pal20	&	\cite{Pal20}	&	Sio90	&	\cite{Sio90}	\\
Lep09	&	\cite{Lep09}	&	Pal23	&	\cite{Pal23}	&	Ski18	&	\cite{Ski18}	\\
Lep13	&	\cite{Lep13}	&	Pas18	&	\cite{Pas18}	&	Smi14	&	\cite{Smi14}	\\
Lim13	&	\cite{Lim13}	&	Pau06	&	\cite{Pau06}	&	Sot21	&	\cite{Sot21}	\\
Lim15	&	\cite{Lim15}	&	Pec13	&	\cite{Pec13}	&	Sou18	&	\cite{Sou18}	\\
Lod05	&	\cite{Lod05}	&	Per17	&	\cite{Per17}	&	Soub08	&	\cite{Soub08}	\\
Lod19	&	\cite{Lod19}	&	Per19	&	\cite{Per19}	&	Spe16	&	\cite{Spe16}	\\
Lop06	&	\cite{Lop06}	&	Pes97	&	\cite{Pes97}	&	Spe19	&	\cite{Spe19}	\\
Lot98	&	\cite{Lot98}	&	Pet84	&	\cite{Pet84}	&	Spi21	&	\cite{Spi21}	\\
											
Sta19	&	\cite{Sta19}	&	Tok18	&	\cite{Tok18}	&	War15	&	\cite{War15}	\\
Sta97	&	\cite{Sta97}	&	Tok22	&	\cite{Tok22}	&	Wei16	&	\cite{Wei16}	\\
Ste86a	&	\cite{Ste86a}	&	Tok97	&	\cite{Tok97}	&	Wes11	&	\cite{Wes11}	\\
Ste86b	&	\cite{Ste86b}	&	Tol21	&	\cite{Tol21}	&	Wes93	&	\cite{Wes93}	\\
Stef20	&	\cite{Stef20}	&	Tor02	&	\cite{Tor02}	&	Whi01	&	\cite{Whi01}	\\
Stei20	&	\cite{Stei20}	&	Tor06	&	\cite{Tor06}	&	Whi07	&	\cite{Whi07}	\\
Sto20a	&	\cite{Sto20a}	&	Tor09	&	\cite{Tor09}	&	Wil53	&	\cite{Wil53}	\\
Sto20b	&	\cite{Sto20b}	&	Tor10	&	\cite{Tor10}	&	Win17	&	\cite{Win17}	\\
Ston20	&	\cite{Ston20}	&	Tor22	&	\cite{Tor22}	&	Win19b	&	\cite{Win19b}	\\
Str03	&	\cite{Str03}	&	TP86	&	\cite{TP86}	&	Win20	&	\cite{Win20}	\\
Str93	&	\cite{Str93}	&	Tre20	&	\cite{Tre20}	&	Win22	&	\cite{Win22}	\\
Str94	&	\cite{Str94}	&	Tri18	&	\cite{Tri18}	&	Wri16	&	\cite{Wri16}	\\
Sua17a	&	\cite{Sua17a}	&	Tri21	&	\cite{Tri21}	&	Xua24	&	\cite{Xua24}	\\
Sua17b	&	\cite{Sua17b}	&	Tuo14	&	\cite{Tuo14}	&	Yi14	&	\cite{Yi14}	\\
Sua23	&	\cite{Sua23}	&	Udr07	&	\cite{Udr07}	&	Zac10	&	\cite{Zac10}	\\
Tab17	&	\cite{Tab17}	&	van07	&	\cite{van07}	&	Zac12	&	\cite{Zac12}	\\
Tam06	&	\cite{Tam06}	&	van09	&	\cite{van09}	&	Zak22	&	\cite{Zak22}	\\
Tam08	&	\cite{Tam08}	&	Van20	&	\cite{Van20}	&	Zak79	&	\cite{Zak79}	\\
Tau20	&	\cite{Tau20}	&	van95	&	\cite{van95}	&	Zap04	&	\cite{Zap04}	\\
Ter15a	&	\cite{Ter15a}	&	Vog02	&	\cite{Vog02}	&	Zbo98	&	\cite{Zbo98}	\\
Ter15b	&	\cite{Ter15b}	&	Vog05	&	\cite{Vog05}	&	Zec19	&	\cite{Zec19}	\\
Tin19	&	\cite{Tin19}	&	Vri20	&	\cite{Vri20}	&	Zon20	&	\cite{Zon20}	\\
Tok08	&	\cite{Tok08}	&	Wan22	&	\cite{Wan22}	&		&		\\
\end{longtable}
}

\end{appendix}
\end{document}